\documentclass[11pt]{article}

\usepackage{amssymb}
\usepackage{amsmath}
\usepackage{dsfont}
\usepackage{xcolor}
\usepackage{youngtab}
\usepackage{tikz}
\usepackage{pgfplots}
\pgfplotsset{compat = newest}
\usetikzlibrary{intersections}
\usetikzlibrary{decorations.markings}
\newcommand{\vertex}{\node[fill,circle,inner sep=0pt,minimum size=0pt]}
\newcommand{\coord}[1]{({sin(#1)},{cos(#1)})}

\usepackage{slashed}
\usepackage{mathrsfs}
\usepackage{mathrsfs}
\usepackage{xcolor}
\definecolor{darkblue}{rgb}{0.1,0.1,.7}
\definecolor{darkgreen}{rgb}{0,.5,0}
\usepackage[pdftex,breaklinks,colorlinks,linkcolor=darkblue,citecolor=darkgreen]{hyperref}

\csname @addtoreset\endcsname{equation}{section}

\usepackage{float}
\usepackage{appendix}

\newcommand{\nn}{\nonumber}
\newcommand{\be}{\begin{equation}}
\newcommand{\ee}{\end{equation}}

\newcommand{\pr}{\partial}
\newcommand{\cG}{{\mathcal G}}
\newcommand{\rmd}{{\rm d}}
\newcommand{\vep}{\varepsilon}
\newcommand{\HH}{\mathcal H}
\newcommand{\ts}{\textstyle}
\newcommand{\tlam}{{\tilde \lambda}}
\newcommand{\vphi}{\varphi}

\newcommand{\F}{{\cal F}}
\newcommand{\D}{{\cal D}}

\renewcommand{\L}{{\cal L}}

\newcommand \teta{\bar\eta}
\newcommand \tG{\bar G}

\begin{document}
\hypersetup{pageanchor=false}
\begin{titlepage}
\thispagestyle{empty}
\begin{flushright}
\small
\today \\
\normalsize
\end{flushright}
\vspace{1cm} 
\begin{center}
{\LARGE {$\phi^6$ at $6$ (and some $8$) loops in $3d$}}\\
\end{center}
\vspace{1cm}
\begin{center}
{ \Large Ian Jack$^a$ and Hugh Osborn$^b$}

\vskip 1.5cm
${}^a$Department of Mathematical Sciences, University of Liverpool,\\
 {Liverpool L69 3BX}
\vskip 6pt

${}^b$Department of Applied Mathematics and Theoretical Physics, Wilberforce Road, Cambridge CB3 0WA
\end{center}

\date{latest update: \today}
\vskip 2cm
\begin{abstract}
We recalculate the contributions of  individual six  loop graphs to the $\beta$-function for a three
dimensional scalar theory with an arbitrary sextic scalar potential. Previously this was calculated
by Hager who specialised to a theory with maximal $O(N)$ symmetry. Our results differ in some
contributions to the overall $\beta$-function but agree with a recent calculation \cite{Kompaniets2}. 
At large $N$ three eight loop diagrams which are
relevant are calculated. At the $O(N)$ fixed point some critical exponents
are determined to $\rm O(\varepsilon^3)$. Imposing that the $\beta$-function satisfies a gradient flow
equation is shown to require linear relations between some $\beta$-function coefficients. 
The curvature for the associated metric is also determined. Detailed results for the Feynman integrals
are described in the appendices.
 \end{abstract}

\thispagestyle{empty}
\end{titlepage}

\hypersetup{pageanchor=true}
\pagenumbering{roman}
\newpage

\pagenumbering{arabic}

\setcounter{footnote}{0}

\section{Introduction}

Scalar field theories have proved a laboratory for extending our calculational abilities 
in quantum field theories to higher loops. Results 
for individual graphs for $\phi^4$ theories in four dimensions
were given more than 20 years ago in \cite{KleinertB} at five loops and have now quite recently been extended to
six loops in \cite{Bednyakov} and even seven in \cite{Schnetz7}. With this information it is possible  to find 
critical exponents in the $\vep$-expansion \cite{Kleinert,Panzer,Sixloop,Shalaby} with greatly improved accuracy
and with resummation techniques give results close to bootstrap calculations.
Similarly renormalisable $\phi^3$ theories in six dimensions have been extensively investigated, at three loops
in  \cite{Fei3loop}, four loops in \cite{Gracey4loop} and now at five loops in \cite{Kompaniets,Borinsky}
and six loops in \cite{Schnetz6}.

In addition to the above theories renormalisable $\phi^6$ theories in three dimensions have been
of particular interest since the early days of the Wilsonian revolution \cite{Tricrit,Lewis}. Such theories
have IR fixed points in $3-\vep$ dimensions which describe tricritical behaviour where critical
exponents can be calculated in the $\vep$-expansion.
For theories with $N$ scalar fields forming an $O(N)$ vector then $O(N)$ symmetry ensures there is a 
single coupling. This theory has been extensively discussed previously, in particular
 the large $N$ limit has been considered with varied conclusions,  both in the pre archive era \cite{Townsend0,Townsend1,Townsend2,Appelquist,Appelquist2,Gudmundsdottir1A,Gudmundsdottir2,
Pisarski,Pisarski2,Bardeen,David} 
and more recently \cite{Giombi,Giombi2,Delamotte,Semenoff,Sakhi,Kapoor,Jepsen,TSteudtner}. 
For the $O(N)$ symmetric theory four loop calculations in \cite{Pisarski}, which show there is a 
possible UV fixed point,  were extended to six loops in \cite{Hager0,Hager}.  A CFT approach 
to obtaining $\vep$-expansion results for the $O(N)$ $\phi^6$ theory was described in \cite{Basu},
albeit at a low order. A four loop calculation outside the dimensional regularisation paradigm 
is contained in \cite{Kharuk}.
A very helpful review is to be found in \cite{HenrikssonRev}.

More recent higher loop calculations for the $\phi^6$ theory are described in \cite{Bednyakov2,Kompaniets2}.
The six loop results in \cite{Kompaniets2} for a single scalar disagree in places with \cite{Hager}, in accord
with our independent calculations.

In the interests of transparency, and  possible independent checking, we list the the necessary 
counterterms for each four and six loop graph contributing to the $\beta$-function. 
This should allow corresponding six loop results to be obtained for other symmetry groups than $O(N)$.

In the next section we present detailed results for the 2,4 and 6 loop graphs for the $\phi^6$ theory
and also give the $N$ factors which arise for each graph when there are $N$ scalars with an $O(N)$ symmetry.
In section 3 we briefly describe  the corresponding large $N$ limit which reduces to an effective $\varphi^3$
theory with a non local $\varphi^2$ interaction. In section 4 results for critical exponents at the $\vep$-fixed 
point are given to $\rm O(\vep^3)$. In section 5 it is shown how there are linear relations between different
contributions to the $\beta$-function obtained by requiring gradient flow. This requires a metric for the space
of couplings and the associated curvature is obtained in section 6. Some further details are given in various
appendices. Appendix A gives the $Z$'s for the scalar field and composite operators in the $O(N)$ case.
Appendix B describes in detail how the contributions to the various $Z$ factors corresponding
to individual Feynman graphs may be obtained. Appendix C discusses in some detail  results for a  two
loop master integral recapitulating and extending results of \cite{Kotikov1}. Appendix D is concerned
with operator mixing issues.

\section{Detailed Results}

For a  renormalisable  scalar theory in three dimensions with $N$ real scalars we may consider a potential
\be
V(\phi) = \tfrac{1}{6!} \, \lambda_{ijklmn}
\phi_i \phi_j \phi_k \phi_l \phi_m \phi_n \, ,
\label{V3D}
\ee
for an arbitrary symmetric tensor coupling $ \lambda_{ijklmn}$.The associated $\beta$-function in $d=3-\vep$ dimensions is  expressible in the form
\be
\beta_V(\phi) = \vep   V(\phi)+ \phi_i( - \vep\, \tfrac12 \, \delta_{ij} + \gamma_{ij} ) V_j (\phi) 
+  {\tilde \beta}_V (\phi) \, , \quad V_j (\phi) =\pr_j V(\phi) \, .
\label{Bone}
\ee
with ${\tilde \beta}_V (\phi) $ determined by one particle irreducible, 1PI, and one vertex 
irreducible, 1VI,  vertex graphs. Correspondingly $\gamma_{ij}$ is given in terms of 1PI, 1VI, 
graphs  with two external lines.  For maximal $O(N)$ symmetry 
the potential is  restricted to
\be
V_\lambda(\phi) = \lambda \, \tfrac{1}{48}  (\phi^2)^3 \, ,
\label{VO6}
\ee
with just a single coupling $\lambda$ and then 
$\beta_V(\phi)_{V\to V_\lambda} = \beta_\lambda \, V_1(\phi)$ and $ \gamma_{ij} = \gamma_\phi \,\delta_{ij}$.

With dimensional regularisation in $3-\vep$ dimensions divergences arise just at even loops. 
For $2,4,6$ loops results are given, after rescaling $V\to(8\pi)^2 V$, in tables below.
Here $V_n$ denotes contributions involving $n$, $n=2,3,4,5$, derivatives of $V(\phi)$ and so with 
$6-n$ external $\phi$ lines.
For the general potential \eqref{V3D} $V_n \to V_{i_1 \dots i_n} = \pr_{i_1}\dots \pr_{i_n}V$. For each vertex 
graph $\cG_v$ then $V(\cG_v)$ denotes the associated monomial of degree 6 in $\phi$ defined by $\cG_v$ where the $V_n$ and $\lambda$ are contracted  according to the dictates of the graph, thus for the two loop
vertex graph $\cG_2$, $\tikz[baseline=(vert_cent.base),scale=0.2]{
  \node (vert_cent) {\hspace{-13pt}$\phantom{-}$};
  \draw (0,0) circle [radius=1.2];
  \draw (-1.2,0)-- (1.2,0);
 \node at (-2.1,0) {$\scriptstyle{V_3}$};
 \node at (2.1,0) {$\scriptstyle {V_3}$};
 \fill[black]  (1.2,0) circle [radius = 0.1cm];
\fill[black]  (-1.2,0) circle [radius = 0.1cm];
}$,
$V(\cG_{2}) = V_{ijk}V_{ijk}$. In general
\be
V(\cG_v) \sim V_2{\!}^r V_3{\!}^sV_4{\!}^tV_5{\!}^u \lambda^v \, , \quad 4r+3s+2t+u =6 \, , \quad
l_{\cG_v} =1+ \tfrac12s +t +\tfrac32u + 2 v\, , 
\label{VGv}
\ee
with $l_{\cG_v} $ the number of loops in $\cG_v$.
For a propagator graph $\cG_\phi$ then $\Gamma_{ij}(\cG_\phi)$ is the 
corresponding tensor formed by contracting the coupling indices as required by  $\cG_\phi$.
 For us each Feynman integral is defined with a factor $\pi^{-\frac12 d}$ for each loop momentum. 
Assuming minimal subtraction with all necessary counterterms subtracted
$ Z_{\cG_v},  \,  Z_{\cG_\phi} $ are the divergent parts of the Feynman integrals associated with the 
graphs $\cG_v, \, \cG_\phi $ containing  just poles in $\vep$,  although for convenience we factor off 
$\pi^{\frac12 l_{\cG_v}}$  or $\pi^{\frac12 l_{\cG_\phi}}$.
 No symmetry  or additional sign factors are included for  either 
$ Z_{\cG_v}, \, Z_{\cG_\phi} $. Corresponding to a vertex or propagator graph $\cG_v,\cG_\phi$
there is an associated contribution to the $\beta$-function or anomalous dimension $\gamma$
\be
\beta_V(\phi)\big |_{\cG_v} =  c_{\cG_v}\,  V(\cG_v)(\phi)\, ,\qquad
2 \, \gamma_{\phi,ij} \big |_{\cG_{\phi}} =  d_{\cG_\phi}\,  \Gamma_{ij}(\cG_{\phi}) \, ,
\ee
where for graphs of even loop order $l_{\cG_V}$, $l_{\cG_\phi} $, 
\be
c_{\cG_{v}}^{\vphantom g} = (-1)^{\frac12 l_{\cG_v}+1}  \frac{l_{\cG_v}}{S_{\cG_{v}}  E_{\cG_{v}} }
\mathop{\mathrm{Res}}_{\vep = 0} \,  Z_{\cG_{v}}  \, , \qquad
d_{\cG_{\phi}} = (-1)^{\frac12 l_{\cG_\phi} }\frac{l_{\cG_\phi}}{S_{\cG_{\phi}} }\mathop{\mathrm{Res}}_{\vep = 0} Z_{\cG_{\phi}}   \, , 
\ee
with $ S_{\cG_v}, \, S_{\cG_\phi}$  the usual symmetry factors arising from permutations of internal lines and 
vertices involving $\lambda$ leaving $\cG_v, \, \cG_\phi $ invariant while
$E_{\cG_v}$ is the order of the permutation group of equivalent vertices involving the same $V_n$, $n=3,4,5$,
in $V(\cG_v)$. From \eqref{VGv} we may also define
\be
P_{\cG_v} = \frac{720}{4!^r \, 3!^s\, 2!^t} \, ,
\ee
where $P_{\cG_v}/E_{\cG_v}$ is the number of terms arising in $\pr_i\pr_j\pr_k\pr_l\pr_m\pr_n V(\cG_v)$
which are necessary to form a symmetric $6$-tensor. 

The results for each graph $\cG$ are dependent on the subgraphs $\cG^\prime\in \cG$ for which it is necessary 
to introduce counterterms and in consequence there are  higher order poles in $\vep$. To elucidate this structure
we define for all connected  1PI graphs with superficial degree of divergence $D=0,2$ forming a set 
$\HH_{\rm 1PI} $,  
which is extended to include their disconnected products $\HH$, the reduced Hopf coproduct 
$\Delta : \HH\to \HH\otimes \HH$ by, for $\cG\in \HH_{\rm 1PI}$,
\be
\Delta\, (\cG)  = \begin{cases} {\ts \sum_{\cG^\prime \in \HH , \, \cG^\prime \subsetneqq G, \, 
\cG^\prime \ne {\mathds {1}}}} \; 
\cG^\prime \otimes \cG/\cG^\prime  \, , \quad \cG, \, \cG/\cG^\prime  \in  \HH \, ,
\\ \ \ \emptyset  \, , \quad \mbox{if} \ \cG \ \mbox{has no proper subgraphs in} \ \HH_{\rm 1PI} \, , \end{cases}
\ee
with $ {\mathds {1}}$ the empty graph.
Here  $\cG^\prime$ may include disconnected 1PI graphs
$\cG^\prime = \cG_1 \cup  \cG_2  \cup \dots $, $\cG_i \in \HH_{\rm 1PI}$ and  
$\cG_i \subsetneqq \cG$ for all $i$ and also 
non overlapping, so that  $\cG_i  \cap \cG_j  = \emptyset$ for all $i \ne j$, then in $\cG/\cG^\prime$ each subgraph 
$\cG_i \in \cG$ is contracted  to a single vertex or line, $\cG/ {\mathds {1}} = \cG$ and $\cG/ \cG = {\mathds {1}}$.
The extension for $\cG$ formed from disconnected 
products is straightforward but not relevant here. We restrict here just to the cases where if $\cG/\cG^\prime$ 
is 1VR it does not involve tadpole graphs where removing a vertex generates disconnected graphs one
of which is a vacuum graph. With dimensional regularisation tadpole graphs do not contribute.
In the case when $\cG$ has no proper subgraphs $\cG^\prime  \in \HH_{\rm 1PI}$,
$\cG$ is primitive, $\Delta(\cG) = \emptyset $, and generates just a simple pole in $\vep$.  Graphs with degree of divergence $D=1,3$,
such as 
\tikz[baseline=(vert_cent.base),scale=0.7]{
  \node (vert_cent) {\hspace{-13pt}$\phantom{-}$};
  \node at  (0,0)  (C) {};
   \node at  (1,0)  (D) {};
    \draw [bend left = 80]  (C.center) to (D.center);
   \draw [bend left = 25]  (C.center) to (D.center);
   \draw [bend right= 25] (C.center) to (D.center);
     \draw [bend right= 80] (C.center) to (D.center);
     \draw (0,0) --+ (140:0.45) ;
      \draw (0,0) --+ (220:0.45) ;
        \draw (1,0) --+ (40:0.45) ;
      \draw (1,0) --+ (320:0.45) ;
  \fill[black]  (C) circle [radius = 0.08cm];
 \fill[black]  (D) circle [radius = 0.08cm];
}, and which are present for odd numbers of loops, do not generate poles in $\vep$ and are therefore not
included in $\cG^\prime$.

Otherwise the necessary counterterms to
obtain $Z_{\cG_V} $ are determined by the terms arising in the coproduct $\Delta\, (\cG_V)$.

At two loops there is just one divergent graph
\begin{align}
\label{2loopZ}
\text{
\begin{tabular}{  c   c   c   c  c  c  }
$\cG$ label &  $~\cG ~$ & $~\Delta(\cG) ~$ & $ S_\cG,E_\cG,P_\cG $ &  $~~~Z_\cG~$  & $~~~c_{\cG}~~$\\
\noalign {\vskip 2pt}
\hline
\noalign {\vskip 4pt}
$2$ & 
\tikz[baseline=(vert_cent.base),scale=0.4]{
  \node (vert_cent) {\hspace{-13pt}$\phantom{-}$};
  \draw (0,0) circle [radius=1.2];
  \draw (-1.2,0)-- (1.2,0);
 \node at (-1.8,0) {$\scriptstyle{V_3}$};
 \node at (1.8,0) {$\scriptstyle {V_3}$};
 \fill[black]  (1.2,0) circle [radius = 0.1cm];
\fill[black]  (-1.2,0) circle [radius = 0.1cm];
}
& $  \emptyset $ & $ 6,2,20$   & ~ $\tfrac{2}{\vep} $  &~ $ \tfrac13 $\\ 
\noalign {\vskip 2pt}
\end{tabular} \, .
}
\end{align}

At four loops there is one propagator graph and four vertex graphs
\begin{align}
\label{4loop}
\text{
\begin{tabular}{  c   c   c   c  c  c }
$\cG$ label &  $~\cG ~$ & $~\Delta(\cG) ~$ & $ S_\cG,E_\cG,P_\cG$ &  $~~~Z_\cG~$  & $~~~d_{\cG},c_{\cG}~~$\\
\noalign {\vskip 2pt}
\hline
\noalign {\vskip 4pt}
$4\gamma $ & \hskip 0.5cm
\tikz[baseline=(vert_cent.base),scale=0.4]{
  \node (vert_cent) {\hspace{-13pt}$\phantom{-}$};
  \draw (0,0) ellipse [x radius=1.4, y radius= 1];
  \draw (-1.4,0)-- (1.4,0);
  \draw [bend left =40] (-1.4,0) to (1.4,0);
  \draw [bend right =40] (-1.4,0) to (1.4,0);
  \draw  (1.4,0) --(2.2,0) ;
   \draw  (-1.4,0) --(-2.2,0) ;
       \node at (1.6,-0.4) {$\scriptstyle {\lambda}$};
       \node at (-1.6,-0.4) {$\scriptstyle {\lambda}$};
  \node at (-2.4,0) {$\scriptstyle {i}$};
  \node at (2.4,0) {$\scriptstyle {j}$};
 \fill[black]  (1.4,0) circle [radius = 0.1cm];
\fill[black]  (-1.4,0) circle [radius = 0.1cm];
}
& $  \emptyset $ & $ 120 ~ $ & ~ $-\tfrac{2}{3\vep} $  &  $ \tfrac{1}{45} $  \\ 
 \noalign {\vskip 4pt} 
$4a$ & \tikz[baseline=(vert_cent.base),scale=0.4]{
  \node (vert_cent) {\hspace{-13pt}$\phantom{-}$};
  \draw (0,0) circle [radius=1.2];
  \node at  (70:1.2)  (A) {};
   \node at  (-70:1.2)  (B) {};
   \draw (A.center) -- (B.center);
    \draw [bend left = 38] (A.center) to (B.center);
     \draw [bend right =38] (A.center) to (B.center);
 \node at (-1.8,0) {$\scriptstyle{V_2}$};
 \node at (70:1.7) {$\scriptstyle {V_5}$};
  \node at (-70:1.7) {$\scriptstyle {V_5}$};
 \fill[black]  (70:1.2) circle [radius = 0.1cm];
\fill[black]  (-70:1.2) circle [radius = 0.1cm];
 \fill[black]  (-1.2,0) circle [radius = 0.1cm];
}
& $  \emptyset $ & $ 24,2,30$ & ~ $- \tfrac{2}{\vep} $  &$ \tfrac16 $\\ 
\noalign {\vskip 0pt} 
$4b$ & \tikz[baseline=(vert_cent.base),scale=0.4]{
  \node (vert_cent) {\hspace{-13pt}$\phantom{-}$};
  \draw (0,0) circle [radius=1.2];
  \node at  (55:1.2)  (A) {};
   \node at  (-55:1.2)  (B) {};
   \draw (A.center) -- (B.center);
   \draw [bend right = 30] (-1.2,0) to (A.center);
     \draw [bend right =50] (A.center) to (B.center);
 \node at (-1.8,0) {$\scriptstyle{V_3}$};
 \node at (55:1.7) {$\scriptstyle {V_5}$};
  \node at (-55:1.75) {$\scriptstyle {V_4}$};
 \fill[black]  (55:1.2) circle [radius = 0.1cm];
\fill[black]  (-55:1.2) circle [radius = 0.1cm];
 \fill[black]  (-1.2,0) circle [radius = 0.1cm];
}
& $ \cG_{2}\otimes \cG_{2} $ & $ 12,1,60$& $- \tfrac{2}{\vep^2}\big (1 -  2 \, \vep\big )\ $  & $- \tfrac43 $\\ 
\noalign {\vskip 0pt} 
$4c$ & 
\tikz[baseline=(vert_cent.base),scale=0.4]{
  \node (vert_cent) {\hspace{-13pt}$\phantom{-}$};
  \draw (0,0) circle [radius=1.2];
  \node at  (60:1.2)  (A) {};
   \node at  (-60:1.2)  (B) {};
   \draw [bend right = 28] (-1.2,0) to (A.center);
     \draw [bend left = 28] (-1.2,0) to (B.center);
     \draw [bend right =48] (A.center) to (B.center);
 \node at (-1.8,0) {$\scriptstyle{V_4}$};
 \node at (60:1.7) {$\scriptstyle {V_4}$};
  \node at (-60:1.7) {$\scriptstyle {V_4}$};
 \fill[black]  (60:1.2) circle [radius = 0.1cm];
\fill[black]  (-60:1.2) circle [radius = 0.1cm];
 \fill[black]  (-1.2,0) circle [radius = 0.1cm];
}
& $  \emptyset $ & $ 8,6,90$ & ~ $ \tfrac{\pi^2}{\vep}\ $  &$- \tfrac{\pi^2}{12} $
\end{tabular} \, .
}
\end{align}
The graphs $\cG_{4a}, \, \cG_{4c}$ have no subgraphs which would generate $\vep$-poles and
are primitive.
There  is also one 1VR divergent graph which contributes to overall RG factors but is
not relevant for the $\beta$-function itself
\begin{align}
\label{4loopR}
\text{
\begin{tabular}{  c   c   c   c  c  c }
$\cG$ label &  $~\cG ~$ & $~\Delta(\cG) ~$ & $ S_\cG,E_\cG,P_\cG$ &  $~~~Z_\cG~$  & \\
\noalign {\vskip 2pt}
\hline
\noalign {\vskip 4pt}
$ 4R  $ & \hskip 0.5cm
\tikz[baseline=(vert_cent.base),scale=0.45]{
  \node (vert_cent) {\hspace{-13pt}$\phantom{-}$};
  \draw (0,0) circle [radius=1];
  \draw (2,0) circle [radius=1];
  \draw (1,0) -- (3,0);
  \draw (-1,0) -- (1,0);
    \node at (3.5,0) {$\scriptstyle {V_3}$};
  \node at (1,-0.8) {$\scriptstyle {\lambda}$};
   \node at (-1.5,0) {$\scriptstyle {V_3}$};
 \fill[black]  (-1,0) circle [radius = 0.1cm];
\fill[black]  (1,0) circle [radius = 0.1cm];
\fill[black]  (3,0) circle [radius = 0.1cm];
} \ .
& $ 2\, \cG_2\otimes \cG_2 $ & $ 36,2,20 $ & ~ $-\tfrac{4}{\vep^2 } $  &  
\end{tabular} \, .
}
\end{align}

From the two and four loop results in \eqref{2loopZ} and \eqref{4loop} one may read off
\be
\beta_V(\phi)^{(2)}  =   \tfrac13 \, V_{ijk}(\phi) V_{ijk}(\phi) \, ,
\ee
and
\begin{align}
2\gamma_{ij}{\!}^{(4)} = {}& \tfrac{1}{45} \, \lambda_{iklmnp} \lambda_{jklmnp} \, , \nn \\
\noalign{\vskip 2pt}
{\tilde \beta}_V(\phi)^{(4)} = {}&  \tfrac16 \,
V_{ij}(\phi) V_{iklmn}(\phi) V_{jklmn}(\phi) - \tfrac43 \,
V_{ijk}(\phi)  V_{ilmn}(\phi) V_{jklmn}(\phi ) \nn \\
\noalign{\vskip 0pt}
&{} - \tfrac{\pi^2}{12} \, V_{ijkl}(\phi)  V_{klmn}(\phi) V_{ijmn}(\phi ) \, ,
\label{Bfour}
\end{align}
which of course are in accord  with previous calculations \cite{Pisarski,Hager0,Dwyer} and the form given in \eqref{Bfour}
was exhibited in \cite{Seeking}.

At six loops there is still just one propagator graph but 26 1PI, 1VI vertex graphs. Here we also
include the labelling of graphs given by Hager \cite{Hager}.
\begin{align}
\label{6loop}
\text{
\begin{tabular}{  c   c   c   c  c  c }
$\cG$ label &  $~\cG ~$ & $~\Delta(\cG) ~$ & $ S_\cG,E_\cG,P_\cG$  & $~~~Z_\cG~$  & $~~~d_{\cG},c_{\cG}~~$\\
\noalign {\vskip 2pt}
\hline
 \noalign {\vskip 4pt} 
$6a,a1$ &\hskip- 0.5cm \tikz[baseline=(vert_cent.base),scale=0.45]{
  \node (vert_cent) {\hspace{-13pt}$\phantom{-}$};
  \draw (0,0) circle [radius=1.2];
  \node at  (100:1.2)  (A) {};
   \node at  (-100:1.2)  (B) {};
    \draw (A.center) to (B.center);
   \draw [bend left = 28] (1.2,0) to (A.center);
     \draw [bend left = 48] (B.center) to (1.2,0);
        \draw [bend left = -10] (B.center) to (1.2,0);
  \draw (-100:1.2) arc(230:395:0.953) ;
 \node at (-1.8,0) {$\scriptstyle{V_2}$};
 \node at (95:1.7) {$\scriptstyle {V_4}$};
  \node at (-95:1.7) {$\scriptstyle {\lambda}$};
   \node at (1.7,0.1) {$\scriptstyle {\lambda}$};
 \fill[black]  (100:1.2) circle [radius = 0.1cm];
\fill[black]  (-100:1.2) circle [radius = 0.1cm];
 \fill[black]  (1.2,0) circle [radius = 0.1cm];
 \fill[black]  (-1.2,0) circle [radius = 0.1cm];
}
& $  \emptyset $ & $ 48,1,15$  & $0 $  &$\ 0 $\\ 
\noalign {\vskip 0pt} 
$6b,a2$ & \hskip -0.5cm
\tikz[baseline=(vert_cent.base),scale=0.4]{
  \node (vert_cent) {\hspace{-13pt}$\phantom{-}$};
  \draw (0,0) circle [radius=1.2];
  \node at  (95:1.2)  (A) {};
   \node at  (-95:1.2)  (B) {};
   \draw [bend left = 25] (1.2,0) to (A.center);
    \draw [bend right = 25] (1.2,0) to (B.center);
     \draw [bend right =70] (A.center) to (B.center);
      \draw [bend right =20] (A.center) to (B.center);
      \draw [bend left =20] (A.center) to (B.center);
 \node at (-1.8,0) {$\scriptstyle{V_2}$};
 \node at (95:1.7) {$\scriptstyle {\lambda}$};
  \node at (-95:1.7) {$\scriptstyle {\lambda}$};
   \node at (1.7,0) {$\scriptstyle {V_4}$};
 \fill[black]  (95:1.2) circle [radius = 0.1cm];
\fill[black]  (-95:1.2) circle [radius = 0.1cm];
 \fill[black]  (-1.2,0) circle [radius = 0.1cm];
  \fill[black]  (1.2,0) circle [radius = 0.1cm];
}
& $ \cG_{2}\otimes \cG_{4Ra } $ & $ 48,1,15$ & $  -\tfrac{4\pi^2}{3\vep} $  & $- \tfrac{\pi^2}{6} $\\ 
\noalign {\vskip 0pt} 
$6c,b1$ & \hskip -0.5cm
\tikz[baseline=(vert_cent.base),scale=0.45]{
  \node (vert_cent) {\hspace{-13pt}$\phantom{-}$};
  \draw (0,0) circle [radius=1.2];
  \node at  (100:1.2)  (A) {};
   \node at  (-100:1.2)  (B) {};
    \draw (A.center) to (B.center);
   \draw [bend right = 48]  (A.center) to (1.2,0);
   \draw [bend left = 5]  (A.center) to (1.2,0);
     \draw [bend left = 48] (B.center) to (1.2,0);
        \draw [bend right = 5] (B.center) to (1.2,0);
  \node at (-1.8,0) {$\scriptstyle{V_2}$};
 \node at (95:1.7) {$\scriptstyle {V_5}$};
  \node at (-95:1.7) {$\scriptstyle {V_5}$};
   \node at (1.7,0.1) {$\scriptstyle {\lambda}$};
 \fill[black]  (100:1.2) circle [radius = 0.1cm];
\fill[black]  (-100:1.2) circle [radius = 0.1cm];
 \fill[black]  (1.2,0) circle [radius = 0.1cm];
 \fill[black]  (-1.2,0) circle [radius = 0.1cm];
}
&\hskip -0cm $  2\, \cG_{2}\otimes \cG_{4a} $ & 
$ 36, 2,30$  & ~ $ \tfrac{8}{3\vep^2}(1-2\vep)\, $  &$- \tfrac{4}{9} $ \\
\noalign {\vskip 0pt} 
$6d,b2$ & \hskip -0.5cm
\tikz[baseline=(vert_cent.base),scale=0.45]{
  \node (vert_cent) {\hspace{-13pt}$\phantom{-}$};
  \draw (0,0) circle [radius=1.2];
  \node at  (100:1.2)  (A) {};
   \node at  (-100:1.2)  (B) {};
    \draw [bend right = 40]  (A.center) to (B.center);
    \draw [bend left = 25]  (A.center) to (B.center);
   \draw [bend right = 38]  (A.center) to (1.2,0);
     \draw [bend left = 38] (B.center) to (1.2,0);
        \draw [bend right = 5] (B.center) to (1.2,0);
  \node at (-1.8,0) {$\scriptstyle{V_2}$};
 \node at (95:1.7) {$\scriptstyle {V_5}$};
  \node at (-95:1.7) {$\scriptstyle {\lambda}$};
   \node at (1.8,0) {$\scriptstyle {V_5}$};
 \fill[black]  (100:1.2) circle [radius = 0.1cm];
\fill[black]  (-100:1.2) circle [radius = 0.1cm];
 \fill[black]  (1.2,0) circle [radius = 0.1cm];
 \fill[black]  (-1.2,0) circle [radius = 0.1cm];
}
&\hskip -0.3cm $  \ \ \cG_{2}\otimes \cG_{4a }  $ & 
 $ 24,1,30$ & ~ $  \tfrac{4}{3\vep^2}(1-6\vep) $  &$- 2 $
\end{tabular} \, .
}
\end{align}
Here $\cG_{6a}$ is primitive.
 Results  for $Z_\cG$ for these graphs can be found in \cite{Hager0}. Graphs containing 
two subgraphs with degree of divergence $D=0$ and which then  generate
a leading singular contribution of order $\vep^{-3}$ are
\begin{align}
\label{6loop2}
\hskip -0.5cm
\text{
\begin{tabular}{  c   c   c   c  c  c}
$\cG$ label &  $\cG $ & $~\Delta(\cG) ~$ & $ S_\cG,E_\cG,P_\cG $ & $~~~Z_\cG~$  & $~~~c_{\cG}~~$\\
\noalign {\vskip 2pt}
\hline
\noalign {\vskip 4pt}
\hskip -0.2cm $6e,d5$ &\hskip- 0.5cm 
\tikz[baseline=(vert_cent.base),scale=0.45]{
  \node (vert_cent) {\hspace{-13pt}$\phantom{-}$};
  \draw (0,0) circle [radius=1.2];
  \node at  (100:1.2)  (A) {};
   \node at  (-100:1.2)  (B) {};
   \draw [bend left = 50]  (A.center) to (-1.2,0);
   \draw [bend right = 48]  (A.center) to (1.2,0);
   \draw [bend left = 5]  (A.center) to (1.2,0);
     \draw [bend left = 48] (B.center) to (1.2,0);
        \draw [bend right = 5] (B.center) to (1.2,0);
  \node at (-1.8,0) {$\scriptstyle{V_3}$};
 \node at (95:1.7) {$\scriptstyle {V_5}$};
  \node at (-95:1.7) {$\scriptstyle {V_4}$};
   \node at (1.7,0.1) {$\scriptstyle {\lambda}$};
 \fill[black]  (100:1.2) circle [radius = 0.1cm];
\fill[black]  (-100:1.2) circle [radius = 0.1cm];
 \fill[black]  (1.2,0) circle [radius = 0.1cm];
 \fill[black]  (-1.2,0) circle [radius = 0.1cm];
}
& $\hskip -0.5cm 2\, \cG_2 \otimes \cG_{4b} + \cG_ 2\cG_2 \otimes \cG_2$ & $72,1,60$ & $\tfrac{8}{3\vep^3}
\big (1-2\vep + \tfrac13 (\pi^2 -6) \vep^2\big )  $  &$\tfrac{2}{27}(\pi^2-6) $\\ 
\noalign {\vskip 4pt} 
\hskip -0.2cm $6f,g1a$ & \hskip -0.5cm
\tikz[baseline=(vert_cent.base),scale=0.45]{
  \node (vert_cent) {\hspace{-13pt}$\phantom{-}$};
  \draw (0,0) circle [radius=1.2];
  \node at  (40:1.2)  (A) {};
   \node at  (140:1.2)  (B) {};
   \draw [bend left = 50]  (A.center) to (B.center);
     \draw [bend left = 0]  (A.center) to (B.center);
        \node at  (220:1.2)  (C) {};
   \node at  (320:1.2)  (D) {};
   \draw [bend left = 50]  (C.center) to (D.center);
   \draw [bend left = 0]  (C.center) to (D.center);
       \draw [bend left= 48] (D.center) to (A.center);
  \node at (145:1.8) {$\scriptstyle{V_4}$};
   \node at (220:1.7) {$\scriptstyle{V_4}$};
 \node at (40:1.7) {$\scriptstyle {V_5}$};
  \node at (-40:1.8) {$\scriptstyle {V_5}$};
 \fill[black]  (A) circle [radius = 0.1cm];
\fill[black]  (B) circle [radius = 0.1cm];
 \fill[black]  (C) circle [radius = 0.1cm];
 \fill[black]  (D) circle [radius = 0.1cm];
}
& $\hskip -0.5cm 2\, \cG_2 \otimes \cG_{4b} + \cG_ 2\cG_2 \otimes \cG_2$ & $ 72, 2,180$ & $\tfrac{8}{3\vep^3}
\big (1-2\vep + \tfrac13 (\pi^2 -6) \vep^2\big )  $  &$\tfrac{1}{27}(\pi^2-6) $\\ 
\noalign {\vskip 4pt}
\hskip -0.2cm $6g,d8$ &\hskip- 0.5cm 
\tikz[baseline=(vert_cent.base),scale=0.45]{
  \node (vert_cent) {\hspace{-13pt}$\phantom{-}$};
  \draw (0,0) circle [radius=1.2];
  \node at  (90:1.2)  (A) {};
   \node at  (-90:1.2)  (B) {};
  \draw [bend left = 0]  (A.center) to (1.2,0);
   \draw [bend right = 50]  (A.center) to (1.2,0);
  \draw [bend left = 50]  (A.center) to (-1.2,0);
     \draw [bend left = 50] (B.center) to (1.2,0);
     \draw (A.center) -- (B.center) ;     
  \node at (-1.8,0) {$\scriptstyle{V_3}$};
 \node at (90:1.7) {$\scriptstyle {\lambda}$};
  \node at (-85:1.7) {$\scriptstyle {V_4}$};
   \node at (1.7,0) {$\scriptstyle {V_5}$};
 \fill[black]  (90:1.2) circle [radius = 0.1cm];
\fill[black]  (-90:1.2) circle [radius = 0.1cm];
 \fill[black]  (1.2,0) circle [radius = 0.1cm];
 \fill[black]  (-1.2,0) circle [radius = 0.1cm];
}
& $ \cG_{2}\otimes \cG_{4b }+ \cG_{4b} \otimes \cG_2   $ & $ 24,1,60$ & $\tfrac{4}{3\vep^3}
\big (1-6\vep - \tfrac23 (\pi^2 - 24) \vep^2\big )$  & $ -\tfrac{2}{9}(\pi^2-24) $\\ 
\noalign {\vskip 4pt}
\hskip -0.2cm $6h,e3$ &\hskip- 0.5cm 
\tikz[baseline=(vert_cent.base),scale=0.45]{
  \node (vert_cent) {\hspace{-13pt}$\phantom{-}$};
  \draw (0,0) circle [radius=1.2];
  \node at  (90:1.2)  (A) {};
   \node at  (-90:1.2)  (B) {};
  \draw [bend left = 0]  (B.center) to (1.2,0);
   \draw [bend right = 50]  (A.center) to (1.2,0);
  \draw [bend left = 50]  (A.center) to (-1.2,0);
     \draw [bend left = 50] (B.center) to (1.2,0);
     \draw (A.center) -- (B.center) ;     
  \node at (-1.8,0) {$\scriptstyle{V_3}$};
 \node at (90:1.7) {$\scriptstyle {V_5}$};
  \node at (-90:1.7) {$\scriptstyle {V_5}$};
   \node at (1.7,0) {$\scriptstyle {V_5}$};
 \fill[black]  (90:1.2) circle [radius = 0.1cm];
\fill[black]  (-90:1.2) circle [radius = 0.1cm];
 \fill[black]  (1.2,0) circle [radius = 0.1cm];
 \fill[black]  (-1.2,0) circle [radius = 0.1cm];
}
& $ \cG_{2}\otimes \cG_{4b } + \cG_{4b} \otimes \cG_2 $ & $ 24,1,120 $ & $\tfrac{4}{3\vep^3}
\big (1-6\vep - \tfrac23 (\pi^2 - 24) \vep^2\big )$  & $ -\tfrac{2}{9}(\pi^2-24) $\\ 
\noalign {\vskip 4pt} 
\hskip -0.2cm $6i,d1$ & \hskip -0.5cm
\tikz[baseline=(vert_cent.base),scale=0.45]{
  \node (vert_cent) {\hspace{-13pt}$\phantom{-}$};
  \draw (0,0) circle [radius=1.2];
  \node at  (-0.17,0)  (A) {};
   \node at  (0:1.2)  (B) {};
      \node at  (90:1.2)  (C) {};
     \node at  (270:1.2)  (D) {};
   \draw [bend right = 50]  (B.center) to (D.center);
   \draw   (A.center) to (B.center);
   \draw [bend left = 5]  (B.center) to (C.center);
     \draw [bend left = 45] (B.center) to (C.center);
     \draw [bend left = 15] (D.center) to (C.center) ;     
  \node at (-0.64,0) {$\scriptstyle{V_3}$};
 \node at (90:1.7) {$\scriptstyle {V_5}$};
  \node at (-90:1.7) {$\scriptstyle {V_4}$};
   \node at (1.7,0) {$\scriptstyle {\lambda}$};
 \fill[black]  (A) circle [radius = 0.1cm];
\fill[black]  (D) circle [radius = 0.1cm];
 \fill[black]  (B) circle [radius = 0.1cm];
 \fill[black]  (C) circle [radius = 0.1cm];
}
&\hskip -0cm $  \cG_{2}\otimes \cG_{4b }+ \cG_{4b} \otimes \cG_2  $ & 
$ 12, 1,60 $ & $  \tfrac{4}{3\vep^3}
\big (1-6\vep - \tfrac43 (\pi^2 - 12) \vep^2\big ) $  &$  -\tfrac{8}{9}(\pi^2-12)  $ \\
\noalign {\vskip 4pt} 
\noalign {\vskip 0pt} 
\hskip -0.2cm $6j,g0a$ & \hskip -0.5cm
\tikz[baseline=(vert_cent.base),scale=0.45]{
  \node (vert_cent) {\hspace{-13pt}$\phantom{-}$};
  \draw (0,0) circle [radius=1.2];
  \node at  (90:1.2)  (A) {};
   \node at  (-90:1.2)  (B) {};
    \draw (A.center) to (B.center);
   \draw [bend right = 48]  (A.center) to (1.2,0);
   \draw [bend left = 5]  (A.center) to (1.2,0);
     \draw [bend right = 48] (B.center) to (-1.2,0);
        \draw [bend left = 5] (B.center) to (-1.2,0);
  \node at (-1.8,0) {$\scriptstyle{V_4}$};
 \node at (90:1.7) {$\scriptstyle {V_5}$};
  \node at (-90:1.7) {$\scriptstyle {V_5}$};
   \node at (1.7,0) {$\scriptstyle {V_4}$};
 \fill[black]  (A) circle [radius = 0.1cm];
\fill[black]  (B) circle [radius = 0.1cm];
 \fill[black]  (1.2,0) circle [radius = 0.1cm];
 \fill[black]  (-1.2,0) circle [radius = 0.1cm];
}
&\hskip -0.5cm $ \  \ 2\, \cG_{2}\otimes \cG_{4b} + \cG_2\cG_2 \otimes \cG_2 $ & 
$ 36, 2,180 $ & ~ $ \tfrac{8}{3\vep^3}
\big (1 -  2 \vep -2 \vep^2\big ) $  &$ -\tfrac{4}{9} $\\
\hskip -0.2cm $6k,d6$ & \hskip -0.5cm
\tikz[baseline=(vert_cent.base),scale=0.45]{
  \node (vert_cent) {\hspace{-13pt}$\phantom{-}$};
  \draw (0,0) circle [radius=1.2];
  \node at  (180:1.2)  (A) {};
   \node at  (0:1.2)  (B) {};
   \node at  (90:1.2)  (C) {};
\node at  (270:1.2)  (D) {};
   \draw [bend right = 37]  (B.center) to (D.center);
   \draw [bend right = 37]  (A.center) to (C.center);
     \draw [bend right = 23]   (D.center) to (C.center) ;   
       \draw [bend left = 23]   (D.center) to (C.center) ;  
   \draw   (D.center) to (B.center) ;    
  \node at (-1.8,0) {$\scriptstyle{V_3}$};
 \node at (90:1.7) {$\scriptstyle {V_5}$};
  \node at (-90:1.7) {$\scriptstyle {\lambda}$};
   \node at (1.7,0) {$\scriptstyle {V_4}$};
 \fill[black]  (90:1.2) circle [radius = 0.1cm];
\fill[black]  (-90:1.2) circle [radius = 0.1cm];
 \fill[black]  (1.2,0) circle [radius = 0.1cm];
 \fill[black]  (-1.2,0) circle [radius = 0.1cm];
}
&\hskip -0.5cm $  \ \ \ \cG_{2}\otimes \cG_{4b}  + \cG_{4b} \otimes \cG_2 $ & 
$ 24, 1,60  $ & ~ $ \tfrac{4}{3\vep^3}
\big (1 -  6 \vep + 16 \vep^2\big ) $  &$ \tfrac{16}{3} $ \\
\noalign {\vskip 0pt} 
\hskip -0.2cm $6l,c0b$ & \hskip -0.5cm
\tikz[baseline=(vert_cent.base),scale=0.45]{
  \node (vert_cent) {\hspace{-13pt}$\phantom{-}$};
  \draw (0,0) circle [radius=1.2];
  \node at  (180:1.2)  (A) {};
   \node at  (0:1.2)  (B) {};
   \node at  (90:1.2)  (C) {};
\node at  (270:1.2)  (D) {};
   \draw [bend right = 37]  (B.center) to (D.center);
   \draw [bend right = 37]  (A.center) to (C.center);
     \draw [bend right = 23]   (D.center) to (C.center) ;   
       \draw [bend left = 23]   (D.center) to (C.center) ;  
        \draw   (D.center) to (C.center) ;    
  \node at (-1.8,0) {$\scriptstyle{V_3}$};
 \node at (90:1.7) {$\scriptstyle {\lambda}$};
  \node at (-90:1.7) {$\scriptstyle {\lambda}$};
   \node at (1.75,0) {$\scriptstyle {V_3}$};
 \fill[black]  (90:1.2) circle [radius = 0.1cm];
\fill[black]  (-90:1.2) circle [radius = 0.1cm];
 \fill[black]  (1.2,0) circle [radius = 0.1cm];
 \fill[black]  (-1.2,0) circle [radius = 0.1cm];
}
&\hskip -0.5cm $  \ \ \ \cG_{2}\otimes \cG_{4R}  + 2\, \cG_{4b} \otimes \cG_2  $ & 
$ 24, 2 ,20 $ & ~ $ \tfrac{8}{3\vep^3}
\big (1 -  4 \vep + 2 \vep^2\big ) $  &$ \tfrac{2}{3} $\\
\end{tabular} \, .
}
\end{align}
In the coproducts besides $\cG_R$ in \eqref{4loopR} there is another  1VR non tadpole  graph
labelled by $\cG_{4Ra}$,
\tikz[baseline=(vert_cent.base),scale=0.3]{
  \node (vert_cent) {\hspace{-13pt}$\phantom{-}$};
  \draw (0,0) circle [radius=1];
  \draw (2,0) circle [radius=1];
  \draw [bend left = 45] (1,0) to (3,0);
   \draw [bend right = 45] (1,0) to (3,0);
    \node at (3.7,0) {$\scriptstyle {V_4}$};
  \node at (1,-1.1) {$\scriptstyle {\lambda}$};
   \node at (-1.7,0) {$\scriptstyle {V_2}$};
 \fill[black]  (-1,0) circle [radius = 0.1cm];
\fill[black]  (1,0) circle [radius = 0.1cm];
\fill[black]  (3,0) circle [radius = 0.1cm];
}\hskip -0.1cm ,
where the associated integral has no poles in $\vep$.

The 6 loop contributions which involve $\vep^{-2}$ singular contributions are given by
\begin{align}
\label{6loop3}
\text{
\begin{tabular}{  c   c   c   c  c  c}
$\cG$ label &  $~\cG ~$ & $~\Delta(\cG) ~$ & $ S_\cG,E_\cG,P_\cG$ & $~~~Z_\cG~$  & $~~~d_{\cG},c_{\cG}~~$\\
\noalign {\vskip 2pt}
\hline
\noalign {\vskip 4pt}
\noalign {\vskip 4pt}
$6\gamma $ & \hskip- 0.5cm 
\tikz[baseline=(vert_cent.base),scale=0.45]{
  \node (vert_cent) {\hspace{-13pt}$\phantom{-}$};
  \draw (0,0) ellipse [x radius=1.4, y radius= 1.1];
  \draw (-1.4,0)-- (1.4,0);
  \draw [bend left =70] (-1.4,0) to (0,0);
  \draw [bend right =70] (-1.4,0) to (0,0);
  \draw [bend left =70] (0,0) to (1.4,0);
  \draw [bend right =70] (0,0) to (1.4,0);
   \draw  (1.4,0) --(2.2,0) ;
   \draw  (-1.4,0) --(-2.2,0) ;
  \node at (-2.4,0) {$\scriptstyle {i}$};
  \node at (2.4,0) {$\scriptstyle {j}$};
   \node at (0,-0.5) {$\scriptstyle {\lambda}$};
     \node at (1.55,-0.4) {$\scriptstyle {\lambda}$};
       \node at (-1.55,-0.4) {$\scriptstyle {\lambda}$};
 \fill[black]  (1.4,0) circle [radius = 0.1cm];
\fill[black]  (-1.4,0) circle [radius = 0.1cm];
\fill[black]  (0,0) circle [radius = 0.1cm];
}
& \hskip -0.1cm $  2\,\cG_{2} \otimes \cG_{4\gamma}$ & $ 72  ~  $ & 
$ \tfrac{8}{9\vep^2}\big (1 - \tfrac23 \, \vep\big )$ & $-\tfrac{4}{81} $  \\ 
\noalign {\vskip 6pt}
\hskip -0.2cm $6m,c2$ & \hskip -0.5cm
\tikz[baseline=(vert_cent.base),scale=0.45]{
  \node (vert_cent) {\hspace{-13pt}$\phantom{-}$};
 \draw (0,0) ellipse [x radius=1.3, y radius= 1.1];
    \draw (0,0) ellipse [x radius=0.72, y radius= 0.6];
       \draw [bend left = 45] (-0.72,0) to (0.72,0);
         \draw [bend right = 45](-0.72,0) to (0.72,0);
  \draw (-1.2,0)-- (1.2,0) ;
 \node at (-1.8,0) {$\scriptstyle{V_3}$};
 \node at (1.8 ,0) {$\scriptstyle{V_3}$};
 \node at (0.88,-0.35) {$\scriptstyle {\lambda}$};
 \node at (-0.88,-0.35) {$\scriptstyle {\lambda}$};
 \fill[black]  (1.3,0) circle [radius = 0.1cm];
\fill[black]  (-1.3,0) circle [radius = 0.1cm];
 \fill[black]  (0.72,0) circle [radius = 0.1cm];
  \fill[black]  (-0.72,0) circle [radius = 0.1cm];
}
&\hskip -0.5cm $\  \cG_{\gamma4} \otimes \cG_2 $ & 
$ 240,2, 20 $ & ~ $ \tfrac{8}{9\vep^2}\big ( 1- \tfrac{10}{3} \vep \big )  $  &  $- \tfrac{1}{27} $ \\
\noalign {\vskip 4pt}
\hskip -0.2cm $6n,c1b$ & \hskip -0.5cm
\tikz[baseline=(vert_cent.base),scale=0.45]{
  \node (vert_cent) {\hspace{-13pt}$\phantom{-}$};
  \draw (0,0) circle [radius=1.2];
    \draw (0,0.6) ellipse [x radius=0.5, y radius= 0.6];
    \draw [bend left = 40] (0,0) to (0,1.2);
    \draw [bend right = 40](0,0) to (0,1.2);
  \draw (-1.2,0)-- (1.2,0) ;
 \node at (-1.8,0) {$\scriptstyle{V_3}$};
 \node at (1.75,0) {$\scriptstyle{V_3}$};
 \node at (0.0,1.7) {$\scriptstyle {\lambda}$};
 \node at (0,-0.5) {$\scriptstyle {\lambda}$};
 \fill[black]  (1.2,0) circle [radius = 0.1cm];
\fill[black]  (-1.2,0) circle [radius = 0.1cm];
 \fill[black]  (0.0,1.2) circle [radius = 0.1cm];
  \fill[black]  (0,0) circle [radius = 0.1cm];
}
&\hskip -0.5cm   \ \ $ 2\,\cG_{4a}\otimes \cG_{2}  $ & 
$ 48,2,20  $ & ~ $\tfrac{16}{3\vep^2}(1- 2 \vep) $  &   $- \tfrac23 $ \\
\noalign {\vskip 4pt}
\hskip -0.2cm $6o,d4$ & \hskip -0.5cm
\tikz[baseline=(vert_cent.base),scale=0.45]{
  \node (vert_cent) {\hspace{-13pt}$\phantom{-}$};
  \draw (0,0) circle [radius=1.2];
  \node at  (95:1.2)  (A) {};
   \node at  (-95:1.2)  (B) {};
    \draw (A.center) to (B.center);
   \draw [bend right =45] (-1.2,0) to (A.center);
     \draw [bend left = 55] (B.center) to (1.2,0);
        \draw [bend left = -1] (B.center) to (1.2,0);
  \draw (-95:1.2) arc(230:395:0.9) ;
 \node at (-1.8,0) {$\scriptstyle{V_3}$};
 \node at (95:1.7) {$\scriptstyle {V_4}$};
  \node at (-95:1.7) {$\scriptstyle {\lambda}$};
   \node at (1.7,0) {$\scriptstyle {V_5}$};
 \fill[black]  (95:1.2) circle [radius = 0.1cm];
\fill[black]  (-95:1.2) circle [radius = 0.1cm];
 \fill[black]  (1.2,0) circle [radius = 0.1cm];
 \fill[black]  (-1.2,0) circle [radius = 0.1cm];
}
&\hskip -0.5cm   \ \ $ \cG_{4a}\otimes \cG_{2}  $ & 
$ 48,1, 60 $ & ~ $ \tfrac{8}{3\vep^2}(1-4 \vep) $  &   $- \tfrac43 $ \\
\noalign {\vskip 4pt}
\hskip -0.2cm $6p,g3$ & \hskip -0.5cm
\tikz[baseline=(vert_cent.base),scale=0.45]{
  \node (vert_cent) {\hspace{-13pt}$\phantom{-}$};
  \draw (0,0) circle [radius=1.2];
  \node at  (40:1.2)  (A) {};
   \node at  (140:1.2)  (B) {};
   \draw [bend left = 50]  (A.center) to (B.center);
     \draw [bend left = 0]  (A.center) to (B.center);
        \node at  (220:1.2)  (C) {};
   \node at  (320:1.2)  (D) {};
   \draw [bend left = 60]  (C.center) to (D.center);
   \draw [bend left = 20]  (C.center) to (D.center);
   \draw [bend right= 18] (C.center) to (D.center);
  \node at (140:1.75) {$\scriptstyle{V_4}$};
   \node at (220:1.7) {$\scriptstyle{V_5}$};
 \node at (40:1.7) {$\scriptstyle {V_4}$};
  \node at (-40:1.8) {$\scriptstyle {V_5}$};
 \fill[black]  (A) circle [radius = 0.1cm];
\fill[black]  (B) circle [radius = 0.1cm];
 \fill[black]  (C) circle [radius = 0.1cm];
 \fill[black]  (D) circle [radius = 0.1cm];
}
&\hskip -0.5cm   \ \ $ \cG_{2}\otimes \cG_{4a}  $ & 
$ 144,2,180 $ & ~ $ \tfrac{4}{3\vep^2} $  &   $ ~ 0 $ \\
\noalign {\vskip 4pt}
\hskip -0.2cm $6q,f2$ & \hskip -0.5cm
\tikz[baseline=(vert_cent.base),scale=0.45]{
  \node (vert_cent) {\hspace{-13pt}$\phantom{-}$};
  \draw (0,0) circle [radius=1.2];
  \node at  (180:1.2)  (A) {};
   \node at  (0:1.2)  (B) {};
      \node at  (90:1.2)  (C) {};
     \node at  (270:1.2)  (D) {};
   \draw [bend right = 50]  (B.center) to (D.center);
   \draw [bend right = 48]  (A.center) to (C.center);
   \draw [bend left = 5]  (A.center) to (C.center);
     \draw [bend left = 45] (B.center) to (C.center);
     \draw (D.center) -- (C.center) ;     
  \node at (-1.7,0) {$\scriptstyle{V_4}$};
 \node at (90:1.7) {$\scriptstyle {\lambda}$};
  \node at (-90:1.7) {$\scriptstyle {V_4}$};
   \node at (1.7,0) {$\scriptstyle {V_4}$};
 \fill[black]  (A) circle [radius = 0.1cm];
\fill[black]  (D) circle [radius = 0.1cm];
 \fill[black]  (B) circle [radius = 0.1cm];
 \fill[black]  (C) circle [radius = 0.1cm];
}
&\hskip -0.5cm $  \ \ \cG_{2}\otimes \cG_{4c }  $ & 
$ 24,1,90  $ & ~ $ - \tfrac{2\pi^2}{3\vep^2}
\big (1 +   (2\ln 2 - 5) \vep\big )\ $  &$- \tfrac{\pi^2}{6}(2 \ln 2 -5) $ \\
\noalign {\vskip 4pt}
\hskip -0.2cm $6r,g2a$ & \hskip -0.5cm
\tikz[baseline=(vert_cent.base),scale=0.45]{
  \node (vert_cent) {\hspace{-13pt}$\phantom{-}$};
  \draw (0,0) circle [radius=1.2];
  \node at  (50:1.2)  (A) {};
   \node at  (-50:1.2)  (B) {};
    \node at  (135:1.2)  (C) {};
   \node at  (-135:1.2)  (D) {};
   \draw [bend right = 45] (C.center)to (A.center);
     \draw [bend left = 45] (D.center) to (B.center);
     \draw [bend right =45] (A.center) to (B.center);
       \draw [bend left =1] (A.center) to (B.center);
    \draw [bend left =45] (C.center) to (D.center);
 \node at (135:1.8) {$\scriptstyle{V_4}$};
 \node at (-135:1.7) {$\scriptstyle{V_4}$};
 \node at (50:1.7) {$\scriptstyle {V_5}$};
 \node at (-50:1.8) {$\scriptstyle {V_5}$};
 \fill[black]  (A) circle [radius = 0.1cm];
\fill[black]  (B) circle [radius = 0.1cm];
 \fill[black]  (C) circle [radius = 0.1cm];
  \fill[black]  (D) circle [radius = 0.1cm];
}
&\hskip -0.5cm $  \ \ \cG_{2}\otimes \cG_{4c }  $ & 
$ 48,2,180  $ & ~ $ - \tfrac{2\pi^2}{3\vep^2}
\big (1 +   (2\ln 2 - 5) \vep\big )\ $  &$- \tfrac{\pi^2}{24}(2 \ln 2 -5) $ \\
\noalign {\vskip 4pt} 
\hskip -0.2cm $6s,d7$ & \hskip -0.5cm
\tikz[baseline=(vert_cent.base),scale=0.45]{
  \node (vert_cent) {\hspace{-13pt}$\phantom{-}$};
  \draw (0,0) circle [radius=1.2];
  \node at  (90:1.2)  (A) {};
   \node at  (-90:1.2)  (B) {};
   \draw [bend left = 37]   (A.center) to (-1.2,0);
   \draw [bend right = 37]   (A.center) to (1.2,0);
     \draw [bend left = 37] (B.center) to (1.2,0);
     \draw [bend right = 23]   (A.center) to (B.center) ;   
       \draw [bend left = 23]   (A.center) to (B.center) ;    
  \node at (-1.8,0) {$\scriptstyle{V_3}$};
 \node at (90:1.7) {$\scriptstyle {\lambda}$};
  \node at (-90:1.7) {$\scriptstyle {V_5}$};
   \node at (1.7,0) {$\scriptstyle {V_4}$};
 \fill[black]  (90:1.2) circle [radius = 0.1cm];
\fill[black]  (-90:1.2) circle [radius = 0.1cm];
 \fill[black]  (1.2,0) circle [radius = 0.1cm];
 \fill[black]  (-1.2,0) circle [radius = 0.1cm];
}
&\hskip -0.5cm \ \ $\cG_{4c}\otimes \cG_{2}  $ & 
$ 16,1,60  $ & ~ $ - \tfrac{4\pi^2}{3\vep^2}
\big (1 -   (\ln 2  + \tfrac32) \vep\big ) $  & $ \tfrac{\pi^2}{4}(2 \ln 2 +3) $ \\
\noalign {\vskip 4pt} 
\hskip -0.2cm $6t,e1$ & \hskip -0.5cm
\tikz[baseline=(vert_cent.base),scale=0.50]{
  \node (vert_cent) {\hspace{-13pt}$\phantom{-}$};
  \draw (0,0) circle [radius=1.2];
 \node at  (60:1.2)  (A) {};
  \node at  (-60:1.2)  (B) {};
  \node at  (-1.2,0)  (C) {};
    \draw [bend right = 20] (C.center) to (A.center);
     \draw [bend left = 20] (C.center) to (B.center);
    \draw [bend right = 20] (A.center) to (B.center);
  \draw (C.center)-- (0,0) ;
   \draw (A.center)-- (0,0) ;
    \draw (B.center)-- (0,0) ;
     \fill[white]  (0.4,0) circle [radius = 0.23cm];
 \node at (-1.7,0) {$\scriptstyle{V_5}$};
 \node at (0.5,0) {$\scriptstyle{V_3}$};
 \node at (60:1.7) {$\scriptstyle {V_5}$};
 \node at (-60:1.7) {$\scriptstyle {V_5}$};
 \fill[black]  (A) circle [radius = 0.1cm];
\fill[black]  (B) circle [radius = 0.1cm];
 \fill[black]  (0,0) circle [radius = 0.1cm];
  \fill[black]  (C) circle [radius = 0.1cm];
}
&\hskip -0.5cm \ \ $\cG_{4c}\otimes \cG_{2}  $ & 
$ 8,6,120  $ & $- \tfrac{4\pi^2}{3 \vep^2} \big (1 - 3 (  \ln 2 + \tfrac12 ) \vep \big )
 - \tfrac{28\, \zeta_3}{\vep}  $  & $ \tfrac{\pi^2}{4} (2\ln 2 + 1) - \tfrac72 \zeta_3$  \\
\noalign {\vskip 4pt} 
\hskip -0.2cm $6u,g2b$ & \hskip -0.5cm
\tikz[baseline=(vert_cent.base),scale=0.55]{
  \node (vert_cent) {\hspace{-13pt}$\phantom{-}$};
  \draw (0,0) circle [radius=1.2];
 \node at  (50:1.2)  (A) {};
  \node at  (-50:1.2)  (B) {};
  \node at  (0,0)  (C) {};
    \draw [bend left = 40] (C.center) to (A.center);
    \draw [bend right = 40](C.center) to (B.center);
    \draw [bend right = 16] (A.center) to (B.center);
  \draw (-1.2,0)-- (C.center) ;
   \draw [bend left = 65] (-1.2,0) to  (C.center) ;
    \draw [bend right = 56] (-1.2,0) to (C.center) ;
 \node at (-1.7,0) {$\scriptstyle{V_5}$};
 \node at (0.35,0) {$\scriptstyle{V_5}$};
 \node at (50:1.7) {$\scriptstyle {V_4}$};
 \node at (-50:1.7) {$\scriptstyle {V_4}$};
 \fill[black]  (A) circle [radius = 0.1cm];
\fill[black]  (B) circle [radius = 0.1cm];
 \fill[black]  (-1.2,0) circle [radius = 0.1cm];
  \fill[black]  (C) circle [radius = 0.1cm];
}
&\hskip -0.2cm \ \ $\cG_{2}\otimes \cG_{4c}  $ & 
$ 12,4,180  $ & $ - \tfrac{2\pi^2}{3 \vep^2} \big (1 -  ( 2 \ln 2 + 5 ) \vep \big )
 - \tfrac{28\, \zeta_3}{\vep} $ & $ \tfrac{\pi^2} {12}( 2\ln 2 +5) - \tfrac{7}{2} \zeta_3$ \\
\end{tabular} \, .
}
\end{align}

For primitive graphs when $\Delta(\cG) =\emptyset$
\begin{align}
\label{6loopP}
\text{
\begin{tabular}{  c   c   c   c  c  c}
$\cG$ label &  $\cG ~$ &  $ S_\cG,E_\cG,P_\cG$  &$~~~Z_\cG~$  & $~~~c_{\cG}~~$\\
\noalign {\vskip 2pt}
\hline
\noalign {\vskip 4pt}
\noalign {\vskip 4pt} 
\hskip -0cm $6v,c1a$ & \hskip -0.2cm
\tikz[baseline=(vert_cent.base),scale=0.45]{
  \node (vert_cent) {\hspace{-13pt}$\phantom{-}$};
  \draw (0,0) circle [radius=1.2];
  \node at  (50:1.2)  (A) {};
   \node at  (-50:1.2)  (B) {};
    \node at  (135:1.2)  (C) {};
   \node at  (-135:1.2)  (D) {};
   \draw [bend right = 28] (C.center)to (A.center);
     \draw [bend left = 28] (D.center) to (B.center);
     \draw [bend right =55] (A.center) to (B.center);
       \draw [bend right =20] (A.center) to (B.center);
         \draw [bend left = 15] (A.center) to (B.center);
 \node at (138:1.8) {$\scriptstyle{V_3}$};
 \node at (-135:1.8) {$\scriptstyle{V_3}$};
 \node at (50:1.7) {$\scriptstyle {\lambda}$};
 \node at (-50:1.7) {$\scriptstyle {\lambda}$};
 \fill[black]  (A) circle [radius = 0.1cm];
\fill[black]  (B) circle [radius = 0.1cm];
 \fill[black]  (C) circle [radius = 0.1cm];
  \fill[black]  (D) circle [radius = 0.1cm];
}
& $ 96, 2,20$  &  $- \tfrac{4\pi^2}{3\vep}$ & $-\tfrac{\pi^2}{24} $
 \\
 \noalign {\vskip 4pt} 
 \hskip -0cm $6w,d3$ & \hskip -0.2cm
 \tikz[baseline=(vert_cent.base),scale=0.45]{
  \node (vert_cent) {\hspace{-13pt}$\phantom{-}$};
  \draw (0,0) circle [radius=1.2];
  \node at  (45:1.2)  (A) {};
   \node at  (-45:1.2)  (B) {};
    \node at  (138:1.2)  (C) {};
   \node at  (-138:1.2)  (D) {};
   \draw [bend right = 55] (D.center)to (C.center);
     \draw [bend left = 28] (D.center) to (B.center);
     \draw [bend right =55] (A.center) to (B.center);
       \draw [bend right =20] (A.center) to (B.center);
         \draw [bend left = 15] (A.center) to (B.center);
 \node at (140:1.8) {$\scriptstyle{V_3}$};
 \node at (-140:1.8) {$\scriptstyle{V_4}$};
 \node at (50:1.7) {$\scriptstyle {V_5}$};
 \node at (-48:1.65) {$\scriptstyle {\lambda}$};
 \fill[black]  (A) circle [radius = 0.1cm];
\fill[black]  (B) circle [radius = 0.1cm];
 \fill[black]  (C) circle [radius = 0.1cm];
  \fill[black]  (D) circle [radius = 0.1cm];
}
&
$ 96, 1,60$  & ~ $-\tfrac{4\pi^2}{3\vep}$ & $-\tfrac{\pi^2}{12} $ \\
 \noalign {\vskip 4pt} 
\hskip -0cm $6x,f1$ & \hskip -0.2cm
\tikz[baseline=(vert_cent.base),scale=0.45]{
  \node (vert_cent) {\hspace{-13pt}$\phantom{-}$};
  \draw (0,0) circle [radius=1.2];
   \node at  (60:1.2)  (A) {};
   \node at  (-60:1.2)  (B) {};
    \node at  (180:1.2)  (C) {};
   \node at  (0,0)  (D) {};
 \fill[white]  (0.23,0.2) circle [radius = 0.17cm];  
         \fill[white]  (0.23,-0.25) circle [radius = 0.2cm];  
   \draw [bend right = 45]  (A.center) to (D.center);
    \draw [bend left = 45]  (A.center) to (D.center);
     \draw [bend right = 45]  (B.center) to (D.center);
    \draw [bend left = 45]  (B.center) to (D.center);
 \draw [bend right = 45]  (C.center) to (D.center);
    \draw [bend left = 45]  (C.center) to (D.center);
   \node at (-1.7,0) {$\scriptstyle{V_4}$};
 \node at (60:1.7) {$\scriptstyle {V_4}$};
  \node at (-60:1.7) {$\scriptstyle {V_4}$};
   \fill[white]  (0.17,0.15) circle [radius = 0.1cm];
    \fill[white]  (0.2,-0.15) circle [radius = 0.1cm];
   \node at (0.23,0) {$\scriptstyle {\lambda}$};
 \fill[black]  (A.center) circle [radius = 0.1cm];
\fill[black]  (B.center) circle [radius = 0.1cm];
 \fill[black]  (C.center) circle [radius = 0.1cm];
 \fill[black]  (D.center) circle [radius = 0.1cm];
}
& $ 8,6,90 $  &  ~ $\tfrac{2\pi^4}{3\vep}$ & $\tfrac{\pi^4}{12} $ 
\\
 \noalign {\vskip 4pt} 
\hskip -0cm $6y,g0b$ & \hskip -0.2cm
\tikz[baseline=(vert_cent.base),scale=0.45]{
  \node (vert_cent) {\hspace{-13pt}$\phantom{-}$};
  \draw (0,0) circle [radius=1.2];
  \node at  (90:1.2)  (A) {};
   \node at  (-90:1.2)  (B) {};
   \draw [bend right = 50]  (B.center) to (-1.2,0);
   \draw [bend right = 50]  (A.center) to (1.2,0);
  \draw [bend left = 50]  (A.center) to (-1.2,0);
     \draw [bend left = 45] (B.center) to (1.2,0);
     \draw (A.center) -- (B.center) ;     
  \node at (-1.8,0) {$\scriptstyle{V_4}$};
 \node at (90:1.7) {$\scriptstyle {V_5}$};
  \node at (-90:1.7) {$\scriptstyle {V_5}$};
   \node at (1.7,0) {$\scriptstyle {V_4}$};
 \fill[black]  (90:1.2) circle [radius = 0.1cm];
\fill[black]  (-90:1.2) circle [radius = 0.1cm];
 \fill[black]  (1.2,0) circle [radius = 0.1cm];
 \fill[black]  (-1.2,0) circle [radius = 0.1cm];
}
&
$ 16,4,180$  & ~ $\tfrac{2\pi^4}{3\vep}$ & $\tfrac{\pi^4}{16} $ \\
 \noalign {\vskip 4pt} 
\hskip -0cm $6z,g1b$ & \hskip -0.2cm
\tikz[baseline=(vert_cent.base),scale=0.45]{
  \node (vert_cent) {\hspace{-13pt}$\phantom{-}$};
  \draw (0,0) circle [radius=1.2];
   \node at  (45:1.2)  (A) {};
   \node at  (-45:1.2)  (B) {};
    \node at  (138:1.2)  (C) {};
   \node at  (-138:1.2)  (D) {};
    \draw [bend left = 40] (A.center) to (C.center);
    \draw [bend left = 40] (B.center) to (A.center);
      \draw [bend left = 40] (D.center) to (B.center);
  \draw (C.center)-- (B.center) ;
  \fill[white]  (-0.06,0) circle [radius = 0.12cm];
   \draw (A.center)-- (D.center) ;
 \node at (138:1.8) {$\scriptstyle{V_4}$};
 \node at (45:1.7) {$\scriptstyle{V_5}$};
 \node at (-43:1.8) {$\scriptstyle {V_5}$};
 \node at (-138:1.7) {$\scriptstyle {V_4}$};
 \fill[black]  (A) circle [radius = 0.1cm];
\fill[black]  (B) circle [radius = 0.1cm];
 \fill[black]  (C) circle [radius = 0.1cm];
  \fill[black]  (D) circle [radius = 0.1cm];
}
&
$ 8,2,180$  & $\tfrac{4C}{3\vep} $& $ \tfrac12C$ \\
\end{tabular} \, .
}
\end{align}
Here $C$ is a new irrational constant appearing for this primitive 6 loop graph
\be
C = 32\,  \beta_4 + \tfrac43 \,\pi^2 \beta_2 \, ,   \qquad \beta_k = {\ts {\sum_{n \ge 0}}} \, \tfrac{(-1)^n}{(2n+1)^k}\, , 
\label{Cres}
\ee
with $\beta_k$ the Dirichlet $\beta$-function, $\beta_2=G$ the Catalan number.

There remain three  1VR graphs which generate poles in $\vep$ 
\begin{align}
\label{6loopR}
\hskip -1cm
\text{
\begin{tabular}{  c   c   c   c  c  c}
$\cG$ label &  $\cG ~$ & $~\Delta(\cG) ~$ & $ S_\cG,E_\cG,P_\cG$  &$~~~Z_\cG~$  & \\
\noalign {\vskip 2pt}
\hline
 \noalign {\vskip 8pt} 
 \hskip -0cm $6Ra,c0a$ & \hskip -0.4cm
\tikz[baseline=(vert_cent.base),scale=0.45]{
  \node (vert_cent) {\hspace{-13pt}$\phantom{-}$};
  \draw (0,0) circle [radius=1];
  \draw (2,0) circle [radius=1];
   \draw (4 ,0) circle [radius=1];
  \draw (1,0) -- (3,0);
  \draw (-1,0) -- (1,0);
  \draw (3,0) -- (5,0);
    \node at (5.5,0) {$\scriptstyle {V_3}$};
  \node at (1,-0.8) {$\scriptstyle {\lambda}$};
    \node at (3,-0.8) {$\scriptstyle {\lambda}$};
   \node at (-1.5,0) {$\scriptstyle {V_3}$};
 \fill[black]  (-1,0) circle [radius = 0.1cm];
\fill[black]  (1,0) circle [radius = 0.1cm];
\fill[black]  (3,0) circle [radius = 0.1cm];
\fill[black]  (5,0) circle [radius = 0.1cm];
} 
& \hskip - 0.2cm  $ 3\, \cG_2 \otimes \cG_{4R} +  2\, \cG_{4R} \otimes \cG_2 + \cG_2\cG_2 \otimes \cG_2$  
& $ 216, 2,20$  &  $ \tfrac{8}{\vep^3}$ &
 \\
 \noalign {\vskip 4pt} 
 \hskip -0cm $6Rb,d2$ & \hskip -0.2cm
\tikz [baseline=(vert_cent.base),scale=0.45]{
  \node (vert_cent) {\hspace{-13pt}$\phantom{-}$};
 \draw (-2,0) circle [radius=1];
   \draw (-3,0) -- (-1,0);
  \node at (-1,-0.8) {$\scriptstyle {\lambda}$};
 \fill[black]  (-1,0) circle [radius = 0.1cm];
\fill[black]  (-3,0) circle [radius = 0.1cm];
  \draw (0,0) circle [radius=1];
  \node at  (60:1)  (A) {};
   \node at  (-60:1)  (B) {};
   \draw (A.center) -- (B.center);
   \draw [bend right = 30] (-1,0) to (A.center);
     \draw [bend right =50] (A.center) to (B.center);
 \node at (-3.5,0) {$\scriptstyle{V_3}$};
 \node at (60:1.5) {$\scriptstyle {V_5}$};
  \node at (-60:1.55) {$\scriptstyle {V_4}$};
 \fill[black]  (60:1) circle [radius = 0.1cm];
\fill[black]  (-60:1) circle [radius = 0.1cm];
 \fill[black]  (-1,0) circle [radius = 0.1cm];
 }
&   \hskip - 0.5cm  $  \cG_2 \otimes \cG_{4R} +   \cG_2 \otimes \cG_{4b} + \cG_{4b} \otimes \cG_2 
+ \cG_2\cG_2 \otimes \cG_2$  &
$ 72,1,60$  & $\tfrac{4}{\vep^3}(1-2\vep)$ &\\
 \noalign {\vskip 4pt} 
\hskip -0cm $6Rc,e2$ & \hskip -0.2cm
\tikz[baseline=(vert_cent.base),scale=0.45]{
  \node (vert_cent) {\hspace{-13pt}$\phantom{-}$};
 \draw (-2,0) circle [radius=1];
   \draw (-3,0) -- (-1,0);
  \node at (-1,-1) {$\scriptstyle {V_5}$};
 \fill[black]  (-1,0) circle [radius = 0.1cm];
\fill[black]  (-3,0) circle [radius = 0.1cm];
  \draw (0,0) circle [radius=1];
  \node at  (70:1)  (A) {};
   \node at  (-70:1)  (B) {};
   \draw[bend left  = 38]  (A.center) to (B.center);
   \draw [bend right = 38] (A.center) to (B.center);
     \draw [bend right =0] (A.center) to (B.center);
 \node at (-3.5,0) {$\scriptstyle{V_3}$};
 \node at (68:1.5) {$\scriptstyle {V_5}$};
  \node at (-68:1.55) {$\scriptstyle {V_5}$};
 \fill[black]  (70:1) circle [radius = 0.1cm];
\fill[black]  (-70:1) circle [radius = 0.1cm];
 \fill[black]  (-1,0) circle [radius = 0.1cm];
 }
& $  \cG_2 \otimes \cG_{4a} +   \cG_{4a} \otimes \cG_{2} $   &
$ 144,2,120$  & ~ $\tfrac{4}{\vep^2}$ & \\
 \end{tabular} \, .
}
\end{align}
Results for $Z_\cG$ for the  individual six loop graphs
$\cG_{6e}, \, \cG_{6g}, \, \cG_{6i}$ (where the overall sign in the original version
should be changed), $\cG_{6k}, \, \cG_{6l}, \, \cG_{6q}, \, \cG_{6s}$ and $\cG_{6x}$ were  recently obtained in \cite{JackJones} and the results matched to semiclassical results
for large charge in three dimensional theories with a $U(1)$ symmetry \cite{Badel}.
Their results also encompass $\cG_{6j}$. For vertex graphs, where the
superficial divergence is logarithmic,  the incoming momenta on all but two external vertices can be set to zero, 
though it is necessary to be careful about potential IR divergences. 
A consistency check is that the same result is obtained irrespective of which
pair of vertices is chosen. Methods for evaluating the Feynman integrals with massless propagators
associated with the six loop graphs, involving integration by parts and
IR rearrangement, were described in \cite{JackJones} and these techniques can be extended to many of the
remaining integrals necessary for the full 6 loop calculations required here. The graphs
$\cG_{6i},\, \cG_{6n},\, \cG_{6t}, \, \cG_{6u}, \, \cG_{6x}, \, \cG_{6z}$ are tetrahedral in that all vertices
are linked. The $\vep$-poles arising in Feynman integrals from $\cG_{6i}, \, \cG_{6t}, \, \cG_{6x}$   can be 
determined by using
the  integration by parts  techniques on a basic two loop graph as described in \cite{JackJones}.

From \eqref{6loop3} 
\be
2\gamma_{\phi,ij}{\!\!}^{(6)} = -\tfrac{4}{81} \, \lambda_{iklmnp}\lambda_{jklrst}\lambda_{mnprst}\,,
\ee
but we refrain from writing the lengthy expression for ${\tilde \beta}_V(\phi)^{(6)}$ since this is easily 
recovered from the above tables \eqref{6loop}, \eqref{6loop2}, \eqref{6loop3}, \eqref{6loopP}.

Imposing $O(N)$ symmetry as in \eqref{VO6} and for $\cG_v, \, \cG_\phi $ of loop order $l=2,4,6$ 
\begin{align}
V({\cG_v})\big |_{V\to V_\lambda}={}& N_{\cG_v} \lambda^{\frac12 l_{\cG_v}+1 }\, V_1\, , \quad 
 \beta_V\big |_{\cG_v,V\to V_\lambda} = \beta_\lambda \big |_{\cG_v} V_1 \,  ,\quad
\beta_\lambda \big |_{\cG_v} =  c_{\cG_v}   N_{\cG_v}  \lambda^{\frac12 l_{\cG_v}+1 } \, , \nn \\
&  \Gamma_{ij} (\cG_\phi) \big |_{V\to V_\lambda}= N_{\cG_\phi} \lambda^{\frac12 l_{\cG_\phi} }\, \delta_{ij}\, , \quad \  \
 2 \, \gamma_\phi \big |_{\cG_\phi} = d_{\cG_\phi}  N_{\cG_\phi}  \lambda^{\frac12 l_{\cG_\phi} }\, ,
\end{align}
so that
\be
\beta_\lambda = -2\vep \, \lambda
+ \big ( {\ts \sum_{\cG_v}} \,  c_{\cG_v}   N_{\cG_v}  \lambda^{\frac12 l_{\cG_v} } 
+ 6\, \gamma_\phi \big )  \lambda \, , \qquad
2\, \gamma_\phi  =  {\ts \sum_{\cG_\phi}} \,  d_{\cG_\phi}   N_{\cG_\phi}  \lambda^{\frac12 l_{\cG_\phi} } \, .
\ee
The results here are easily extended to lower dimension operators by taking 
\be
V_{\lambda,\sigma,\tau,\nu,\kappa}(\phi) =  \lambda\, \tfrac{1}{48}  (\phi^2)^3 + \sigma \, \tfrac12 \phi^2+ 
\tau_i\, \, \tfrac12 \phi_i \phi^2+ \nu \, \tfrac18 (\phi^2)^2 +  \kappa_i \, \tfrac18 \phi_i (\phi^2)^2
\ee
so that for a $l$-loop $\cG_v$, $l$ even,
\begin{align}
V({\cG_v})\big |_{V\to V_{\lambda,\sigma,\tau,\nu,\kappa}} = {}&  \lambda^{\frac12 l_{\cG_v}} \big ( 
N_{\cG_v}\, \lambda \, \tfrac{1}{48}  (\phi^2)^3 
+ N_{\cG_v ,2} \, \sigma\, \tfrac12 \phi^2   
+  N_{\cG_v ,3} \, \tau_i \, \tfrac12 \phi_i  \phi^2 \nn \\
&\hskip 1cm + N_{\cG_v ,4}  \, \nu\,  \tfrac18 (\phi^2)^2   
+ N_{\cG_v ,5}  \, \kappa_i\,  \tfrac18 \phi_i(\phi^2)^2 \big ) 
+ {\rm O}(\sigma \nu,\nu^2,\tau^2,\kappa^2) \, . 
\end{align}
The associated anomalous dimensions are  then given by
\be
\gamma_n = {\ts \sum_{\cG_v}} \,  c_{\cG_v}   N_{\cG_v,n} \,  \lambda^{\frac12 l_{\cG_v} }
+n \, \gamma_\phi \, , \qquad n= 2,3,4,5\, .
\label{anom}
\ee

It remains to determine $N_{\cG_v,n}$ for each $n$ at $2,4,6$ loops.
In general
\be
N_{\cG_v,5} = N_{\cG_v}  \, ,
\label{gam5}
\ee
which follows just by differentiating $V({\cG_v})\big |_{V\to V_\lambda}$.
At lowest two loop order $N_{\cG,2}=0$ and  
\begin{align}
\label{2loopN}
\text{
\begin{tabular}{  c   c   c   c  c  c  }
$\cG$  & $~ N_{\cG} ~$&  $~ {\genfrac{}{}{0 pt}{2}{N_{\cG,3}}{N_{\cG,4}}} ~$ &\\
\noalign {\vskip 2pt}
\hline
\noalign {\vskip 4pt}
$2$ & $~ 12(3N+22)~$ & $~ {\genfrac{}{}{0 pt}{2}{6(N+4)}{24(N+4)}} ~$  &\\ 
\noalign {\vskip 2pt}
\end{tabular}
}
\end{align}

At four loops the results are
\begin{align}
\label{4loopN}
\hskip -0.5cm
\text{
\begin{tabular}{  c   c   c   c  c  c }
$\cG$  &  $~N_\cG ~$ &  $~N_{\cG,2} ~$ & $~ {\genfrac{}{}{0 pt}{2}{N_{\cG,3}}{N_{\cG,4}}} ~$  \\
\noalign {\vskip 2pt}
\hline
\noalign {\vskip 4pt}
$4\gamma $ & \hskip 0.5cm $15(N+2)(N +4)$  &  \\ 
 \noalign {\vskip 4pt} 
$4a$ &  $54(N+4)(3N+22) $ &   $ 30(N+2)(N+4) $ &$~ {\genfrac{}{}{0 pt}{2}{6(N+4)(7N+38)}{12(N+4)(7N+38)}} ~$    \\
\noalign {\vskip 4pt} 
$4b$ & $36(5N^2+78N+292) $ & &$~ {\genfrac{}{}{0 pt}{2}{15(N+4)(N+8)}{48(N+4)(2N+13)}} ~$  \\ 
\noalign {\vskip 4pt} 
$4c$ & \hskip 0.5cm  $6(N^3+34N^2+620N+2720)$ & & $~ {\genfrac{}{}{0 pt}{2}{0}{6(N+4)(N^2+18N+116)}} ~$   & \\
\noalign {\vskip 4pt} 
$4R$ & \hskip 0.5cm  $36(3N^2+ 30N+ 92)$ & &$~ {\genfrac{}{}{0 pt}{2}{9(N+4)^2}{72(N+4)^2} }~$ & \\
\end{tabular} 
}
\end{align}

At six loops the results corresponding to the tables above are
\begin{align}
\label{6loopN}
\hskip - 2cm
\text{
\begin{tabular}{  c   c   c   c  c  c }
$\cG $ &  $~N_\cG ~$ & $N_{\cG,2}$ &  $~ {\genfrac{}{}{0 pt}{2}{N_{\cG,3}}{N_{\cG,4}}} ~$ &\\
\noalign {\vskip 2pt}
\hline
 \noalign {\vskip 4pt} 
$6a,a1$ &  $9(N+4)^2(N+14)^2$ & $ 3 (N+2)(N+4)^2(N+14) $ &  
$~ {\genfrac{}{}{0 pt}{2}{3(N+4)^2(N+8)(N+14)}{9(N+4)^2(N+6)(N+14)} }~$  & \\
\noalign {\vskip 4pt} 
$6b,a2$ &  $9(N+4)(N+8)(N^2+16N+108) $ & $ 3(N+2) (N+4)^2 (N+14)$ &   
$~ {\genfrac{}{}{0 pt}{2}{3(N+4)(N^3+22N^2+164N+488)}{3(N+4)(3N^3+64N^2+420N+1088)} }~$ & \\ 
\noalign {\vskip 4pt} 
$6c,b1$ & $ 54(N+4)(7N^2+84N+284)$  &   $18(N+2)(N+4)(3N+22) $ &  
$~ {\genfrac{}{}{0 pt}{2}{18(N+4)(5N^2+56 N + 164)}{36(N+4)(5N^2+56 N + 164)} }~$    &   \\
\noalign {\vskip 4pt} 
$6d,b2$ &  $ 18(N+4)(13N^2+228N+884) $ &   $18(N+2)(N+4)(3N+22) $ 
& $~ {\genfrac{}{}{0 pt}{2}{6(N+4)(11N^2+156N+508)}{12(N+4)(11N^2+156N+508)} }~$ & \\
\noalign{\vskip 4pt}
\end{tabular} \, .
}
\end{align}

\begin{align}
\label{6loop2N}
\text{
\begin{tabular}{  c   c   c   c  c  c}
$\cG$  &  $N_\cG $ & $~ {\genfrac{}{}{0 pt}{2}{N_{\cG,3}}{N_{\cG,4}}} ~$ &\\
\noalign {\vskip 2pt}
\hline
\noalign {\vskip 4pt}
\hskip -0.2cm $6e,d5$ &\hskip  0.2cm $108(5N^3+90N^2 +556N+1224) $ & 
$~ {\genfrac{}{}{0 pt}{2}{9(N+4)(5N^2 +52N+168)}{ 144(N+4)(2N^2+19N+54) }}~$   &
\\ 
\noalign {\vskip 4pt} 
\hskip -0.2cm $6f,g1a$ & \hskip 0.2cm $108(11N^3 +218 N^2 +1564N+3832) $  &
$~ {\genfrac{}{}{0 pt}{2}{0 }{360 (N+4)^2(N+8) } }~$   &\\
\noalign {\vskip 4pt}
\hskip -0.2cm $6g,d8$ &\hskip 0.2cm $36(11 N^3 +218 N^2 + 1564 N +3832) $ &
$~ {\genfrac{}{}{0 pt}{2}{3(N+4)(11N^2+144N+520)}{ 48 (N+4) (4N^2 +51 N +170)} }~$  & \\ 
\noalign {\vskip 4pt}
\hskip -0.2cm $6h,e3$ &\hskip  0.2cm  $72(7N^3 +190 N^2 +1532N+3896) $ &
$~ {\genfrac{}{}{0 pt}{2}{6(N+4)((7N^2+132N+536)}{24(N+4)(7N^2 +132 N + 536) } }~$ &  \\ 
\noalign {\vskip 4pt} 
\hskip -0.2cm $6i,d1$ & \hskip  0.2cm   $ 36(7N^3+190 N^2+1532N +3896) $ &
$~ {\genfrac{}{}{0 pt}{2}{3(N+4)(7N^2+132N+536)}{ 144(N+4) (N^2+16N +58) } }~$  &  \\
 \noalign {\vskip 4pt}  
\hskip -0.2cm $6j,g0a$ & \hskip 0.2cm  $108(7N^3+190 N^2 +1532 N +3896)$ &
$~ {\genfrac{}{}{0 pt}{2}{0}{360(N+4)^2 (N+8)} }~$  & \\
\noalign {\vskip 4pt} 
\hskip -0.2cm $6k,d6$ & \hskip  0.2cm   $ 36(11 N^3 +218 N^2 + 1564 N +3832) $ &
$~ {\genfrac{}{}{0 pt}{2}{3(N+4)(11N^2+144N+520)}{48(N+4)(5N^2+54N+166) } }~$  & \\
\noalign {\vskip 4pt} 
\hskip -0.2cm $6l,c0b$ & \hskip 0.2cm  $36(5N^3+90N^2+556 N +1224) $  &
$~ {\genfrac{}{}{0 pt}{2}{30(N+4)^2(N+8)}{ 120(N+4)^2 (N+8)  } }~$   &\\
\end{tabular} \, .
}
\end{align}

\begin{align}
\label{6loop3N}
\text{
\begin{tabular}{  c   c   c   c  c  c}
$\cG$ label &  $~N_\cG ~$ & $~ {\genfrac{}{}{0 pt}{2}{N_{\cG,3}}{N_{\cG,4}}} ~$ & \\
\noalign {\vskip 2pt}
\hline
\noalign {\vskip 4pt}
$6\gamma $ & \hskip 0.2cm $9(N+2)(N+4)(3N+22)$
&   \\ 
\noalign {\vskip 4pt}
\hskip -0.2cm $6m,c2$ & \hskip 0.2cm  $180(N+2)(N+4)(3N+22)$
& $~ {\genfrac{}{}{0 pt}{2}{90(N+2)(N+4)^2}{360(N+2)(N+4)^2} }~$ &\\
\noalign {\vskip 4pt}
\hskip -0.2cm $6n,c1b$ & \hskip 0.2cm  $36(N+4)(7N^2+84N+284)$
&$~ {\genfrac{}{}{0 pt}{2}{6(N+4)^2(7N+38)}{24(N+4)^2(7N+38) } }~$  & \\
\noalign {\vskip 4pt}
\hskip -0.2cm $6o,d4$ & \hskip 0.2cm  $ 36(N+4)(13N^2+228N+884)$
& $~ {\genfrac{}{}{0 pt}{2}{3(N+4)^2(13N+122)}{ 240(N+4)^2 (N+8)  } }~$ & \\
\noalign {\vskip 4pt}
\hskip -0.2cm $6p,g3$ & \hskip 0.2cm  $108(N+4)(13 N^2 +228N+884)$
& $~ {\genfrac{}{}{0 pt}{2}{0}{72(N+4)^2(7N+38)} } ~$ &  \\
\noalign {\vskip 4pt}
\hskip -0.2cm $6q,f2$ & \hskip 0.2cm  $18(N^4+34N^3  + 584 N^2 +4448 N+11808 )$
&$~ {\genfrac{}{}{0 pt}{2}{0}{ 6(N+4)(3N^3 +62N^2  +504N +1456) } }~$ &\\
\noalign {\vskip 4pt}
\hskip -0.2cm $6r,g2a$ & \hskip 0.2cm $36(N^4 +34 N^3+584N^2+4448N+11808)$
& $~ {\genfrac{}{}{0 pt}{2}{0}{24(N+4)(N^3+18N^2+152N+504) } }~$ & \\
\noalign {\vskip 4pt} 
\hskip -0.2cm $6s,d7$ & \hskip 0.2cm  $12(N^4+34N^3+584N^2+4448N+11808) $
& $~ {\genfrac{}{}{0 pt}{2}{(N+4)(N^3+28N^2+404N+1592)}{  8(N+4)(N^3+25N^2 +296N+1028)} }~$  &\\
\noalign {\vskip 4pt} 
\hskip -0.2cm $6t,e1$ & \hskip 0.2cm  $ 24(11N^3+428N^2+4228 N+12208)$
& $~ {\genfrac{}{}{0 pt}{2}{18(N+4)(N^2+36N+188)}{ 72(N+4) (N^2+36N+188)  } }~$  &\\
\noalign {\vskip 4pt} 
\hskip -0.2cm $6u,g2b$ & \hskip 0.2cm   $ 36 (11N^3+428N^2+4228 N+12208)$
&$~ {\genfrac{}{}{0 pt}{2}{0}{72(N+4)(N+8)(3N+22) } }~$  &\\
\end{tabular} \, .
}
\end{align}

\begin{align}
\label{6loopPN}  \hskip -1cm
\text{
\begin{tabular}{  c   c   c   c  }
$\cG$  &  $N_\cG ~$ & $~ {\genfrac{}{}{0 pt}{2}{N_{\cG,3}}{N_{\cG,4}}} ~$& \\
\noalign {\vskip 2pt}
\hline
\noalign {\vskip 4pt}
\hskip -0cm $6v,c1a$ & \hskip 0.2cm $36(N+4)(N^3+14N^2+88N+272)$
& $~ {\genfrac{}{}{0 pt}{2}{6(N+4)^2(N^2+8N+36)}{24(N+4)^2(N^2+8N+ 36)}}~$  & \\
 \noalign {\vskip 4pt} 
 \hskip -0cm $6w,d3$ & \hskip 0.2cm $36(N+4)(N+8)(N^2+16N+108)$
& $~ {\genfrac{}{}{0 pt}{2}{3(N+4)^2(N^2+14N+120)}{24(N+4)^2(N^2+ 11 N+ 78)} }~$    & \\
 \noalign {\vskip 4pt} 
\hskip -0cm $6x,f1$ & \hskip 0.2cm  $ 6(N^4+ 64 N^3 +1352N^2 + 12248 N +36960)$
& $~ {\genfrac{}{}{0 pt}{2}{0}{ 6(N+4)(N^3+40N^2+440N+1544)} }~$  & \\
 \noalign {\vskip 4pt} 
\hskip -0cm $6y,g0b$ & \hskip 0.2cm  $12(N^4+64N^3  + 1352 N^2 + 12248 N+ 36960 )$
& $~ {\genfrac{}{}{0 pt}{2}{0}{8(N+4)(N^3+40N^2+440N+1544)} }~$ & \\
 \noalign {\vskip 4pt} 
\hskip -0cm $6z,g1b$ & \hskip 0.2cm $12(31 N^3 + 1126 N^2 +11876 N +37592)$
 & $~ {\genfrac{}{}{0 pt}{2}{0}{24(N+4)(7N^2 +132N+536)} }~$  &  \\
\end{tabular} \, .
}
\end{align}

\begin{align}
\label{6loopRN}
\text{
\begin{tabular}{  c   c   c   c  c  c}
$\cG$  &  $N_\cG ~$ &   $~ {\genfrac{}{}{0 pt}{2}{N_{\cG,3}}{N_{\cG,4}}} ~$ & \\
\noalign {\vskip 2pt}
\hline
\noalign {\vskip 4pt}
\hskip -0cm $6Ra,c0a$ & \hskip 0.2cm $108(3N^3+42N^2+204N+376)$
&$~ {\genfrac{}{}{0 pt}{2}{54(N+4)^3}{216(N+4)^3} }~$ &  \\
 \noalign {\vskip 4pt} 
 \hskip -0cm $6Rb,d2$ & \hskip 0.2cm $108(5N^3+90N^2+556N+1224)$
& $~ {\genfrac{}{}{0 pt}{2}{45(N+4)^2(N+8)}{144(N+4)^2(2N+13)} }~$ &\\
 \noalign {\vskip 4pt} 
\hskip -0cm $6Rc,e2$ & \hskip 0.2cm  $ 216(N+4)(7N^2 + 84 N + 284)$
& $~ {\genfrac{}{}{0 pt}{2}{18(N+4)^2(7N+38)}{72(N+4)^2(7N+38)} }~$ & \\
\end{tabular} \, .
}
\end{align}
In these tables  with our normalisation $N_{\cG_v}\big |_{N=1} = 15^{\frac12 l_{\cG_v} } P_{\cG_v}$.

 For the anomalous dimensions to ${\rm O}(\lambda^3)$ from \eqref{anom}
\begin{align}
2\gamma_{\phi} ={}&  \tfrac{1}{3} \, (N+2)(N+4) \lambda^2 -\tfrac{4}{9} \,(N+2)(N+4)(3N+22) \lambda^3\,  ,\nn \\
\gamma_{2} ={}&  \tfrac{16}{3} \, (N+2)(N+4) \lambda^2 -\tfrac{400}{9} \,(N+2)(N+4)(3N+22) \lambda^3 \nn \\
\noalign{\vskip -1pt}
&{} - \pi^2 \ \tfrac12 \,(N+2)(N+4)^2(N+14) \lambda^3\,  ,\nn \\
\gamma_{3} ={}& 2(N+4) \lambda -  \tfrac12(N+4)  (25N+ 242)  \lambda^2
\nn \\
&{}+ (N+4) \big ( \tfrac43 (407N^2 +  7327 N + 30666) \nn \\
&\hskip 1.8cm{} - \pi^2 \ \tfrac{1}{12}( 3 N^3  + 300 N^2 + 3208 N + 8864)  \nn \\
&\hskip 1.8cm{} + \pi^2 \ln 2 \ \tfrac12( N^3  + 46 N^2 + 1052 N + 4976)  \nn \\
&\hskip 1.8cm{} - \zeta_3 \ 63 (N^2 + 36 N +188)  \big ) \lambda^3\, , 
\nn \\
\gamma_{4} ={}& 8(N+4) \lambda - (N+4) \big ( \tfrac43 (85N+566) + \pi^2\, \tfrac12 ( N^2 + 18N + 116) \big ) \lambda^2
\nn \\
&{}+ (N+4) \big ( \tfrac89 (4113N^2 + 62522N +233440) \nn \\
&\hskip 1.8cm{} + \pi^2 \ \tfrac16( 129 N^3  + 2464 N^2 + 21772 N + 73360)  \nn \\
&\hskip 1.8cm{} - \pi^2 \ln 2 \ 4( N^3 -3N^2 - 430 N- 2268)  \nn \\
&\hskip 1.8cm{} - \zeta_3 \, 504 (N+14)(2N+13) + \pi^4\,  (N^3 + 40 N^2+ 440 N+ 1544) \nn \\
&\hskip 1.8cm{}+  C\, 12  ( 7 N^2 + 132 N+ 536 ) \big ) \lambda^3\, .
\label{gam6}
\end{align}
To ${\rm O}(\lambda^2)$ these results coincide with those given in \cite{Seeking}.
For the $\beta$-function to ${\rm O}(\lambda^4)$
\begin{align}
\beta_\lambda =&{} - 2\vep \, \lambda + 4(3N+22) \lambda^2  \nn\\
&{} - 4 (53N^2 + 858 N + 3304) \lambda^3
- \pi^2 \ \tfrac12 (N^3 + 34 N^2 +620 N +2720) \lambda^3 \nn \\
&{}+ 4 ( 1857 N^3 + 45976 N^2 +367716N +  950576 ) \lambda^4  \nn \\
&{}+ \pi^2 \, \tfrac12(51 N^4 + 1618 N^3 + 32804 N^2 +288968 N + 837184) \lambda^4 \nn \\
&{} - \pi^2 \ln 2 \ 3(N^4 - 32N^3 -1984N^2 -20920N-61440) \lambda^4 \nn \\
& - \zeta_3 \, 210(11N^3 + 428 N^2 +4228 N + 12208) \lambda^4 \nn \\
&{}+ \pi^4\,  \tfrac54(N^4+ 64N^3 + 1352N^2+12248N+ 36960) \lambda^4  \nn \\
&{}+ C\, 6( 31N^3 + 1126N^2 +11876N+ 37592 )\lambda^4\, .
\label{bet6}
\end{align}  
The results here agree almost exactly with Hager \cite{Hager}.\footnote{The correspondence can be made
by taking $\lambda = {\bar w}_R/30 $ with $30\beta_\lambda(\lambda) = \beta({\bar w}_R)_{\rm{Hager}}$.}  
The terms involving $\pi^2$ and $\pi^2 \ln 2$ at order $\lambda^4$ arising from 
applying  his results for the $\beta_\lambda$ would have $N$ polynomials 
which are $36N^4 + 1607 N^3 + \dots$
and $- 6(N^4 + N^3 + \dots)$ respectively. Similar issues arises for $\gamma_4$. As far as the $\pi^2 \ln 2$ 
terms in  $\beta_\lambda$ and $\gamma_4$ are concerned the difference can be isolated to the
contribution corresponding to the graph $\cG_{6s}$. Hager associates the irrational coefficient $C$, involving
a polynomial $31N^3 + \dots$, to his graph  $g2b$, our $6u$, but it must in fact correspond to his $g1b$, our 
$6z$.\footnote{Hager in his result for the $\pi^2 \ln 2$ in $Z_4$ has a polynomial ending in $-3508$, in his expression
for $\gamma_u$, our $\gamma_4$, the corresponding term ends in $-3805$.}

As a  consequence of \eqref{gam5} to all orders in the perturbative expansion
\be
\gamma_5 = 2 \vep + \beta_\lambda/ \lambda  - \gamma_\phi \, .
\label{D5}
\ee
This result is discussed further in appendix \ref{eom}.

\section{Large $N$}

As commented in the introduction, there have been many discussions of the $O(N)$ $\phi^6$ theory 
at large $N$, primarily in the context of discussing the effective potential.
If we consider the rescaling
\be
{\tilde \lambda} = N^2 \lambda \, , \qquad \beta \raisebox{-1.5 pt}{$\scriptstyle {\tilde \lambda}$}
= N^2 \, \beta_\lambda \big |_{\lambda  \to {\tilde \lambda}/N^2  } \, , 
\label{lN1}
\ee
then at large $N$ from \eqref{gam6} and \eqref{bet6}
\begin{align}
2 \, \gamma_\phi ={}&\frac{\tlam^2}{3\, N^2} + {\rm O} \Big ( \frac{1}{N^3} \Big ) \, , \nn \\
\beta \raisebox{-1.5 pt}{$\scriptstyle {\tilde \lambda}$}
 = {}& - 2\vep \, \tlam + \frac{1}{N} \Big (  12 \, \tlam^2 - \tfrac12 \pi^2 \tlam^3 \Big ) \nn \\
&{} + \frac{1}{N^2} \Big ( 88\, \tlam^2 - \big (212+17\pi^2\big ) \tlam^3 
+ \pi^2 \big ( \tfrac{51}{2} - 3 \ln2 + \tfrac54 \pi^2 \big )\tlam^4 \Big)
+ {\rm O} \Big ( \frac{1}{N^3} \Big ) \, .
\label{largeN}
\end{align}
There is then an apparent UV fixed point arising just from the four loop contributions at 
${\tilde \lambda} = 24/\pi^2 + {\rm O}(N\vep,1/N)$ as was discussed in \cite{Pisarski}.
To the next to leading order in an expansion in $1/N$, as shown in \eqref{largeN},
a corresponding result was given in  \cite{Sakhi} based on an equivalent large $N$ effective 
theory.\footnote{In terms  of the notation there $\tlam = \eta/(8\pi^2)$.}  The details of the calculation
are unclear to us but the expression quoted in \cite{Sakhi} agrees with that which would be 
obtained by taking the appropriate limit of the Hager result (and so disagrees with \eqref{largeN}
for the $\pi^2 \tlam^4$ and $\pi^2 \ln 2 \, \tlam^4$ terms).
 
For $O(N)$ symmetric theories, with the scalar fields $\phi$ forming a $N$-dimensional vector,
the dominant contributions at large $N$ are formed from bubble graphs. For $\phi^4$ theory in 
$4-\vep$ dimensions then the leading large $N$ contributions arise from strings of bubble graphs.
For the propagator graph with one string of bubbles the basic graphs are of the form
\tikz[baseline=(vert_cent.base),scale=0.5]{
  \node (vert_cent) {\hspace{-13pt}$\phantom{-}$};
  \draw (-0.5,0)--(0.1,0)
        (0.7,0) ++(0:0.6cm and 0.4cm) arc (0:28:0.6cm and 0.4cm) node (n1)
        {}
        (0.7,0) ++(28:0.6cm and 0.4cm) arc (30:72:0.6cm and 0.4cm) node (n2)
        {}
         (0.7,0) ++(72:0.6cm and 0.4cm) arc (72:108:0.6cm and 0.4cm) node (n3)
         {}
          (0.7,0) ++(108:0.6cm and 0.4cm) arc (108:152:0.6cm and 0.4cm) node (n4)
          {}
                    (0.7,0) ++(152:0.6cm and 0.4cm) arc (152:360:0.6cm and 0.4cm)       
               (1.3,0)--(1.9,0);
                \filldraw [white](n1.base) circle [radius =5.3pt];
                 \filldraw [white](n2.base) circle [radius =5.3pt];
                 \filldraw [white](n3.base) circle [radius =5.3pt];
                 \filldraw [white](n4.base) circle [radius =5.5pt];
                 \draw (n1.base) circle [radius = 5.3pt];
                   \draw (n2.base) circle [radius = 5.3pt];
                   \draw (n3.base) circle [radius = 5.3pt];
                   \draw (n4.base) circle [radius = 5.5pt];
                   },
as shown  for just 4 bubbles.  For an interaction  $\frac18 \lambda \, (\phi^2)^2$ then after taking 
 $\lambda\to (4\pi)^2 \lambda$ and defining here $\tlam = N \lambda$ then in a
$ \overline{M\!S}$ scheme the leading large $N$ contribution to the anomalous dimension 
 is just  $2\gamma_\phi =\frac{1}{N} \, \frac{\tlam^2\, \Gamma(2- \tlam)}{\Gamma(1-\frac12\tlam)^2\,
  \Gamma(3-\frac12\tlam) \, \Gamma(1+\frac12\tlam)} + {\rm O}(\frac{1}{N^2})$.  At the
  Heisenberg  fixed point  to leading order $\tlam_* = \vep$ and applying this gives the well known result 
  for the critical exponent $\eta$ at large $N$ as a function  of the dimension $d$.
  
  The $O(N)$ $\phi^6$ theory is rather different. In this case three bubbles join at a vertex and hence
   the dominant contributions correspond to a theory with cubic vertices. At lowest order the leading 
   large $N$ result is associated with 
   the 4 loop graph $\cG_{4c}$ in \eqref{4loop} which, as shown in \eqref{4loopN}, is proportional to $N^3$. 
  The six loop graph contributions are sub leading but at eight loops the following graphs, corresponding to
  two loop graphs for a $\varphi^3$ theory, are relevant
  \be
   \tikz[baseline=(vert_cent.base),scale=0.35]{
  \node (vert_cent) {\hspace{-13pt}$\phantom{-}$};
  \node at  (0,1.4)  (A) {};
   \node at  (1.8,-0.5)  (B) {};
    \node at  (-1.8,-0.5)  (C) {};
       \node at  (-1.1,-2.5)  (D) {};
          \node at  (1.1,-2.5)  (E) {};
      \draw [bend right = 48] (B.center) to (A.center);
   \draw [bend left = 40] (B.center) to (A.center);
     \draw [bend left = 48] (C.center) to (A.center);
     \draw [bend right = 40] (C.center) to (A.center);
         \draw [bend left = 48] (D.center) to (C.center);
          \draw [bend right = 48] (D.center) to (C.center);
           \draw [bend left = 48] (E.center) to (B.center);
          \draw [bend right = 48] (E.center) to (B.center);
           \draw [bend left = 65] (D.center) to (E.center);
          \draw [bend right = 65] (D.center) to (E.center);
            \draw [bend left = 20] (D.center) to (E.center);
          \draw [bend right = 20] (D.center) to (E.center);
 \node at (-2.5,-0.5) {$\scriptstyle{V_4}$};
 \node at (0,2.1) {$\scriptstyle {V_4}$};
  \node at (2.5,-0.5) {$\scriptstyle {V_4}$};
    \node at (1.2,-3) {$\scriptstyle{\lambda}$};
      \node at (-1.2,-3) {$\scriptstyle{\lambda}$};
 \fill[black]  (A) circle [radius = 0.1cm];
\fill[black]  (B) circle [radius = 0.1cm];
 \fill[black]  (C) circle [radius = 0.1cm];
  \fill[black]  (D) circle [radius = 0.1cm];
    \fill[black]  (E) circle [radius = 0.1cm];
     \node at  (-4.4,0)  {$\cG_{8A}$};
}
\qquad
\tikz[baseline=(vert_cent.base),scale=0.35]{
  \node (vert_cent) {\hspace{-13pt}$\phantom{-}$};
  \node at  (0,1.8)  (A) {};
   \node at  (1.6,-0.5)  (B) {};
    \node at  (-1.6,-0.5)  (C) {};
       \node at  (-1.1,-2.5)  (D) {};
          \node at  (1.1,-2.5)  (E) {};
      \draw [bend right = 48] (B.center) to (A.center);
   \draw [bend left = 40] (B.center) to (A.center);
     \draw [bend left = 48] (C.center) to (A.center);
     \draw [bend right = 40] (C.center) to (A.center);
      \draw [bend right = 30] (C.center) to (B.center);
    \draw [bend left =30] (C.center) to (B.center);
       \draw [bend left = 48] (D.center) to (C.center);
          \draw [bend right = 48] (D.center) to (C.center);
           \draw [bend left = 48] (E.center) to (B.center);
          \draw [bend right = 48] (E.center) to (B.center);
           \draw [bend left = 52] (D.center) to (E.center);
          \draw [bend right = 52] (D.center) to (E.center);
 \node at (-2,-2.6) {$\scriptstyle{V_4}$};
 \node at (0,2.3) {$\scriptstyle {V_4}$};
  \node at (2,-2.6) {$\scriptstyle {V_4}$};
   \node at (-2.25,-0.5) {$\scriptstyle{\lambda}$};
     \node at (2.25,-0.5) {$\scriptstyle{\lambda}$};
 \fill[black]  (A) circle [radius = 0.1cm];
\fill[black]  (B) circle [radius = 0.1cm];
 \fill[black]  (C) circle [radius = 0.1cm];
  \fill[black]  (D) circle [radius = 0.1cm];
    \fill[black]  (E) circle [radius = 0.1cm];
     \node at  (-4.2,0)  {$\cG_{8B}$};
}
\qquad
     \tikz[baseline=(vert_cent.base),scale=0.4]{
  \node (vert_cent) {\hspace{-13pt}$\phantom{-}$};
  \node at  (0,1.7)  (A) {};
   \node at  (1.8,-0.4)  (B) {};
    \node at  (-1.8,-0.4)  (C) {};
       \node at  (0,-0.4)  (D) {};
          \node at  (0,-2.5)  (E) {};
      \draw [bend right = 30] (B.center) to (A.center);
   \draw [bend left = 30] (B.center) to (A.center);
     \draw [bend left = 30] (C.center) to (A.center);
     \draw [bend right = 30] (C.center) to (A.center);
      \draw [bend right = 30] (D.center) to (A.center);
          \draw [bend left = 30] (D.center) to (A.center);
               \fill[white]  (0,-0.9) circle [radius = 0.2cm];
            \draw [bend left = 30] (E.center) to (B.center);
          \draw [bend right = 30] (E.center) to (B.center);
           \draw [bend left = 30] (E.center) to (C.center);
          \draw [bend right = 30] (E.center) to (C.center);
           \draw [bend left =30] (E.center) to (D.center);
       \draw [bend right = 30] (E.center) to (D.center);
 \node at (-2.4,-0.4) {$\scriptstyle{V_4}$};
 \node at (-0.6,-0.4) {$\scriptstyle {V_4}$};
  \node at (2.4,-0.4) {$\scriptstyle {V_4}$};
  \node at (0,-3) {$\scriptstyle{\lambda}$};
     \node at (0,2.2) {$\scriptstyle{\lambda}$};
 \fill[black]  (A) circle [radius = 0.1cm];
\fill[black]  (B) circle [radius = 0.1cm];
 \fill[black]  (C) circle [radius = 0.1cm];
  \fill[black]  (D) circle [radius = 0.1cm];
    \fill[black]  (E) circle [radius = 0.1cm];
     \node at  (-4.2,0)  {$\cG_{8C}$};
}
\ee
For these three graphs we have
\begin{align}
\label{6loopRT}
\hskip -1.5cm
\text{
\begin{tabular}{  c   c   c   c  c  c}
$\cG$  &   $~\Delta(\cG) ~$ & $ S_\cG,E_\cG,P_\cG$  &$~~~Z_\cG~$  & $c_\cG $ & $ ~~~~ N_\cG~~~~~$ \\
\noalign {\vskip 2pt}
\hline
 \noalign {\vskip 8pt} 
$8A$ &   $ \emptyset $  
& $ 384, 2,90$  &  $ - \tfrac{\pi^4}{\vep}$ & $ ~ \tfrac{\pi^4}{96} $ &
$\scriptstyle{18(N+4)(N^5+32N^4+ 440N^3 + 3144N^2+ 14112N+32896)}$
 \\
 \noalign {\vskip 4pt} 
 $8B$ &   $  \cG_{4c} \otimes \cG_{4c}$  &
$ 64,2,90$  & $- \tfrac{\pi^4}{2\vep^2}(1-4\vep)$ &$ - \tfrac{\pi^4}{8} $   &
$ \scriptstyle{6(N^6 + 38N^5 + 640 N^4 + 6600 N^3+49376 N^2+ 232960N+ 469760)}$\\
 \noalign {\vskip 4pt} 
$8C$ 
& $\emptyset $   &
$ 128,6,90$  & ~ $\tfrac{\pi^6}{2\vep}$ &$ -\tfrac{\pi^6}{192} $  &
$ \scriptstyle{6(N^6 + 30N^5 + 472 N^4 + 5160 N^3+42160 N^2+ 220800N+ 490752)}$\\
 \end{tabular} \,
}
\end{align}
The leading contributions are then $N^6 \lambda^5$ while the four loop graph $4c$  gives 
$N^3 \lambda^3$ and at six loops we have at most $N^4 \lambda^4$.  For graphs corresponding
to a $\varphi^3$ theory with 3 external lines then at $L$ loops there $V=2L+1$  vertices and $I=3L$ internal
lines. The corresponding $N$ dependence is then obtained by  a factor $N$ arising from a bubble
for each internal line so that overall  there is a contribution
$N^{3L} \lambda^{2L+1}$ or $(N \tlam)^V N^{-I-3}$.
For the  $\phi^6$ theory the associated number of loops is $l=4L$. If instead  of \eqref{lN1} we take
\be
{\hat \lambda} = \pi N^{\frac32} \lambda \, , \qquad \beta \raisebox{-1.5 pt}{$\scriptstyle {\hat \lambda}$}
=\pi  N^{\frac32} \, \beta_\lambda \big |_{\lambda  \to {\hat \lambda}/(\pi N^{\frac32})  } \, , 
\label{lN2}
\ee
then the leading large $N$ behaviour is given by
\be
\beta \raisebox{-1.5 pt}{$\scriptstyle {\hat \lambda}$} 
\approx - 2\vep\, {\hat \lambda} -\tfrac12  {\hat \lambda}{}^3 
-  \tfrac{1}{32}\big ( 18  + \pi^2  \big ) {\hat \lambda}{}^5 \, , \qquad \gamma_4 =2\vep + 
\beta \raisebox{-1.5 pt}{$\scriptstyle {\hat \lambda}$} /{\hat \lambda} \, .
\label{largeNP}
\ee
In this large $N$ limit there is a corresponding equivalent 
cubic theory which is apparently asymptotically free \cite{Pisarski3A} even when $\vep \to 0$.
In  \cite{Pisarski3A}  results are quoted which should correspond to \eqref{largeNP}.\footnote{In \cite{Pisarski3A}
a wave function renormalisation is introduced as would correspond to a counterterm proportional to
the kinetic term. From our perspective since this is non local no such counterterm is required, see
\cite{Behan} (especially in 4.1). This contribution in \cite {Pisarski3A} accounts for the difference between the
result quoted for the $\beta$-function  there and the result obtained in \eqref{largeNP}.}

\section{Results at the $O(N)$ Fixed Point}

The  expressions for the $\beta$-function and anomalous dimensions are of course scheme dependent.
We consider the  restriction to a fixed point where such issues go away. For the simple $O(N)$ theory
then solving $\beta_\lambda=0$ from \eqref{bet6} gives
\begin{align}
\lambda_* = {}& \tfrac{1}{2(3N+22)} \, \vep +  \tfrac{1}{4(3N+22)^3} \big ( 53N^2 + 858 N + 3304
+ \tfrac18 \, \pi^2 ( N^3+ 34 N^2 +620 N +2720 ) \big )\vep^2 \nn \\
&{}+   \tfrac{1}{8(3N+22)^5} \big ( 47N^4 + 3114 N^3 + 58156 N^2 +397848N + 920160 \big ) \vep^3\nn \\
&{}+   \tfrac{1}{64(3N+22)^5}\, \pi^2 \big ( 59 N^5 + 4664 N^4 + 127336 N^3 + 1565232 N^2  
+ 8660112N + 17529472 \big ) \vep^3\nn \\
&{}+   \tfrac{1}{256(3N+22)^5}\, \pi^4 \big ( N^6 + 38 N^5 + 256 N^4 -7040 N^3 -95520 N^2  
-430560 N  - 732800 \big ) \vep^3\nn \\
&{}+   \tfrac{3}{32(3N+22)^4}\, \pi^2  \ln 2\big ( N^4 -32   N^3 - 1984 N^2  - 20920N   -61440 \big ) \vep^3\nn \\
&{}+   \tfrac{105}{16(3N+22)^4}\, \zeta_3 \big ( 11 N^3 +428  N^2  + 4228 N   + 12208 \big ) \vep^3\nn \\
&{} -   \tfrac{3}{16(3N+22)^4}\, C \big ( 31N^3 + 1126 N^2  
+ 11876N   + 37592 \big ) \vep^3 + {\rm O}(\vep^4) \, .
\end{align}

Hence to ${\rm O}(\vep^3)$
\begin{align}
2\, \gamma_{\phi *}= {}& \tfrac{(N+2)(N+4)}{12(3N+22)^2} \, \vep^2 \nn \\
&{} + \tfrac{(N+2)(N+4)}{12(3N+22)^4} \big ( \tfrac13(141 N^2 + 2310 N + 8944) +
 \tfrac18\pi^2 ( N^3+ 34 N^2 +620 N +2720 ) \big ) \vep^3 \, .
\end{align} 
From \eqref{D5} we have then
\be
\Delta_{5*}= d -  \Delta_{\phi *} \, ,
\ee
perhaps contrary to naive expectation.  

Further results for anomalous dimensions  at the fixed point to ${\rm O}(\vep^3)$ with uninspiring coefficients are
\begin{align}
\gamma_{2*}= {}&\tfrac{4(N+2)(N+4)}{3(3N+22)^2}\, \vep^2\nn \\
&{}+ \tfrac{(N+2)(N+4)}{(3N+22)^4}\big ( \tfrac{2}{9} ( 93 N^2+1848 N+7724) - \tfrac{1}{48} \pi^2
(N^3- 44N^2 -3268N-18064) \big ) \vep^3\, , \nn \\
\noalign{\vskip 4pt}
\gamma_{3*}= {}& \tfrac{N+4}{3N+22} \, \vep \nn \\
&{} + \tfrac{N+4}{(3N+22)^3} \big ( \tfrac18(137N^2 + 2156 N + 7892) 
 + \tfrac{1}{16} \pi^2(N^3 + 34 N^2 +620  N + 2720) \big ) \vep^2 \nn \\
 &{} + \tfrac{N+4}{(3N+22)^5} \big ( \tfrac{1}{24}( 3009N^4+ 101418 N^3 + 1235180 N^2 +  6412120 N
+ 12118848) \nn \\
\noalign{\vskip -2pt}
& \hskip 1cm {}
+ \tfrac{1}{192}\pi^2  (75N^5 + 10314 N^4 + 338544 N^3 + 4566120 N^2  +  26200432 N 
+ 53152640)\nn \\
\noalign{\vskip -2pt}
& \hskip 1cm {}
+ \tfrac{1}{128}\pi^4( N^6 + 38 N^5+ 256 N^4 -7040 N^3 - 95520  N^2 - 430560 N
 - 732800) \big ) \vep^3 \nn \\
&{} + \tfrac{N+4}{(3N+22)^4} \big ( \tfrac18 \pi^2 \ln 2 (3 N^4 + 32 N^3  - 892 N^2 -12344 N -37424) \nn \\
\noalign{\vskip -2pt}
& \hskip 2cm {} + \tfrac{21}{4}\zeta_3 ( 23 N^3 + 875 N^2 + 8536 N+ 24316) \nn \\
\noalign{\vskip -2pt}
& \hskip 2cm {} - \tfrac38 \,C ( 31N^3 + 1126 N^2 + 11876 N+ 37592 ) \big ) \vep^3 \, , \nn \\
\noalign{\vskip 4pt}
\gamma_{4*}= {}& \tfrac{4(N+4)}{3N+22} \, \vep \nn \\
&{} + \tfrac{N+4}{(3N+22)^3} \big ( \tfrac13(63N^2 + 1580 N + 7372) 
 - \tfrac18 \pi^2(N^3 + 8 N^2 -496 N -2888) \big ) \vep^2 \nn \\
&{} - \tfrac{N+4}{(3N+22)^5} \big ( \tfrac19( 3105N^4 + 90042N^3 + 823500N^2 + 2762104 N
+ 2157824) \nn \\
\noalign{\vskip -2pt}
& \hskip 1cm {}
- \tfrac{1}{48}\pi^2  (51N^5 + 3100 N^4 + 76608 N^3 + 1311584 N^2  +  9444688 N 
+ 22353344)\nn \\
\noalign{\vskip -2pt}
& \hskip 1cm {}
+ \tfrac{1}{64}\pi^4( N^6 + 30 N^5+ 740 N^4 +  19416 N^3 + 215120  N^2 + 1132896 N
 + 2428672) \big ) \vep^3 \nn \\
&{} - \tfrac{N+4}{(3N+22)^4} \big ( \tfrac14 \pi^2 \ln 2 (3 N^4 +122 N^3 +3240 N^2 + 30232 N + 84528) \nn \\
\noalign{\vskip -2pt}
& \hskip 2cm {} - \tfrac{21}{2}\zeta_3 ( 19N^3 + 1138 N^2 + 12452 N+ 37016) \nn \\
\noalign{\vskip -2pt}
& \hskip 2cm {} +3 \,C ( 5N^3 + 288 N^2 + 3682 N+ 12900 ) \big ) \vep^3 \, .
\end{align} 
The ${\rm O}(\vep)$ results are as found in \cite{Basu}.
In addition
\begin{align} 
\beta_\lambda{}^\prime(\lambda_*) = {}& 2 \vep
- \tfrac{1}{(3N+22)^2} \big ( 53N^2 + 858 N + 3304
  +\tfrac18 \pi^2(N^3 + 34 N^2 +620 N +2720) \big ) \vep^2 \nn \\
&{} + \tfrac{1}{(3N+22)^4} \big ( 2( 1381N^4 + 43917N^3 + 514116N^2 + 2635908 N
+ 4998128) \nn \\
\noalign{\vskip -2pt}
& \hskip 0.5cm {}
+ \tfrac{1}{8}\pi^2  (47N^5 + 656 N^4 + 3336 N^3 + 11680 N^2  +  104368 N 
+ 444288)\nn \\
\noalign{\vskip -2pt}
& \hskip 0.5cm {}
- \tfrac{1}{64}\pi^4( N^6 + 8 N^5 -1884 N^4 -61680 N^3 -760400  N^2 -4233920 N
 - 8864000) \big ) \vep^3 \nn \\
 &{} - \tfrac{1}{(3N+22)^3} \big ( \tfrac34 \pi^2 \ln 2 ( N^4-32  N^3 -1984  N^2 -20920 N -61440) \nn \\
\noalign{\vskip -2pt}
& \hskip 2cm {} + \tfrac{105}{2}\zeta_3 ( 11N^3 + 428 N^2 + 4228 N+ 12208) \nn \\
\noalign{\vskip -2pt}
& \hskip 2cm {} - \tfrac32\,C ( 31 N^3 + 1126 N^2 + 11876 N+ 37592 ) \big ) \vep^3 \, ,
\end{align}
where $\gamma_{6*}= 2\vep+ \beta_\lambda{}^\prime(\lambda_*)$ and  from \eqref{D5}
$\gamma_{5*} = 2\vep - \gamma_{\phi*}$.
In general the  scaling dimension  $\Delta_k = \tfrac12k(d-2) + \gamma_{k*}$.
As commented by Hager the expansion coefficients grow rapidly, more so than for $\phi^4$ in the 
usual $\vep$ expansion.

For $N=1$ the results reduce to
\begin{align}
2\, \gamma_{\phi *}= {}& \tfrac{1}{500}\, \vep^2 + \tfrac{1}{15000}\big (\tfrac{4558}{25} + \tfrac{81}{4} \pi^2 \big ) \vep^3
\, , \nn \\
\noalign{\vskip 4pt}
\gamma_{2*}= {}&  \tfrac{4}{125}\, \vep^2+
\tfrac{1}{625}\big (\tfrac{3866}{75} + \tfrac{171}{16} \pi^2 \big ) \vep^3 \, , \nn\\
\noalign{\vskip 4pt}
\gamma_{3*}= {}& \tfrac15 \, \vep +\tfrac{1}{200}\big (\tfrac{2037}{25}+ \tfrac{27}{2} \pi^2\big )  \vep^2\nn \\
&\qquad {} +  \tfrac{1}{1250} \big ( \tfrac{264941}{500}
+ \tfrac{44943}{160}\pi^2 - \tfrac{405}{4}\pi^2  \ln2  -  \tfrac{405}{64}\pi^4 +  2835 \zeta_3 - \tfrac{1215}{4}C \big ) \vep^3\, , \nn \\
\noalign{\vskip 4pt}
\gamma_{4*}= {}& \tfrac45 \, \vep +\tfrac{1}{25}\big( \tfrac{601}{25} + \tfrac{27}{8} \pi^2 \big )\vep^2 \nn \\
&\qquad {} -   \tfrac{1}{250} \big ( \tfrac{155642}{1875}
- \tfrac{17701}{200}\pi^2 + \tfrac{189}{2}\pi^2  \ln2  +  \tfrac{243}{32}\pi^4 -  1701 \zeta_3 + 162C \big ) \vep^3\, , \nn \\
\noalign{\vskip 4pt}
\beta_\lambda{}^\prime(\lambda_*) = {}& 2 \vep -  \tfrac{1}{25}\big( \tfrac{843}{5} + \tfrac{135}{8}\pi^2 \big )\vep^2 \nn \\
&\qquad {} +  \tfrac{1}{50} \big ( \tfrac{1310952}{625}
+ \tfrac{903}{100}\pi^2 + \tfrac{405}{2}\pi^2  \ln2  +  \tfrac{891}{32}\pi^4-  2835 \zeta_3 + 243C \big ) \vep^3\, .
\end{align}
The results given here for $\beta_\lambda{}^\prime(\lambda_*)$, as well as those for $\gamma_2(\lambda),
\gamma_4(\lambda),\beta_\lambda(\lambda)$ when $N=1$ given in \eqref{gam6} and \eqref{bet6},
 agree precisely with corresponding expressions in \cite{Kompaniets2}
reassuringly confirming our disagreement with \cite{Hager} as far as the $\pi^2, \pi^2 \ln 2$ terms are concerned.

\section{Gradient Flow}

A constraint on results for the perturbative $\beta$-function for renormalisable  theories
with general dimensionless couplings $g^I$ 
is that $\beta^I(g)$  can be expressed as a gradient flow in terms of a scalar $A(g)$. Such possibilities
were first explored for $\phi^4 $ in four dimensions by Wallace and Zia \cite{WallaceG}. In the context
of the $a$-theorem a perturbative proof was given in four dimensions for general renormalisable 
theories in \cite{Analogs,Weyl}.
 
 The flow equation can be expressed  in the form
\be
\rmd A(g) = T_{IJ}(g) \,  \rmd g^I \beta^J (g) \, ,
\label{Athm}
\ee
where for gradient flow $T_{IJ}$ is symmetric and if positive then $A$ decreases under RG flow giving 
a strong version of the $a$-theorem. In four dimensions 
at a fixed point, where the $\beta$-function vanishes, $A$  reduces to the  $a$ anomaly, the coefficient of 
the Euler density in the energy momentum tensor trace on a curved background.
The tensor $T_{IJ}$  can further  be defined in terms  of quantities necessary for renormalisation on a curved
background but  may also be obtained by virtue of various consistency relations  in terms of flat space calculations 
such that the integrability conditions necessary for \eqref{Athm}  to hold are valid. 
$T_{IJ}$ in general need not be symmetric but a further constraint is that the antisymmetric
part is expressible as
\be
T_{[IJ]} = \pr_I W_J -  \pr_J W_I \, .
\label{TAW}
\ee
Away from fixed points $A$ is not unique since \eqref{Athm} is invariant, for any fixed $\beta^I$, under the changes
\be
\delta A = g_{IJ} \, \beta^I \beta^J \, , \quad \delta \, T_{IJ} = {\mathcal L}_\beta \, g_{IJ}
+ \pr_I \big (g_{JK} \beta^K \big ) - \pr_J\big  (g_{IK} \beta^K\big ) \, , \quad \delta W_J = g_{JK} \beta^K \, , 
\label{Avar}
\ee
for any symmetric  $g_{IJ}(g)$. This may be used to remove at least some antisymmetric pieces from $T_{IJ}$. 
The existence of $A(g)$ imposes constraints on the $\beta$-function which may be verified using perturbative
results. 

Besides four dimensions gradient flow properties have been explored in six dimensions in general $\phi^3$
theories in \cite{Jack6d} and in three dimensions for theories with fermions and gauge fields 
in \cite{Jack3d,Jack3d2,Jack3d3}. For four dimensional scalar
theories constraints on the $\beta$-function  at four loops were obtained in \cite{Jack4d} and more
recently extended to higher loops in \cite{Pannell1} and the associated scalar curvature corresponding to the metric
on the space of couplings considered in \cite{Pannell2}. For purely scalar $\phi^6$ theories in three dimensions
results corresponding to the four loop $\beta$-function were given in \cite{Kapoor} but there are no constraints
at this order.

Here we consider applications to three dimensional $\phi^6$ theory using our six loop results 
and demonstrate that there are non trivial constraints which provide some check on the perturbative
calculations. Expressions   for $A, \, T_{IJ}$ for $\phi^6$ theory are realised in terms of graphs with 
no external lines, vacuum diagrams, with all vertices  of degree six and the expression for $A(\lambda)$
obtained by associating $\lambda_{ijklmn}\equiv \lambda_I$ with each vertex with indices contracted as
required by the graph. For $n$ vertices these graphs 
have $2n+1$ loops. Results for $T_{IJ}(\lambda) \rmd \lambda^I \rmd \lambda^J$ correspond to two vertices
being identified in a vacuum graph.
For six loop $\beta$-functions it is necessary to consider for $A(g)$ vacuum graphs with five vertices.

To illustrate the possible vacuum graphs we consider connected simple graphs, where only one line
can link any two vertices, and then attach numbers to each edge so that they add to six at every vertex.
Such graphs are then relevant for vacuum diagrams in $\phi^6$ theory with 6 lines linked to
the coupling $\lambda^I$ at each vertex.
For $n=2,3$ there is only one graph in each case
\be
\tikz[baseline=(vert_cent.base)]{
\node (vert_cent) {\hspace{-13pt}$\phantom{-}$};
\node at  (0:0.8)  (A) {};
\node at  (180:0.8)  (B) {};
\draw (A.center) -- (B.center);
\fill[red]  (A) circle [radius = 0.1cm];
	\fill[red]  (B) circle [radius = 0.1cm];
		\node at  (-1.9,0) {$\cG_{2}$}  ;
		\node at  (90:0.2)  {6};
			\node at  (0:1.1)  {$a$};
	\node at  (180:1.1)  {$b$};
	}
\qquad
\tikz[baseline=(vert_cent.base)]{
\node (vert_cent) {\hspace{-13pt}$\phantom{-}$};
\node at  (90:0.9)  (A) {};
\node at  (210:0.9)  (B) {};
\node at  (330:0.9)  (C) {};
\draw (A.center) -- (B.center);
\draw (B.center)--(C.center);
\draw (C.center)--(A.center);
\fill[red]  (A) circle [radius = 0.1cm];
	\fill[red]  (B) circle [radius = 0.1cm];
	\fill[red]  (C) circle [radius = 0.1cm];
	\node at  (30:0.75)  {3};
		\node at  (270:0.75)  {3};
	\node at  (150:0.75)  {3};
		\node at  (90:1.2)  {$a$};
	\node at  (330:1.2)  {$b$};
	\node at (210:1.2)  {$c$};
	\node at  (-1.7,0) {$\cG_{3}$}  ;
	}	
\ee
All vertices and pairs of vertices are equivalent and in the expansion of $A(\lambda)$ 
the coefficients are $A_2, \, A_3$
and for the single corresponding contributions to $T_{IJ}(\lambda)$ the coefficients are just $T_2, \, T_3$
and with a choice of normalisation $T_2=1$.
When $n=4$ there are 6 possible graphs
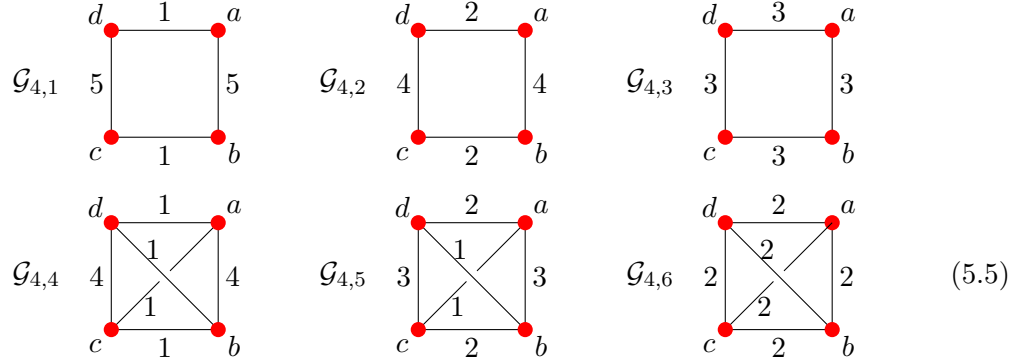
\begin{align}
 \tikz[baseline=(vert_cent.base)]{
\node (vert_cent) {\hspace{-13pt}$\phantom{-}$};
\node at  (45:1)  (A) {};
\node at  (135:1)  (B) {};
\node at  (225:1)  (C) {};
\node at  (315:1)  (D) {};
\draw (A.center) -- (B.center);
\draw (B.center)--(C.center);
\draw (C.center)--(D.center);
\draw (D.center)--(A.center);
\fill[red]  (A) circle [radius = 0.1cm];
	\fill[red]  (B) circle [radius = 0.1cm];
	\fill[red]  (C) circle [radius = 0.1cm];
	\fill[red]  (D) circle [radius = 0.1cm];
	\node at  (0:0.9)  {5};
	\node at  (90:0.95)  {1};
	\node at  (180:0.9)  {5};
	\node at  (270:0.95)  {1};
	\node at  (45:1.3)  {$a$};
	\node at  (315:1.3)  {$b$};
	\node at  (225:1.3)  {$c$};
	\node at  (135:1.3)  {$d$};
	\node at  (-1.7,0) {$\cG_{4,1}$}  ;
	}
	\qquad
	 \tikz[baseline=(vert_cent.base)]{
\node (vert_cent) {\hspace{-13pt}$\phantom{-}$};
\node at  (45:1)  (A) {};
\node at  (135:1)  (B) {};
\node at  (225:1)  (C) {};
\node at  (315:1)  (D) {};
\draw (A.center) -- (B.center);
\draw (B.center)--(C.center);
\draw (C.center)--(D.center);
\draw (D.center)--(A.center);
\fill[red]  (A) circle [radius = 0.1cm];
	\fill[red]  (B) circle [radius = 0.1cm];
	\fill[red]  (C) circle [radius = 0.1cm];
	\fill[red]  (D) circle [radius = 0.1cm];
	\node at  (0:0.9)  {4};
	\node at  (90:0.95)  {2};
	\node at  (180:0.9)  {4};
	\node at  (270:0.95)  {2};
		\node at  (45:1.3)  {$a$};
	\node at  (315:1.3)  {$b$};
	\node at  (225:1.3)  {$c$};
	\node at  (135:1.3)  {$d$};
	\node at  (-1.7,0) {$\cG_{4,2}$}  ;
	}
\qquad
 \tikz[baseline=(vert_cent.base)]{
\node (vert_cent) {\hspace{-13pt}$\phantom{-}$};
\node at  (45:1)  (A) {};
\node at  (135:1)  (B) {};
\node at  (225:1)  (C) {};
\node at  (315:1)  (D) {};
\draw (A.center) -- (B.center);
\draw (B.center)--(C.center);
\draw (C.center)--(D.center);
\draw (D.center)--(A.center);
\fill[red]  (A) circle [radius = 0.1cm];
	\fill[red]  (B) circle [radius = 0.1cm];
	\fill[red]  (C) circle [radius = 0.1cm];
	\fill[red]  (D) circle [radius = 0.1cm];
	\node at  (0:0.9)  {3};
	\node at  (90:0.95)  {3};
	\node at  (180:0.9)  {3};
	\node at  (270:0.95)  {3};
		\node at  (45:1.3)  {$a$};
	\node at  (315:1.3)  {$b$};
	\node at  (225:1.3)  {$c$};
	\node at  (135:1.3)  {$d$};
	\node at  (-1.7,0) {$\cG_{4,3}$}  ;
	}   \nn \\
 \tikz[baseline=(vert_cent.base)]{
\node (vert_cent) {\hspace{-13pt}$\phantom{-}$};
\node at  (45:1)  (A) {};
\node at  (135:1)  (B) {};
\node at  (225:1)  (C) {};
\node at  (315:1)  (D) {};
\draw (A.center) -- (B.center);
\draw (B.center)--(C.center);
\draw (C.center)--(D.center);
\draw (D.center)--(A.center);
\draw (A.center)--(C.center);
\fill[white]  (0,0) circle [radius = 0.1cm];
\draw (B.center)--(D.center);
\fill[red]  (A) circle [radius = 0.1cm];
	\fill[red]  (B) circle [radius = 0.1cm];
	\fill[red]  (C) circle [radius = 0.1cm];
	\fill[red]  (D) circle [radius = 0.1cm];
	\node at  (0:0.9)  {4};
	\node at  (90:0.95)  {1};
	\node at  (180:0.9)  {4};
	\node at  (270:0.95)  {1};
	\node at  (114:0.38)  {1};
	\node at  (245:0.45)  {1};
	\node at  (45:1.3)  {$a$};
	\node at  (315:1.3)  {$b$};
	\node at  (225:1.3)  {$c$};
	\node at  (135:1.3)  {$d$};
	\node at  (-1.7,0) {$\cG_{4,4}$}  ;
	}
	\qquad
	 \tikz[baseline=(vert_cent.base)]{
\node (vert_cent) {\hspace{-13pt}$\phantom{-}$};
\node at  (45:1)  (A) {};
\node at  (135:1)  (B) {};
\node at  (225:1)  (C) {};
\node at  (315:1)  (D) {};
\draw (A.center) -- (B.center);
\draw (B.center)--(C.center);
\draw (C.center)--(D.center);
\draw (D.center)--(A.center);
\draw (A.center)--(C.center);
\fill[white]  (0,0) circle [radius = 0.1cm];
\draw (B.center)--(D.center);
\fill[red]  (A) circle [radius = 0.1cm];
	\fill[red]  (B) circle [radius = 0.1cm];
	\fill[red]  (C) circle [radius = 0.1cm];
	\fill[red]  (D) circle [radius = 0.1cm];
	\node at  (0:0.9)  {3};
	\node at  (90:0.95)  {2};
	\node at  (180:0.9)  {3};
	\node at  (270:0.95)  {2};
	\node at  (114:0.38)  {1};
	\node at  (245:0.45)  {1};
		\node at  (45:1.3)  {$a$};
	\node at  (315:1.3)  {$b$};
	\node at  (225:1.3)  {$c$};
	\node at  (135:1.3)  {$d$};
	\node at  (-1.7,0) {$\cG_{4,5}$}  ;
	}
\qquad
 \tikz[baseline=(vert_cent.base)]{
\node (vert_cent) {\hspace{-13pt}$\phantom{-}$};
\node at  (45:1)  (A) {};
\node at  (135:1)  (B) {};
\node at  (225:1)  (C) {};
\node at  (315:1)  (D) {};
\draw (A.center) -- (B.center);
\draw (B.center)--(C.center);
\draw (C.center)--(D.center);
\draw (D.center)--(A.center);
\fill[red]  (A) circle [radius = 0.1cm];
\draw (A.center)--(C.center);
\fill[white]  (0,0) circle [radius = 0.1cm];
\draw (B.center)--(D.center);
	\fill[red]  (B) circle [radius = 0.1cm];
	\fill[red]  (C) circle [radius = 0.1cm];
	\fill[red]  (D) circle [radius = 0.1cm];
	\node at  (0:0.9)  {2};
	\node at  (90:0.95)  {2};
	\node at  (180:0.9)  {2};
	\node at  (270:0.95)  {2};
	\node at  (114:0.38)  {2};
	\node at  (245:0.45)  {2};
		\node at  (45:1.3)  {$a$};
	\node at  (315:1.3)  {$b$};
	\node at  (225:1.3)  {$c$};
	\node at  (135:1.3)  {$d$};
	\node at  (-1.7,0) {$\cG_{4,6}$}  ;
	}
	\label{Vfour}	
\end{align}
Each graph is vertex transitive and the inequivalent contributions  to $T_{IJ}$ are all symmetric and labelled
by $T_{4,p\,ab}, \, T_{4,p\,ac}, T_{4,p\,ad}$, $p=1,2,5$,  $T_{4,p\, ab}, \, T_{4,p\, ac}, \ p=3,4$ and 
$T_{4,6\, ab}$. The graphs $\cG_{4,2}$ and $\cG_{4,4}$ have no subgraphs which would generate 
$\vep$-poles  and are primitive.

For $n=5$ there are more graphs and the associated vertices are not all equivalent. With colour coding for
equivalent vertices
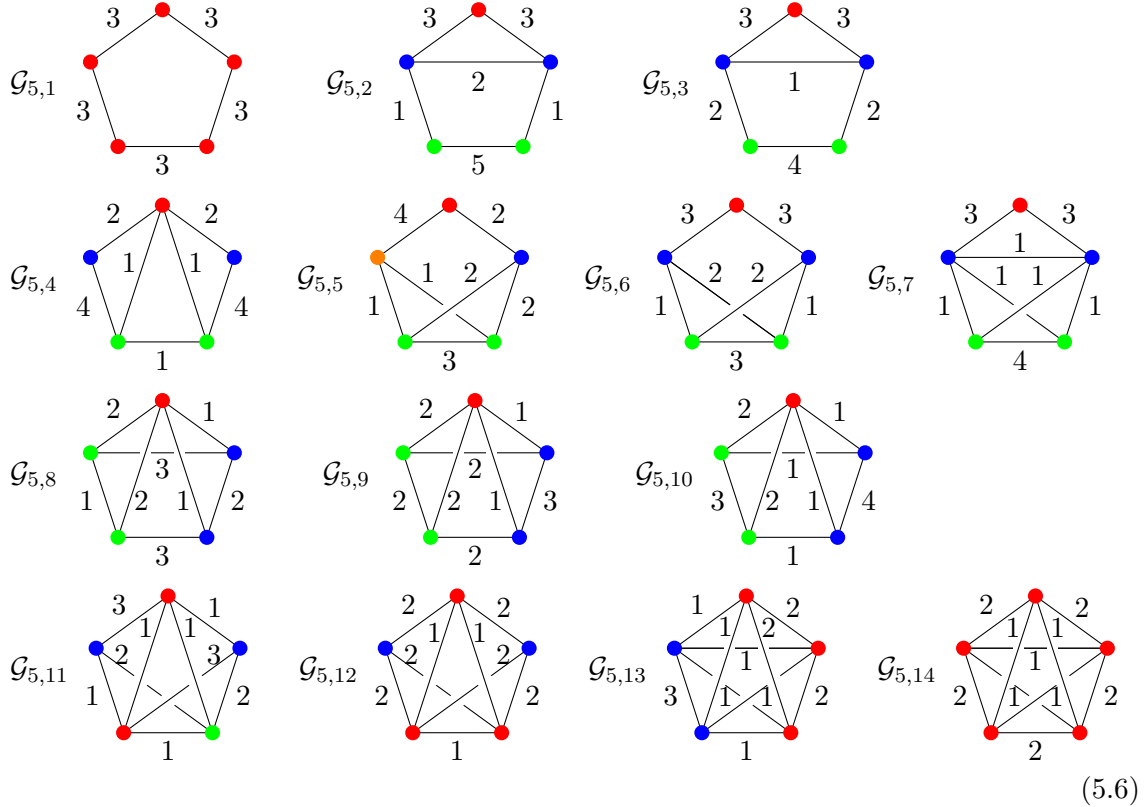
\begin{align}
& \tikz[baseline=(vert_cent.base)]{
\node (vert_cent) {\hspace{-13pt}$\phantom{-}$};
\node at  (90:1)  (A) {};
\node at  (18:1)  (B) {};
\node at  (306:1)  (C) {};
\node at  (234:1)  (D) {};
\node at  (162:1)  (E) {};
\draw (A.center) -- (B.center);
\draw (B.center)--(C.center);
\draw (C.center)--(D.center);
\draw (D.center)--(E.center);
\draw (E.center) -- (A.center);
\fill[red]  (A) circle [radius = 0.1cm];
	\fill[red]  (B) circle [radius = 0.1cm];
	\fill[red]  (C) circle [radius = 0.1cm];
	\fill[red]  (D) circle [radius = 0.1cm];
	\fill[red]  (E) circle [radius = 0.1cm];
	\node at  (54:1.1)  {3};
	\node at  (126:1.1)  {3};
	\node at  (198:1.1)  {3};
	\node at  (270:1.05)  {3};
	\node at  (342:1.1)  {3};
	\node at  (-1.7,0) {$\cG_{5,1}$}  ;
	}
	\qquad
	\tikz[baseline=(vert_cent.base)]{
\node (vert_cent) {\hspace{-13pt}$\phantom{-}$};
\node at  (90:1)  (A) {};
\node at  (18:1)  (B) {};
\node at  (306:1)  (C) {};
\node at  (234:1)  (D) {};
\node at  (162:1)  (E) {};
\draw (A.center) -- (B.center);
\draw (B.center)--(C.center);
\draw (C.center)--(D.center);
\draw (D.center)--(E.center);
\draw (E.center) -- (A.center);
\draw (B.center)--(E.center);
\fill[red]  (A) circle [radius = 0.1cm];
	\fill[blue]  (B) circle [radius = 0.1cm];
	\fill[green]  (C) circle [radius = 0.1cm];
	\fill[green]  (D) circle [radius = 0.1cm];
	\fill[blue]  (E) circle [radius = 0.1cm];
	\node at  (54:1.1)  {3};
	\node at  (126:1.1)  {3};
	\node at  (198:1.1)  {1};
	\node at  (270:1.05)  {5};
	\node at  (342:1.1)  {1};
	\node at  (90:0.05)  {2};
	\node at  (-1.7,0) {$\cG_{5,2}$}  ;
	}
	\qquad
\tikz[baseline=(vert_cent.base)]{
\node (vert_cent) {\hspace{-13pt}$\phantom{-}$};
\node at  (90:1)  (A) {};
\node at  (18:1)  (B) {};
\node at  (306:1)  (C) {};
\node at  (234:1)  (D) {};
\node at  (162:1)  (E) {};
\draw (A.center) -- (B.center);
\draw (B.center)--(C.center);
\draw (C.center)--(D.center);
\draw (D.center)--(E.center);
\draw (E.center) -- (A.center);
\draw (B.center)--(E.center);
\fill[red]  (A) circle [radius = 0.1cm];
	\fill[blue]  (B) circle [radius = 0.1cm];
	\fill[green]  (C) circle [radius = 0.1cm];
	\fill[green]  (D) circle [radius = 0.1cm];
	\fill[blue]  (E) circle [radius = 0.1cm];
	\node at  (54:1.1)  {3};
	\node at  (126:1.1)  {3};
	\node at  (198:1.1)  {2};
	\node at  (270:1.05)  {4};
	\node at  (342:1.1)  {2};
	\node at  (90:0.05)  {1};
	\node at  (-1.7,0) {$\cG_{5,3}$}  ;	
}
\nn \\
&
\tikz[baseline=(vert_cent.base)]{
\node (vert_cent) {\hspace{-13pt}$\phantom{-}$};
\node at  (90:1)  (A) {};
\node at  (18:1)  (B) {};
\node at  (306:1)  (C) {};
\node at  (234:1)  (D) {};
\node at  (162:1)  (E) {};
\draw (A.center) -- (B.center);
\draw (B.center)--(C.center);
\draw (C.center)--(D.center);
\draw (D.center)--(E.center);
\draw (E.center) -- (A.center);
\draw (A.center)--(C.center);
\draw (A.center)--(D.center);
\fill[red]  (A) circle [radius = 0.1cm];
	\fill[blue]  (B) circle [radius = 0.1cm];
	\fill[green]  (C) circle [radius = 0.1cm];
	\fill[green]  (D) circle [radius = 0.1cm];
	\fill[blue]  (E) circle [radius = 0.1cm];
	\node at  (54:1.1)  {2};
	\node at  (126:1.1)  {2};
	\node at  (198:1.1)  {4};
	\node at  (270:1.05)  {1};
	\node at  (342:1.1)  {4};
	\node at  (28:0.5)  {1};
	\node at  (152:0.5)  {1};
	\node at  (-1.7,0) {$\cG_{5,4}$}  ;
	}
	\quad
	\tikz[baseline=(vert_cent.base)]{
\node (vert_cent) {\hspace{-13pt}$\phantom{-}$};
\node at  (90:1)  (A) {};
\node at  (18:1)  (B) {};
\node at  (306:1)  (C) {};
\node at  (234:1)  (D) {};
\node at  (162:1)  (E) {};
\draw (A.center) -- (B.center);
\draw (B.center)--(C.center);
\draw (C.center)--(D.center);
\draw (D.center)--(E.center);
\draw (E.center) -- (A.center);
\draw [white] [name path=bd] (B.center)--(D.center);
\draw  [name path=ce] (C.center)--(E.center);
\path [name intersections={of=bd and ce,by= bdce}];
 	\node[fill=white, inner sep=3pt, rotate=45] at (bdce) {};
\draw (B.center) -- (D.center); 
\fill[red]  (A) circle [radius = 0.1cm];
	\fill[blue]  (B) circle [radius = 0.1cm];
	\fill[green]  (C) circle [radius = 0.1cm];
	\fill[green]  (D) circle [radius = 0.1cm];
	\fill[orange]  (E) circle [radius = 0.1cm];
	\node at  (54:1.1)  {2};
	\node at  (126:1.1)  {4};
	\node at  (198:1.05)  {1};
	\node at  (270:1.05)  {3};
	\node at  (342:1.1)  {2};
	\node at  (20:0.3)  {2};
	\node at  (160:0.3)  {1};
	\node at  (-1.7,0) {$\cG_{5,5}$}  ;
	}
	\quad
\tikz[baseline=(vert_cent.base)]{
\node (vert_cent) {\hspace{-13pt}$\phantom{-}$};
\node at  (90:1)  (A) {};
\node at  (18:1)  (B) {};
\node at  (306:1)  (C) {};
\node at  (234:1)  (D) {};
\node at  (162:1)  (E) {};
\draw (A.center) -- (B.center);
\draw (B.center)--(C.center);
\draw (C.center)--(D.center);
\draw (D.center)--(E.center);
\draw (E.center) -- (A.center);
\draw (C.center)--(E.center);
\draw [white] [name path=bd] (B.center)--(D.center);
\draw  [name path=ce] (C.center)--(E.center);
\path [name intersections={of=bd and ce,by= bdce}];
 	\node[fill=white, inner sep=3pt, rotate=45] at (bdce) {};
\draw (B.center) -- (D.center); 
\fill[red]  (A) circle [radius = 0.1cm];
	\fill[blue]  (B) circle [radius = 0.1cm];
	\fill[green]  (C) circle [radius = 0.1cm];
	\fill[green]  (D) circle [radius = 0.1cm];
	\fill[blue]  (E) circle [radius = 0.1cm];
	\node at  (54:1.1)  {3};
	\node at  (126:1.1)  {3};
	\node at  (198:1.05)  {1};
	\node at  (270:1.05)  {3};
	\node at  (342:1.05)  {1};
	\node at  (20:0.3)  {2};
	\node at  (160:0.3)  {2};
	\node at  (-1.7,0) {$\cG_{5,6}$}  ;
} 
\quad
\tikz[baseline=(vert_cent.base)]{
\node (vert_cent) {\hspace{-13pt}$\phantom{-}$};
\node at  (90:1)  (A) {};
\node at  (18:1)  (B) {};
\node at  (306:1)  (C) {};
\node at  (234:1)  (D) {};
\node at  (162:1)  (E) {};
\draw (A.center) -- (B.center);
\draw (B.center)--(C.center);
\draw (C.center)--(D.center);
\draw (D.center)--(E.center);
\draw (E.center) -- (A.center);
\draw [white] [name path=bd] (B.center)--(D.center);
\draw  [name path=ce] (C.center)--(E.center);
\path [name intersections={of=bd and ce,by= bdce}];
 	\node[fill=white, inner sep=3pt, rotate=45] at (bdce) {};
\draw (B.center) -- (D.center); 
\draw (B.center)--(E.center);
\fill[red]  (A) circle [radius = 0.1cm];
	\fill[blue]  (B) circle [radius = 0.1cm];
	\fill[green]  (C) circle [radius = 0.1cm];
	\fill[green]  (D) circle [radius = 0.1cm];
	\fill[blue]  (E) circle [radius = 0.1cm];
	\node at  (54:1.1)  {3};
	\node at  (126:1.1)  {3};
	\node at  (198:1.05)  {1};
	\node at  (270:1.05)  {4};
	\node at  (342:1.05)  {1};
	\node at  (16:0.25)  {1};
	\node at  (164:0.25)  {1};
	\node at  (90:0.5)  {1} ;
	\node at  (-1.7,0) {$\cG_{5,7}$}  ;
}
\nn \\
&
\tikz[baseline=(vert_cent.base)]{
\node (vert_cent) {\hspace{-13pt}$\phantom{-}$};
\node at  (90:1)  (A) {};
\node at  (18:1)  (B) {};
\node at  (306:1)  (C) {};
\node at  (234:1)  (D) {};
\node at  (162:1)  (E) {};
\draw (A.center) -- (B.center);
\draw (B.center)--(C.center);
\draw (C.center)--(D.center);
\draw (D.center)--(E.center);
\draw (E.center) -- (A.center);
\draw [white] [name path=ac] (A.center)--(C.center);
\draw [white] [name path=ad] (A.center)--(D.center);
\draw [name path=be] (B.center)--(E.center);
\path [name intersections={of=ad and be,by= adbe}];
\path [name intersections={of=ac and be,by= acbe}];
 	\node[fill=white, inner sep=2pt, rotate=90] at (adbe) {};
	\node[fill=white, inner sep=2pt, rotate=90] at (acbe) {};
\draw (A.center)--(C.center);
\draw (A.center)--(D.center);
\fill[red]  (A) circle [radius = 0.1cm];
	\fill[blue]  (B) circle [radius = 0.1cm];
	\fill[blue]  (C) circle [radius = 0.1cm];
	\fill[green]  (D) circle [radius = 0.1cm];
	\fill[green]  (E) circle [radius = 0.1cm];
	\node at  (54:1.05)  {1};
	\node at  (126:1.1)  {2};
	\node at  (198:1.05)  {1};
	\node at  (270:1.05)  {3};
	\node at  (342:1.05)  {2};
	\node at  (-50:0.43)  {1};
	\node at  (230:0.43)  {2};
	\node at  (90:0.1)  {3};
	\node at  (-1.7,0) {$\cG_{5,8}$}  ;
}
\qquad
\tikz[baseline=(vert_cent.base)]{
\node (vert_cent) {\hspace{-13pt}$\phantom{-}$};
\node at  (90:1)  (A) {};
\node at  (18:1)  (B) {};
\node at  (306:1)  (C) {};
\node at  (234:1)  (D) {};
\node at  (162:1)  (E) {};
\draw (A.center) -- (B.center);
\draw (B.center)--(C.center);
\draw (C.center)--(D.center);
\draw (D.center)--(E.center);
\draw (E.center) -- (A.center);
\draw [white] [name path=ac] (A.center)--(C.center);
\draw [white] [name path=ad] (A.center)--(D.center);
\draw [name path=be] (B.center)--(E.center);
\path [name intersections={of=ad and be,by= adbe}];
\path [name intersections={of=ac and be,by= acbe}];
 	\node[fill=white, inner sep=2pt, rotate=90] at (adbe) {};
	\node[fill=white, inner sep=2pt, rotate=90] at (acbe) {};
\draw (A.center)--(C.center);
\draw (A.center)--(D.center);
\fill[red]  (A) circle [radius = 0.1cm];
	\fill[blue]  (B) circle [radius = 0.1cm];
	\fill[blue]  (C) circle [radius = 0.1cm];
	\fill[green]  (D) circle [radius = 0.1cm];
	\fill[green]  (E) circle [radius = 0.1cm];
	\node at  (54:1.05)  {1};
	\node at  (126:1.1)  {2};
	\node at  (198:1.05)  {2};
	\node at  (270:1.05)  {2};
	\node at  (342:1.05)  {3};
	\node at  (-50:0.43)  {1};
	\node at  (230:0.43)  {2};
	\node at  (90:0.1)  {2};
	\node at  (-1.7,0) {$\cG_{5,9}$}  ;
}
\qquad
\tikz[baseline=(vert_cent.base)]{
\node (vert_cent) {\hspace{-13pt}$\phantom{-}$};
\node at  (90:1)  (A) {};
\node at  (18:1)  (B) {};
\node at  (306:1)  (C) {};
\node at  (234:1)  (D) {};
\node at  (162:1)  (E) {};
\draw (A.center) -- (B.center);
\draw (B.center)--(C.center);
\draw (C.center)--(D.center);
\draw (D.center)--(E.center);
\draw (E.center) -- (A.center);
\draw [white] [name path=ac] (A.center)--(C.center);
\draw [white] [name path=ad] (A.center)--(D.center);
\draw [name path=be] (B.center)--(E.center);
\path [name intersections={of=ad and be,by= adbe}];
\path [name intersections={of=ac and be,by= acbe}];
 	\node[fill=white, inner sep=2pt, rotate=90] at (adbe) {};
	\node[fill=white, inner sep=2pt, rotate=90] at (acbe) {};
\draw (A.center)--(C.center);
\draw (A.center)--(D.center);
\fill[red]  (A) circle [radius = 0.1cm];
	\fill[blue]  (B) circle [radius = 0.1cm];
	\fill[blue]  (C) circle [radius = 0.1cm];
	\fill[green]  (D) circle [radius = 0.1cm];
	\fill[green]  (E) circle [radius = 0.1cm];
	\node at  (54:1.05)  {1};
	\node at  (126:1.1)  {2};
	\node at  (198:1.05)  {3};
	\node at  (270:1.05)   {1};
	\node at  (342:1.05)  {4};
	\node at  (-50:0.43)  {1};
	\node at  (230:0.43)  {2};
	\node at  (90:0.1)  {1};
	\node at  (-1.7,0) {$\cG_{5,10}$}  ;
}
\nn \\
&
\tikz[baseline=(vert_cent.base)]{
\node (vert_cent) {\hspace{-13pt}$\phantom{-}$};
\node at  (90:1)  (A) {};
\node at  (18:1)  (B) {};
\node at  (306:1)  (C) {};
\node at  (234:1)  (D) {};
\node at  (162:1)  (E) {};
\draw (A.center) -- (B.center);
\draw (B.center)--(C.center);
\draw (C.center)--(D.center);
\draw (D.center)--(E.center);
\draw (E.center) -- (A.center);
\draw [white] [name path=bd] (B.center)--(D.center);
\draw  [name path=ce] (C.center)--(E.center);
\path [name intersections={of=bd and ce,by= bdce}];
 	\node[fill=white, inner sep=3pt, rotate=45] at (bdce) {};
\draw (B.center)--(D.center);
\draw [white] [name path=ad] (A.center)--(D.center);
\path [name intersections={of=ad and ce,by= adce}];
 	\node[fill=white, inner sep=3pt, rotate=45] at (adce) {};
	\draw [white] [name path=ac] (A.center)--(C.center);
\path [name intersections={of=ac and bd,by= acbd}];
 	\node[fill=white, inner sep=3pt, rotate=45] at (acbd) {};
\draw (A.center)--(D.center);
\draw (A.center)--(C.center);
\fill[red]  (A) circle [radius = 0.1cm];
	\fill[blue]  (B) circle [radius = 0.1cm];
	\fill[green]  (C) circle [radius = 0.1cm];
	\fill[red]  (D) circle [radius = 0.1cm];
	\fill[blue]  (E) circle [radius = 0.1cm];
	\node at  (54:1.05)  {1};
	\node at  (126:1.1)  {3};
	\node at  (198:1.05)  {1};
	\node at  (270:1.05)  {1};
	\node at  (342:1.05)  {2};
	\node at  (20:0.65)  {3};
	\node at  (160:0.65)  {2};
	\node at  (63:0.65)  {1};
	\node at  (117:0.65)  {1};
	\node at  (-1.7,0) {$\cG_{5,11}$}  ;
}
\quad
\tikz[baseline=(vert_cent.base)]{
\node (vert_cent) {\hspace{-13pt}$\phantom{-}$};
\node at  (90:1)  (A) {};
\node at  (18:1)  (B) {};
\node at  (306:1)  (C) {};
\node at  (234:1)  (D) {};
\node at  (162:1)  (E) {};
\draw (A.center) -- (B.center);
\draw (B.center)--(C.center);
\draw (C.center)--(D.center);
\draw (D.center)--(E.center);
\draw (E.center) -- (A.center);
\draw [white] [name path=bd] (B.center)--(D.center);
\draw  [name path=ce] (C.center)--(E.center);
\path [name intersections={of=bd and ce,by= bdce}];
 	\node[fill=white, inner sep=3pt, rotate=45] at (bdce) {};
\draw (B.center)--(D.center);
\draw [white] [name path=ad] (A.center)--(D.center);
\path [name intersections={of=ad and ce,by= adce}];
 	\node[fill=white, inner sep=3pt, rotate=45] at (adce) {};
	\draw [white] [name path=ac] (A.center)--(C.center);
\path [name intersections={of=ac and bd,by= acbd}];
 	\node[fill=white, inner sep=3pt, rotate=45] at (acbd) {};
\draw (A.center)--(D.center);
\draw (A.center)--(C.center);
\fill[red]  (A) circle [radius = 0.1cm];
	\fill[blue]  (B) circle [radius = 0.1cm];
	\fill[red]  (C) circle [radius = 0.1cm];
	\fill[red]  (D) circle [radius = 0.1cm];
	\fill[blue]  (E) circle [radius = 0.1cm];
	\node at  (54:1.05)  {2};
	\node at  (126:1.1)  {2};
	\node at  (198:1.05)  {2};
	\node at  (270:1.05)  {1};
	\node at  (342:1.05)  {2};
	\node at  (20:0.65)  {2};
	\node at  (160:0.65)  {2};
	\node at  (60:0.6)  {1};
	\node at  (120:0.6)  {1};
	\node at  (-1.7,0) {$\cG_{5,12}$}  ;
}
\quad
\tikz[baseline=(vert_cent.base)]{
\node (vert_cent) {\hspace{-13pt}$\phantom{-}$};
\node at  (90:1)  (A) {};
\node at  (18:1)  (B) {};
\node at  (306:1)  (C) {};
\node at  (234:1)  (D) {};
\node at  (162:1)  (E) {};
\draw (A.center) -- (B.center);
\draw (B.center)--(C.center);
\draw (C.center)--(D.center);
\draw (D.center)--(E.center);
\draw (E.center) -- (A.center);
\draw [white] [name path=bd] (B.center)--(D.center);
\draw  [name path=ce] (C.center)--(E.center);
\path [name intersections={of=bd and ce,by= bdce}];
 	\node[fill=white, inner sep=3pt, rotate=45] at (bdce) {};
\draw (B.center) -- (D.center); 
\draw (B.center)--(E.center);
\draw [white] [name path=ac] (A.center)--(C.center);
\draw [white] [name path=ad] (A.center)--(D.center);
\draw [name path=be] (B.center)--(E.center);
\path [name intersections={of=ad and be,by= adbe}];
\path [name intersections={of=ac and be,by= acbe}];
 	\node[fill=white, inner sep=2pt, rotate=90] at (adbe) {};
	\node[fill=white, inner sep=2pt, rotate=90] at (acbe) {};
	\draw [white] [name path=ad] (A.center)--(D.center);
\path [name intersections={of=ad and ce,by= adce}];
 	\node[fill=white, inner sep=3pt, rotate=45] at (adce) {};
	\draw [white] [name path=ac] (A.center)--(C.center);
\path [name intersections={of=ac and bd,by= acbd}];
 	\node[fill=white, inner sep=3pt, rotate=45] at (acbd) {};
\draw (A.center)--(D.center);
\draw (A.center)--(C.center);
\fill[red]  (A) circle [radius = 0.1cm];
	\fill[red]  (B) circle [radius = 0.1cm];
	\fill[red]  (C) circle [radius = 0.1cm];
	\fill[blue]  (D) circle [radius = 0.1cm];
	\fill[blue]  (E) circle [radius = 0.1cm];
	\node at  (54:1.05)  {2};
	\node at  (126:1.1)  {1};
	\node at  (198:1.05)  {3};
	\node at  (270:1.05)   {1};
	\node at  (342:1.05)  {2};
	\node at  (-50:0.43)  {1};
	\node at  (230:0.43)  {1};
	\node at  (90:0.14)  {1};
	\node at  (115:0.65)  {1};
	\node at  (63:0.65)  {2};
	\node at  (-1.7,0) {$\cG_{5,13}$}  ;
}
\quad
\tikz[baseline=(vert_cent.base)]{
\node (vert_cent) {\hspace{-13pt}$\phantom{-}$};
\node at  (90:1)  (A) {};
\node at  (18:1)  (B) {};
\node at  (306:1)  (C) {};
\node at  (234:1)  (D) {};
\node at  (162:1)  (E) {};
\draw (A.center) -- (B.center);
\draw (B.center)--(C.center);
\draw (C.center)--(D.center);
\draw (D.center)--(E.center);
\draw (E.center) -- (A.center);
\draw [white] [name path=bd] (B.center)--(D.center);
\draw  [name path=ce] (C.center)--(E.center);
\path [name intersections={of=bd and ce,by= bdce}];
 	\node[fill=white, inner sep=3pt, rotate=45] at (bdce) {};
\draw (B.center) -- (D.center); 
\draw (B.center)--(E.center);
\draw [white] [name path=ac] (A.center)--(C.center);
\draw [white] [name path=ad] (A.center)--(D.center);
\draw [name path=be] (B.center)--(E.center);
\path [name intersections={of=ad and be,by= adbe}];
\path [name intersections={of=ac and be,by= acbe}];
 	\node[fill=white, inner sep=2pt, rotate=90] at (adbe) {};
	\node[fill=white, inner sep=2pt, rotate=90] at (acbe) {};
	\draw [white] [name path=ad] (A.center)--(D.center);
\path [name intersections={of=ad and ce,by= adce}];
 	\node[fill=white, inner sep=3pt, rotate=45] at (adce) {};
	\draw [white] [name path=ac] (A.center)--(C.center);
\path [name intersections={of=ac and bd,by= acbd}];
 	\node[fill=white, inner sep=3pt, rotate=45] at (acbd) {};
\draw (A.center)--(D.center);
\draw (A.center)--(C.center);
\fill[red]  (A) circle [radius = 0.1cm];
	\fill[red]  (B) circle [radius = 0.1cm];
	\fill[red]  (C) circle [radius = 0.1cm];
	\fill[red]  (D) circle [radius = 0.1cm];
	\fill[red]  (E) circle [radius = 0.1cm];
	\node at  (54:1.05)  {2};
	\node at  (126:1.1)  {2};
	\node at  (198:1.05)  {2};
	\node at  (270:1.05)   {2};
	\node at  (342:1.05)  {2};
	\node at  (-50:0.43)  {1};
	\node at  (230:0.43)  {1};
	\node at  (90:0.14)  {1};
	\node at  (117:0.65)  {1};
	\node at  (63:0.65)  {1};
	\node at  (-1.7,0) {$\cG_{5,14}$}  ;
}
\label{Vfive}
\end{align}
Here the graphs $\cG_{5,4}$, $\cG_{5,12}$  and $\cG_{5,14}$ are primitive with $\cG_{5,14}$ vertex transitive.

For the three and four vertex vacuum graphs imposing \eqref{Athm} leads to one relation for each graph
\be
3\, A_3 = 2\, c_2 \, , 
\ee
and
\begin{align}
& 4 \, A_{4,1} = 3 \, d_4 \, ,  && 4\, A_{4,2} = 0 \, ,   && 4 \, A_{4,3} = 2\, T_3 \, c_2 \, , \nn \\
& 4 \, A_{4,4}= 30 \, c_{4a} \, ,   && 4\, A_{4,5} = 60 \, c_{4b} + 18 \, T_3 \, c_2 \, , &&
4\, A_{4,6} = 90\, c_{4c} \, .
\label{A4C}
\end{align}
Clearly there are no constraints on the $\beta$-function at this order. \eqref{A4C} is invariant under
\be
\delta A_{4,3} = 40\, g_2 \, c_2{\!}^2 \, , \qquad \delta A_{4,5} = 360\, g_2 \,  c_2{\!}^2 \, , \qquad
\delta T_3 = 80\, g_2 \, c_2 \, ,
\label{A3var}
\ee
reflecting the freedom in \eqref{Avar}.

Non trivial relations arise when considering the 5 vertex vacuum graphs. Applying \eqref{Athm} gives
for each vacuum graph and for each inequivalent vertex in turn
\begin{align}
5 \, A_{5,1} = {}& 2(T_{4,3ab}+T_{4,3ac}) c_2 \, , \nn \\
\noalign{\vskip 1pt}
  A_{5,2} = {}& 20 \, c_{6m} + \tfrac32 \, T_3 \, d_{4\gamma}
\, , &&   \hskip -7.5cm 2 \, A_{5,2} =  20 \, T_{4,1ab} \, c_2 + \tfrac32 \, T_3 \, d_{4\gamma}
= 20(T_{4,1ac}+T_{4,1ad}) c_2 \, , \nn \\
\noalign{\vskip 1pt}
A_{5,3} = {}&  20 \, c_{6v} \, ,  && \hskip -7.5cm 
2 \, A_{5,3} =  8 \, T_{4, 2ab}\, c_2 = 8 (T_{4,2ac}+ T_{4,2ad} ) c_2 \, , \nn \\
  A_{5,4} = {}& 0  \, ,  && \hskip -7.5cm  2 \, A_{5,4} = 15\, c_{6a}  = 0 \, , \nn \\
  \noalign{\vskip 1pt}
  A_{5,5} = {}& 15 \, c_{6b} + 12 \, T_{4,2ac} \, c_2 = 12 \, T_{4,2ab} \, c_2= 12 \, T_{4,2ad} \, c_2 \, , \qquad
 2 \, A_{5,5} = 60\, c_{6w} \, , \nn \\
 \noalign{\vskip 1pt}
  A_{5,6} = {}& 20\, c_{6l}   + 18 \, T_{4,3ac}\, c_2 + 3\, T_3 \, c_{4b} \, , \nn \\
\noalign{\vskip -2pt}
  2 \, A_{5,6} = {}& 18 \, T_{4,3ab} \, c_2 +  2 \, T_{4,5ab}\, c_2  + 3\, T_3 \, c_{4b}
=  60\, c_{6e}+  2 \,( T_{4,5ac}+ T_{4,5ad}) c_2 \, , \nn \\
\noalign{\vskip 1pt}
   A_{5,7} = {}& 20\, c_{6n} + 6\, T_3 \, c_{4a}  \, , \nn \\
\noalign{\vskip -2pt}
  2 \, A_{5,7} = {}& 8 \, T_{4,4ab} \, c_2 +  6\, T_3 \, c_{4a} 
= 30\, c_{6c} +  8 \, T_{4,4ac} \, c_2 \, , \nn \\
\noalign{\vskip 1pt}
   A_{5,8} = {}& 180\, c_{6f} +  6 \, T_{4,5ab}\, c_2  \, , \nn \\ 
\noalign{\vskip -2pt}
  2 \, A_{5,8} = {}& 60\, c_{6g}  +  6 \, T_{4,5ac}\, c_2 +  9\, T_3 \, c_{4b} 
= 60\, c_{6k} +  6 \, T_{4,5ad}\, c_2 +  9\, T_3 \, c_{4b} \, , \nn \\
\noalign{\vskip 1pt}
  A_{5,9} = {}&180\, c_{6r} + 4\, T_{4,6ab} \, c_2  \, , && \hskip -7.5cm 
 2 \, A_{5,9} =  90\, c_{6q} + 8\, T_{4,6ab} \, c_2 = 60\, c_{6s} +  54\, T_3 \, c_{4c} \, , \nn \\
 \noalign{\vskip 1pt}
  A_{5,10} = {}& 180\, c_{6p}  +  12 \, T_{4,4ab}\, c_2 \, , \nn \\
\noalign{\vskip -2pt}
   2 \, A_{5,10} = {}&  60\, c_{6o} + 18\, T_3  \, c_{4a}   
= 30\, c_{6d} +  12 \, T_{4,4ac}\, c_2  \, , \nn \\
\noalign{\vskip 1pt}
   A_{5,11} = {}& 180\, c_{6j}  +  12 \, T_{4,5ab}\, c_2 \, ,\nn \\
\noalign{\vskip -2pt}
 2 \, A_{5,11}=  {}& 120\, c_{6h} +   12 \, T_{4,5ad} \, c_2 +  18\, T_3\, c_{4b} = 
60\, c_{6i} +   12 \, T_{4,5ac}\, c_2 +  18\, T_3\, c_{4b}  \, , \nn \\
\noalign{\vskip 1pt}
2\, A_{5,12}=  {}& 90\, c_{6x}   && \hskip -7.5cm 3\, A_{5,12} = 180 \, c_{6y} \, ,\nn \\
\noalign{\vskip 1pt}
2\, A_{5,13}=  {}& 120\, c_{6t}   +  36\, T_3 \, c_{4c} \, , && \hskip -7.5cm 3\, A_{5,13} = 180 \, c_{6u} 
+ 8\, T_{4,6ab} \, c_2\, ,\nn \\
\noalign{\vskip 1pt}
5\, A_{5,14} = {}& 180\, c_{6z} \, .
\label{A5C}
\end{align}
Each of the 14 vacuum graphs $\cG_5$
corresponds to a unique $N$ polynomial $N_\cG$ given, up to an overall constant, by those
 listed in tables \eqref{6loopN}, \eqref{6loop2N},  \eqref{6loop3N} and \eqref{6loopPN}. These
 polynomials  are the same
 for every vertex graph associated on the right hand sides of \eqref{A5C} for each $A_{5}$.

Along with $\delta T_3$ from  \eqref{A3var}, \eqref{A5C} are invariant under the variations
\begin{align}
&\delta A_{5,1} = 4\, g_3 c_2{\!}^2\, , \quad 
\delta A_{5,2} = 120\, g_2 c_2 d_{4\gamma} \, , \quad 
\delta A_{5,6} = 240\, g_2 c_2 c_{4b} + 72\, g_3 c_2{\!}^2\ \, , \nn \\
& \delta A_{5,7} = 480\, g_2 c_2 c_{4a} \, , \quad
\delta A_{5,8} = 720\, g_2 c_2 c_{4b} + 108\, g_3 c_2{\!}^2 \, , \quad
\delta A_{5,9} = 2160\, g_2 c_2 c_{4c} \, , \nn \\
& \delta A_{5,10} = 720\, g_2 c_2 c_{4a} \, , \quad
 \delta A_{5,11} = 1440\, g_2 c_2 c_{4b} + 216 \, g_3 c_2{\!}^2 \, ,  \qquad 
\delta A_{5,13} =1440\, g_2 c_2 c_{4c}\nn  \\
& \delta T_{4,1ab} = \delta T_{4,1ac}=\delta T_{4,1ad} = 6\, g_2 d_{4\gamma}\, \qquad
\delta T_{4,3ab} =  6\, g_3 c_2\, , \quad \delta T_{4,3ac} =  4\, g_3 c_2  \, , \nn \\
&\delta T_{4,4ab} = 60\, g_2  c_{4a} \, , \quad \delta T_{4,4ac} = 120\, g_2  c_{4a} \, , \nn \\
& \delta T_{4,5ab} = 120\, g_2 c_{4b} + 18 \, g_3 c_2\, ,  
\qquad \delta T_{4,5ac}=\delta T_{4,5ad} = 120\, g_2 c_{4b} + 36 \, g_3 c_2\, , \nn \\
&  \delta T_{4,6ab} = 540\, g_2  c_{4c} \, .
\label{A4var}
\end{align}

For the $\beta$-function coefficients corresponding to primitive graphs from \eqref{A5C} 
\be
c_{6a} = 0 \, , \qquad 2\, c_{6v} -c_{6w} =0 \, ,  \qquad 2\, c_{6w} -c_{6b} =0 \, , \qquad 3\, c_{6x} - 4\, c_{6y} = 0 \, .
\label{rel1}
\ee
A further 2 term relation is
\be
4\, c_{6r} - c_{6q}= 0 \, ,
\label{rel2}
\ee
and also there are four 4 term linear relations and one 3 term relation,
\begin{align}
& 6\, c_{6f} + c_{6h} - 3 \, c_{6j} - c_{6k}=0 \, ,  &&  6\, c_{6r} + 3\, c_{6t} - c_{6s} - 3\, c_{6u} = 0 \, ,\nn \\
& 2(c_{6k} - c_{6h} - c_{6g} ) + c_{6i} =0\, ,   && 3\, c_{6c} + 4\, c_{6n}  + 24 \, c_{6p} - 2\, c_{6d}= 0\, ,  \nn\\
& 2\, c_{6n} + 6 \, c_{6p} - c_{6o} =  0 \, .
\label{rel34}
\end{align}
All these relations are satisfied by the calculated  results given in the tables for $c_{6a}, \dots,c_{6z}$
and provide constraints on 22 of the coefficients. Manifestly changing the contribution coresponding
to  one graph, such as $c_{6s}$, would vitiate the agreement satisfied by our results. The relation between
$c_{6x}$ and $c_{6y}$ in \eqref{rel1} is a consequence of the arguments of \cite{SchnetzPeriod} when joining
the external lines of a primitive graphs with an extra vertex generates a common vacuum graph.

The coefficients appearing in \eqref{rel2} and \eqref{rel34}  reflect the various  symmetry factors
since
\begin{align}
\Delta( \cG_{6r}- \cG_{6q} ) ={}&  \Delta(\cG_{6f} + \cG_{6h}-\cG_{6j}-\cG_{6k})\nn \\
= {}&\Delta(\cG_{6r} + \cG_{6t}-\cG_{6s}-\cG_{6u}) = \Delta(\cG_{6i} + \cG_{6k}-\cG_{6g}-\cG_{6h})
=\emptyset  \, , 
\end{align}
and
\begin{align}
 \Delta(\cG_{6c} -\cG_{6d}+\cG_{6n}+\cG_{6p}) = {}&2( \cG_2\otimes \cG_{4a} + \cG_{4a}\otimes \cG_2)\,, \nn \\
  \Delta(\cG_{6n} +\cG_{6p}-\cG_{6o}) = {}&  \cG_2\otimes \cG_{4a} + \cG_{4a}\otimes \cG_2 \, .
\end{align}
The symmetric or null coproducts ensure that the corresponding results  in \eqref{rel34} are scheme
invariant.

After eliminating the various $A_5{}^\prime$s from \eqref{A5C} there remain 22 equations. These entail the 10
consistency relations in \eqref{rel1}, \eqref{rel2} and \eqref{rel34} leaving  12 equations  constraining 
the 14 possible $T_{5}{}^\prime$s,
\begin{align}
& c_2 \, T_{4,1ab}- \tfrac{3}{40} \, d_{4\gamma} \, T_3 
= c_2 ( T_{4,1ac} +T_{4,1ad} ) - \tfrac{3}{20} \, d_{4\gamma} \, T_3  = 2\, c_{6m} \, , \nn \\
& c_2\, T_{4,2ab}= c_2\, T_{4,2ad} = 5\, c_{6v} \,, \qquad \quad \ \ c_2\, T_{4,2ac} = 0  \,,  \nn \\
& 4c_2\, T_{4,4ab} -  3\, c_{4a}\, T_3 = 20\, c_{6n} \, , \qquad \quad
2c_2\, T_{4,4ac} -  3\, c_{4a}\, T_3 = 10\, c_{6o} - 5\, c_{6d} \, , \nn \\
& 2c_2 ( 2\, T_{4,5ab} - T_{4,5ac} ) -  3\, c_{4b} \, T_3 = 20 \, c_{6g} -120\, c_{6f}\, , \nn \\
& 2c_2 ( 2\, T_{4,5ab} - T_{4,5ad} ) -  3\, c_{4b} \, T_3 = 20 \, c_{6k} -120\, c_{6f}\, , \nn \\
& 4\, c_2 \, T_{4,6ab} - 27 \, c_{4c}\, T_3 = 90 (c_{6t} - c_{6u} ) \, , \nn \\
& 18\, c_2 \, T_{4,3ab} + 2 c_2 ( T_{4,5ab}- T_{4,5ac} - T_{4,5ad} ) + 3\, c_{4b}\, T_3 = 60\, c_{6e} \, , \nn \\
& 18\, c_2 \, T_{4,3ac} - c_2 (  T_{4,5ac} + T_{4,5ad} ) + 3\, c_{4b}\, T_3 = 30\, c_{6e} -20\, c_{6l} \, .
\label{Trel}
\end{align}
The remaining ambiguity  arises from the variations in  \eqref{A3var} and \eqref{A4var} since the 
left hand sides of \eqref{Trel} are independent of $g_2,g_3$ parameterising the variations in \eqref{A3var}.

\section{Results for Curvature}

A symmetric $T_{IJ}$ plays the role of a metric on the manifold defined by the couplings $\lambda^I$ as coordinates.
Given a general metric it is natural to calculate the associated Riemann tensor and scalar curvature. For
a general $\phi^4$ theory in four dimensions this was carried out in \cite{Pannell2}.

An expansion of the metric
\be
G_{\mu\nu}\, \rmd x_\mu \rmd x_\nu = \rmd x_\mu \rmd x_\mu + A_{\mu\nu\sigma} \,  \rmd x_\mu \rmd x_\nu \, x_\sigma 
+ B_{\mu\nu\sigma\rho} \,  \rmd x_\mu \rmd x_\nu \,  x_\sigma x_\rho + \dots \, , 
\label{met1}
\ee
is sufficient to determine the Riemann tensor at $x=0$. Here $A_{\mu\nu\sigma} = A_{(\mu\nu)\sigma}, \
 B_{\mu\nu\sigma\rho} = B_{(\mu\nu)(\sigma\rho)}$ and they have the decomposition into irreducible tensorial
 components given by 
 $  \raisebox{0pt}{ {\tiny  \yng(2) } }\times  \raisebox{0pt}{ {\tiny  \yng(1) } } \simeq
 \raisebox{0pt}{ \tiny  \yng(3)  } + 
 \raisebox{-6pt}{\tiny  \yng(2,1) } 
 $
 and
 $  \raisebox{0pt}{ {\tiny  \yng(2) } }\times  \raisebox{0pt}{ \tiny  \yng(2) } \simeq
 \raisebox{0pt}{ {\tiny  \yng(4)   } } + 
 \raisebox{-6pt}{ {\tiny  \yng(3,1)} } + \raisebox{-6pt}{ {\tiny  \yng(2,2)} } 
 $.  The Riemann tensor corresponds to the $\raisebox{-6pt}{ {\tiny  \yng(2,2)} } $ contribution.
 This can be isolated by choosing normal coordinates. For a change of variable
 \be
 x_\mu = y_\mu + \tfrac 12\, a_{\mu\alpha\beta}\, y_\alpha y_\beta + 
  \tfrac 13\, b_{\mu\alpha\beta\gamma}\, y_\alpha y_\beta  y_\gamma + \dots \, ,
  \ee
  with $a_{\mu\alpha\beta} = a_{\mu(\alpha\beta)}, \ b_{\mu(\alpha\beta\gamma)}$ and where the $b$ tensor
 is decomposed as $\raisebox{0pt}{ {\tiny  \yng(3) } }\times  \raisebox{0pt}{ {\tiny  \yng(1) } } \simeq
  \raisebox{0pt}{ {\tiny  \yng(4)   } } + \raisebox{-6pt}{ {\tiny  \yng(3,1)} }$. 
  Defining
  \be
  G_{\mu\nu}\, \rmd x_\mu \rmd x_\nu = {\tilde G}_{\mu\nu}\, \rmd y_\mu \rmd y_\nu \, ,
  \ee
  then if 
  \be
  a_{\mu \sigma \nu} + a_{\nu \sigma \mu} + A_{\mu\nu\sigma}=0 \quad \Rightarrow \quad
  a_{\sigma\mu\nu} = -\tfrac12 ( A_{\nu \sigma \nu}+  A_{\mu \sigma \mu}-  A_{\mu\nu   \sigma} )\, ,
  \label{asol}
  \ee
  the expansion in \eqref{met1} takes the form
  \be
{\tilde G}_{\mu\nu}\, \rmd y_\mu \rmd y_\nu = \rmd y_\mu \rmd y_\mu 
+ {\tilde B}_{\mu\nu\sigma\rho} \,  \rmd y_\mu \rmd y_\nu \,  y_\sigma y_\rho + \dots \, , 
\label{met2}
\ee
with
\begin{align}
  {\tilde B}_{\mu\nu\sigma\rho} ={}& B_{\mu\nu\sigma\rho}  + b_{\mu\nu\sigma \rho} +  b_{\nu\mu\sigma \rho} 
  + \tfrac12( a_{\alpha \mu \sigma}\, a_{\alpha \nu \rho} 
  + a_{\alpha \mu \rho} \, a_{\alpha \nu \sigma} )  + \tfrac12\, A_{\mu\nu\alpha} \,a_{\alpha \sigma\rho } \nn \\
  &{}+ \tfrac12 ( A_{\mu\alpha\sigma} \,a_{\alpha \nu\rho } + A_{\mu\alpha\rho} \,a_{\alpha \nu\sigma} 
  + A_{\nu\alpha\sigma} \,a_{\alpha \mu\rho }+ A_{\nu\alpha\rho} \,a_{\alpha \mu\sigma } ) \, ,
\end{align}
where $a$ may be eliminated using \eqref{asol}. For simplicity assuming $A_{\mu\nu\sigma}$ is totally symmetric,
as is relevant later, 
\be
{\tilde B}_{\mu\nu\sigma\rho} =  B_{\mu\nu\sigma\rho} + b_{\mu\nu\sigma \rho} +  b_{\nu\mu\sigma \rho} 
 -\tfrac14 \, A_{\mu\nu \alpha}\, A_{\sigma\rho\alpha} 
- \tfrac38 ( A_{\mu\sigma \alpha}\, A_{\nu\rho\alpha} + A_{\mu\rho \alpha}\, A_{\nu \sigma\alpha} )\, .
\ee

In general we may decompose ${\tilde B}$ as
\begin{align}
{\tilde B}_{\mu(\nu\sigma\rho)} + {\tilde B}_{\nu(\mu\sigma\rho)} - \tfrac12\big ( {\tilde B}_{\sigma(\mu\nu\rho)} 
+ {\tilde B}_{\rho(\mu\nu\sigma)}\big  ) -\tfrac23 \big ( {\tilde B}_{[\mu[\rho\, \sigma]\nu]} 
+ {\tilde B}_{[\mu[\sigma\, \rho]\nu]} \big )\, ,
\end{align}
with the terms involving antisymmetrising more explicity given by taking
\be
{\tilde B}_{[\mu[\rho\, \sigma]\nu]} = \tfrac14 \big ( {\tilde B}_{\mu\rho\sigma\nu} - {\tilde B}_{\sigma\rho\nu\nu} 
-  {\tilde B}_{\mu\nu\sigma\rho} +  {\tilde B}_{\sigma\nu\mu\rho} \big ) \, .
\ee
Choosing $b$ appropriately reduces the metric to standard normal coordinate form where
\begin{align}
& {\tilde G}_{\mu\nu}\, \rmd y_\mu \rmd y_\nu = \rmd y_\mu \rmd y_\mu 
-\tfrac13\, R _{\mu\sigma\nu\rho} \,  \rmd y_\mu \rmd y_\nu \,  y_\sigma y_\rho + \dots \, , \nn \\
 & R _{\mu\sigma\nu\rho} +  R _{\mu\rho\nu\sigma} = 4 \big ( {\tilde B}_{[\mu[\rho\, \sigma]\nu]} 
+ {\tilde B}_{[\mu[\sigma\, \rho]\nu]} \big ) \, ,
 \end{align}
 or
 \be
 R_{\mu\sigma\rho\nu} = {\tilde B}_{\mu\nu\,\sigma\rho} + {\tilde B}_{\sigma\rho\,\mu\nu}
 - {\tilde B}_{\mu\rho\,\sigma\nu} - {\tilde B}_{\sigma\nu\,\mu\rho} \, .
 \ee
 It is easy to verify that this satisfies the necessary symmetries for the Riemann tensor.
 
For application to $\phi^6$ theory
 \be
A_{IJK}  = T_3 
\raisebox{-4pt}{\tikz[baseline=(vert_cent.base),scale=0.4]{
	\node (vert_cent) {\hspace{-13pt}$\phantom{-}$};
	\vertex at \coord{0} (A) {};
	\vertex at \coord{-120} (B) {};
	\vertex at \coord{120} (C) {};
	\draw (A) to (B)   (B) to  (C)  (C) to (A);
	\node at (0,1.5) {$\scriptstyle I$};
	\node at (1.2,-0.5) {$\scriptstyle J$};
	\node at (-1.4,-0.5) {$\scriptstyle K$};
	\node at (30:0.95) {$\scriptstyle 3$};
	\node at (150:0.95) {$\scriptstyle 3$};
	\node at (270:0.95) {$\scriptstyle 3$};
		\fill[black]  (A) circle [radius = 0.08cm];
	\fill[black]  (B) circle [radius = 0.08cm];
	\fill[black]  (C) circle [radius = 0.08cm];
	} } \, , 
\ee
which is clearly symmetric. For determining  ${\tilde B}$ we require
\be
A_{KLM} A_{MIJ} \lambda^K \lambda^L \rmd \lambda^I  \rmd \lambda^J = T_3{}^2
\tikz[baseline=(vert_cent.base),scale=0.4]{
  \coordinate (A1) at (3.5,1);
  \coordinate (B1) at (3.5,-1);
  \node at (4,0) {$\scriptstyle 3$};
   \draw (B1) to (A1);
       \coordinate (A2) at (0.5,1);
  \coordinate (B2) at (0.5,-1);
  \draw (B2) to (A2) ;
   \node at (0,0) {$\scriptstyle 3$};
  \draw (2.5,-1.1) -- (2.5,1.1)
        (2.5,1.1) -- (1.5,1.1)
        (1.5,1.1) -- (1.5,-1.1)
        (1.5,-1.1) -- (2.5, -1.1);
  \coordinate (R) at (2.5,0.95);
  \coordinate (S) at (2.5,0.1);
  \coordinate (Q) at (2.5,0.55);
  \coordinate (T) at (2.5,-0.1);
  \coordinate (U) at (2.5,-0.95);
   \coordinate (V) at (2.5,-0.55);
  \coordinate (R1) at  (1.5,0.95);
  \coordinate (S1) at (1.5,0.1);
  \coordinate (Q1) at (1.5,0.55);
  \coordinate (T1) at (1.5,-0.1);
  \coordinate (U1) at (1.5,-0.95);
    \coordinate (V1) at (1.5,-0.55);
  \draw  [bend right= 5] (A1) to (R);
  \draw [bend  left = 23](A1) to (S);
   \draw [bend  left = 12](A1) to (Q);
 \draw [bend right= 23] (B1) to (T);
   \draw  [bend left = 5]  (B1) to (U);
    \draw [bend  left = 12](B1) to (V);
   \draw  [bend left= 5] (A2) to (R1);
  \draw [bend  right = 23](A2) to (S1);
    \draw [bend  right = 12](A2) to (Q1);
 \draw [bend left= 23] (B2) to (T1);
   \draw  [bend right= 5]  (B2) to (U1);
      \draw [bend  right = 12](B2) to (V1);
\node at (3.7,1.4) {$\scriptstyle \rmd \lambda$};
\node at (3.7,-1.4) {$\scriptstyle  \rmd \lambda$};
\node at (0.3,1.4) {$\scriptstyle \lambda$};
\node at (0.3,-1.4) {$\scriptstyle \lambda $};
\fill[black]  (A1) circle [radius = 0.08cm];
	\fill[black]  (B1) circle [radius = 0.08cm];
	\fill[black]  (A2) circle [radius = 0.08cm];
	\fill[black]  (B2) circle [radius = 0.08cm];
 \node at (2,0) {$\delta$};
  }
  =  \ \frac{1}{10} \, T_3{}^2  \left ( 
\tikz[baseline=(vert_cent.base),scale=0.5]{
	\node (vert_cent) {\hspace{-13pt}$\phantom{-}$};
	\vertex at \coord{-45} (A) {};
	\vertex at \coord{-135} (B) {};
	\vertex at \coord{135} (C) {};
	\vertex at \coord{45} (D) {};
	\draw  (A) to (B)  (B) to (C)  (C) to (D)  (D) to (A); 
	\node at (1.25,0.85) {$\scriptstyle \rmd \lambda$};
	\node at (1.2,-0.85) {$\scriptstyle \rmd \lambda$};
	\node at (-1.1,-0.85) {$\scriptstyle \lambda $};
	\node at (-1.1,0.85) {$\scriptstyle \lambda$};
	\node at (1.1,0) {$\scriptstyle 3 $};
	\node at (-1.1,0) {$\scriptstyle 3 $};
	\node at (0,1.1) {$\scriptstyle 3 $};
	\node at (0,-1.1) {$\scriptstyle 3 $};
	\fill[black]  (A) circle [radius = 0.08cm];
	\fill[black]  (B) circle [radius = 0.08cm];
	\fill[black]  (C) circle [radius = 0.08cm];
	\fill[black]  (D) circle [radius = 0.08cm];
} 
+ 9 \tikz[baseline=(vert_cent.base),scale=0.5]{
	\node (vert_cent) {\hspace{-13pt}$\phantom{-}$};
	\vertex at \coord{-45} (A) {};
	\vertex at \coord{-135} (B) {};
	\vertex at \coord{135} (C) {};
	\vertex at \coord{45} (D) {};
	\draw  (A) to (B)  (B) to (C)  (C) to (D)  (D) to (A); 
	\draw (B)--(D); 
	\fill[white]  (0,0) circle [radius = 0.1cm];
	\draw (C)--(A); 
	\node at (1.25,0.85) {$\scriptstyle \rmd \lambda$};
	\node at (1.2,-0.85) {$\scriptstyle \rmd \lambda$};
	\node at (-1.1,-0.85) {$\scriptstyle \lambda $};
	\node at (-1.1,0.85) {$\scriptstyle \lambda$};
	\node at (1.1,0) {$\scriptstyle 3 $};
	\node at (-1.1,0) {$\scriptstyle 3 $};
	\node at (0,1.1) {$\scriptstyle 2 $};
	\node at (0,-1.1) {$\scriptstyle 2 $};
	\node at (-0.1,0.46) {$\scriptstyle 1$};
	\node at (-0.1,-0.46) {$\scriptstyle 1 $};		
	\fill[black]  (A) circle [radius = 0.08cm];
	\fill[black]  (B) circle [radius = 0.08cm];
	\fill[black]  (C) circle [radius = 0.08cm];
	\fill[black]  (D) circle [radius = 0.08cm];
} 
\right )  \, ,
\ee
where $
\raisebox{-0.4cm}{\tikz[baseline=(vert_cent.base),scale=0.75]{
   \draw (0.5,0) -- (0.5,1)
        (0.5,1) -- (0,1)
        (0,1) -- (0,0)
        (0,0) -- (0.5, 0);
  \coordinate (R) at (0.5,0.02);
  \coordinate (S) at (0.5,0.22);
  \coordinate (T) at (0.5,0.42);
  \coordinate (U) at (0.5,0.62);
    \coordinate (V) at (0.5,0.82);
      \coordinate (W) at (0.5,0.98);
  \draw (R) --+ (0:0.4);
  \draw (S) --+ (0:0.4);
  \draw (T) --+ (0:0.4);
  \draw (U) --+ (0:0.4);
  \draw (V) --+ (0:0.4);
  \draw (W) --+ (0:0.4);
  \coordinate (R1) at  (0,0.02);
  \coordinate (S1) at (0,0.22);
  \coordinate (T1) at (0,0.42);
  \coordinate (U1) at (0,0.62);
  \coordinate (V1) at (0,0.82);
  \coordinate (W1) at (0,0.98);
  \draw (R1) --+ (180:0.4);
  \draw (S1) --+ (180:0.4);
  \draw (T1) --+ (180:0.4);
  \draw (U1) --+ (180:0.4);
   \draw (V1) --+ (180:0.4);
  \draw (W1) --+ (180:0.4);
 \node at (0.25,0.5) {$\delta$};
  }}
  \simeq \delta_{ijklmn,pqrstu}
$
acts as the identity for symmetric 6 index tensors with total dimension
$\delta_{ijklmn,ijklmn} = \frac{1}{720}N(N+1)(N+2)(N+3)(N+4)(N+5)$.
Starting from the graphs in \eqref{Vfour} which determine the possible  contributions to $T$
we then have
\begin{align}
{\tilde B}_{IJKL}  \lambda^K \lambda^L \rmd \lambda^I  \rmd \lambda^J 
= {}& T_{4,1ab} \tikz[baseline=(vert_cent.base),scale=0.5]{
	\node (vert_cent) {\hspace{-13pt}$\phantom{-}$};
	\vertex at \coord{-45} (A) {};
	\vertex at \coord{-135} (B) {};
	\vertex at \coord{135} (C) {};
	\vertex at \coord{45} (D) {};
	\draw  (A) to (B)  (B) to (C)  (C) to (D)  (D) to (A); 
	\node at (1.25,0.85) {$\scriptstyle \rmd \lambda$};
	\node at (1.2,-0.85) {$\scriptstyle \rmd \lambda$};
	\node at (-1.1,-0.85) {$\scriptstyle \lambda $};
	\node at (-1.1,0.85) {$\scriptstyle \lambda$};
	\node at (1.1,0) {$\scriptstyle 5 $};
	\node at (-1.1,0) {$\scriptstyle 5 $};
	\node at (0,1.1) {$\scriptstyle 1 $};
	\node at (0,-1.1) {$\scriptstyle 1 $};
	\fill[black]  (A) circle [radius = 0.08cm];
	\fill[black]  (B) circle [radius = 0.08cm];
	\fill[black]  (C) circle [radius = 0.08cm];
	\fill[black]  (D) circle [radius = 0.08cm];
} 
+ T_{4,1ac} \tikz[baseline=(vert_cent.base),scale=0.5]{
	\node (vert_cent) {\hspace{-13pt}$\phantom{-}$};
	\vertex at \coord{-45} (A) {};
	\vertex at \coord{-135} (B) {};
	\vertex at \coord{135} (C) {};
	\vertex at \coord{45} (D) {};
	\draw  (A) to (B)  (B) to (C)  (C) to (D)  (D) to (A); 
	\node at (1.25,0.85) {$\scriptstyle \rmd \lambda$};
	\node at (1.2,-0.85) {$\scriptstyle  \lambda$};
	\node at (-1.1,-0.85) {$\scriptstyle \rmd \lambda $};
	\node at (-1.1,0.85) {$\scriptstyle \lambda$};
	\node at (1.1,0) {$\scriptstyle 5 $};
	\node at (-1.1,0) {$\scriptstyle 5 $};
	\node at (0,1.1) {$\scriptstyle 1 $};
	\node at (0,-1.1) {$\scriptstyle 1 $};
	\fill[black]  (A) circle [radius = 0.08cm];
	\fill[black]  (B) circle [radius = 0.08cm];
	\fill[black]  (C) circle [radius = 0.08cm];
	\fill[black]  (D) circle [radius = 0.08cm];
} 
+
 T_{4,1ad} \tikz[baseline=(vert_cent.base),scale=0.5]{
	\node (vert_cent) {\hspace{-13pt}$\phantom{-}$};
	\vertex at \coord{-45} (A) {};
	\vertex at \coord{-135} (B) {};
	\vertex at \coord{135} (C) {};
	\vertex at \coord{45} (D) {};
	\draw  (A) to (B)  (B) to (C)  (C) to (D)  (D) to (A); 
	\node at (1.25,0.85) {$\scriptstyle \rmd \lambda$};
	\node at (1.2,-0.85) {$\scriptstyle  \lambda$};
	\node at (-1.1,-0.85) {$\scriptstyle \lambda $};
	\node at (-1.2,0.85) {$\scriptstyle \rmd \lambda$};
	\node at (1.1,0) {$\scriptstyle 5 $};
	\node at (-1.1,0) {$\scriptstyle 5 $};
	\node at (0,1.1) {$\scriptstyle 1 $};
	\node at (0,-1.1) {$\scriptstyle 1 $};
	\fill[black]  (A) circle [radius = 0.08cm];
	\fill[black]  (B) circle [radius = 0.08cm];
	\fill[black]  (C) circle [radius = 0.08cm];
	\fill[black]  (D) circle [radius = 0.08cm];
} \nn \\
\noalign {\vskip -6pt}
&{}+ T_{4,2ab} \tikz[baseline=(vert_cent.base),scale=0.5]{
	\node (vert_cent) {\hspace{-13pt}$\phantom{-}$};
	\vertex at \coord{-45} (A) {};
	\vertex at \coord{-135} (B) {};
	\vertex at \coord{135} (C) {};
	\vertex at \coord{45} (D) {};
	\draw  (A) to (B)  (B) to (C)  (C) to (D)  (D) to (A); 
	\node at (1.25,0.85) {$\scriptstyle \rmd \lambda$};
	\node at (1.2,-0.85) {$\scriptstyle \rmd \lambda$};
	\node at (-1.1,-0.85) {$\scriptstyle \lambda $};
	\node at (-1.1,0.85) {$\scriptstyle \lambda$};
	\node at (1.1,0) {$\scriptstyle 4 $};
	\node at (-1.1,0) {$\scriptstyle 4 $};
	\node at (0,1.1) {$\scriptstyle 2 $};
	\node at (0,-1.1) {$\scriptstyle 2 $};
	\fill[black]  (A) circle [radius = 0.08cm];
	\fill[black]  (B) circle [radius = 0.08cm];
	\fill[black]  (C) circle [radius = 0.08cm];
	\fill[black]  (D) circle [radius = 0.08cm];
} 
+ T_{4,2ac} \tikz[baseline=(vert_cent.base),scale=0.5]{
	\node (vert_cent) {\hspace{-13pt}$\phantom{-}$};
	\vertex at \coord{-45} (A) {};
	\vertex at \coord{-135} (B) {};
	\vertex at \coord{135} (C) {};
	\vertex at \coord{45} (D) {};
	\draw  (A) to (B)  (B) to (C)  (C) to (D)  (D) to (A); 
	\node at (1.25,0.85) {$\scriptstyle \rmd \lambda$};
	\node at (1.2,-0.85) {$\scriptstyle  \lambda$};
	\node at (-1.1,-0.85) {$\scriptstyle \rmd \lambda $};
	\node at (-1.1,0.85) {$\scriptstyle \lambda$};
	\node at (1.1,0) {$\scriptstyle 4 $};
	\node at (-1.1,0) {$\scriptstyle 4 $};
	\node at (0,1.1) {$\scriptstyle 2 $};
	\node at (0,-1.1) {$\scriptstyle 2 $};
	\fill[black]  (A) circle [radius = 0.08cm];
	\fill[black]  (B) circle [radius = 0.08cm];
	\fill[black]  (C) circle [radius = 0.08cm];
	\fill[black]  (D) circle [radius = 0.08cm];
} 
+
 T_{4,2ad} \tikz[baseline=(vert_cent.base),scale=0.5]{
	\node (vert_cent) {\hspace{-13pt}$\phantom{-}$};
	\vertex at \coord{-45} (A) {};
	\vertex at \coord{-135} (B) {};
	\vertex at \coord{135} (C) {};
	\vertex at \coord{45} (D) {};
	\draw  (A) to (B)  (B) to (C)  (C) to (D)  (D) to (A); 
	\node at (1.25,0.85) {$\scriptstyle \rmd \lambda$};
	\node at (1.2,-0.85) {$\scriptstyle  \lambda$};
	\node at (-1.1,-0.85) {$\scriptstyle \lambda $};
	\node at (-1.2,0.85) {$\scriptstyle \rmd \lambda$};
	\node at (1.1,0) {$\scriptstyle 4 $};
	\node at (-1.1,0) {$\scriptstyle 4 $};
	\node at (0,1.1) {$\scriptstyle 2 $};
	\node at (0,-1.1) {$\scriptstyle 2 $};
	\fill[black]  (A) circle [radius = 0.08cm];
	\fill[black]  (B) circle [radius = 0.08cm];
	\fill[black]  (C) circle [radius = 0.08cm];
	\fill[black]  (D) circle [radius = 0.08cm];
} \nn \\
\noalign {\vskip -6pt}
&{}+ {\hat T}_{4,3ab} \tikz[baseline=(vert_cent.base),scale=0.5]{
	\node (vert_cent) {\hspace{-13pt}$\phantom{-}$};
	\vertex at \coord{-45} (A) {};
	\vertex at \coord{-135} (B) {};
	\vertex at \coord{135} (C) {};
	\vertex at \coord{45} (D) {};
	\draw  (A) to (B)  (B) to (C)  (C) to (D)  (D) to (A); 
	\node at (1.25,0.85) {$\scriptstyle \rmd \lambda$};
	\node at (1.2,-0.85) {$\scriptstyle \rmd \lambda$};
	\node at (-1.1,-0.85) {$\scriptstyle \lambda $};
	\node at (-1.1,0.85) {$\scriptstyle \lambda$};
	\node at (1.1,0) {$\scriptstyle 3 $};
	\node at (-1.1,0) {$\scriptstyle 3 $};
	\node at (0,1.1) {$\scriptstyle 3 $};
	\node at (0,-1.1) {$\scriptstyle 3 $};
	\fill[black]  (A) circle [radius = 0.08cm];
	\fill[black]  (B) circle [radius = 0.08cm];
	\fill[black]  (C) circle [radius = 0.08cm];
	\fill[black]  (D) circle [radius = 0.08cm];
} 
+ {\hat T}_{4,3ac} \tikz[baseline=(vert_cent.base),scale=0.5]{
	\node (vert_cent) {\hspace{-13pt}$\phantom{-}$};
	\vertex at \coord{-45} (A) {};
	\vertex at \coord{-135} (B) {};
	\vertex at \coord{135} (C) {};
	\vertex at \coord{45} (D) {};
	\draw  (A) to (B)  (B) to (C)  (C) to (D)  (D) to (A); 
	\node at (1.25,0.85) {$\scriptstyle \rmd \lambda$};
	\node at (1.2,-0.85) {$\scriptstyle  \lambda$};
	\node at (-1.1,-0.85) {$\scriptstyle \rmd \lambda $};
	\node at (-1.1,0.85) {$\scriptstyle \lambda$};
	\node at (1.1,0) {$\scriptstyle 3$};
	\node at (-1.1,0) {$\scriptstyle 3 $};
	\node at (0,1.1) {$\scriptstyle 3 $};
	\node at (0,-1.1) {$\scriptstyle 3 $};
	\fill[black]  (A) circle [radius = 0.08cm];
	\fill[black]  (B) circle [radius = 0.08cm];
	\fill[black]  (C) circle [radius = 0.08cm];
	\fill[black]  (D) circle [radius = 0.08cm];
} 
\nn \\
\noalign {\vskip -4pt}
&{}+ T_{4,4ab} \tikz[baseline=(vert_cent.base),scale=0.5]{
	\node (vert_cent) {\hspace{-13pt}$\phantom{-}$};
	\vertex at \coord{-45} (A) {};
	\vertex at \coord{-135} (B) {};
	\vertex at \coord{135} (C) {};
	\vertex at \coord{45} (D) {};
	\draw  (A) to (B)  (B) to (C)  (C) to (D)  (D) to (A); 
	\draw (B)--(D); 
	\fill[white]  (0,0) circle [radius = 0.1cm];
	\draw (C)--(A); 
	\node at (1.25,0.85) {$\scriptstyle \rmd \lambda$};
	\node at (1.2,-0.85) {$\scriptstyle \rmd \lambda$};
	\node at (-1.1,-0.85) {$\scriptstyle \lambda $};
	\node at (-1.1,0.85) {$\scriptstyle \lambda$};
	\node at (1.1,0) {$\scriptstyle 4 $};
	\node at (-1.1,0) {$\scriptstyle 4 $};
	\node at (0,1.1) {$\scriptstyle 1 $};
	\node at (0,-1.1) {$\scriptstyle 1 $};
	\node at (-0.1,0.46) {$\scriptstyle 1$};
	\node at (-0.1,-0.46) {$\scriptstyle 1 $};		
	\fill[black]  (A) circle [radius = 0.08cm];
	\fill[black]  (B) circle [radius = 0.08cm];
	\fill[black]  (C) circle [radius = 0.08cm];
	\fill[black]  (D) circle [radius = 0.08cm];
} 
+ T_{4,4ac}
\tikz[baseline=(vert_cent.base),scale=0.5]{
	\node (vert_cent) {\hspace{-13pt}$\phantom{-}$};
	\vertex at \coord{-45} (A) {};
	\vertex at \coord{-135} (B) {};
	\vertex at \coord{135} (C) {};
	\vertex at \coord{45} (D) {};
	\draw  (A) to (B)  (B) to (C)  (C) to (D)  (D) to (A); 
	\draw (B)--(D); 
	\fill[white]  (0,0) circle [radius = 0.1cm];
	\draw (C)--(A); 
	\node at (1.25,0.85) {$\scriptstyle \rmd \lambda$};
	\node at (1.2,-0.85) {$\scriptstyle  \lambda$};
	\node at (-1.1,-0.85) {$\scriptstyle \rmd \lambda $};
	\node at (-1.1,0.85) {$\scriptstyle \lambda$};
	\node at (1.1,0) {$\scriptstyle 4 $};
	\node at (-1.1,0) {$\scriptstyle 4 $};
	\node at (0,1.1) {$\scriptstyle 1 $};
	\node at (0,-1.1) {$\scriptstyle 1 $};
	\node at (-0.1,0.46) {$\scriptstyle 1$};
	\node at (-0.1,-0.46) {$\scriptstyle 1 $};		
	\fill[black]  (A) circle [radius = 0.08cm];
	\fill[black]  (B) circle [radius = 0.08cm];
	\fill[black]  (C) circle [radius = 0.08cm];
	\fill[black]  (D) circle [radius = 0.08cm];
} \nn \\
\noalign {\vskip -4pt}
&{}+ {\hat T}_{4,5ab} \tikz[baseline=(vert_cent.base),scale=0.5]{
	\node (vert_cent) {\hspace{-13pt}$\phantom{-}$};
	\vertex at \coord{-45} (A) {};
	\vertex at \coord{-135} (B) {};
	\vertex at \coord{135} (C) {};
	\vertex at \coord{45} (D) {};
	\draw  (A) to (B)  (B) to (C)  (C) to (D)  (D) to (A); 
	\draw (B)--(D); 
	\fill[white]  (0,0) circle [radius = 0.1cm];
	\draw (C)--(A); 
	\node at (1.25,0.85) {$\scriptstyle \rmd \lambda$};
	\node at (1.2,-0.85) {$\scriptstyle \rmd \lambda$};
	\node at (-1.1,-0.85) {$\scriptstyle \lambda $};
	\node at (-1.1,0.85) {$\scriptstyle \lambda$};
	\node at (1.1,0) {$\scriptstyle 3 $};
	\node at (-1.1,0) {$\scriptstyle 3 $};
	\node at (0,1.1) {$\scriptstyle 2 $};
	\node at (0,-1.1) {$\scriptstyle 2 $};
	\node at (-0.1,0.46) {$\scriptstyle 1$};
	\node at (-0.1,-0.46) {$\scriptstyle 1 $};		
	\fill[black]  (A) circle [radius = 0.08cm];
	\fill[black]  (B) circle [radius = 0.08cm];
	\fill[black]  (C) circle [radius = 0.08cm];
	\fill[black]  (D) circle [radius = 0.08cm];
} 
+ {\hat T}_{4,5ac}
\tikz[baseline=(vert_cent.base),scale=0.5]{
	\node (vert_cent) {\hspace{-13pt}$\phantom{-}$};
	\vertex at \coord{-45} (A) {};
	\vertex at \coord{-135} (B) {};
	\vertex at \coord{135} (C) {};
	\vertex at \coord{45} (D) {};
	\draw  (A) to (B)  (B) to (C)  (C) to (D)  (D) to (A); 
	\draw (B)--(D); 
	\fill[white]  (0,0) circle [radius = 0.1cm];
	\draw (C)--(A); 
	\node at (1.25,0.85) {$\scriptstyle \rmd \lambda$};
	\node at (1.2,-0.85) {$\scriptstyle  \lambda$};
	\node at (-1.1,-0.85) {$\scriptstyle \rmd \lambda $};
	\node at (-1.1,0.85) {$\scriptstyle \lambda$};
	\node at (1.1,0) {$\scriptstyle 3 $};
	\node at (-1.1,0) {$\scriptstyle 3 $};
	\node at (0,1.1) {$\scriptstyle 2 $};
	\node at (0,-1.1) {$\scriptstyle 2 $};
	\node at (-0.1,0.46) {$\scriptstyle 1$};
	\node at (-0.1,-0.46) {$\scriptstyle 1 $};		
	\fill[black]  (A) circle [radius = 0.08cm];
	\fill[black]  (B) circle [radius = 0.08cm];
	\fill[black]  (C) circle [radius = 0.08cm];
	\fill[black]  (D) circle [radius = 0.08cm];
} 
+ {\hat T}_{4,5ad}
\tikz[baseline=(vert_cent.base),scale=0.5]{
	\node (vert_cent) {\hspace{-13pt}$\phantom{-}$};
	\vertex at \coord{-45} (A) {};
	\vertex at \coord{-135} (B) {};
	\vertex at \coord{135} (C) {};
	\vertex at \coord{45} (D) {};
	\draw  (A) to (B)  (B) to (C)  (C) to (D)  (D) to (A); 
	\draw (B)--(D); 
	\fill[white]  (0,0) circle [radius = 0.1cm];
	\draw (C)--(A); 
	\node at (1.25,0.85) {$\scriptstyle \rmd \lambda$};
	\node at (1.2,-0.85) {$\scriptstyle  \lambda$};
	\node at (-1.1,-0.85) {$\scriptstyle  \lambda $};
	\node at (-1.2,0.85) {$\scriptstyle \rmd \lambda$};
	\node at (1.1,0) {$\scriptstyle 3 $};
	\node at (-1.1,0) {$\scriptstyle 3 $};
	\node at (0,1.1) {$\scriptstyle 2 $};
	\node at (0,-1.1) {$\scriptstyle 2 $};
	\node at (-0.1,0.46) {$\scriptstyle 1$};
	\node at (-0.1,-0.46) {$\scriptstyle 1 $};		
	\fill[black]  (A) circle [radius = 0.08cm];
	\fill[black]  (B) circle [radius = 0.08cm];
	\fill[black]  (C) circle [radius = 0.08cm];
	\fill[black]  (D) circle [radius = 0.08cm];
} \nn \\
\noalign {\vskip -4pt}
&{}+ T_{4,6ab} \tikz[baseline=(vert_cent.base),scale=0.5]{
	\node (vert_cent) {\hspace{-13pt}$\phantom{-}$};
	\vertex at \coord{-45} (A) {};
	\vertex at \coord{-135} (B) {};
	\vertex at \coord{135} (C) {};
	\vertex at \coord{45} (D) {};
	\draw  (A) to (B)  (B) to (C)  (C) to (D)  (D) to (A); 
	\draw (B)--(D); 
	\fill[white]  (0,0) circle [radius = 0.1cm];
	\draw (C)--(A); 
	\node at (1.25,0.85) {$\scriptstyle \rmd \lambda$};
	\node at (1.2,-0.85) {$\scriptstyle \rmd \lambda$};
	\node at (-1.1,-0.85) {$\scriptstyle \lambda $};
	\node at (-1.1,0.85) {$\scriptstyle \lambda$};
	\node at (1.1,0) {$\scriptstyle 2 $};
	\node at (-1.1,0) {$\scriptstyle 2 $};
	\node at (0,1.1) {$\scriptstyle 2 $};
	\node at (0,-1.1) {$\scriptstyle 2 $};
	\node at (-0.1,0.46) {$\scriptstyle 2$};
	\node at (-0.1,-0.46) {$\scriptstyle 2 $};		
	\fill[black]  (A) circle [radius = 0.08cm];
	\fill[black]  (B) circle [radius = 0.08cm];
	\fill[black]  (C) circle [radius = 0.08cm];
	\fill[black]  (D) circle [radius = 0.08cm];
} 
\end{align}
for 
\begin{align}
& {\hat T}_{4,3ab} = T_{4,3ab} - \tfrac{5}{80} \, T_3{\!}^2 \, , \qquad
{\hat T}_{4,3ac} = T_{4,3ac} - \tfrac{3}{80} \, T_3{\!}^2 \, , \nn \\
& {\hat T}_{4,5ab} = T_{4,5ab} - \tfrac{9}{40} \, T_3{\!}^2 \, , \quad
{\hat T}_{4,5ac} = T_{4,5ac} - \tfrac{27}{80} \, T_3{\!}^2 \, ,  \quad
{\hat T}_{4,5ad} = T_{4,5ad} - \tfrac{27}{80} \, T_3{\!}^2 \, .
\end{align}
The Riemann tensor is then
\begin{align}
R_{IKLJ}= {}&  \big (T_{4,1ab}-  T_{4,1ac}) 
\,  \bigg ( 
\tikz[baseline=(vert_cent.base),scale=0.5]{
	\node (vert_cent) {\hspace{-13pt}$\phantom{-}$};
	\vertex at \coord{-45} (A) {};
	\vertex at \coord{-135} (B) {};
	\vertex at \coord{135} (C) {};
	\vertex at \coord{45} (D) {};
	\draw  (A) to (B)  (B) to (C)  (C) to (D)  (D) to (A); 
	\node at (1.25,0.85) {$\scriptstyle I $};
	\node at (1.2,-0.85) {$\scriptstyle J $};
	\node at (-1.1,-0.85) {$\scriptstyle L  $};
	\node at (-1.1,0.85) {$\scriptstyle K $};
	\node at (1.1,0) {$\scriptstyle 5 $};
	\node at (-1.1,0) {$\scriptstyle 5 $};
	\node at (0,1.1) {$\scriptstyle 1 $};
	\node at (0,-1.1) {$\scriptstyle 1 $};
	\fill[black]  (A) circle [radius = 0.08cm];
	\fill[black]  (B) circle [radius = 0.08cm];
	\fill[black]  (C) circle [radius = 0.08cm];
	\fill[black]  (D) circle [radius = 0.08cm];
} 
 -
\tikz[baseline=(vert_cent.base),scale=0.5]{
	\node (vert_cent) {\hspace{-13pt}$\phantom{-}$};
	\vertex at \coord{-45} (A) {};
	\vertex at \coord{-135} (B) {};
	\vertex at \coord{135} (C) {};
	\vertex at \coord{45} (D) {};
	\draw  (A) to (B)  (B) to (C)  (C) to (D)  (D) to (A); 
	\node at (1.25,0.85) {$\scriptstyle I $};
	\node at (1.2,-0.85) {$\scriptstyle L $};
	\node at (-1.1,-0.85) {$\scriptstyle J $};
	\node at (-1.1,0.85) {$\scriptstyle K $};
	\node at (1.1,0) {$\scriptstyle 5 $};
	\node at (-1.1,0) {$\scriptstyle 5 $};
	\node at (0,1.1) {$\scriptstyle 1 $};
	\node at (0,-1.1) {$\scriptstyle 1 $};
	\fill[black]  (A) circle [radius = 0.08cm];
	\fill[black]  (B) circle [radius = 0.08cm];
	\fill[black]  (C) circle [radius = 0.08cm];
	\fill[black]  (D) circle [radius = 0.08cm];
} 
\bigg  )  
+  
 \big (T_{4,1ad}-  T_{4,1ab}) 
\,  \bigg ( 
\tikz[baseline=(vert_cent.base),scale=0.5]{
	\node (vert_cent) {\hspace{-13pt}$\phantom{-}$};
	\vertex at \coord{-45} (A) {};
	\vertex at \coord{-135} (B) {};
	\vertex at \coord{135} (C) {};
	\vertex at \coord{45} (D) {};
	\draw  (A) to (B)  (B) to (C)  (C) to (D)  (D) to (A); 
	\node at (1.25,0.85) {$\scriptstyle I $};
	\node at (1.2,-0.85) {$\scriptstyle L $};
	\node at (-1.1,-0.85) {$\scriptstyle K  $};
	\node at (-1.1,0.85) {$\scriptstyle J $};
	\node at (1.1,0) {$\scriptstyle 5 $};
	\node at (-1.1,0) {$\scriptstyle 5 $};
	\node at (0,1.1) {$\scriptstyle 1 $};
	\node at (0,-1.1) {$\scriptstyle 1 $};
	\fill[black]  (A) circle [radius = 0.08cm];
	\fill[black]  (B) circle [radius = 0.08cm];
	\fill[black]  (C) circle [radius = 0.08cm];
	\fill[black]  (D) circle [radius = 0.08cm];
} 
 -
\tikz[baseline=(vert_cent.base),scale=0.5]{
	\node (vert_cent) {\hspace{-13pt}$\phantom{-}$};
	\vertex at \coord{-45} (A) {};
	\vertex at \coord{-135} (B) {};
	\vertex at \coord{135} (C) {};
	\vertex at \coord{45} (D) {};
	\draw  (A) to (B)  (B) to (C)  (C) to (D)  (D) to (A); 
	\node at (1.25,0.85) {$\scriptstyle I $};
	\node at (1.2,-0.85) {$\scriptstyle J $};
	\node at (-1.1,-0.85) {$\scriptstyle K $};
	\node at (-1.1,0.85) {$\scriptstyle L $};
	\node at (1.1,0) {$\scriptstyle 5 $};
	\node at (-1.1,0) {$\scriptstyle 5 $};
	\node at (0,1.1) {$\scriptstyle 1 $};
	\node at (0,-1.1) {$\scriptstyle 1 $};
	\fill[black]  (A) circle [radius = 0.08cm];
	\fill[black]  (B) circle [radius = 0.08cm];
	\fill[black]  (C) circle [radius = 0.08cm];
	\fill[black]  (D) circle [radius = 0.08cm];
} 
\bigg  )  
\nn \\
&{}+  \big (T_{4,1ac}-  T_{4,1ad})  \bigg ( 
\tikz[baseline=(vert_cent.base),scale=0.5]{
	\node (vert_cent) {\hspace{-13pt}$\phantom{-}$};
	\vertex at \coord{-45} (A) {};
	\vertex at \coord{-135} (B) {};
	\vertex at \coord{135} (C) {};
	\vertex at \coord{45} (D) {};
	\draw  (A) to (B)  (B) to (C)  (C) to (D)  (D) to (A); 
	\node at (1.25,0.85) {$\scriptstyle I $};
	\node at (1.2,-0.85) {$\scriptstyle K $};
	\node at (-1.1,-0.85) {$\scriptstyle J  $};
	\node at (-1.1,0.85) {$\scriptstyle L $};
	\node at (1.1,0) {$\scriptstyle 5 $};
	\node at (-1.1,0) {$\scriptstyle 5 $};
	\node at (0,1.1) {$\scriptstyle 1 $};
	\node at (0,-1.1) {$\scriptstyle 1 $};
	\fill[black]  (A) circle [radius = 0.08cm];
	\fill[black]  (B) circle [radius = 0.08cm];
	\fill[black]  (C) circle [radius = 0.08cm];
	\fill[black]  (D) circle [radius = 0.08cm];
} 
 -
\tikz[baseline=(vert_cent.base),scale=0.5]{
	\node (vert_cent) {\hspace{-13pt}$\phantom{-}$};
	\vertex at \coord{-45} (A) {};
	\vertex at \coord{-135} (B) {};
	\vertex at \coord{135} (C) {};
	\vertex at \coord{45} (D) {};
	\draw  (A) to (B)  (B) to (C)  (C) to (D)  (D) to (A); 
	\node at (1.25,0.85) {$\scriptstyle I $};
	\node at (1.2,-0.85) {$\scriptstyle K $};
	\node at (-1.1,-0.85) {$\scriptstyle L $};
	\node at (-1.1,0.85) {$\scriptstyle J $};
	\node at (1.1,0) {$\scriptstyle 5 $};
	\node at (-1.1,0) {$\scriptstyle 5 $};
	\node at (0,1.1) {$\scriptstyle 1 $};
	\node at (0,-1.1) {$\scriptstyle 1 $};
	\fill[black]  (A) circle [radius = 0.08cm];
	\fill[black]  (B) circle [radius = 0.08cm];
	\fill[black]  (C) circle [radius = 0.08cm];
	\fill[black]  (D) circle [radius = 0.08cm];
} 
\bigg  )  \nn \\
 &{}  +  \big (T_{4,2ab}-  T_{5,2ac}) 
\,  \bigg ( 
\tikz[baseline=(vert_cent.base),scale=0.5]{
	\node (vert_cent) {\hspace{-13pt}$\phantom{-}$};
	\vertex at \coord{-45} (A) {};
	\vertex at \coord{-135} (B) {};
	\vertex at \coord{135} (C) {};
	\vertex at \coord{45} (D) {};
	\draw  (A) to (B)  (B) to (C)  (C) to (D)  (D) to (A); 
	\node at (1.25,0.85) {$\scriptstyle I $};
	\node at (1.2,-0.85) {$\scriptstyle J $};
	\node at (-1.1,-0.85) {$\scriptstyle L  $};
	\node at (-1.1,0.85) {$\scriptstyle K $};
	\node at (1.1,0) {$\scriptstyle 4 $};
	\node at (-1.1,0) {$\scriptstyle 4 $};
	\node at (0,1.1) {$\scriptstyle 2 $};
	\node at (0,-1.1) {$\scriptstyle 2 $};
	\fill[black]  (A) circle [radius = 0.08cm];
	\fill[black]  (B) circle [radius = 0.08cm];
	\fill[black]  (C) circle [radius = 0.08cm];
	\fill[black]  (D) circle [radius = 0.08cm];
} 
 -
\tikz[baseline=(vert_cent.base),scale=0.5]{
	\node (vert_cent) {\hspace{-13pt}$\phantom{-}$};
	\vertex at \coord{-45} (A) {};
	\vertex at \coord{-135} (B) {};
	\vertex at \coord{135} (C) {};
	\vertex at \coord{45} (D) {};
	\draw  (A) to (B)  (B) to (C)  (C) to (D)  (D) to (A); 
	\node at (1.25,0.85) {$\scriptstyle I $};
	\node at (1.2,-0.85) {$\scriptstyle L $};
	\node at (-1.1,-0.85) {$\scriptstyle J $};
	\node at (-1.1,0.85) {$\scriptstyle K $};
	\node at (1.1,0) {$\scriptstyle 4 $};
	\node at (-1.1,0) {$\scriptstyle 4 $};
	\node at (0,1.1) {$\scriptstyle 2 $};
	\node at (0,-1.1) {$\scriptstyle 2 $};
	\fill[black]  (A) circle [radius = 0.08cm];
	\fill[black]  (B) circle [radius = 0.08cm];
	\fill[black]  (C) circle [radius = 0.08cm];
	\fill[black]  (D) circle [radius = 0.08cm];
} 
\bigg  )  
+  
 \big (T_{4,2ad}-  T_{4,2ab}) 
\,  \bigg ( 
\tikz[baseline=(vert_cent.base),scale=0.5]{
	\node (vert_cent) {\hspace{-13pt}$\phantom{-}$};
	\vertex at \coord{-45} (A) {};
	\vertex at \coord{-135} (B) {};
	\vertex at \coord{135} (C) {};
	\vertex at \coord{45} (D) {};
	\draw  (A) to (B)  (B) to (C)  (C) to (D)  (D) to (A); 
	\node at (1.25,0.85) {$\scriptstyle I $};
	\node at (1.2,-0.85) {$\scriptstyle L $};
	\node at (-1.1,-0.85) {$\scriptstyle K  $};
	\node at (-1.1,0.85) {$\scriptstyle J $};
	\node at (1.1,0) {$\scriptstyle 4 $};
	\node at (-1.1,0) {$\scriptstyle 4 $};
	\node at (0,1.1) {$\scriptstyle 2 $};
	\node at (0,-1.1) {$\scriptstyle 2 $};
	\fill[black]  (A) circle [radius = 0.08cm];
	\fill[black]  (B) circle [radius = 0.08cm];
	\fill[black]  (C) circle [radius = 0.08cm];
	\fill[black]  (D) circle [radius = 0.08cm];
} 
 -
\tikz[baseline=(vert_cent.base),scale=0.5]{
	\node (vert_cent) {\hspace{-13pt}$\phantom{-}$};
	\vertex at \coord{-45} (A) {};
	\vertex at \coord{-135} (B) {};
	\vertex at \coord{135} (C) {};
	\vertex at \coord{45} (D) {};
	\draw  (A) to (B)  (B) to (C)  (C) to (D)  (D) to (A); 
	\node at (1.25,0.85) {$\scriptstyle I $};
	\node at (1.2,-0.85) {$\scriptstyle J $};
	\node at (-1.1,-0.85) {$\scriptstyle K $};
	\node at (-1.1,0.85) {$\scriptstyle L $};
	\node at (1.1,0) {$\scriptstyle 4 $};
	\node at (-1.1,0) {$\scriptstyle 4 $};
	\node at (0,1.1) {$\scriptstyle 2 $};
	\node at (0,-1.1) {$\scriptstyle 2 $};
	\fill[black]  (A) circle [radius = 0.08cm];
	\fill[black]  (B) circle [radius = 0.08cm];
	\fill[black]  (C) circle [radius = 0.08cm];
	\fill[black]  (D) circle [radius = 0.08cm];
} 
\bigg  )  
\nn \\
&{}+  \big (T_{4,2ac}-  T_{4,2ad})  \bigg ( 
\tikz[baseline=(vert_cent.base),scale=0.5]{
	\node (vert_cent) {\hspace{-13pt}$\phantom{-}$};
	\vertex at \coord{-45} (A) {};
	\vertex at \coord{-135} (B) {};
	\vertex at \coord{135} (C) {};
	\vertex at \coord{45} (D) {};
	\draw  (A) to (B)  (B) to (C)  (C) to (D)  (D) to (A); 
	\node at (1.25,0.85) {$\scriptstyle I $};
	\node at (1.2,-0.85) {$\scriptstyle K $};
	\node at (-1.1,-0.85) {$\scriptstyle J  $};
	\node at (-1.1,0.85) {$\scriptstyle L $};
	\node at (1.1,0) {$\scriptstyle 4 $};
	\node at (-1.1,0) {$\scriptstyle 4 $};
	\node at (0,1.1) {$\scriptstyle 2 $};
	\node at (0,-1.1) {$\scriptstyle 2 $};
	\fill[black]  (A) circle [radius = 0.08cm];
	\fill[black]  (B) circle [radius = 0.08cm];
	\fill[black]  (C) circle [radius = 0.08cm];
	\fill[black]  (D) circle [radius = 0.08cm];
} 
 -
\tikz[baseline=(vert_cent.base),scale=0.5]{
	\node (vert_cent) {\hspace{-13pt}$\phantom{-}$};
	\vertex at \coord{-45} (A) {};
	\vertex at \coord{-135} (B) {};
	\vertex at \coord{135} (C) {};
	\vertex at \coord{45} (D) {};
	\draw  (A) to (B)  (B) to (C)  (C) to (D)  (D) to (A); 
	\node at (1.25,0.85) {$\scriptstyle I $};
	\node at (1.2,-0.85) {$\scriptstyle K $};
	\node at (-1.1,-0.85) {$\scriptstyle L $};
	\node at (-1.1,0.85) {$\scriptstyle J $};
	\node at (1.1,0) {$\scriptstyle 4 $};
	\node at (-1.1,0) {$\scriptstyle 4 $};
	\node at (0,1.1) {$\scriptstyle 2 $};
	\node at (0,-1.1) {$\scriptstyle 2 $};
	\fill[black]  (A) circle [radius = 0.08cm];
	\fill[black]  (B) circle [radius = 0.08cm];
	\fill[black]  (C) circle [radius = 0.08cm];
	\fill[black]  (D) circle [radius = 0.08cm];
} 
\bigg  )  \nn \\
&{}  +  \big ({\hat T}_{4,3ab}- 2\,  {\hat T}_{4,3ac}) 
\,  \bigg ( 
\tikz[baseline=(vert_cent.base),scale=0.5]{
	\node (vert_cent) {\hspace{-13pt}$\phantom{-}$};
	\vertex at \coord{-45} (A) {};
	\vertex at \coord{-135} (B) {};
	\vertex at \coord{135} (C) {};
	\vertex at \coord{45} (D) {};
	\draw  (A) to (B)  (B) to (C)  (C) to (D)  (D) to (A); 
	\node at (1.25,0.85) {$\scriptstyle I $};
	\node at (1.2,-0.85) {$\scriptstyle J $};
	\node at (-1.1,-0.85) {$\scriptstyle L  $};
	\node at (-1.1,0.85) {$\scriptstyle K $};
	\node at (1.1,0) {$\scriptstyle 3 $};
	\node at (-1.1,0) {$\scriptstyle 3 $};
	\node at (0,1.1) {$\scriptstyle 3 $};
	\node at (0,-1.1) {$\scriptstyle 3 $};
	\fill[black]  (A) circle [radius = 0.08cm];
	\fill[black]  (B) circle [radius = 0.08cm];
	\fill[black]  (C) circle [radius = 0.08cm];
	\fill[black]  (D) circle [radius = 0.08cm];
} 
 -
\tikz[baseline=(vert_cent.base),scale=0.5]{
	\node (vert_cent) {\hspace{-13pt}$\phantom{-}$};
	\vertex at \coord{-45} (A) {};
	\vertex at \coord{-135} (B) {};
	\vertex at \coord{135} (C) {};
	\vertex at \coord{45} (D) {};
	\draw  (A) to (B)  (B) to (C)  (C) to (D)  (D) to (A); 
	\node at (1.25,0.85) {$\scriptstyle I $};
	\node at (1.2,-0.85) {$\scriptstyle L $};
	\node at (-1.1,-0.85) {$\scriptstyle J $};
	\node at (-1.1,0.85) {$\scriptstyle K $};
	\node at (1.1,0) {$\scriptstyle 3 $};
	\node at (-1.1,0) {$\scriptstyle 3 $};
	\node at (0,1.1) {$\scriptstyle 3 $};
	\node at (0,-1.1) {$\scriptstyle 3 $};
	\fill[black]  (A) circle [radius = 0.08cm];
	\fill[black]  (B) circle [radius = 0.08cm];
	\fill[black]  (C) circle [radius = 0.08cm];
	\fill[black]  (D) circle [radius = 0.08cm];
} 
\bigg  )  \nn \\
&{}  +  \big (2\, T_{4,4ab}-   T_{4,4ac}) 
\,  \bigg ( 
\tikz[baseline=(vert_cent.base),scale=0.5]{
	\node (vert_cent) {\hspace{-13pt}$\phantom{-}$};
	\vertex at \coord{-45} (A) {};
	\vertex at \coord{-135} (B) {};
	\vertex at \coord{135} (C) {};
	\vertex at \coord{45} (D) {};
	\draw  (A) to (B)  (B) to (C)  (C) to (D)  (D) to (A); 
	\draw (B)--(D); 
	\fill[white]  (0,0) circle [radius = 0.1cm];
	\draw (C)--(A); 
	\node at (1.25,0.85) {$\scriptstyle I $};
	\node at (1.2,-0.85) {$\scriptstyle J $};
	\node at (-1.1,-0.85) {$\scriptstyle L $};
	\node at (-1.1,0.85) {$\scriptstyle K $};
	\node at (1.1,0) {$\scriptstyle 4 $};
	\node at (-1.1,0) {$\scriptstyle 4 $};
	\node at (0,1.1) {$\scriptstyle 1 $};
	\node at (0,-1.1) {$\scriptstyle 1 $};
	\node at (-0.1,0.46) {$\scriptstyle 1$};
	\node at (-0.1,-0.46) {$\scriptstyle 1 $};		
	\fill[black]  (A) circle [radius = 0.08cm];
	\fill[black]  (B) circle [radius = 0.08cm];
	\fill[black]  (C) circle [radius = 0.08cm];
	\fill[black]  (D) circle [radius = 0.08cm];
} 
 - 
 \tikz[baseline=(vert_cent.base),scale=0.5]{
	\node (vert_cent) {\hspace{-13pt}$\phantom{-}$};
	\vertex at \coord{-45} (A) {};
	\vertex at \coord{-135} (B) {};
	\vertex at \coord{135} (C) {};
	\vertex at \coord{45} (D) {};
	\draw  (A) to (B)  (B) to (C)  (C) to (D)  (D) to (A); 
	\draw (B)--(D); 
	\fill[white]  (0,0) circle [radius = 0.1cm];
	\draw (C)--(A); 
	\node at (1.25,0.85) {$\scriptstyle I $};
	\node at (1.2,-0.85) {$\scriptstyle L $};
	\node at (-1.1,-0.85) {$\scriptstyle J $};
	\node at (-1.1,0.85) {$\scriptstyle K $};
	\node at (1.1,0) {$\scriptstyle 4 $};
	\node at (-1.1,0) {$\scriptstyle 4 $};
	\node at (0,1.1) {$\scriptstyle 1 $};
	\node at (0,-1.1) {$\scriptstyle 1 $};
	\node at (-0.1,0.46) {$\scriptstyle 1$};
	\node at (-0.1,-0.46) {$\scriptstyle 1 $};		
	\fill[black]  (A) circle [radius = 0.08cm];
	\fill[black]  (B) circle [radius = 0.08cm];
	\fill[black]  (C) circle [radius = 0.08cm];
	\fill[black]  (D) circle [radius = 0.08cm];
} 
\bigg  )  \nn \\
&{}  +  \big ({\hat T}_{4,5ab} - {\hat T}_{4,5ac}) 
\,  \bigg ( 
\tikz[baseline=(vert_cent.base),scale=0.5]{
	\node (vert_cent) {\hspace{-13pt}$\phantom{-}$};
	\vertex at \coord{-45} (A) {};
	\vertex at \coord{-135} (B) {};
	\vertex at \coord{135} (C) {};
	\vertex at \coord{45} (D) {};
	\draw  (A) to (B)  (B) to (C)  (C) to (D)  (D) to (A); 
	\draw (B)--(D); 
	\fill[white]  (0,0) circle [radius = 0.1cm];
	\draw (C)--(A); 
	\node at (1.25,0.85) {$\scriptstyle I $};
	\node at (1.2,-0.85) {$\scriptstyle J$};
	\node at (-1.1,-0.85) {$\scriptstyle L $};
	\node at (-1.1,0.85) {$\scriptstyle K $};
	\node at (1.1,0) {$\scriptstyle 3 $};
	\node at (-1.1,0) {$\scriptstyle 3 $};
	\node at (0,1.1) {$\scriptstyle 2 $};
	\node at (0,-1.1) {$\scriptstyle 2 $};
	\node at (-0.1,0.46) {$\scriptstyle 1$};
	\node at (-0.1,-0.46) {$\scriptstyle 1 $};		
	\fill[black]  (A) circle [radius = 0.08cm];
	\fill[black]  (B) circle [radius = 0.08cm];
	\fill[black]  (C) circle [radius = 0.08cm];
	\fill[black]  (D) circle [radius = 0.08cm];
} 
 - 
 \tikz[baseline=(vert_cent.base),scale=0.5]{
	\node (vert_cent) {\hspace{-13pt}$\phantom{-}$};
	\vertex at \coord{-45} (A) {};
	\vertex at \coord{-135} (B) {};
	\vertex at \coord{135} (C) {};
	\vertex at \coord{45} (D) {};
	\draw  (A) to (B)  (B) to (C)  (C) to (D)  (D) to (A); 
	\draw (B)--(D); 
	\fill[white]  (0,0) circle [radius = 0.1cm];
	\draw (C)--(A); 
	\node at (1.25,0.85) {$\scriptstyle I $};
	\node at (1.2,-0.85) {$\scriptstyle L $};
	\node at (-1.1,-0.85) {$\scriptstyle J $};
	\node at (-1.1,0.85) {$\scriptstyle K $};
	\node at (1.1,0) {$\scriptstyle 3 $};
	\node at (-1.1,0) {$\scriptstyle 3 $};
	\node at (0,1.1) {$\scriptstyle 2 $};
	\node at (0,-1.1) {$\scriptstyle 2 $};
	\node at (-0.1,0.46) {$\scriptstyle 1$};
	\node at (-0.1,-0.46) {$\scriptstyle 1 $};		
	\fill[black]  (A) circle [radius = 0.08cm];
	\fill[black]  (B) circle [radius = 0.08cm];
	\fill[black]  (C) circle [radius = 0.08cm];
	\fill[black]  (D) circle [radius = 0.08cm];
} 
\bigg  ) 
 +  \big ( {\hat T}_{4,5ad} - {\hat T}_{4,5ab}) 
\,  \bigg ( 
\tikz[baseline=(vert_cent.base),scale=0.5]{
	\node (vert_cent) {\hspace{-13pt}$\phantom{-}$};
	\vertex at \coord{-45} (A) {};
	\vertex at \coord{-135} (B) {};
	\vertex at \coord{135} (C) {};
	\vertex at \coord{45} (D) {};
	\draw  (A) to (B)  (B) to (C)  (C) to (D)  (D) to (A); 
	\draw (B)--(D); 
	\fill[white]  (0,0) circle [radius = 0.1cm];
	\draw (C)--(A); 
	\node at (1.25,0.85) {$\scriptstyle I $};
	\node at (1.2,-0.85) {$\scriptstyle L $};
	\node at (-1.1,-0.85) {$\scriptstyle K $};
	\node at (-1.1,0.85) {$\scriptstyle J $};
	\node at (1.1,0) {$\scriptstyle 3 $};
	\node at (-1.1,0) {$\scriptstyle 3 $};
	\node at (0,1.1) {$\scriptstyle 2 $};
	\node at (0,-1.1) {$\scriptstyle 2 $};
	\node at (-0.1,0.46) {$\scriptstyle 1$};
	\node at (-0.1,-0.46) {$\scriptstyle 1 $};		
	\fill[black]  (A) circle [radius = 0.08cm];
	\fill[black]  (B) circle [radius = 0.08cm];
	\fill[black]  (C) circle [radius = 0.08cm];
	\fill[black]  (D) circle [radius = 0.08cm];
} 
 - 
 \tikz[baseline=(vert_cent.base),scale=0.5]{
	\node (vert_cent) {\hspace{-13pt}$\phantom{-}$};
	\vertex at \coord{-45} (A) {};
	\vertex at \coord{-135} (B) {};
	\vertex at \coord{135} (C) {};
	\vertex at \coord{45} (D) {};
	\draw  (A) to (B)  (B) to (C)  (C) to (D)  (D) to (A); 
	\draw (B)--(D); 
	\fill[white]  (0,0) circle [radius = 0.1cm];
	\draw (C)--(A); 
	\node at (1.25,0.85) {$\scriptstyle I $};
	\node at (1.2,-0.85) {$\scriptstyle J $};
	\node at (-1.1,-0.85) {$\scriptstyle K $};
	\node at (-1.1,0.85) {$\scriptstyle L $};
	\node at (1.1,0) {$\scriptstyle 3 $};
	\node at (-1.1,0) {$\scriptstyle 3 $};
	\node at (0,1.1) {$\scriptstyle 2 $};
	\node at (0,-1.1) {$\scriptstyle 2 $};
	\node at (-0.1,0.46) {$\scriptstyle 1$};
	\node at (-0.1,-0.46) {$\scriptstyle 1 $};		
	\fill[black]  (A) circle [radius = 0.08cm];
	\fill[black]  (B) circle [radius = 0.08cm];
	\fill[black]  (C) circle [radius = 0.08cm];
	\fill[black]  (D) circle [radius = 0.08cm];
} 
\bigg  )   \nn \\
&{}  +  \big ({\hat T}_{4,5ac} - {\hat T}_{4,5ad}) 
\,  \bigg ( 
\tikz[baseline=(vert_cent.base),scale=0.5]{
	\node (vert_cent) {\hspace{-13pt}$\phantom{-}$};
	\vertex at \coord{-45} (A) {};
	\vertex at \coord{-135} (B) {};
	\vertex at \coord{135} (C) {};
	\vertex at \coord{45} (D) {};
	\draw  (A) to (B)  (B) to (C)  (C) to (D)  (D) to (A); 
	\draw (B)--(D); 
	\fill[white]  (0,0) circle [radius = 0.1cm];
	\draw (C)--(A); 
	\node at (1.25,0.85) {$\scriptstyle I $};
	\node at (1.2,-0.85) {$\scriptstyle K$};
	\node at (-1.1,-0.85) {$\scriptstyle J $};
	\node at (-1.1,0.85) {$\scriptstyle L $};
	\node at (1.1,0) {$\scriptstyle 3 $};
	\node at (-1.1,0) {$\scriptstyle 3 $};
	\node at (0,1.1) {$\scriptstyle 2 $};
	\node at (0,-1.1) {$\scriptstyle 2 $};
	\node at (-0.1,0.46) {$\scriptstyle 1$};
	\node at (-0.1,-0.46) {$\scriptstyle 1 $};		
	\fill[black]  (A) circle [radius = 0.08cm];
	\fill[black]  (B) circle [radius = 0.08cm];
	\fill[black]  (C) circle [radius = 0.08cm];
	\fill[black]  (D) circle [radius = 0.08cm];
} 
 - 
 \tikz[baseline=(vert_cent.base),scale=0.5]{
	\node (vert_cent) {\hspace{-13pt}$\phantom{-}$};
	\vertex at \coord{-45} (A) {};
	\vertex at \coord{-135} (B) {};
	\vertex at \coord{135} (C) {};
	\vertex at \coord{45} (D) {};
	\draw  (A) to (B)  (B) to (C)  (C) to (D)  (D) to (A); 
	\draw (B)--(D); 
	\fill[white]  (0,0) circle [radius = 0.1cm];
	\draw (C)--(A); 
	\node at (1.25,0.85) {$\scriptstyle I $};
	\node at (1.2,-0.85) {$\scriptstyle K $};
	\node at (-1.1,-0.85) {$\scriptstyle L $};
	\node at (-1.1,0.85) {$\scriptstyle J $};
	\node at (1.1,0) {$\scriptstyle 3 $};
	\node at (-1.1,0) {$\scriptstyle 3 $};
	\node at (0,1.1) {$\scriptstyle 2 $};
	\node at (0,-1.1) {$\scriptstyle 2 $};
	\node at (-0.1,0.46) {$\scriptstyle 1$};
	\node at (-0.1,-0.46) {$\scriptstyle 1 $};		
	\fill[black]  (A) circle [radius = 0.08cm];
	\fill[black]  (B) circle [radius = 0.08cm];
	\fill[black]  (C) circle [radius = 0.08cm];
	\fill[black]  (D) circle [radius = 0.08cm];
} 
\bigg  ) \, .
\label{Riem}
 \end{align} 
 It is easy to verify that this satisfies the necessary symmetries of the Riemann tensor. 
 
The corresponding scalar curvature $R=R_{IJIJ}$ so that we require from \eqref{Riem} 
\begin{align}
&\tikz[baseline=(vert_cent.base),scale=0.5]{
	\node (vert_cent) {\hspace{-13pt}$\phantom{-}$};
	\vertex at \coord{-45} (A) {};
	\vertex at \coord{-135} (B) {};
	\vertex at \coord{135} (C) {};
	\vertex at \coord{45} (D) {};
	\draw  (A) to (B)  (B) to (C)  (C) to (D)  (D) to (A); 
	\node at (1.25,0.85) {$\scriptstyle I $};
	\node at (1.2,-0.85) {$\scriptstyle J $};
	\node at (-1.1,-0.85) {$\scriptstyle I  $};
	\node at (-1.1,0.85) {$\scriptstyle J $};
	\node at (1.1,0) {$\scriptstyle 5 $};
	\node at (-1.1,0) {$\scriptstyle 5 $};
	\node at (0,1.1) {$\scriptstyle 1 $};
	\node at (0,-1.1) {$\scriptstyle 1 $};
	\fill[black]  (A) circle [radius = 0.08cm];
	\fill[black]  (B) circle [radius = 0.08cm];
	\fill[black]  (C) circle [radius = 0.08cm];
	\fill[black]  (D) circle [radius = 0.08cm];
} 
= \tfrac{1}{6\, 6!} N (N+2)_4 (N+11)\, , \quad
\tikz[baseline=(vert_cent.base),scale=0.5]{
	\node (vert_cent) {\hspace{-13pt}$\phantom{-}$};
	\vertex at \coord{-45} (A) {};
	\vertex at \coord{-135} (B) {};
	\vertex at \coord{135} (C) {};
	\vertex at \coord{45} (D) {};
	\draw  (A) to (B)  (B) to (C)  (C) to (D)  (D) to (A); 
	\node at (1.25,0.85) {$\scriptstyle I $};
	\node at (1.2,-0.85) {$\scriptstyle I$};
	\node at (-1.1,-0.85) {$\scriptstyle J  $};
	\node at (-1.1,0.85) {$\scriptstyle J $};
	\node at (1.1,0) {$\scriptstyle 5 $};
	\node at (-1.1,0) {$\scriptstyle 5 $};
	\node at (0,1.1) {$\scriptstyle 1 $};
	\node at (0,-1.1) {$\scriptstyle 1 $};
	\fill[black]  (A) circle [radius = 0.08cm];
	\fill[black]  (B) circle [radius = 0.08cm];
	\fill[black]  (C) circle [radius = 0.08cm];
	\fill[black]  (D) circle [radius = 0.08cm];
} 
= \tfrac{1}{6!^2} N (N+1)_5{}^2\, , \nn \\
\noalign{\vskip -6pt}
&\tikz[baseline=(vert_cent.base),scale=0.5]{
	\node (vert_cent) {\hspace{-13pt}$\phantom{-}$};
	\vertex at \coord{-45} (A) {};
	\vertex at \coord{-135} (B) {};
	\vertex at \coord{135} (C) {};
	\vertex at \coord{45} (D) {};
	\draw  (A) to (B)  (B) to (C)  (C) to (D)  (D) to (A); 
	\node at (1.25,0.85) {$\scriptstyle I $};
	\node at (1.2,-0.85) {$\scriptstyle J $};
	\node at (-1.1,-0.85) {$\scriptstyle J  $};
	\node at (-1.1,0.85) {$\scriptstyle I $};
	\node at (1.1,0) {$\scriptstyle 5 $};
	\node at (-1.1,0) {$\scriptstyle 5 $};
	\node at (0,1.1) {$\scriptstyle 1 $};
	\node at (0,-1.1) {$\scriptstyle 1 $};
	\fill[black]  (A) circle [radius = 0.08cm];
	\fill[black]  (B) circle [radius = 0.08cm];
	\fill[black]  (C) circle [radius = 0.08cm];
	\fill[black]  (D) circle [radius = 0.08cm];
} 
= \tfrac{1}{6\, 6!} (N)_5 (N+5)^2\, , \nn \\
&\tikz[baseline=(vert_cent.base),scale=0.5]{
	\node (vert_cent) {\hspace{-13pt}$\phantom{-}$};
	\vertex at \coord{-45} (A) {};
	\vertex at \coord{-135} (B) {};
	\vertex at \coord{135} (C) {};
	\vertex at \coord{45} (D) {};
	\draw  (A) to (B)  (B) to (C)  (C) to (D)  (D) to (A); 
	\node at (1.25,0.85) {$\scriptstyle I $};
	\node at (1.2,-0.85) {$\scriptstyle J $};
	\node at (-1.1,-0.85) {$\scriptstyle I  $};
	\node at (-1.1,0.85) {$\scriptstyle J $};
	\node at (1.1,0) {$\scriptstyle 4 $};
	\node at (-1.1,0) {$\scriptstyle 4 $};
	\node at (0,1.1) {$\scriptstyle 2 $};
	\node at (0,-1.1) {$\scriptstyle 2 $};
	\fill[black]  (A) circle [radius = 0.08cm];
	\fill[black]  (B) circle [radius = 0.08cm];
	\fill[black]  (C) circle [radius = 0.08cm];
	\fill[black]  (D) circle [radius = 0.08cm];
} 
= \tfrac{1}{§15\, 6!} N (N+2)(N+4)(N+5)(N+9)  (N+11)\, , \quad
\tikz[baseline=(vert_cent.base),scale=0.5]{
	\node (vert_cent) {\hspace{-13pt}$\phantom{-}$};
	\vertex at \coord{-45} (A) {};
	\vertex at \coord{-135} (B) {};
	\vertex at \coord{135} (C) {};
	\vertex at \coord{45} (D) {};
	\draw  (A) to (B)  (B) to (C)  (C) to (D)  (D) to (A); 
	\node at (1.25,0.85) {$\scriptstyle I $};
	\node at (1.2,-0.85) {$\scriptstyle I$};
	\node at (-1.1,-0.85) {$\scriptstyle J  $};
	\node at (-1.1,0.85) {$\scriptstyle J $};
	\node at (1.1,0) {$\scriptstyle 4 $};
	\node at (-1.1,0) {$\scriptstyle 4 $};
	\node at (0,1.1) {$\scriptstyle 2 $};
	\node at (0,-1.1) {$\scriptstyle 2 $};
	\fill[black]  (A) circle [radius = 0.08cm];
	\fill[black]  (B) circle [radius = 0.08cm];
	\fill[black]  (C) circle [radius = 0.08cm];
	\fill[black]  (D) circle [radius = 0.08cm];
} 
= \tfrac{2}{6!^2}( N)_2 (N+2)_4{}^2\, , \nn \\
\noalign{\vskip -6pt}
&\tikz[baseline=(vert_cent.base),scale=0.5]{
	\node (vert_cent) {\hspace{-13pt}$\phantom{-}$};
	\vertex at \coord{-45} (A) {};
	\vertex at \coord{-135} (B) {};
	\vertex at \coord{135} (C) {};
	\vertex at \coord{45} (D) {};
	\draw  (A) to (B)  (B) to (C)  (C) to (D)  (D) to (A); 
	\node at (1.25,0.85) {$\scriptstyle I $};
	\node at (1.2,-0.85) {$\scriptstyle J $};
	\node at (-1.1,-0.85) {$\scriptstyle J  $};
	\node at (-1.1,0.85) {$\scriptstyle I $};
	\node at (1.1,0) {$\scriptstyle 4 $};
	\node at (-1.1,0) {$\scriptstyle 4 $};
	\node at (0,1.1) {$\scriptstyle 2 $};
	\node at (0,-1.1) {$\scriptstyle 2 $};
	\fill[black]  (A) circle [radius = 0.08cm];
	\fill[black]  (B) circle [radius = 0.08cm];
	\fill[black]  (C) circle [radius = 0.08cm];
	\fill[black]  (D) circle [radius = 0.08cm];
} 
= \tfrac{1}{30\, 6!} (N)_4 (N+4)_2{}^2 \, , \nn \\
&\tikz[baseline=(vert_cent.base),scale=0.5]{
	\node (vert_cent) {\hspace{-13pt}$\phantom{-}$};
	\vertex at \coord{-45} (A) {};
	\vertex at \coord{-135} (B) {};
	\vertex at \coord{135} (C) {};
	\vertex at \coord{45} (D) {};
	\draw  (A) to (B)  (B) to (C)  (C) to (D)  (D) to (A); 
	\node at (1.25,0.85) {$\scriptstyle I $};
	\node at (1.2,-0.85) {$\scriptstyle J $};
	\node at (-1.1,-0.85) {$\scriptstyle I  $};
	\node at (-1.1,0.85) {$\scriptstyle J $};
	\node at (1.1,0) {$\scriptstyle 3 $};
	\node at (-1.1,0) {$\scriptstyle 3 $};
	\node at (0,1.1) {$\scriptstyle 3 $};
	\node at (0,-1.1) {$\scriptstyle 3 $};
	\fill[black]  (A) circle [radius = 0.08cm];
	\fill[black]  (B) circle [radius = 0.08cm];
	\fill[black]  (C) circle [radius = 0.08cm];
	\fill[black]  (D) circle [radius = 0.08cm];
} 
= \tfrac{1}{5!^2} N (N+2)(N+4)(N+7)(N+9)(N+11)\, , \quad
\tikz[baseline=(vert_cent.base),scale=0.5]{
	\node (vert_cent) {\hspace{-13pt}$\phantom{-}$};
	\vertex at \coord{-45} (A) {};
	\vertex at \coord{-135} (B) {};
	\vertex at \coord{135} (C) {};
	\vertex at \coord{45} (D) {};
	\draw  (A) to (B)  (B) to (C)  (C) to (D)  (D) to (A); 
	\node at (1.25,0.85) {$\scriptstyle I $};
	\node at (1.2,-0.85) {$\scriptstyle I$};
	\node at (-1.1,-0.85) {$\scriptstyle J  $};
	\node at (-1.1,0.85) {$\scriptstyle J $};
	\node at (1.1,0) {$\scriptstyle 3 $};
	\node at (-1.1,0) {$\scriptstyle 3 $};
	\node at (0,1.1) {$\scriptstyle 3 $};
	\node at (0,-1.1) {$\scriptstyle 3 $};
	\fill[black]  (A) circle [radius = 0.08cm];
	\fill[black]  (B) circle [radius = 0.08cm];
	\fill[black]  (C) circle [radius = 0.08cm];
	\fill[black]  (D) circle [radius = 0.08cm];
} 
= \tfrac{1}{5!6!}( N)_3 (N+3)_3{}^2\, , \nn \\
& \tikz[baseline=(vert_cent.base),scale=0.5]{
	\node (vert_cent) {\hspace{-13pt}$\phantom{-}$};
	\vertex at \coord{-45} (A) {};
	\vertex at \coord{-135} (B) {};
	\vertex at \coord{135} (C) {};
	\vertex at \coord{45} (D) {};
	\draw  (A) to (B)  (B) to (C)  (C) to (D)  (D) to (A); 
	\draw (B)--(D); 
	\fill[white]  (0,0) circle [radius = 0.1cm];
	\draw (C)--(A); 
	\node at (1.25,0.85) {$\scriptstyle I $};
	\node at (1.2,-0.85) {$\scriptstyle J $};
	\node at (-1.1,-0.85) {$\scriptstyle I $};
	\node at (-1.1,0.85) {$\scriptstyle J $};
	\node at (1.1,0) {$\scriptstyle 4 $};
	\node at (-1.1,0) {$\scriptstyle 4 $};
	\node at (0,1.1) {$\scriptstyle 1 $};
	\node at (0,-1.1) {$\scriptstyle 1 $};
	\node at (-0.1,0.46) {$\scriptstyle 1$};
	\node at (-0.1,-0.46) {$\scriptstyle 1 $};		
	\fill[black]  (A) circle [radius = 0.08cm];
	\fill[black]  (B) circle [radius = 0.08cm];
	\fill[black]  (C) circle [radius = 0.08cm];
	\fill[black]  (D) circle [radius = 0.08cm];
}   = \tfrac{4}{5!6!} N (N+2)_3(N+5)^2 (N+9) \, , \quad
\tikz[baseline=(vert_cent.base),scale=0.5]{
	\node (vert_cent) {\hspace{-13pt}$\phantom{-}$};
	\vertex at \coord{-45} (A) {};
	\vertex at \coord{-135} (B) {};
	\vertex at \coord{135} (C) {};
	\vertex at \coord{45} (D) {};
	\draw  (A) to (B)  (B) to (C)  (C) to (D)  (D) to (A); 
	\draw (B)--(D); 
	\fill[white]  (0,0) circle [radius = 0.1cm];
	\draw (C)--(A); 
	\node at (1.25,0.85) {$\scriptstyle I $};
	\node at (1.2,-0.85) {$\scriptstyle I $};
	\node at (-1.1,-0.85) {$\scriptstyle J $};
	\node at (-1.1,0.85) {$\scriptstyle J $};
	\node at (1.1,0) {$\scriptstyle 4 $};
	\node at (-1.1,0) {$\scriptstyle 4 $};
	\node at (0,1.1) {$\scriptstyle 1 $};
	\node at (0,-1.1) {$\scriptstyle 1 $};
	\node at (-0.1,0.46) {$\scriptstyle 1$};
	\node at (-0.1,-0.46) {$\scriptstyle 1 $};		
	\fill[black]  (A) circle [radius = 0.08cm];
	\fill[black]  (B) circle [radius = 0.08cm];
	\fill[black]  (C) circle [radius = 0.08cm];
	\fill[black]  (D) circle [radius = 0.08cm];
}   = \tfrac{1}{6!^2} N (N+2)_4{}^2 \, , \nn \\
& \tikz[baseline=(vert_cent.base),scale=0.5]{
	\node (vert_cent) {\hspace{-13pt}$\phantom{-}$};
	\vertex at \coord{-45} (A) {};
	\vertex at \coord{-135} (B) {};
	\vertex at \coord{135} (C) {};
	\vertex at \coord{45} (D) {};
	\draw  (A) to (B)  (B) to (C)  (C) to (D)  (D) to (A); 
	\draw (B)--(D); 
	\fill[white]  (0,0) circle [radius = 0.1cm];
	\draw (C)--(A); 
	\node at (1.25,0.85) {$\scriptstyle I $};
	\node at (1.2,-0.85) {$\scriptstyle J$};
	\node at (-1.1,-0.85) {$\scriptstyle I$};
	\node at (-1.1,0.85) {$\scriptstyle J $};
	\node at (1.1,0) {$\scriptstyle 3 $};
	\node at (-1.1,0) {$\scriptstyle 3 $};
	\node at (0,1.1) {$\scriptstyle 2 $};
	\node at (0,-1.1) {$\scriptstyle 2 $};
	\node at (-0.1,0.46) {$\scriptstyle 1$};
	\node at (-0.1,-0.46) {$\scriptstyle 1 $};		
	\fill[black]  (A) circle [radius = 0.08cm];
	\fill[black]  (B) circle [radius = 0.08cm];
	\fill[black]  (C) circle [radius = 0.08cm];
	\fill[black]  (D) circle [radius = 0.08cm];
} = \tfrac{2}{5!6!}N(N+2)(N+4)(N+5)^2(N+7)(N+9) \, , \quad
 \tikz[baseline=(vert_cent.base),scale=0.5]{
	\node (vert_cent) {\hspace{-13pt}$\phantom{-}$};
	\vertex at \coord{-45} (A) {};
	\vertex at \coord{-135} (B) {};
	\vertex at \coord{135} (C) {};
	\vertex at \coord{45} (D) {};
	\draw  (A) to (B)  (B) to (C)  (C) to (D)  (D) to (A); 
	\draw (B)--(D); 
	\fill[white]  (0,0) circle [radius = 0.1cm];
	\draw (C)--(A); 
	\node at (1.25,0.85) {$\scriptstyle I $};
	\node at (1.2,-0.85) {$\scriptstyle I$};
	\node at (-1.1,-0.85) {$\scriptstyle J$};
	\node at (-1.1,0.85) {$\scriptstyle J $};
	\node at (1.1,0) {$\scriptstyle 3 $};
	\node at (-1.1,0) {$\scriptstyle 3 $};
	\node at (0,1.1) {$\scriptstyle 2 $};
	\node at (0,-1.1) {$\scriptstyle 2 $};
	\node at (-0.1,0.46) {$\scriptstyle 1$};
	\node at (-0.1,-0.46) {$\scriptstyle 1 $};		
	\fill[black]  (A) circle [radius = 0.08cm];
	\fill[black]  (B) circle [radius = 0.08cm];
	\fill[black]  (C) circle [radius = 0.08cm];
	\fill[black]  (D) circle [radius = 0.08cm];
} = \tfrac{2}{6!^2}N(N+2)(N+3)_3{}^2 \, , \nn \\
&\tikz[baseline=(vert_cent.base),scale=0.5]{
	\node (vert_cent) {\hspace{-13pt}$\phantom{-}$};
	\vertex at \coord{-45} (A) {};
	\vertex at \coord{-135} (B) {};
	\vertex at \coord{135} (C) {};
	\vertex at \coord{45} (D) {};
	\draw  (A) to (B)  (B) to (C)  (C) to (D)  (D) to (A); 
	\draw (B)--(D); 
	\fill[white]  (0,0) circle [radius = 0.1cm];
	\draw (C)--(A); 
	\node at (1.25,0.85) {$\scriptstyle I $};
	\node at (1.2,-0.85) {$\scriptstyle J$};
	\node at (-1.1,-0.85) {$\scriptstyle J$};
	\node at (-1.1,0.85) {$\scriptstyle I$};
	\node at (1.1,0) {$\scriptstyle 3 $};
	\node at (-1.1,0) {$\scriptstyle 3 $};
	\node at (0,1.1) {$\scriptstyle 2 $};
	\node at (0,-1.1) {$\scriptstyle 2 $};
	\node at (-0.1,0.46) {$\scriptstyle 1$};
	\node at (-0.1,-0.46) {$\scriptstyle 1 $};		
	\fill[black]  (A) circle [radius = 0.08cm];
	\fill[black]  (B) circle [radius = 0.08cm];
	\fill[black]  (C) circle [radius = 0.08cm];
	\fill[black]  (D) circle [radius = 0.08cm];
} = \tfrac{1}{5!6!}N(N+2)_2(N+4)_2{}^2(N+7) \, .
\end{align}
The scalar curvature is not fully determined due to ambiguities in obtaining $T_{IJ}$. However
at  large $N$
\be
R \sim  \tfrac{1}{6!^2 }\, N^{11} \big ( T_{4,1ac} + T_{5,1ad}-  2\, T_{4,1ab}  \big ) = -\tfrac{1}{6!^2} \,N^{11} \,2c_{6m}/c_2 \, ,
\quad - 2 c_{2m}/c_2 = \tfrac29\, , 
\ee
using the relations in \eqref{Trel}.

\section{Conclusion}

The somewhat lengthy gestation of this paper has for us been an autopedagogic exercise. 
For $\phi^6$ at 6 loops
the number of diagrams is relatively modest and their evaluation can be undertaken by hand without the use
of sophisticated software packages. In this  respect the present work is perhaps somewhat dated but we have
presented the results so that they may be used for symmetry groups other than the $O(N)$ case  that was considered here. 
The $\phi^6$ theory for $O(N)$ has two relevant operators and so has tricritical behaviour. Tricritical fixed
points are  relevant in some physical situations. The $\vep$-expansion is very far from convergent but may have
some relevance in bootstrap studies.

\vskip 24pt
\noindent{\bf Acknowledgements}

\vskip 2pt

We are both grateful to Johan Henriksson and Andy Stergiou for stimulating discussions relating to this
work. In particular the recent review by Johan Henriksson \cite{HenrikssonRev} was very much a trigger for the calculations undertaken here. We are also indebted to Oliver Schnetz for helpful advice.
Despite in the end there being some disagreements the paper by Johannes Hager 
\cite{Hager} was crucial in understanding and checking many aspects of our investigation.
We thank Alexander Bednyakov and Alexander Trenogin
for pointing out a couple of errors in version 1 in some formulae when $N=1$. We are further
acknowledge Johan for discovering an error in our results for $\gamma_3$.

\newpage
\begin{appendices}
\section{Z results} 
With $O(N)$ symmetry the bare couplings $g_{0,I}$ are related to the finite couplings 
$g_I =(\sigma,\tau_i,\nu,\lambda)$, $I=2,3,4,6$, by
\be
g_{0,I} = \mu^{\frac12 (I-2)\vep} g_I \, {\mathcal Z}_I \, , \qquad {\mathcal Z}_I = 1 - \sum_{\cG_{v}} \ 
(-1)^{\frac12 l_{\cG_v} }\lambda^{\frac12 l_{\cG_v} }   N_{\cG_{v},I}\,
\frac{1}{S_{\cG_{v}}  E_{\cG_{v}} }\, Z_{\cG_{v} } \, .
\ee

To ${\rm O}(\lambda^3)$ our results give
\be
{\mathcal Z}_\phi = 1-  \lambda^2  (N+2)(N+4)\,  \tfrac{1}{12\,\vep} 
- \lambda^3  (N+2)(N+4)^2(3N+22)\, \tfrac{1}{9\,\vep^2} \big ( 1- \tfrac23 \, \vep \big ) 
\, .
\ee
\begin{align}
{\mathcal Z}_2= {}& 1+ \lambda^2  (N+2)(N+4)\,  \tfrac{5}{4\, \vep} 
+ \lambda^2  (N+2)(N+4)(3N+22)\, \tfrac{5}{3\,\vep^2} \big ( 1- \tfrac{22}{5} \, \vep \big ) \nn \\
 &{} - \lambda^3\, \pi^2  (N+2)(N+4)^2(N+14)\,  \tfrac{1}{12\, \vep} \, .
\end{align}
\begin{align}
{\mathcal Z}_3= {}& 1+ \lambda\,  (N+4)\,  \tfrac{1}{\vep} 
+ \lambda^2 \, (N+4)\big ( (3N+22)\, \tfrac{1}{\vep^2} - (13N+122)\, \tfrac{1}{4\, \vep} \big ) \nn \\
\noalign{\vskip 2pt}
 &{} +  \lambda^3\,   (N+4)\big ( 21N^2+284N+970)\,  \tfrac{2}{3\, \vep^3} 
 - (107N^2 + 1682N+6440)\,  \tfrac{5}{12\, \vep^2} \nn \\
 \noalign{\vskip -2pt}
  &\hskip 2.5cm {}+  (649N^2+11514N+ 48512)\, \tfrac{1}{9\, \vep} \big ) \nn \\
  \noalign{\vskip 2pt} 
  &{} - \lambda^3 \, \pi^2 \,  (N+4) \big ( N^3 + 34N^2 +620N + 2720 )\,\tfrac{1}{12\, \vep^2}\nn \\
  \noalign{\vskip -2pt}
  &\hskip 2.8cm {}  + (3 N^3 +244N^2 +2152 N + 4576 ) \, \tfrac{1}{72\, \vep} \big  )
  \nn \\
  \noalign{\vskip  2pt}
 &{}  + \lambda^3\, \pi^2 \ln 2 \, (N+4)(N^3 +46 N^2 +1052 N +4976 )\,  \tfrac{1}{12\, \vep} 
 \nn \\
 \noalign{\vskip 2pt}
 &{}  -  \lambda^3 \, \zeta_3 \, (N+4) ( N^2+36N+188) \, \tfrac{21}{2\, \vep}  \, .
\end{align}
\begin{align}
{\mathcal Z}_4= {}& 1+ \lambda\,  (N+4)\,  \tfrac{4}{\vep} 
+ \lambda^2 \, (N+4)\big ( (N+6)\, \tfrac{20}{\vep^2} - (19N+126)\, \tfrac{3}{2\, \vep} \big ) \nn \\
\noalign{\vskip 2pt}
&{} - \lambda^2 \, \pi^2 (N+4)(N^2+18N+126)\, \tfrac{1}{8\, \vep}\nn \\
\noalign{\vskip 2pt}
 &{} +  \lambda^3\,   (N+4)\big ( (N+6)(2N+13)\,  \tfrac{160}{3\, \vep^3} -
 (277N^2 + 3819N+13034)\,  \tfrac{4}{3\, \vep^2} \nn \\
 \noalign{\vskip -2pt}
  &\hskip 2.5cm {}+  (686N^2+10425N+38914)\, \tfrac{8}{9\, \vep} \big ) \nn \\
  \noalign{\vskip 2pt} 
  &{} - \lambda^3 \, \pi^2 \,  (N+4) \big ( 4N^3 + 105N^2 +1274N +4692)\,\tfrac{1}{3\, \vep^2}\nn \\
  \noalign{\vskip -2pt}
  &\hskip 2.8cm {}  - (129 N^3 +2464N^2 +21772 N + 73360 ) \, \tfrac{1}{36\, \vep} \big  )
  \nn \\
  \noalign{\vskip  2pt}
 &{}  - \lambda^3\, \pi^2 \ln 2 \, (N+4)(N^3 -3 N^2 - 430 N - 2268 )\,  \tfrac{2}{3\, \vep} 
 \nn \\
  \noalign{\vskip 2pt}
  &{}+  \lambda^3 \, \pi^4 \, (N+4) ( N^3 + 40N^2 +440N +1544) \, \tfrac{1}{6\, \vep} 
  \nn \\
  \noalign{\vskip 2pt}
 &{}  -  \lambda^3 \, \zeta_3 \, (N+4) ( N+14) (2N+13) \, \tfrac{84}{\vep}  \nn \\
 \noalign{\vskip 2pt}
 &{}+ \lambda^3 \, C \, (N+4)(7N^2+ 132 N+ 536 )\, \tfrac{2}{ \vep}   \, .
\end{align}
\begin{align}
{\mathcal Z}_6 = {}& 1+ \lambda \, (3N+22)\,  \tfrac{2}{\vep} 
+ \lambda^2 \, \big ( (3N+22)^2\, \tfrac{4}{\vep^2} - (71N^2+1146N+4408)\, \tfrac{3}{4\, \vep} \big ) \nn \\
\noalign{\vskip 2pt}
&{} - \lambda^2 \, \pi^2 (N^3 + 34N^2+ 620N+2720)\, \tfrac{1}{8\, \vep}\nn \\
\noalign{\vskip 2pt}
 &{} +  \lambda^3\,  \big ( (3N+22)^3\,  \tfrac{8}{\vep^3} -
 (3N+22) (1489N^2 + 24054N+ 92552)\,  \tfrac{1}{6\, \vep^2} \nn \\
 \noalign{\vskip -1pt}
  &\hskip 2.7cm {}+  (2787N^3+ 68984N^2+551652N+1425952)\, \tfrac{4}{9\, \vep} \big ) \nn \\
  \noalign{\vskip 2pt} 
  &{} - \lambda^3 \, \pi^2 \,\big( (3N+22)  (N^3 + 34N^2 +620N + 2720)\,\tfrac{7}{12\, \vep^2} \nn \\
  \noalign{\vskip -2pt}
  &\hskip 2.7cm {}  - (51N^4+1618 N^3 + 32804N^2 +288968 N + 837184) \, \tfrac{1}{12\, \vep} \big  )
  \nn \\
  \noalign{\vskip  2pt}
 &{}  - \lambda^3\, \pi^2 \ln 2 \, (N^4-32 N^3 - 1984 N^2 - 20920 N - 61440 )\,  \tfrac{1}{2\, \vep} 
 \nn \\
  \noalign{\vskip 2pt}
  &{}+  \lambda^3 \, \pi^4 \,  ( N^4+ 64 N^3 + 1352N^2 +12248N + 36960) \, \tfrac{5}{24\, \vep} 
  \nn \\
  \noalign{\vskip 2pt}
 &{}  -  \lambda^3 \, \zeta_3 \, (11N^3 +428 N^2 +4228 N +12208)\, \tfrac{35}{\vep} \nn \\
 \noalign{\vskip 2pt}
 &{}+ \lambda^3 \, C \, (31 N^3 +1126N^2 + 11876N + 37592) \, \tfrac{1}{\vep} \, .
\end{align}

\section{Feynman Integrals}
\label{sec:int}

Most of the results given here can be obtained by considering a basic $l$ loop integral with two vertices
joined by $l+1$ propagators, corresponding to what are sometimes called sunset diagrams. 
The relevant integral  is then
\begin{align}
&\frac{1}{\pi^{\frac12 l \, d } }\int \rmd^d k_1 \dots \rmd^dk_l \;
\frac{1}{(k_1{}^2)^{\eta_1}_{\vphantom d} \dots (k_l{}^2)^{\eta_l }_{\vphantom d}\, 
((p-k_1- \dots -k_l)^2)^{\eta_{l+1}}_{\vphantom d}}
= L_l(\eta_1,\dots,\eta_{l+1})\, (p^2)^{-{\bar \eta}_{l+1}}_{\vphantom d} \, , \nn \\
& \hskip 2cm {\bar \eta}_{l+1}= \eta_1 + \dots + \eta_{l+1} - \tfrac12 l \, d\, ,
\end{align} 
where explicitly
\be
 L_l(\eta_1,\dots,\eta_{l+1}) = g(\eta_1)\dots g(\eta_{l+1}) \frac{1}{g({\bar \eta}_{l+1})}\, , \qquad
 g(\eta)= \frac{\Gamma(\frac12d - \eta)}{\Gamma(\eta)} \, .
 \label{defLg}
 \ee
 This is easy to obtain inductively starting from the well known result for $l=1$ and
 using the recurrence relation
 \be
  L_l(\eta_1,\dots,\eta_{l+1}) L_1( {\bar \eta}_{l+1}, \eta_{l+2}) =   L_{l+1}(\eta_1,\dots,\eta_{l+2})\, .
  \ee
  
  For $d=3-\vep$
  \be
  L_2(1,1,1) \sim \frac{2\pi}{\vep}\, , \qquad L_4(1,1,1,1,1) \sim-  \frac{2\pi^2}{3\vep}\, ,
  \ee
  which correspond to $Z_2$ and $Z_{4\gamma}$.
  
  For vertex graphs to determine the divergent part it is sufficient to consider a single momentum $p$
  entering  and leaving between one pair of vertices.\footnote{This method was followed for $\phi^4$
  theory in four dimensions at five loops in \cite{KleinertB}.}
   The  same result should be obtained for any particular
  pair although in most cases it is desirable to make a judicious choice to avoid any resulting IR sub divergences.
  In two cases this does not seem to be possible.
  Where feasible we have checked different momentum routings give the same result.
  After taking into account the necessary counterterms contributions the poles in $\vep$ should be independent of
  \be
  P =  e^{- \gamma_E}\frac{p^2}{\mu^2} \, , 
  \label{Pchoice}
  \ee
  which is a useful consistency check. $\gamma_E$ is the Euler constant and with \eqref{Pchoice} individual 
  contributions to each $Z$ are independent of $\gamma_E$.
  
  At four loops, with $\sim$ denoting the restriction to just $\vep$-pole terms,
  \begin{align}
&   L_ 4(1,1,1,1,2) P^{-2\vep} \sim \pi^2 Z_{4a}\, , \nn \\
&   L_2(1,1,1)\,L_2(1,1,1+\vep) P^{-2\vep} - \pi Z_2\, L_2(1,1,1) P^{-\vep}\sim \pi^2 Z_{4b}\,, \nn \\
 &  L_1(1,1)^2  L_2(1,1,1+\vep)P^{-2\vep}\sim \pi^2 Z_{4c} \, .
 \end{align}
 At six loops the necessary counterterms are dictated by the Hopf algebra coproducts.  For \eqref{6loop}
 \begin{align}
& L_1(1,1)L_3(1,1,1,1) L_2(2,1,2\vep) P^{-3\vep}  \sim 0 \, , \nn \\
& L_1(1,1)^2L_3(2,1,1,1) L_1(1+\vep,\tfrac12 + \tfrac32\vep) P^{-3\vep}  \sim \pi^3 Z_{6b} \, , \nn \\ 
& L_2(1,1,1)^2  L_2(2,1,2\vep) P^{-3\vep} - 2 \pi Z_2 \, L_4(1,1,1,1,2)  P^{-2\vep}   \sim \pi^3 Z_{6c} \, , \nn \\ 
 & L_1(1,1)L_2(1,1,1)  L_2(2,1,1)  L_1(1+\vep,\tfrac12 + \tfrac32\vep) P^{-3\vep} 
 -  \pi Z_2 \, L_4(1,1,1,1,2)  P^{-2\vep}   \sim \pi^3 Z_{6d } \, .
\end{align}
For \eqref{6loop2}, excluding $6i,6l$ which are calculated later,
 and noting that $6e,6f$ and $6g,6h$ can both be treated similarly,
\begin{align}
& L_1(1,1)L_2(1,1,1)^2  L_1(1,\tfrac12 + \tfrac52\vep) P^{-3\vep} 
 -  2\pi Z_2 \, L_1(1,1)L_2(1,1,1)L_1(1,\tfrac12 + \tfrac32\vep)  P^{-2\vep}  \nn \\
 \noalign{\vskip -2pt}
& \qquad  {}  +  ( \pi Z_2)^2 \, L_2(1,1,1) P^{-\vep} 
  \sim \pi^3 Z_{6e } \, , \nn \\
  & L_1(1,1)^2L_2(1,1,1)  L_1(1,\tfrac12 + \tfrac32\vep)  L_1(1,\tfrac12 + \tfrac52\vep) P^{-3\vep} 
 -  \pi Z_2 \, L_1(1,1)L_2(1,1,1)L_1(1,\tfrac12 + \tfrac32\vep)  P^{-2\vep}  \nn \\
 \noalign{\vskip -2pt}
& \qquad  {} -  \pi^2 Z_{4b} \, L_2(1,1,1) P^{-\vep} 
  \sim \pi^3 Z_{6g } \, , \nn \\
  & L_2(1,1,1)^2  L_2(1,1+\vep,1+\vep) P^{-3\vep} 
 -  2\pi Z_2 \, L_2(1,1,1)L_2(1,1, 1+ \vep)  P^{-2\vep}  \nn \\
 \noalign{\vskip -2pt}
& \qquad  {}  +  ( \pi Z_2)^2 \, L_2(1,1,1) P^{-\vep} 
  \sim \pi^3 Z_{6j } \, , \nn \\
  &L_1(1,1) L_2(1,1,1) L_2(1,1,1+\vep) L_1(1,\tfrac12+ \tfrac52\vep)P^{-3\vep} 
 -  \pi Z_2 \, L_1(1,1) L_2(1,1,1)L_1(1,\tfrac12+ \tfrac32 \vep)  P^{-2\vep}  \nn \\
 \noalign{\vskip -2pt}
& \qquad  {}  -   \pi^2 Z_{4b} \, L_2(1,1,1) P^{-\vep} 
  \sim \pi^3 Z_{6k } \, .
\end{align}
For \eqref{6loop3}, excluding $6n,6t,6u$ considered below,
\begin{align}
& L_2(1,1,1)^2  L_2(1,1,2\vep) P^{-3\vep} - 2\pi Z_2 \, L_4(1,1,1,1,1)P^{-2\vep}  \sim \pi^3 Z_{6\gamma} \, , \nn \\
& L_4(1,1,1,1,1)  L_2(1,1,1+2\vep) P^{-3\vep} - \pi^2 Z_{4\gamma} \, L_2(1,1,1)P^{-\vep}  
\sim \pi^3 Z_{6m} \, , \nn \\
& L_1(1,1) L_3(1,1,1,1) L_2(1,\tfrac12+\tfrac32\vep) L_1(1,\tfrac12+\tfrac52\vep) P^{-3\vep} 
- \pi^2 Z_{4a} \, L_2(1,1,1)P^{-\vep}  
\sim \pi^3 Z_{6o} \, , \nn \\
& L_2(1,1,1) L_3(1,1,1,1) L_1(2+\vep,-\tfrac12+\tfrac32\vep) P^{-3\vep} 
- \pi Z_{2} \, L_4(2,1,1,1,1)P^{-2\vep}  
\sim \pi^3 Z_{6p} \, , \nn \\ 
& L_1(1,1)^2 L_2(1,1,1) L_2(1,1+\vep,1+\vep) P^{-3\vep} 
- \pi Z_{2} \, L_1(1,1)^2 L_2(1,1,1+\vep)P^{-2\vep}  
\sim \pi^3 Z_{6q} \, , \nn \\
& L_1(1,1)^3 L_2(1,1,1) L_1(1+\vep,\tfrac12+\tfrac32\vep) P^{-3\vep} 
- \pi Z_{2} \, L_1(1,1)^3 L_1(1+\vep,\tfrac12 +\tfrac12 \vep)P^{-2\vep}  
\sim \pi^3 Z_{6r} \, , \nn \\
& L_1(1,1)^4  L_1(1+\vep,\tfrac12+\tfrac12\vep) L_1(1,\tfrac12+\tfrac52\vep)P^{-3\vep} 
- \pi^2 Z_{4c} \,  L_2(1,1,1)P^{-\vep}  
\sim \pi^3 Z_{6s} \, .
\end{align}
There remain $6v,6w$, which have an equivalent reduction, and $6y$ that allow a similar treatment
\begin{align}
& L_1(1,1)^2   L_3(1,1,1,1) L_1(1,\tfrac12+\tfrac52\vep)P^{-3\vep} \sim \pi^3 Z_{6v} \, ,\nn \\
& L_1(1,1)^4   L_2(1,1+\vep,1+\vep)P^{-3\vep} \sim \pi^3 Z_{6y} \, .
\end{align}

For $Z_{6l}$ if we consider momentum flowing through the two vertices labelled by $\lambda$ in \eqref{6loop2}
then IR subtractions are  necessary. 	In this case we may use
\be
(p^2)^{-{\frac12}+{\frac12}\delta} \sim \frac{\mu^\delta}{\delta} \, 
S_d  \, \delta^d(p)   \, , \qquad S_d = \frac{2\pi^{\frac12 d}}{\Gamma(\frac12 d)} \, , 
\label{sing}
\ee
to define
\be
\frac{1}{\pi^{\frac12 d}} \, L_1(1,1) \, (p^2)^{-{\frac32}-{\frac12}\vep} +\frac{2\pi}{\vep}  \, \delta^d(p) \, ,
\ee
as the IR regulated expression corresponding to the subgraph 
\tikz[baseline=(vert_cent.base),scale=0.45]{
  \node (vert_cent) {\hspace{-13pt}$\phantom{-}$};
  \node at  (1.2,0)  (A) {};
   \node at  (0,0)  (B) {};
    \node at  (-1,0)  (C) {};
       \draw [bend left = 70] (B.center) to (A.center);
       \draw [bend  right = 70] (B.center) to (A.center);
        \draw (B.center) to (C.center);
 \fill[black]  (A.center) circle [radius = 0.1cm];
\fill[black]  (B.center) circle [radius = 0.1cm];
 \fill[black]  (C.center) circle [radius = 0.1cm];
}. With this result we may obtain
\begin{align}
& L_1(1,1)^2 L_4 (1,1,1,\tfrac32 + \tfrac12 \vep,\tfrac32 + \tfrac12 \vep)P^{-3\vep} 
+ 2\,  \tfrac{2\pi}{\vep}\,  L_1(1,1)  L_3 (1,1,1,\tfrac32 + \tfrac12 \vep)P^{-2\vep}  \nn \\
& \qquad {} +\big (\tfrac{2\pi}{\vep}\big )^2 L_2(1,1,1)P^{-\vep}  -  \big (\tfrac{2\pi}{\vep}\big )^2 \pi Z_2
- \tfrac{2\pi}{\vep} \; 2 \,  \pi^2 Z_{4b} \sim \pi^3 Z_{6l} \, .
\end{align}

\subsection{Tetrahedral Graphs}

The divergent contributions arising from graphs with tetrahedral topology can be reduced to analysing 
the integral for a two loop master diagram. In the
simplest case it is necessary to consider just one propagator with a general dimension $\eta$.
For just the central line having an exponent $\eta$ different from $1$
\be
 \tikz[baseline=(vert_cent.base)]{
  \node (vert_cent) {\hspace{-13pt}$\phantom{-}$};
 \draw (-0.4,0)--(0.1,0);
    \draw (0.7,0) ++(0:0.6cm and 0.4cm) arc (0:90:0.6cm and 0.4cm) node(n1) {}
     (0.7,0) ++(90:0.6cm and 0.4cm) arc (90:180:0.6cm and 0.4cm) 
             (0.7,0) ++(180:0.6cm and 0.4cm) arc (180:270:0.6cm and 0.4cm) node(n2){}
             (0.7,0) ++(270:0.6cm and 0.4cm) arc (270:360:0.6cm and 0.4cm) ;
              \draw [line  width = 1](n1.base) to (n2.base);
             \draw (1.3,0) -- (1.8,0); 
             \node at (0.84,-0.02) {$\scriptstyle\eta$};         
          }\ \ \rightarrow \ \ G_1(\eta) (p^2)^{d-4-\eta} \, ,  
 \label{central}         
\ee
the essential result used here was obtained by Kotikov \cite{Kotikov1}
\begin{align}
G_1(\eta) ={}& F(\eta) + H(\eta)\,
 {}_3F_2 \Big (
{\genfrac{}{}{0pt}{3}{\scriptstyle{1,\ d-2, \ 2 -\frac12d + \eta}}
{\scriptstyle{\eta+1,\ 3-\frac12d + \eta }} } ; 1 \Big ) \, , \nn \\
F(\eta) = {}& - 2 \pi \, \frac{\Gamma(\frac12 d-1)\, \Gamma(\frac12 d-1-\eta)\, \Gamma(3- d+\eta)^2 \,
\Gamma( d-2 -\eta)}
{\Gamma( d-2)\, \Gamma( d-\frac32 -\eta)\,\Gamma(\frac52 - d + \eta)}\, , \nn \\
H(\eta) = {}&  2 \, \frac{\Gamma(\frac12 d-1)^2\, \Gamma(\frac12 d-2-\eta)\, \Gamma(3- d+\eta)}
{\Gamma( \frac32 d- 4 -\eta)\,\Gamma( \eta+1)}\, , 
\label{Kotikov}
\end{align}
where $G_1(0) = L_1(1,1)^2$.
The expansion of ${}_3F_2 \Big (
{\genfrac{}{}{0pt}{3}{\scriptstyle{a,\ b, \ c}} {\scriptstyle{e,\ f }} }; 1 \Big ) $ is convergent for $e+f>a+b+c$. 
Using \eqref{Kotikov} this fails
for the cases of interest here but the problem can be avoided with the aid of the symmetry relation\footnote{This may be 
obtained directly from \eqref{Kotikov} using the identity
$${}_3F_2 \big ({\genfrac{}{}{0pt}{3}{\scriptstyle{1,\ b, \ c}} {\scriptstyle{e,\ c+1 }} }\big )
= -\tfrac{c(1-e)}{(b-c)(1+b-e)} \,{}_3F_2 \big ( {\genfrac{}{}{0pt}{3}{\scriptstyle{1,\ b, \ 1+b-e}} {\scriptstyle{1+b-c,\ 2+b-e }} }\big )
+  \tfrac{c\,\Gamma(1+c-e)}{\Gamma(b)} \big (\tfrac{\Gamma(b-e)\, \Gamma(1-b+e)\, \Gamma(b-c)}{\Gamma(1-e)}
- \tfrac{\Gamma(c)\, \Gamma(1-c)\, \Gamma(e)}{\Gamma(1-b+c)}\big )\,. $$}
\be
G_1(\eta)= G_1(\tilde \eta)\, , \qquad \tilde \eta= \tfrac32 d - 4 -  \eta \, .
\ee

With this result, taking $d=3-\vep, \ \eta \to {\tilde \eta}$ in \eqref{Kotikov},
\begin{align}
&  L_3(1,1,1,1) G_1(1-3 \vep) L_1(1,\tfrac12+ \tfrac52\vep)P^{-3\vep}  
- 2\pi^2 Z_{4a} L_2(1,1,1) P^{-\vep} \sim \pi^3 Z_{6n} \, , \nn \\
&  L_2(1,1,1) G_1(\tfrac12-\tfrac52\vep) L_2(1,1, 1+2\vep)P^{-3\vep}  
- \pi Z_{2} L_1(1,1)^2 L_2(1,1, 1+\vep)P^{-2\vep} \sim \pi^3 Z_{6u} \, .
\end{align}
For $Z_{6n}$  we need only
\be
{}_3F_2 \Big (
{\genfrac{}{}{0pt}{3}{\scriptstyle{1,\ 1- \vep , \ \frac32 - \frac52\vep}}
{\scriptstyle{2 - 3 \vep,\ \frac52  - \frac52 \vep }} }; 1 \Big ) \sim 6(1-\ln 2)
 \, ,  \qquad e^{\gamma_E \vep} G_1( 1-3 \vep)  \sim 2\pi \big ( \tfrac1\vep + 1 \big ) \, , 
\ee
while  for $Z_{6u}$ the first two terms in an $\vep$ expansion are necessary
\begin{align}
{}_3F_2 \Big (
{\genfrac{}{}{0pt}{2}{\scriptstyle{1,\ 1, \ 1 - 2\vep}}
{\scriptstyle{ \frac32 -\frac52\vep,\ 2  - 2\vep }} }; 1 \Big ) 
\sim {}& \sum_{n\ge 0} \frac{n!}{(\frac32)_n } \frac{1}{n+1} \bigg (  1-2\vep +\frac{2\vep}{n+1} + \sum_{m=1}^n
\Big ( \frac52\, \frac{1}{m+\frac12} -  \frac{1}{m} \Big ) \vep \bigg ) \nn \\
={}& \tfrac{1}{4} \, \pi^2 + \big ({ - \tfrac74}\, \pi^2 + \pi^2 \ln 2 +\tfrac{21}{2}\, \zeta_3 \big )\vep \, ,  \nn \\
e^{\gamma_E \vep} G_1(\tfrac12-\tfrac52\vep) \sim {}& \pi^3 + \pi \big ( 4\, \pi^2 \ln 2 - 21\, \zeta_3 \big )\vep \, .
 \end{align}

 Two other two loop results are relevant here
 \begin{align}
 & \tikz[baseline=(vert_cent.base)]{
  \node (vert_cent) {\hspace{-13pt}$\phantom{-}$};
 \draw (-0.4,0)--(0.1,0);
    \draw (0.7,0) ++(0:0.6cm and 0.4cm) arc (0:90:0.6cm and 0.4cm) node(n1) {};
   \draw [line width =1]     (0.7,0) ++(90:0.6cm and 0.4cm) arc (90:180:0.6cm and 0.4cm) 
          (0.7,0) ++(180:0.6cm and 0.4cm) arc (180:270:0.6cm and 0.4cm) node(n2){};
\draw             (0.7,0) ++(270:0.6cm and 0.4cm) arc (270:360:0.6cm and 0.4cm) ;
              \draw (n1.base) to (n2.base);
             \draw (1.3,0) -- (1.8,0); 
             \node at (0.4,0.52) {$\scriptstyle\eta_1$};    
              \node at (0.4,-0.52) {$\scriptstyle\eta_2$};       
          } \  \rightarrow \  G_2(\eta_1,\eta_2) (p^2)^{d - \eta_1-\eta_2 -3}\, ,  \quad
             \tikz[baseline=(vert_cent.base)]{
  \node (vert_cent) {\hspace{-13pt}$\phantom{-}$};
 \draw (-0.4,0)--(0.1,0);
    \draw  [line width =1]  (0.7,0) ++(0:0.6cm and 0.4cm) arc (0:90:0.6cm and 0.4cm) node(n1) {}
     (0.7,0) ++(90:0.6cm and 0.4cm) arc (90:180:0.6cm and 0.4cm) ;
   \draw       (0.7,0) ++(180:0.6cm and 0.4cm) arc (180:270:0.6cm and 0.4cm) node(n2){};
\draw             (0.7,0) ++(270:0.6cm and 0.4cm) arc (270:360:0.6cm and 0.4cm) ;
              \draw (n1.base) to (n2.base);
             \draw (1.3,0) -- (1.8,0); 
             \node at (0.35,0.52) {$\scriptstyle\eta_1$};    
              \node at (1.14,0.48) {$\scriptstyle \eta_2$};       
          }         \ \rightarrow \ {\tilde G}_2(\eta_1,\eta_2)(p^2)^{d - \eta_1-\eta_2 -3}\, .
          \label{Gtwo}
 \end{align}
 These satisfy 
 \begin{align}
 G_2(\eta_1,\eta_2) ={}&  G_2(\eta_2,\eta_1) = G_2(\eta_6,\eta_2) \, , \qquad  \quad \eta_6=\tfrac32 d - \eta_1 - \eta_2 - 3  \, , \nn \\
 G_2(\eta_1,\eta_2)  ={}&  g(\eta_1 + 2 - \tfrac12 d) \, g(\eta_2 + 2 - \tfrac12 d) \, g(\eta_6+ 2 - \tfrac12 d) 
\, g(d-3) \, {\tilde G}_2(\eta_1, \eta_6)  \, .
\label{central3}   
 \end{align}
 In this case integration by parts techniques can be used to find explicit expressions  just in terms of  $\Gamma$  functions
 \begin{align}
 G_2(\eta_1,\eta_2)= {}& \Big ( \tfrac{(d-{\eta_6} -3) ( \eta_6+2 -\frac12 d)}{(\frac12 d-1 -\eta_1)(\frac12 d-1 -\eta_2)}\, L_1(\eta_1,\eta_2) 
 + \tfrac{(d-\eta_1-3) ( \eta_1+2 -\frac12 d)}{(\frac12 d-1 -\eta_2)( \frac12 d-1- \eta_6 )} \, L_1(\eta_2,\eta_6) \nn \\
 \noalign{\vskip - 2pt}
 &\hskip 4cm {}
 + \tfrac{(d-\eta_2-3) ( \eta_2+2 -\frac12 d)}{(\frac12 d-1 -\eta_1)(\frac12 d-1- \eta_6 )}\,  L _1(\eta_1,\eta_6)  \Big )\, L_1(1,1) \, ,
 \label{G2}
 \end{align}
 and
 \begin{align}
{\tilde G}_2(\eta_1, \eta_2 )
=&{} - \tfrac{1}{(d-3) (\frac12 d-2) }
\big (\eta_1 \eta_6 \,  L_1(\eta_1+1,1) L_1(\eta_6+1 ,1) +  \eta_2 \eta_6\,  L_1(\eta_2+1,1) L_1(\eta_6+1 ,1)\nn \\
& \hskip 2.5cm {} + \eta_1 \eta_2 \, L_1(\eta_1+1,1)L_1(\eta_2+1,1)\big )   \, .
\label{G2t}
\end{align}

With $d=3-\vep$ in \eqref{G2}
\begin{align}
& L_1(1,1)L_2(1,1,1)G_2(\tfrac12 +\tfrac12 \vep ,\vep)L_1(1,\tfrac12 + \tfrac52\vep)P^{-3\vep}
- \pi Z_2 L_2(1,1,1) L_2 (1,1,1+\vep) P^{-2\vep}\nn \\
\noalign{\vskip 2pt}
&\qquad {}- \pi^2 Z_{4b} L_2(1,1,1) P^{-\vep} \sim \pi^3 Z_{6i} \, , \nn \\
&L_1(1,1)^3G_2(\tfrac12 +\tfrac12 \vep,\tfrac12 +\tfrac12 \vep )L_1(1,\tfrac12 + \tfrac12\vep)
L_1(1+2\vep,\tfrac12 + \tfrac12 \vep) P^{-3\vep} \sim \pi^3 Z_{6x} \, .
\end{align}

To obtain $Z_{6t}$ we consider the reduction to restricting the momentum to flow between just two vertices
labelled by $V_5$ in the associated graph in \eqref{6loop3}. However there are then also IR divergences 
which generate poles in $\vep$ and need to be subtracted, corresponding to the $R^*$ operation described
in \cite{KleinertB}. For the $Z_{6t}$  case the IR contribution is
obtained by using, with $d=3-\vep$,
\be
\frac{1}{\pi^d} \, \frac{1}{k^2 \, l^2 \, (k-l)^2} \sim -  \frac{2\pi}{\vep}\, \delta^d(k) \delta^d(l) \, ,
\ee
and hence the subgraph corresponding to ${\tilde G}_2(\eta_1,\eta_2)$ is regulated by taking
\be
{\tilde G}_2(\eta_1,\eta_2) (p^2)^{-\vep -\eta_1-\eta_2} +  \frac{2\pi}{\vep}\,  (p^2)^{-\eta_1-\eta_2}\, .
\ee
Similar considerations apply to the counterterm involving $Z_{4c}$. Naively, with the momentum flow
choice made here, the remaining integral $\frac{1}{\pi^d} \int \rmd^d k\, \rmd^d l \,\frac{1}{k^2 \, l^2 \, (k-l)^2}$
is set to zero using dimensional regularisation but with the necessary  IR subtraction in $3-\vep$ dimensions
there is a contribution $2\pi/\vep$.
Hence, after subtracting IR $\vep$ poles,
\begin{align}
& L_1(1,1)^3 \,{\tilde G}_2(\tfrac12 + \tfrac12 \vep, \tfrac12 + \tfrac12 \vep)
L_1(1+2\vep,\tfrac12 + \tfrac12 \vep) P^{-3\vep} 
+ \tfrac{2\pi}{\vep}\,  L_1(1,1)^3  L_1(1+\vep,\tfrac12 + \tfrac12 \vep )P^{-2\vep}  \nn \\
& \qquad {}- \tfrac{2\pi}{\vep}\,\pi^2 Z_{4c} \sim \pi^3 Z_{6t} \, .
\end{align}

\subsubsection{Primitive Tetrahedral Graph}

The last graph $\cG_{6z}$ is primitive so no subtractions are necessary. It is reducible to the two
loop graph
\be
\tikz[baseline=(vert_cent.base)]{
  \node (vert_cent) {\hspace{-13pt}$\phantom{-}$};
 \draw (-0.4,0)--(0.1,0);
    \draw  [line width =1]  (0.7,0) ++(0:0.6cm and 0.4cm) arc (0:90:0.6cm and 0.4cm) node(n1) {}
     (0.7,0) ++(90:0.6cm and 0.4cm) arc (90:180:0.6cm and 0.4cm) ;
   \draw  [line width =1]       (0.7,0) ++(180:0.6cm and 0.4cm) arc (180:270:0.6cm and 0.4cm) node(n2){};
\draw             (0.7,0) ++(270:0.6cm and 0.4cm) arc (270:360:0.6cm and 0.4cm) ;
              \draw (n1.base) to (n2.base);
             \draw (1.3,0) -- (1.8,0); 
             \node at (0.35,0.52) {$\scriptstyle\eta_1$};    
              \node at (1.14,0.48) {$\scriptstyle \eta_2$};   
               \node at (0.35,-0.52) {$\scriptstyle \eta_3$};       
          }         \ \rightarrow \ {G}_3(\eta_1,\eta_2,\eta_3)(p^2)^{d - \eta_1-\eta_2 -\eta_3-2}\, ,
\ee
with the external lines joined.  Clearly $G_3(\eta_1,1,\eta_2)= G_2(\eta_1,\eta_2), \ 
G_3(\eta_1,\eta_2,1)= {\tilde G}_2(\eta_1,\eta_2)$ in \eqref{Gtwo} but for general $\eta_i$ 
$G_3$ cannot be reduced to $\Gamma$-functions but is expressible in terms of two ${}_3 F_2(1)$
generalised hypergeometric functions.
For $\cG_{6z}$ the three one loop bubbles have a contribution involving
$ {G}_3(\eta_1,\eta_2,\eta_3)$ with each $\eta_i = \tfrac12 + \tfrac12 \vep$. Hence
\be
L_1(1,1)^3 \, G_3\big (\tfrac12 + \tfrac12 \vep,\tfrac12 + \tfrac12 \vep,\tfrac12 + \tfrac12 \vep \big ) \,
  L_1(1,\tfrac12+ \tfrac52\vep)P^{-3\vep}  
\sim \pi^3 Z_{6z} \, ,
\ee
so that
\be
 Z_{6z} = \tfrac{2\pi}{3\, \vep} \, G_3\big (\tfrac12,\tfrac12,\tfrac12\big )\big  |_{d=3} \, , 
 \ee
 where 
 \be
 G_3\big (\tfrac12,\tfrac12,\tfrac12\big )\big  |_{d=3} = \tG_3\big (1,1,1\big )\big  |_{d=3} = \tfrac{2}{\pi}\, C\, ,
\ee
with $\tG_3$ the  corresponding $x$-space integral discussed in appendix \ref{kotikov}.
Using the result obtained in \eqref{Kres} for the two loop master integral this is expressible in terms of a double sum\footnote{For $d=4$
the corresponding result is much simpler  since only $k=0$ contributes and there is a single sum,
$G_3(1,1,1 )\big  |_{d=4} = \tG_3(1,1,1 )\big  |_{d=4}  = 6 \sum_{n\ge 0}\frac{1}{(n+1)^3} = 6\, \zeta_3$.}
\begin{align}
 C ={}& 
\sum_{n,k\ge 0} \, \frac{(\frac12)_k^{\vphantom g}}{k!} \, 
\frac{(n+k)!}{(\frac32)_{n+k}^{\vphantom g}}\, F(n,k) \, , \nn \\
 F (n,k)= {}& \frac{1}{(n+\frac12)(n+k+\frac12)^2_{\vphantom d}} 
 + \frac{2}{(n+\frac12)^2_{\vphantom d}(n+k+\frac12)}\nn \\
 \noalign{\vskip-3pt}
&{} + \frac{1}{(n+\frac12)(n+k+ 1)^2_{\vphantom d}} + \frac{2}{(n+ \frac12)^2_{\vphantom d}(n+k+ 1)} \, .
\end{align} 
The sums here converge rather slowly, cutting off the $k,n$ sums at 1000 gives a result for $C$
only to 5 significant figures. However the sum can be transformed to a double integral on the unit
square giving numerically $C = 43.698497$ which matches \eqref{Cres}.

\section{Two Loop Kotikov Integral}

\label{kotikov}

The   two loop master integral corresponds to the graph
\be
 \tikz[baseline=(vert_cent.base)]{
  \node (vert_cent) {\hspace{-13pt}$\phantom{-}$};
 \draw (-1,0)--(-0.3,0);
    \draw  [line  width = 1] (0.7,0) ++(0:1cm and 0.7cm) arc (0:90:1cm and 0.7cm) node(n1) {}
     (0.7,0) ++(90:1cm and 0.7cm) arc (90:180:1cm and 0.7cm) 
             (0.7,0) ++(180:1cm and 0.7cm) arc (180:270:1cm and 0.7cm) node(n2){}
             (0.7,0) ++(270:1cm and 0.7cm) arc (270:360:1cm and 0.7cm) ;
              \draw [line  width = 1] (n1.base) to (n2.base);
             \draw (1.7,0) -- (2.4,0); 
              \node at (0.3,0.45) {$\scriptstyle\eta_1$}; 
               \node at (0.3,-0.45) {$\scriptstyle\eta_2$};    
                \node at (1.1,0.45) {$\scriptstyle\eta_3$}; 
               \node at (1.1,-0.45) {$\scriptstyle\eta_4$}; 
                \node at (0.93,0) {$\scriptstyle\eta_5$};   
           \node at (-1.2,0) {$\scriptstyle p$};
             \node at (2.6,0) {$\scriptstyle p$};
            \node at (-0.3,0.5) {$\scriptstyle k$}; 
               \node at (-0.5,-0.5) {$\scriptstyle  p-k $};  
                 \node at (1.7,0.5) {$\scriptstyle l$}; 
               \node at (1.8,-0.5) {$\scriptstyle p-l $};   
                \node at (0.35,0) {$\scriptstyle k-l$}; 
          }
   \label{2loop}         
\ee
with each line having an arbitrary exponent 
\begin{align}
\frac{1}{\pi^{d}} & \int \!  {\rm d}^d k \! \int \! {\rm d}^d l\,  \frac{1}{(k^2)^{\eta_1}_{\vphantom d} \,   
((p-k)^2)^{\eta_2}_{\vphantom d}  
\, (l^2)^{\eta_3}_{\vphantom d}  \, ((p-l)^2)^{\eta_4 }_{\vphantom d} \,((k-l)^2)^{\eta_5} _{\vphantom d} } 
 =  G({ \eta_i} ) \,   (p^2)^{ \eta_6 - \frac12 d}_{\vphantom d} \, ,  \nn \\
 & {\ts \sum_{i=1}^6}\,  \eta_i = \tfrac32 d \, , \quad G(\eta_1,\eta_2,\eta_3,\eta_4, \eta_5)=
  G(\eta_2,\eta_1,\eta_4,\eta_3, \eta_5) =  G(\eta_4,\eta_1,\eta_3,\eta_2, \eta_5) \, .
\label{int2}
\end{align}
Equivalently there is a corresponding $x$-space  integral 
\be       
   \tikz[baseline=(vert_cent.base)]{
  \node (vert_cent) {\hspace{-13pt}$\phantom{-}$};
   \node at (- 0.5,0) {$0 $}; 
    \node at (1.9,0) {$ z $}; 
    \fill  (-0.3,0) circle [radius=1.5pt];
    \fill  (1.7,0) circle [radius=1.5pt];
    \fill  (0.7,0.7) circle [radius=1.5pt];
    \fill  (0.7,-0.7) circle [radius=1.5pt];
    \draw  [line  width = 1] (0.7,0) ++(0:1cm and 0.7cm) arc (0:90:1cm and 0.7cm) node(n1) {}
     (0.7,0) ++(90:1cm and 0.7cm) arc (90:180:1cm and 0.7cm) 
             (0.7,0) ++(180:1cm and 0.7cm) arc (180:270:1cm and 0.7cm) node(n2){}
             (0.7,0) ++(270:1cm and 0.7cm) arc (270:360:1cm and 0.7cm) ;
              \draw [line  width = 1] (n1.base) to (n2.base);
                        \node at (0.3,0.45) {$\scriptstyle\eta_1$}; 
               \node at (0.3,-0.45) {$\scriptstyle\eta_3$};    
                \node at (1.1,0.45) {$\scriptstyle\eta_2$}; 
               \node at (1.1,-0.45) {$\scriptstyle\eta_4$}; 
                \node at (0.93,0) {$\scriptstyle\eta_5$};   
            \node at (0.7,0.9) {$x$}; 
               \node at (0.7,-0.9) {$y $};  
                 } \,  
\ee
which has an identical form to \eqref{int2} if $k\to x,\, l\to y,\, p\to z$.
Using Fourier transforms there is then the relation
\be
G ({ \eta_i}^{\vphantom g}  ) = g(\eta_1) g(\eta_2)  g(\eta_3)  g(\eta_4)  g(\eta_5)  g(\eta_6) \, 
 G ( {\teta}_i)\big |_{\eta_2\leftrightarrow \eta_3} \, , \qquad \teta_i = \tfrac12 d - \eta_i \, ,
\ee
with $g$ given by \eqref{defLg}.  An  historical review for results for this  master integral is contained in
\cite{GrozinH}. 

Kotikov \cite{Kotikov1,Kotikov2}\footnote{A  related discussion is given in \cite{Broadhurst} and  related
results are described in \cite{Kotikov3}.} obtained a result when two lines in the momentum space integral had exponent 1  by analysing the corresponding $x$-space integral
\begin{align}
\tG_3(\teta_1,\teta_2,\teta_3)={}& \frac{1}{\pi^{d}} \int \!  {\rm d}^d x \! \int \! {\rm d}^d y\,  
\frac{1}{(x^2)^{\teta_1}_{\vphantom d}\, ((x-z)^2)^{\teta_2}_{\vphantom d}   \, (y^2)^{\teta_3}_{\vphantom d}
((x-y)^2)^{\lambda}_{\vphantom d}  
\,((y-z)^2)^{\lambda}_{\vphantom d} }\, , \nn \\
& \lambda = \tfrac12 d - 1 \, , \quad z^2=1 \, .
\label{G3int}
\end{align}
The integral is free of divergences so long as
\begin{align}
& \teta_1 +\teta_2 >1\,,\quad \teta_3 +\lambda >1\,,\quad \teta_1+\teta_2 +\teta_3 >2 \,, \nn \\
& \teta_1,\teta_2,\teta_3 < \lambda +1 \,,\quad  \teta_1+ \teta_3 < \lambda+2 \,  ,
\label{divG3}
\end{align}
and there is a symmetry
\be
\tG_3(\teta_1,\teta_2,\teta_3)= \tG_3(\teta_6,\teta_2,\teta_3) \, , \quad 
\teta_6= \lambda +3 -\teta_1 -\teta_2 - \teta_3  \, ,
\label{Gsym}
\ee
which relates the conditions in \eqref{divG3}. By using inversions in \eqref{G3int} we may also obtain the  relation
\be
\tG_3(\teta_1,\teta_2,\teta_3)= \tG_3(\lambda + 2-\teta_1-\teta_2 ,\teta_2, 2-\teta_3 )\, .
\label{G3inv}
\ee
More generally linking the external lines in \eqref{2loop}, 
with an associated exponent $\eta_6$, to form a tetrahedron there is a symmetry under permutations of the four vertices. This gives
\be
\tG_3(\teta_1,\teta_2,\teta_3)=G(\eta_1,\eta_6,\lambda,\lambda,\eta_3) = 
G(\lambda,\eta_1,\lambda,\eta_6,\eta_2) = G(\lambda,\eta_1,\eta_6,\lambda,\eta_2) \, ,
\ee
with other relations flowing from the symmetries in \eqref{int2} and \eqref{Gsym}. Hence
\be
G(\lambda,\lambda,\lambda,\lambda,\teta)= \tG_3(\lambda,\teta,3-\lambda-\teta)= 
 \tG_3(\lambda,3-\lambda-\teta,\teta) \, .
 \ee

For completeness and to obtain expressions which are of a convenient form for our discussion
we recapitulate the salient aspects of the derivation in \cite{Kotikov1}.
The analysis depends on an expansion in terms of Gegenbauer polynomials where
\be 
\frac{1}{((x-y)^2)^{\lambda}_{\vphantom d}  } = \sum_{m\ge 0}
\bigg ( \theta(|x| -|y|) \, \frac{|y|^m}{|x|^{2\lambda+m}} +  \theta(|y| -|x|) \, \frac{|x|^m}{|y|^{2\lambda+m} }\bigg )
C^{(\lambda)}_m( {\hat x}\cdot  {\hat y} ) \, ,
\ee 
with a similar expansion for $ ((y-z)^2)^{-\lambda}_{\vphantom d}  $ in terms of 
$C^{(\lambda)}_n( {\hat y}\cdot  {\hat z} ) $. Crucially, taking 
\be
{\rm d}^d y=  {\rm d}|y| \, |y|^{2\lambda +1} \, {\rm d}\Omega_{{\hat y}} \,, 
\ee
 there is the orthogonality relation
\be
\frac{1}{\pi^{\frac12 d} }\int_{S^{d-1}} {\hskip -0.3 cm}  {\rm d}\Omega_{{\hat y}} \; C^{(\lambda)}_m( {\hat x}\cdot  {\hat y} ) \, 
C^{(\lambda)}_n( {\hat y}\cdot  {\hat z} ) = \delta_{mn}  \, \frac{1}{\lambda+n} \frac{2}{\Gamma(\lambda)}\,
C^{(\lambda)}_n( {\hat x}\cdot  {\hat z} ) \,  , \quad C^{(\lambda)}_n(1) = \frac{(2\lambda)_n}{n!} \,.
\label{orthog}
\ee
The integration over $|y|=y$ here becomes
\begin{align}
& \int_0^\infty \!\!\! {\rm d}y \, y^{2\lambda +1-2\teta_3} \,
\bigg ( \theta(|x| -y) \, \frac{y^n}{|x|^{2\lambda+n}} +  \theta(y -|x|) \, \frac{|x|^n}{y^{2\lambda+n} }\bigg )
 \bigg ( \theta(1 -y) \, y^n +  \theta(y - 1) \, \frac{1}{y^{2\lambda+n} }\bigg ) \nn \\
 &= \theta(|x|-1) \frac{\lambda+n}{2(1-\teta_3)} \bigg( \frac{1}{\lambda-1+\teta_3+n}
 \, \frac{1}{|x|^{2\lambda+2\teta_3-2+n} } - \frac{1}{\lambda+1-\teta_3+n}
\,  \frac{1}{|x|^{2\lambda+n} }  \bigg ) \nn \\
 &\ \ {}+  \theta(1-|x|) \frac{\lambda+n}{2(1-\teta_3)} \bigg( \frac{1}{\lambda-1+\teta_3+n}\,
 |x|^{n}  - \frac{1}{2- 2\teta_3+n} \, |x|^{2 -2\teta_3 +n} \bigg ) \,.
\end{align}
There are six potential non zero contributions allowed by the step functions for the $y$ integral 
but these can be simplified to four as above. For the  $x$ integral 
$((x-z)^2)^{-\teta_2}$ is expanded  in terms of $C^{(\teta_2)}_l( {\hat x}\cdot  {\hat z} ) $.
With ${\rm d}^d x=  {\rm d}|x| \, |x|^{2\lambda +1} \, {\rm d}\Omega_{{\hat x}} $ the integral over $|x|=x$ becomes
\begin{align}
& \int_0^\infty \!\!\! {\rm d} x \, x^{2\lambda +1-2\teta_1} \,\nn \\
\noalign{\vskip - 4pt}
& \quad \times\bigg ( \theta(x -1) \, \Big (  \frac{1}{\lambda -1+\teta_3+n }\, 
\frac{1}{x^{2\lambda+ 2\teta_2+2\teta_3-2+l+ n}} 
-  \frac{1}{\lambda +1-\teta_3+n }\, \frac{1}{|x|^{2\lambda+ 2\teta_2  +l+ n}} \Big )
\nn \\ 
& \qquad  \quad  +    \theta(1-x ) \, \Big (  \frac{1}{\lambda -1+\teta_3+n }\,  x^{l+n}
-  \frac{1}{\lambda +1-\teta_3+n } \, x^{2- 2\teta_3  +l+ n} \Big ) \bigg )
\nn \\  
 &=   \frac{1}{2(\lambda-1+\teta_3+n)} \bigg ( \frac{1} {\teta_1+\teta_2+\teta_3-2 +\frac12(l+n)}
 + \frac{1} {\lambda -\teta_1 +1 +\frac12(l+n)} \bigg ) \nn \\
 &\quad - \frac{1}{2(\lambda+1-\teta_3+n)} \bigg ( \frac{1} {\teta_1+\teta_2- 1 +\frac12(l+n)}
 + \frac{1} {\lambda -\teta_1-\teta_3 + 2 +\frac12(l+n)} \bigg ) \, .
 \end{align}
 For the remaining integral involving ${\rm d}\Omega_{{\hat x}} $  it is necessary to expand
 the polynomial $C^{(\teta_2)}_l$ in terms of $C^{(\lambda)}_p$ for $p\le l$,\footnote{
 The formula is just quoted in \cite{Kotikov1} but is contained, with a proof, in \cite{Andrews} as Theorem 7.1.4'.}
 \be
 C^{(\teta_2)}_l = \sum_{k=0}^{\lfloor \frac12 l \rfloor} \frac{ (\teta_2)_{l-k} \ (
 \teta_2-\lambda)_k}{k! \, (\lambda+1)_{l-k} } \ \frac{\lambda+ l - 2k}{\lambda} \ C^{(\lambda)}_{l-  2k}  \,.
 \ee
 Applying \eqref{orthog} then ensures $l= 2k+n$ so there remain two summations. Combining the various
 contributions
 \begin{align}
 \hskip -0.3cm
\tG_3(\teta_1,\teta_2,\teta_3)&{}= \frac{1}{\Gamma(\lambda)\, \Gamma(\lambda+1)} \, 
\frac{1}{1-\teta_3}\,
\sum_{n,k\ge 0} \frac{1}{n!\, k!} \frac{(2\lambda)_n} {(\lambda +1)_{n+k}}\, (\teta_2)_{n+k} \,(\teta_2 -\lambda)_k \nn \\
\noalign{ \vskip -3pt}
&{}\times
 \frac{1}{\lambda-1+\teta_3+n} \bigg ( \frac{1} {\teta_1+\teta_2+\teta_3-2 + n+k}
 + \frac{1} {\lambda -\teta_1 +1 + n+k} \bigg ) \nn \\
 &\quad - \frac{1}{\lambda+1-\teta_3+n} \bigg ( \frac{1} {\teta_1+\teta_2- 1 + n+k}
 + \frac{1} {\lambda -\teta_1-\teta_3 + 2 + n+k} \bigg ) \, .
 \label{Kres}
\end{align}
This result is of course the same as in \cite{Kotikov1}.
 
A special case arises if $\teta_2 =\lambda$. In this case the $k$-sum is unnecessary and we may just set
$k=0$ in \eqref{Kres}. Separating each term in the last two lines of \eqref{Kres} as partial fractions
there are 8 contributions which can each be expressed in terms of
\be
{}_3F_2 \Big ({\genfrac{}{}{0pt}{0}{a,b, c}{b+1, c+1 }} ; 1 \Big ) =\frac{b\, c}{c-b}\, \Gamma(1-a)
\Big ( \frac{\Gamma(b)} {\Gamma(1+b-a)} -  \frac{\Gamma(c)} {\Gamma(1+c-a)} \Big ) \, ,
\label{Fab}
\ee
for $a=2\lambda, \, b=\lambda$ and 6 choices for $c$. With this result the contributions from the first term
in \eqref{Fab} sum to zero leaving the dramatically simplified form
\begin{align}
\tG_3(\teta_1,\lambda,\teta_3) ={}& \frac{\Gamma(1-2\lambda)}{\Gamma(\lambda)^2} \, 
\frac{1}{(1-\teta_1)(1-\teta_3)(1-\teta_6)}\nn \\
& {}\times \bigg ( \frac{\Gamma(\lambda-1+ \teta_1)}{\Gamma(\teta_1 -\lambda)}
-  \frac{\Gamma(\lambda+1- \teta_1)}{\Gamma(2-\lambda -\teta_1)}
+ \frac{\Gamma(\lambda-1+ \teta_3)}{\Gamma(\teta_3 -\lambda)}
-  \frac{\Gamma(\lambda+1- \teta_3)}{\Gamma(2-\lambda -\teta_3)} \nn \\
&\qquad  {} +\frac{\Gamma(\lambda-1+ \teta_6)}{\Gamma(\teta_6 -\lambda)}
-  \frac{\Gamma(\lambda+1-\teta_6)}{\Gamma(2-\lambda -\teta_6)} \bigg )  \nn\\
={}& \frac{1}{\Gamma(\lambda)} \bigg (\frac{1}{(1-\teta_3)(1-\teta_6)}\, L_1(\teta_1,2\lambda) 
+  \frac{1}{(1-\teta_1)(1-\teta_6)}\, L_1(\teta_3,2\lambda) \nn \\
&\hskip 1cm {}+  \frac{1}{(1-\teta_1)(1-\teta_3)}\, L_1(\teta_6,2\lambda) \bigg ) \, , \nn \\
&\teta_6 = 3 - \teta_1 -\teta_3 \, .
\label{spec}
\end{align}
The apparent singularities at $\teta_i=1$ are clearly cancelled. The poles at $\teta_i = 1-\lambda-n$ 
and $\teta_i = 1+\lambda+n$, $i=1,3,6$,  reflect  divergences of the original integral. 
$\tG_3(\lambda,\lambda,\lambda)\big|_{\lambda\to 1} = 6\, \zeta_3$.   The momentum space result
\eqref{G2} may be obtained by
\be
G_2(\eta_1,\eta_2) = g(\eta_1)\,g(\eta_2)\,g(\eta_6)\, g(1)^3\, \tG_3(\teta_1,\lambda,\teta_2) \, , \quad
\eta_6=\lambda+1 -\teta_6\, .
\ee

For the general case we may use
\begin{align}
\hskip -0.5cm
\frac{\Gamma(b)}{\Gamma(e)}\, \frac{1}{c} \, 
{}_3F_2 \Big ({\genfrac{}{}{0pt}{0}{a,b, c}{e, c+1 }} ; 1 \Big ) 
= {}& \Gamma(1-a)\, \frac{\Gamma(b-c)\, \Gamma(c)}{\Gamma(e-c)\, \Gamma(c-a+1)} \nn \\
&{} - \frac{\Gamma(1-a)}{\Gamma(e-b)} \, \frac{\Gamma(b)}{\Gamma(b-a+1)}\, 
 \frac{1}{b-c} \,{}_3F_2 \Big ({\genfrac{}{}{0pt}{0}{b-e+1,b, b-c}{b-a+1, b-c+1 }} ; 1 \Big ) \, , 
\label{F32}
\end{align}
where \eqref{Fab} is a special case, to recast \eqref{Kres} into two parts. The first term in \eqref{F32}
applied to the $k$-sum leads, taking $a=\teta_2-\lambda, \ b= \teta_2+ n, \ e=\lambda+1+n$
and with four choices for $c$, to 
\begin{align}\hskip -0.5cm
\tG_3(\teta_1, \teta_2 & ,\teta_3)_1 =  \frac{1}{\Gamma(\lambda)}\, \frac{1}{1-\teta_3} 
 \nn \\
&{}\times \bigg (  \frac{1} {\lambda-1+\teta_3} \, \bigg ( L_1(\teta_2,\teta_6)\,
{}_3F_2 \Big ({\genfrac{}{}{0pt}{0}{2\lambda,\lambda+1-\teta_6, \lambda-1+\teta_3}
{2\lambda+2-\teta_2-\teta_6, \lambda + \teta_3 }} ; 1 \Big ) \nn \\
& \hskip 2.5cm  +
L_1(\teta_2,\teta_1)  \, 
{}_3F_2 \Big ({\genfrac{}{}{0pt}{0}{2\lambda,\lambda+1-\teta_1, \lambda-1+\teta_3}
{2\lambda+2-\teta_2-\teta_1, \lambda + \teta_3 }} ; 1 \Big ) \bigg )  \nn \\
& \ \  \ {} - \frac{1} {\lambda+1-\teta_3} 
\bigg  ( L_1(\teta_2,\lambda+\teta_1) \,
{}_3F_2 \Big ({\genfrac{}{}{0pt}{0}{2\lambda,\teta_2+\teta_1-1, \lambda+1-\teta_3}
{\lambda+\teta_1, \lambda+2 - \teta_3 }} ; 1 \Big ) \nn \\
& \hskip 2.5cm  +
L_1(\teta_2,\lambda+\teta_6)\, 
{}_3F_2 \Big ({\genfrac{}{}{0pt}{0}{2\lambda,\teta_2+ \teta_6-1, \lambda+1-\teta_3}
{\lambda+\teta_6, \lambda +2 - \teta_3 }} ; 1 \Big ) \bigg ) \bigg ) \, ,
\label{Res1}
\end{align}
with $\teta_6$ here as in \eqref{Gsym}. The other contribution remains a double sum
\begin{align}
\tG_3(\teta_1, \teta_2  ,\teta_3)_2 =&{} -\frac{1}{\Gamma(\lambda)\,\Gamma(\lambda+1)} \, \frac{1}{1-\teta_3}  \nn \\
&{}\times \bigg ( 
C \bigg ( {\genfrac{}{}{0pt}{0}{2\lambda,\lambda-1+\teta_3}{\lambda + \teta_3}} ;  
{\genfrac{}{}{0pt}{0}{\teta_2 -\lambda,\teta_2+\teta_6-\lambda-1}{\teta_2+\teta_6-\lambda}} ; 
 {\genfrac{}{}{0pt}{0}{\teta_2}{\lambda+1}} \bigg ) \nn \\
 &\qquad {}+C \bigg ( {\genfrac{}{}{0pt}{0}{2\lambda,\lambda-1+\teta_3}{\lambda + \teta_3}} ;  
{\genfrac{}{}{0pt}{0}{\teta_2 -\lambda,\teta_2+\teta_1-\lambda-1}{\teta_2+\teta_1-\lambda}} ; 
{ \genfrac{}{}{0pt}{0}{\teta_2}{\lambda+1}} \bigg ) \nn \\
 &\qquad {}- C \bigg ( {\genfrac{}{}{0pt}{0}{2\lambda,\lambda+1-\teta_3}{\lambda +2- \teta_3}} ;  
{\genfrac{}{}{0pt}{0}{\teta_2-\lambda , 1-\teta_1}{2-\teta_1} }; 
 {\genfrac{}{}{0pt}{0}{\teta_2}{\lambda+1}} \bigg ) \nn \\
 &\qquad {}-
C \bigg ( {\genfrac{}{}{0pt}{0}{2\lambda,\lambda+1-\teta_3}{\lambda +2- \teta_3}} ;  
{\genfrac{}{}{0pt}{0}{\teta_2-\lambda,1-\teta_6}{2-\teta_6}} ; 
 {\genfrac{}{}{0pt}{0}{\teta_2}{\lambda+1}} \bigg ) \bigg )\, ,
 \label{G32a}
\end{align}
where
\be
C \bigg ( {\genfrac{}{}{0pt}{0}{a_1,a_2}{c}} ;  {\genfrac{}{}{0pt}{0}{b_1,b_2}{d}} ; 
 {\genfrac{}{}{0pt}{0}{e}{f}} \bigg ) = \frac{1}{a_2\, b_2}\,\sum_{n,m\ge 0} \frac{1}{n!\, m!}\,
 \frac{(a_1)_n (a_2)_n}{(c)_n} \,  \frac{(b_1)_m (b_2)_m}{(d)_m} \, 
  \frac{(e)_{n+m} }{(f)_{n+m}} \, .
\ee

 $C$ is a particular case of a two variable 
Kamp\'{e} de F\'{e}riet function with unit arguments.
For our purposes $c=a_2+1, \ d=b_2+1$ and there is the relation
\footnote{This may be derived by using \eqref{F32} twice, first for
${}_3F_2 \big ({\genfrac{}{}{0pt}{3}{\scriptstyle{\ e+m,\ a_1, \ a_2}} {\scriptstyle{f+m,\ a_2+1 }} }\big )$
and then for 
${}_3F_2 \big ({\genfrac{}{}{0pt}{3}{\scriptstyle{\ e-a_1-n,\ b_1, \ b_2}} {\scriptstyle{f-a_1-n,\ b_2+1 }} }\big )$.
}
\begin{align}
&C \bigg ( {\genfrac{}{}{0pt}{0}{a_1,a_2}{a_2+1}} ;  {\genfrac{}{}{0pt}{0}{b_1,b_2}{b_2+1}} ; 
 {\genfrac{}{}{0pt}{0}{e}{f}} \bigg ) \nn \\
 &{} =  \frac{\Gamma(1-e)\, \Gamma(f)}{\Gamma(1-e+a_1+b_1)\, \Gamma(f-a_1-b_1)} \,
 C \bigg ( {\genfrac{}{}{0pt}{0}{a_1,a_1-a_2}{a_1-a_2+1}} ;  {\genfrac{}{}{0pt}{0}{b_1,b_1-b_2}{b_1-b_2+1}} ; 
  {\genfrac{}{}{0pt}{0}{1-f+a_1+b_1}{1-e+a_1+b_1}} \bigg ) \nn \\
  &\quad {}+   \frac{\Gamma(1-e)\, \Gamma(f)}{\Gamma(1-e+a_2)\, \Gamma(f-a_2)} 
  \frac{\Gamma(a_1-a_2)\,\Gamma(a_2)}{\Gamma(a_1)} \, \frac{1}{b_2} \,
  {}_3F_2 \Big ({\genfrac{}{}{0pt}{0}{e-a_2, b_1,b_2} {f-a_2,\ b_2+1 }} ; 1 \Big ) \nn \\
  &\quad {} -  \frac{\Gamma(1-e)\, \Gamma(f)}{\Gamma(1-e+a_1+b_2)\, \Gamma(\ f-a_1-b_2)} 
  \frac{\Gamma(b_1-b_2)\,\Gamma(b_2)}{\Gamma(b_1)} \nn \\
  &\hskip 4.5cm {}\times  \frac{1}{a_1-a_2} \,
  {}_3F_2 \Big ({\genfrac{}{}{0pt}{0}{1-f +a_1+b_2,a_1,a_1- a_2} {\ 1-e+a_1 +b_2, \ a_1-a_2+1 }} ; 1 \Big ) \, .
\end{align}
There is a related formula obtained by $a_1 \leftrightarrow b_1, \ a_2 \leftrightarrow b_2$.
For our purposes we use a special case with $e+f$ constrained
\begin{align}
&C \bigg ( {\genfrac{}{}{0pt}{0}{a_1,a_2}{a_2+1}} ;  {\genfrac{}{}{0pt}{0}{b_1,b_2}{b_2+1}} ; 
 {\genfrac{}{}{0pt}{0}{e}{f}} \bigg ) - 
 C \bigg ( {\genfrac{}{}{0pt}{0}{a_1,a_1-a_2}{a_1-a_2+1}} ;  {\genfrac{}{}{0pt}{0}{b_1,b_1-b_2}{b_1-b_2+1}} ; 
  {\genfrac{}{}{0pt}{0}{e}{f}} \bigg ) \nn \\
  &\quad {}=   \frac{\Gamma(1-e)\, \Gamma(f)}{\Gamma(1-e+a_2)\, \Gamma(f-a_2)} 
  \frac{\Gamma(a_1-a_2)\,\Gamma(a_2)}{\Gamma(a_1)} \, \frac{1}{b_2} \,
  {}_3F_2 \Big ({\genfrac{}{}{0pt}{0}{e-a_2, b_1,b_2} {f-a_2,\ b_2+1 }} ; 1 \Big ) \nn \\
  &\quad \ \ {} -  \frac{\Gamma(1-e)\, \Gamma(f)}{\Gamma(1-e+a_1+b_2)\, \Gamma(\ f-a_1-b_2)} 
  \frac{\Gamma(b_1-b_2)\,\Gamma(b_2)}{\Gamma(b_1)} \nn \\
  &\hskip 4.5cm {}\times  \frac{1}{a_1-a_2} \,
  {}_3F_2 \Big ({\genfrac{}{}{0pt}{0}{1-f +a_1+b_2,a_1,a_1- a_2} {\ 1-e+a_1 +b_2, \ a_1-a_2+1 }} ; 1 \Big ) \, , \nn \\
  &  \qquad f= a_1+b_1 -e + 1 \, .
 \label{CCrel}
\end{align}
There is an apparent singularity on the right hand side when $e=1$. However this cancels between the two
terms using the Thomae relation
\be
\hskip -0.5cm
\frac{\Gamma(a)}{\Gamma(e) \, \Gamma(f)} \,
{}_3F_2 \Big ({\genfrac{}{}{0pt}{0}{a, b,c} {e,\, f }} ; 1 \Big )
= \frac{\Gamma(s)}{\Gamma(s+b) \, \Gamma(s+c)} \,
{}_3F_2 \Big ({\genfrac{}{}{0pt}{0}{s, e-a ,f- a} {s+b,\, s+c }} ; 1 \Big ) \, ,  \ \
s=e+f-a-b-c\, ,
\label{Thomae}
\ee
for $f=c+1$ and taking $a=b_1, \, b= 1-a_2, \, c= b_2, \, e= a_1+b_1 -a_2$ so that $s=a_1$. 

Applying \eqref{CCrel}  to \eqref{G32a} and also using \eqref{F32}  again gives
\begin{align}
\tG_3(\teta_1, \teta_2  ,\teta_3)_2 ={}&{}  \frac{1}{\Gamma(\lambda)}\,  \frac{1}{1-\teta_3}\,
\frac{\Gamma(1-\teta_2)}{ \Gamma(\teta_2-\lambda)}  \, \frac{1}{\lambda+1-\teta_3}\nn \\
&\qquad {}\times\bigg (\frac{\Gamma(1-\teta_6)\,\Gamma(\teta_2+\teta_6-1-\lambda)}
{\Gamma(\lambda +\teta_6)\, \Gamma(2-\teta_2 -\teta_6)\ }\  
{}_3F_2 \Big ({\genfrac{}{}{0pt}{0}{2\lambda, \teta_2+\teta_6-1,\lambda+1-\teta_3 }
{\lambda+\teta_6,\ \lambda+2-\teta_3}} ; 1 \Big ) \nn \\
&\qquad\ \ \ {} + \frac{\Gamma(1-\teta_1)\,\Gamma(\teta_2+\teta_1-1-\lambda)}
{\Gamma(\lambda +\teta_1)\, \Gamma(2-\teta_2 -\teta_1)}\  
{}_3F_2 \Big ({\genfrac{}{}{0pt}{0}{2\lambda, \teta_2+\teta_1-1,\lambda+1-\teta_3 }
{\lambda+\teta_1,\ \lambda+2-\teta_3}} ; 1 \Big ) \bigg )\nn \\
&  {} -  \frac{1}{\Gamma(\lambda)\, \Gamma(2\lambda)}\, \frac{1}{1-\teta_3}\, 
\Gamma(1-\teta_2)\, \Gamma(\lambda+1-\teta_2)\nn \\
\noalign{\vskip -2pt}
&\hskip 0cm {}\times 
\frac{ \Gamma(\lambda+1-\teta_3)\, \Gamma(\lambda-1+\teta_3) }
{\Gamma(\lambda-\teta_2 +\teta_3) \, \Gamma(\teta_2-\teta_3-\lambda +1)}\, \frac{
\Gamma(\teta_2 + \teta_1 -\lambda-1)\, \Gamma(\teta_2 + \teta_6 -\lambda-1)}{\Gamma(\teta_1) \ 
\Gamma(\teta_6)} \, .
\label{Res2}
\end{align}
Alternatively
\begin{align}
\tG_3(\teta_1, \teta_2  ,\teta_3)_2 ={}&{} - \frac{1}{\Gamma(\lambda)}\,  \frac{1}{1-\teta_3}\,
\frac{\Gamma(1-\teta_2)}{ \Gamma(\teta_2-\lambda)}  \, \frac{1}{\lambda-1+\teta_3}\nn \\
&\qquad {}\times\bigg (\frac{\Gamma(1-\teta_6)\,\Gamma(\teta_2+\teta_6-1-\lambda)}
{\Gamma(\teta_6-\lambda)\, \Gamma(2\lambda+2-\teta_2 -\teta_6)\ }\  
{}_3F_2 \Big ({\genfrac{}{}{0pt}{0}{2\lambda, \lambda+1-\teta_6,\lambda-1+\teta_3 }
{2\lambda+2 -\teta_2-\teta_6,\ \lambda+\teta_3}} ; 1 \Big ) \nn \\
&\qquad\ \ \ {} + \frac{\Gamma(1-\teta_1)\,\Gamma(\teta_2+\teta_1-1-\lambda)}
{\Gamma(\teta_1-\lambda)\, \Gamma(2\lambda+2-\teta_2 -\teta_1)}\  
{}_3F_2 \Big ({\genfrac{}{}{0pt}{0}{2\lambda, \lambda+1-\teta_1,\lambda-1+\teta_3 }
{2\lambda+2-\teta_2-\teta_1,\ \lambda+\teta_3}} ; 1 \Big ) \bigg )\nn \\
&  {} + \frac{1}{\Gamma(\lambda)\, \Gamma(2\lambda)}\,  \frac{1}{1-\teta_3}\, 
\Gamma(1-\teta_2)\, \Gamma(\lambda+1-\teta_2)\nn \\
\noalign{\vskip -2pt}
&\hskip 0cm {}\times 
\frac{ \Gamma(\lambda+1-\teta_3)\, \Gamma(\lambda-1+\teta_3) }
{\Gamma(\lambda+2-\teta_2 -\teta_3) \, \Gamma(\teta_2+\teta_3-\lambda -1)}\, \frac{\Gamma(1-\teta_1) \ 
\Gamma(1-\teta_6)} 
{\Gamma(\lambda +2-\teta_2 -\teta_1)\, \Gamma(\lambda+2-\teta_2 - \teta_6 )}\, .
\label{Res3}
\end{align}
The results in \eqref{Res3} and \eqref{Res2} correspond to (19) and (20) in \cite{Kotikov1}. The two expressions
are related by \eqref{G3inv}.
As a check we may verify consistency with \eqref{spec}, where $\tG_3(\teta_1, \lambda  ,\teta_3)
- \tG_3(\teta_1, \lambda  ,\teta_3)_1 = - \frac{1}{(1-\teta_1)(1-\teta_6)} \,\frac{1}{\Gamma(\lambda)} L_1(\teta_3, 2\lambda)$.

Combining \eqref{Res1} and \eqref{Res2} and using \eqref{F32}\footnote{For combining $\Gamma$-functions
it is sufficient to repeatedly use $\Gamma(x)\, \Gamma(1-x)= \pi / \sin \pi \,x$ and the trigonometric relation
$\sin A \, \sin(A+B+C) + \sin B \sin C = \sin (A+B) \sin (A + C)$.} once more there is a result containing just two
${}_3F_2(1)$ functions
\begin{align}\hskip -0.5cm
\tG_3(\teta_1, \teta_2  ,\teta_3)  =  {}&  
 \frac{1} {\lambda+1-\teta_3} \, \frac{1}{1-\teta_3}
\bigg  ( P(\teta_1,\teta_2)\ {}_3F_2 \Big ({\genfrac{}{}{0pt}{0}{2\lambda,\teta_2+\teta_1-1, \lambda+1-\teta_3}
{\lambda+\teta_1,\ \lambda+2 - \teta_3 }} ; 1 \Big ) \nn \\
& \hskip 3cm  +  P(\teta_6,\teta_2)\
{}_3F_2 \Big ({\genfrac{}{}{0pt}{0}{2\lambda,\teta_2+ \teta_6-1, \lambda+1-\teta_3}
{\lambda+\teta_6, \ \lambda +2 - \teta_3 }} ; 1 \Big ) \bigg )  \nn \\
&  {} +  Q(\teta_1,\teta_2,\teta_3) \, ,
\label{Res4}
\end{align}
with
\begin{align}
P(\teta_1,\teta_2) = {}&- \frac{1}{\Gamma(\lambda)}\, 
 L_1(\teta_1,\lambda+\teta_2)\, \frac{\Gamma(\teta_1-\lambda)}
{\Gamma(\lambda+\teta_1)}\, \frac{\Gamma(\lambda+1-\teta_2)}{\Gamma(1-\lambda-\teta_2)} \, , \nn \\
Q(\teta_1,\teta_2,\teta_3) = {}&- \frac{1}{\Gamma(\lambda)\,\Gamma(2\lambda)}\, \frac{1}{1-\teta_3}\, 
\Gamma(1-\teta_2)\, \Gamma(\lambda+1-\teta_2)\, \, g(\teta_1)\, g(\teta_6)\nn \\
\noalign{\vskip -2pt}
&\hskip 0cm {}\times 
\frac{ \Gamma(\teta_1-\lambda)\,\Gamma(\teta_6-\lambda)}
{\Gamma(\lambda+2-\teta_2 -\teta_1) \, \Gamma(\lambda+2- \teta_2 -\teta_6)}\, 
\frac{\Gamma(\lambda+1-\teta_3)\, \Gamma(\lambda-1+\teta_3) } 
{\Gamma(\lambda +\teta_2 +\teta_3)\, \Gamma(1-\lambda-\teta_2 - \teta_3 )}\, .
\label{ResPQ}
\end{align}
Since
\be
{}_3F_2 \Big ({\genfrac{}{}{0pt}{0}{a,b, c}{e, c+1 }} ; 1 \Big ) =\frac{ \Gamma(e)\, \Gamma(e+1-a-b)}
{\Gamma(e-a)\, \Gamma(e+1-b)}\,{}_3F_2 \Big ({\genfrac{}{}{0pt}{0}{a,c+1-b, 1}{e+1-b, c+1 }} ; 1 \Big ) \, ,
\ee
the result in \eqref{Res4} can then be  written in a form more comparable to that in \cite{Broadhurst}.

Alternative expressions to \eqref{Res4} may be obtained by using \eqref{G3inv} or by using \eqref{Thomae}
so that 
\begin{align}
&\frac{1}{\lambda+1-\teta_3}\, {}_3F_2 \Big ({\genfrac{}{}{0pt}{0}{2\lambda,\teta_2+\teta_1-1, \lambda+1-\teta_3}
{\lambda+\teta_1,\ \lambda+2 - \teta_3 }} ; 1 \Big ) \nn \\
&{}=\frac{\Gamma(\lambda+\teta_1)\, \Gamma(\lambda+1-\teta_3)\, \Gamma(2-\lambda-\teta_2)}
{\Gamma(2\lambda)\, \Gamma(3-\teta_2-\teta_3)\, \Gamma(\teta_1-\lambda)}\, 
 \frac{1}{\teta_1 -\lambda}\, {}_3F_2 \Big ({\genfrac{}{}{0pt}{0}{2-\lambda-\teta_2,2-\lambda-\teta_3, \teta_1-\lambda}
{3 - \teta_2-\teta_3, \ \teta_1-\lambda+1 }} ; 1 \Big ) \, ,
\label{ThomF}
\end{align}
and also for $\teta_1\to \teta_6$. For $\teta_3=\lambda$ then the result obtained by using \eqref{ThomF} allows
a reduction to just $\Gamma$-functions using \eqref{Fab}. After some manipulation
\begin{align} \hskip - 1cm {}
\tG_3(\teta_1,\teta_2,\lambda)= -\frac{1}{(\lambda-1)(2\lambda-1)\Gamma(\lambda) }&  \bigg( 
\frac{1}{(1-\teta_1)(1-\teta_2)} \, g(\lambda-1+\teta_1)\, g(\lambda-1+\teta_2) \, g(\teta_6) \nn \\
&{}+ \frac{1}{(1-\teta_1)(1-\teta_6)} \, g(\lambda-1+\teta_1)\, g(\lambda-1+\teta_6) \, g(\teta_2)  \nn \\
&{}+ \frac{1}{(1-\teta_2)(1-\teta_6)} \, g(\lambda-1+\teta_2)\, g(\lambda-1+\teta_6) \, g(\teta_1) \bigg )\, 
\end{align}
displaying the manifest symmetry under permutations of $\teta_1,\teta_2,\teta_6$. For the momentum space result
in \eqref{G2t} 
\be
{\tilde G}_2(\eta_1,\eta_2) = g(\eta_1)\,  g(\eta_2)\,  g(\eta_6)\, g(1)^3\,  \tG_3(\teta_1,\teta_2,\lambda) \, .
\ee

As a consistency check we may consider limiting cases when $\teta_i\to \lambda+1$ using
\be
\frac{1}{(x^2)^{\eta}_{\vphantom d}} \sim - \frac{1}{\eta-\frac12 d} \, \frac{\pi^{\frac12 d}}{\Gamma(\frac12 d)}\,\delta^d(x)\, .
\ee
At these poles the associated lines are contracted to a point so that
\begin{align}
\mathop{\mathrm{Res}}_{\teta_1 = \lambda+1 }\tG_3(\teta_1, \teta_2  ,\teta_3) ={}&  -\frac{1}{\Gamma(\lambda+1)}\, 
L_1(\teta_3+\lambda,\lambda)\, , \quad
\mathop{\mathrm{Res}}_{\teta_2 = \lambda+1 }\tG_3(\teta_1, \teta_2  ,\teta_3) = -\frac{1}{\Gamma(\lambda+1)}\, 
L_1(2\lambda,\teta_3)\, , \nn \\
\mathop{\mathrm{Res}}_{\teta_3 = \lambda+1 }\tG_3(\teta_1, \teta_2  ,\teta_3) = {}& -\frac{1}{\Gamma(\lambda+1)}\, 
L_1(\teta_1+\lambda,\teta_2)\, ,
\end{align}
which correspond  respectively to the associated contracted diagrams
‘\be
          \tikz[baseline=(vert_cent.base)]{
  \node (vert_cent) {\hspace{-13pt}$\phantom{-}$};
    \draw [line width =1]  (0.7,0) ++(0:0.6cm and 0.4cm) arc (0:180:0.6cm and 0.4cm) 
          (0.7,0) ++(180:0.6cm and 0.4cm) node(n1){};
            \draw   [line width =1]  (0.7,0) ++(180:0.6cm and 0.4cm)  arc (180:270:0.6cm and 0.4cm) node(n2){} ;
\draw             (0.7,0) ++(270:0.6cm and 0.4cm) arc (270:360:0.6cm and 0.4cm)  node (n3) {} ;
\draw     (n1.base) to [out=0, in = 90]  (n2.base)  ;
             \node at (0.7,0.58) {$\scriptstyle{\teta_2}$};    
              \node at (0.2,-0.42) {$\scriptstyle \teta_3$};   
               \node at (-0.05,0.0) {$\scriptstyle{0}$};                  
                \node at (1.45,0.0) {$\scriptstyle{z}$};       
}\, , \qquad
          \tikz[baseline=(vert_cent.base)]{
  \node (vert_cent) {\hspace{-13pt}$\phantom{-}$};
    \draw [line width =1]  (0.7,0) ++(0:0.6cm and 0.4cm) arc (0:180:0.6cm and 0.4cm) 
          (0.7,0) ++(180:0.6cm and 0.4cm) node(n1){};
            \draw   [line width =1]  (0.7,0) ++(180:0.6cm and 0.4cm)  arc (180:270:0.6cm and 0.4cm) node(n2){} ;
\draw             (0.7,0) ++(270:0.6cm and 0.4cm) arc (270:360:0.6cm and 0.4cm)  node (n3) {} ;
\draw     (n3.base) to [out=180, in = 90]  (n2.base)  ;
             \node at (0.7,0.58) {$\scriptstyle{\teta_1}$};    
              \node at (0.2,-0.42) {$\scriptstyle \teta_3$};   
               \node at (-0.05,0.0) {$\scriptstyle{0}$};                  
                \node at (1.45,0.0) {$\scriptstyle{z}$};       
}\, , \qquad
  \tikz[baseline=(vert_cent.base)]{
  \node (vert_cent) {\hspace{-13pt}$\phantom{-}$};
    \draw [line width =1]  (0.7,0) ++(0:0.6cm and 0.4cm) arc (0:90:0.6cm and 0.4cm) 
          (0.7,0) ++(90:0.6cm and 0.4cm) node(n1){};
            \draw   [line width =1]  (0.7,0) ++(90:0.6cm and 0.4cm)  arc (90:180:0.6cm and 0.4cm) node(n2){} ;
\draw             (0.7,0) ++(180:0.6cm and 0.4cm) arc (180:360:0.6cm and 0.4cm)  node (n3) {} ;
\draw     (n2.base) to [out=0, in = 270]  (n1.base)  ;
             \node at (0.3,0.48) {$\scriptstyle{\teta_1}$};    
              \node at (1.1,0.48) {$\scriptstyle \teta_2$};   
               \node at (-0.05,0.0) {$\scriptstyle{0}$};                  
                \node at (1.45,0.0) {$\scriptstyle{z}$};       
}\, .
\ee

In the result \eqref{Res4} with \eqref{ResPQ} there is an apparent singularity at $\teta_3=1$. However
since
$\scriptstyle{  \frac{1}{\Gamma(\lambda+\teta_1)\Gamma(\teta_6-\lambda)}
{}_3F_2 \big ({\genfrac{}{}{0pt}{1}{2\lambda,\teta_2+\teta_1-1, \lambda+1}
{\lambda+\teta_1, \lambda+ 1}} ; 1 \big ) + \frac{1}{\Gamma(\lambda+\teta_6)\Gamma(\teta_1-\lambda)}
{}_3F_2 \big ({\genfrac{}{}{0pt}{1}{2\lambda,\teta_2+ \teta_6-1, \lambda}
{\lambda+\teta_6, \lambda + 1 }} ; 1 \big )= \frac{\Gamma(\lambda)^2}{\Gamma(2\lambda)}  \frac{1}{\Gamma(\teta_1)
\Gamma(\teta_6)}}$, with here $\teta_6 = 2 +\lambda-\teta_1-\teta_2$,
 from \eqref{F32} the overall pole at $\teta_3=1$ is absent. The further absence  of a singularity at $\teta_2=1$ follows 
similarly after using \eqref{ThomF}. Singularities at $\teta_1, \teta_6 =\lambda$ also cancel.

\section{Composite and Equation of Motion Operators\protect\footnote{The discussion contained here was undertaken in
collaboration with Andy Stergiou.}}
\label{eom}

For higher dimensional scalar operators it is necessary to take into account mixing with operators involving 
derivatives. These include operators which vanish on the equations of motion and are  redundant. In previous
literature issues related to equations of motion operators were discussed in \cite{Brezin,Nicoll,Nicoll2,Zhang}.

In this appendix we reconsider 
composite operators and equation of motion in a perturbative expansion  for 
a single scalar field $\phi$ and a basic action in $d$-dimensions
\be
S[\phi,V]  = \int \rmd^d x \: \big ( \tfrac12 (\pr \phi )^2 + V(\phi) \big ) \, .
\label{lag}
\ee
For a  critical  dimension
\be
d_n = \frac{2n}{n-1} \, , \quad n=2,3,\dots \, 
\label{dee}
\ee
the theory is renormalisable for $V(\phi)\equiv V(g,\phi)$ a polynomial of degree $2n$ where  $\{g^i \}$ are the couplings
parameterising the independent monomials in  $V$.  
Extending to dimension $d=d_n-\vep $ the counterterms necessary for finiteness, $S_{\rm{c.t.}}[\phi,V]$, may be taken to have  just poles in $\vep$,
giving a mass independent minimal subtraction regularisation scheme.
For $\mu$ the  regularisation scale necessitated  for $d\ne d_n$ then
\be
\mu^{-\vep}\big (S[\phi,V] + S_{\rm{c.t.}}(\phi,V]\big ) = S_0 = S[\phi_0,V_0] \, ,
\ee 
 and the usual perturbative  $\beta$-functions
and anomalous dimensions are  then defined by
\be
\big ( - \vep + \D_\phi + \D_V \big ) S_0 = 0 \, ,
\label{S0B}
\ee
where
\be
\D_\phi = -  \int \rmd^d x \; {\hat \gamma}_\phi\, \phi\frac{\delta}{\delta \phi}  \, , \quad
\D_V = {\hat \beta}_V \cdot \frac{\pr}{\pr V} =  {\hat \beta}_g{\!}^i \frac{\pr}{\pr g^i} \, ,
 \quad  {\hat \beta}_V(\phi)  = V({\hat \beta}_g,\phi) \, .
\label{RGdef}
\ee
${\hat \gamma}_\phi$ and ${\hat \beta}_V( \phi) $ can be decomposed as
\begin{align}
{\hat \gamma}_\phi
= {}& - \tfrac12 \vep + \gamma_\phi \, , \nn \\ {\hat \beta}_V \! (\phi) ={}&   \vep \big ( V(\phi)   - \tfrac12 \phi \,   V'(\phi)  \big ) +  \beta_V(\phi) =
 \vep \,V(\phi)  + {\hat \gamma}_\phi \,  \phi V'(\phi) +  {\tilde \beta}_V(\phi) \, ,
\label{betaV}
\end{align}
where  ${\tilde \beta}_V(\phi)$ depends just  on products of $V(\phi)$ with two or more derivatives and is a sum of 
contributions, corresponding to graphs in which all lines link different vertices, of the form
 $\prod_{i=1}^{p}  V^{(r_i)}(\phi)$, $p=2,3,\dots$,  with $(p-1) 2n = \sum_{i=1}^p r_i$, $2\le r_i \le 2n$ and $(p-1)(n-1)$ the number of loops.
 Perturbative calculations at leading and next to leading order for general $n$ were undertaken in \cite{Dwyer}. 
 
 Assuming no mass scales other than $\mu$ there is a single dimensionless coupling $\lambda$ and
\be
V(\phi) \to V_\lambda(\phi)= \frac{1}{(2n)!}\, \lambda \, \phi^{2n}  \, ,
\ee
so that
\be
{\hat \beta}_{V_\lambda} (\phi)  =  \frac{1}{(2n)!}\, {\hat \beta}_\lambda(\lambda) \,\phi^{2n}\, , \quad
 {\hat \beta}_\lambda(\lambda) =  - \vep\, (n-1) \lambda + \beta_\lambda(\lambda)\, , \quad  \gamma_\phi = \gamma_\phi(\lambda) \, .
 \label{Vred}
\ee
Perturbatively both $ \beta_\lambda(\lambda), \,  \gamma_\phi(\lambda)$ are ${\rm O}(\lambda^2)$.
As  usual in the $\vep$-expansion, there may be fixed points where 
\be
{\hat \beta}_\lambda(\lambda_*)=0 \, , \qquad  \eta = 2\gamma_\phi(\lambda_*) \, ,
\label{fixp}
\ee
with $\lambda_*$ and the critical exponent $\eta$ expressible perturbatively as a power series in $\vep$. 
At a fixed point $\Delta_\phi = \frac{1}{n-1} + {\hat \gamma}_\phi(\lambda_*) = \frac12(d-2+\eta)$.

For an extension to composite operators we introduce a vector of monomials
\be
v(\phi) = \big ( \phi, \, \tfrac12 \phi^2, \dots , \tfrac{1}{k!} \phi^k , \dots \big ) \, ,
\ee
and allow for an additional contribution to the action
\be
S'[\phi,U,Z,E] =  \int \rmd^d x \:  \big ( v(\phi) \cdot U + \pr^2 \phi \, v(\phi) \cdot Z + \pr^2 v(\phi)\cdot E \big ) \, ,
\label{Sextra}
\ee
where $U_k ,\, Z_k, \, E_k$, $k=1,2,\dots$,  are components of  vectors which are $x$-dependent so that the corresponding local
operators may be defined by functional differentiation. To first order in $U, \, Z,\, E$, 
\be 
S_t[\phi,\lambda , U,Z,E]= S[\phi,V_\lambda]+ S'[\phi,U,Z,E] \, ,
\ee
defines a renormalisable theory so long as  we restrict the additional contributions to the finite range  $U_k ,\, k<3n-1, \ Z_k, \, k< n, \  E_k, \,  k < n+1$ 
so as to avoid the necessity  of including four derivative operators. As previously the necessary counterterms to first order in $U, \, Z,\, E$
may be absorbed by taking $S_0 = S_t[\phi_0,\lambda_0, U_0,Z_0,E_0] $ and then \eqref{S0B}  becomes
\be
\big ( - \vep + \D_\phi + \D_\beta \big ) S_0 = 0 \, ,
\label{S0eq}
\ee
with
\be
 \D_\beta= {\hat \beta}_\lambda(\lambda) \frac{\pr}{\pr \lambda} + \sum_{X=U,Z,  E}
  \int \rmd^d x \:  {\hat \beta}_X \cdot \frac{\delta}{\delta  X} \, .
  \label{Dbeta}
 \ee
 \eqref{S0eq} extends to
  \be
\big (-\vep 
 + \D_\vphi + \D_\beta \big ) \Gamma = 0 \, ,
\ee
  for the finite 1PI generating functional $\Gamma[\vphi,\lambda , U,Z,E]$, which is  linear in $(U,\, Z,\, E)$. Correlation functions involving
  single insertions of the composite operators $\frac{1}{k!} \phi^k, \,  \frac{1}{k!} \phi^k \pr^2 \phi , \, \frac{1}{k!} \pr^2 \phi^k$ 
  are defined by functional differentiation of $\Gamma$ with  respect to $U_k , \, Z_k, \, E_k$.
 
 In \eqref{Dbeta} $ {\hat \beta}_X $ is linear in $X$ so that
 \be
 \begin{pmatrix}  {\hat \beta}_U \\  {\hat \beta}_Z  \\ {\hat \beta}_E \end{pmatrix} = 
  \begin{pmatrix}  {\hat \gamma}_{UU} & \gamma_{UZ}  & 0 \\   \gamma_{ZU}  & {\hat \gamma}_{ZZ}  & 0 \\   \gamma_{EU} &  \gamma_{EZ} &  {\hat \gamma}_{EE}
  \end{pmatrix}
 \begin{pmatrix}  U \\  Z  \\ E  \end{pmatrix} \, .
 \ee
 The anomalous dimensions are further restricted by
\begin{align}
& ( {\hat \gamma}_{UU})_{k'k} = (  {\hat \gamma}_{EE} )_{k'k} = \delta_{k'k} \, {\hat \gamma}_k(\lambda)  \, , \quad &
 (  {\hat \gamma}_{ZZ} )_{k'k} = {}& \delta_{k'k} \, {\hat \gamma}_{Z,k} (\lambda) \, , \nn \\
&  ( \gamma_{UZ} )_{k'k} = \delta_{k'\, k+2n-1} \, { \gamma}_{UZ,k}(\lambda)  \, , \ \  &   ( \gamma_{ZU} )_{k'k} = {}& \delta_{k'\, k-2n+1} \, { \gamma}_{ZU,k}(\lambda)  \, ,\nn  \\
&  ( \gamma_{EU} )_{k'k} = \delta_{k'\, k-2n+2} \, {\gamma}_{EU,k}(\lambda)  \ , \ \  &
  ( \gamma_{EZ} )_{k'k} ={}&  \delta_{k'\, k+1} \, {\gamma}_{EZ,k}(\lambda) \, .
  \label{gamexp}
 \end{align}
 Hence the anomalous dimension matrices are reduced to decoupled matrices for each $k$ of the form
 \be 
 {\hat \gamma}_{k} \, , \ k\le  2n-2 \, , \quad
 \begin{pmatrix}  {\hat \gamma}_{2n-1} & 0 \\
 {\gamma}_{EU,2n-1} & {\hat \gamma}_{1}  \end{pmatrix} \, , \quad
 \begin{pmatrix}  {\hat \gamma}_{k} & \gamma_{UZ,k-2n+1} & 0 \\
 \gamma_{ZU,k} &  {\hat \gamma}_{Z,k-2n+1} & 0 \\
 {\gamma}_{EU,k } &  {\gamma}_{EZ,k-2n+1 }& {\hat \gamma}_{k-2n+2}  \end{pmatrix} \, , \ k\ge 2n \, .
 \label{mat3}
 \ee 
 The dependence on $\vep$ is given by
 \be
  {\hat \gamma}_{k} (\lambda) = - \tfrac12(k-2) \vep +  {\gamma}_{k}  (\lambda) \, , \qquad 
   {\hat \gamma}_{Z,k}  (\lambda)  = - \tfrac12(k-1) \vep +  {\gamma}_{Z,k} (\lambda)  \, .
   \ee
   For $k\le 2n-2$ the scaling dimension for $\phi^k$ is $\Delta_k= \frac{k}{n-1} - \vep + {\hat \gamma}_k(\lambda_*)$.
  Taking $k= 2n-1+l$, $l= 1,\dots , n-1$,
  the three eigenvalues $\omega_{l,s}$ and associated eigenvectors $u_{l,s}$, $s=1,2,3$, of the $3\times 3$ matrix in \eqref{mat3} at the 
  fixed point \eqref {fixp} determine scaling operators $(\phi^{2n-1+l}/(2n+l-1)!, \, \phi^l \pr^2 \phi /l!, \, \pr^2 \phi^{l+1}/(l+1)! ) u_{l,s}$  with 
  scaling dimensions
  \be
  \Delta_{l,s} = d +  \frac{l-1}{n-1}+\omega_{l,s} \, .
  \ee
  The eigenvector $u_{l,3}= (0, \, 0,\,1)$ determines $\pr^2 \phi^{l+1}$ as a scaling operator with $\omega_{l,3}= {\hat \gamma}_{l+1}(\lambda_*)$,
  $\Delta_{l,3} = \Delta_{l+1} + 2$.

 In general 
 \be
 {\hat \beta}_{V_\lambda + \epsilon \, v_k} = {\hat \beta}_{V_\lambda}  + \epsilon \, {\hat \gamma_k}(\lambda) \, v_k + {\rm O}(\epsilon^2) \, , \quad
 v_k(\phi) = \tfrac{1}{k!} \phi^k \, , \quad k <  2n \, .
 \ee
 Since $V_1$ cannot contribute to ${\tilde \beta}_V$ from \eqref{betaV} we must have
 \be
{\hat \gamma}_1(\lambda)=  \vep+  {\hat \gamma}_\phi(\lambda)  \quad \Rightarrow  \quad \Delta_1 = \Delta_\phi \, .
\label{gamma1}
\ee

The action $S[\phi,V_\lambda]+ S'[\phi,U,Z,E] $ has redundancies which are analogous to a gauge  invariance. Extending this to $S_0$ 
gives rise to consistency conditions which constrain the form of the anomalous dimension matrices. As the simplest
illustration corresponding to a shift in $\phi$
\be
\L_\epsilon \big ( S[\phi,V_\lambda]+ S'[\phi,U,Z,E] \big ) = 0 \, ,  \quad
\L_\epsilon = \int \rmd^dx\; \epsilon \bigg ( - \frac{\delta}{\delta \phi} + \lambda  \frac{\delta}{\delta U_{2n-1}} -  \frac{\delta}{\delta E_1} \bigg ) \, ,
\label{Lep}
\ee
for arbitrary $\epsilon(x)$.  With the mass independent regularisation
\be
\L_\epsilon S_0 = 0  \quad \Rightarrow \quad \L_\epsilon  \Gamma =0 \, ,
\label{LSG}
\ee
then it is necessary that $ [ \L_\epsilon, \, \D_\phi + \D_\beta ] $ must also annihilate $S_0$ and $\Gamma$. By direct calculation and using \eqref{gamexp} 
\begin{align}
 \big  [ \L_\epsilon, \, \D_\phi + \D_\beta \big  ] ={}& - \L_{\, \epsilon\, \F}\nn \\
& {} +
 \int \rmd^dx\; \epsilon \bigg ( \big ( {\hat \gamma}_\phi - \F\big )     \frac{\delta}{\delta \phi} 
 + \big (  \lambda\, {\hat \gamma}_{2n-1} - {\hat \beta}_\lambda  +  \lambda \, \F \big )  
 \, \frac{\delta}{\delta U_{2n-1}} \bigg ) \, , \nn \\
 & \F=  \gamma_{EU,2n-1}\lambda - {\hat \gamma}_{1}
\end{align}
The additional terms must vanish, giving $\F= {\hat \gamma}_\phi$, so that with \eqref{gamma1}
\be
\gamma_{EU,2n-1} (\lambda) = 2\gamma_\phi (\lambda)  /\lambda \, , \qquad 
{\hat \gamma}_{2n-1} (\lambda)  = {\hat \beta}_\lambda (\lambda) /\lambda - {\hat \gamma}_\phi (\lambda) \, .
\ee
This coincides with \eqref{D5} when $n=3$.
At the fixed point \eqref{fixp} the anomalous dimension matrix becomes
\be
\begin{pmatrix} \frac12(\vep - \eta) & 0 \\ \eta/\lambda_*  & \frac12(\vep + \eta) \end{pmatrix} \, .
\ee
The eigenvectors determine the scaling operators
\begin{align}
& E_\phi = \lambda_* \tfrac{1}{(2n-1)! }\phi^{2n-1} - \pr^2 \phi \, , &&  \Delta_{E_\phi} = \tfrac12 ( d+2-\eta)= 
d - \Delta_\phi \, , \nn \\
& \pr^2 \phi \, , && \Delta_{\pr^2 \phi} = \tfrac12 ( d+2+\eta) = \Delta_\phi+ 2\, .
\label{Ephi}
\end{align}
The result for $\Delta_{\pr^2 \phi}$ reflects that  $\pr^2\phi$ is a descendant of $\phi$ while 
$\Delta_{E_\phi}$ is necessary for the consistency of the identity
\be
\big  \langle E_\phi (x) \, \phi(x_1 ) \dots \phi(x_r) \dots  \big \rangle =  {\ts \sum_r }\,  \delta^d(x- x_r) 
\langle \dots \phi(x_{r-1})\,  \phi(x_{r+1} ) \dots  \big \rangle \, .
\label{Ephi0}
\ee

\subsection{Higher Order Operators}

To first order in $\epsilon$ and taking $(U,\, Z,\, E)= {\rm O}(\epsilon) $ the action  $S_t$ is  invariant under
  \be
  \delta \phi = - \epsilon \, \frac{1}{l!} \phi^l \, , \quad \delta U_{2n+l-1} =  c_{n,l} \lambda\, \epsilon \, , \quad
  \delta Z_l = - \epsilon \, , \quad l\ge 1 \, , \quad  c_{n,l} = \frac{(2n+l-1)!}{(2n-1)!\, l!} \, ,
  \label{Llambda}
  \ee
  so that
  \be
  \frac{\delta S_t}{\delta U_l}\, \frac {\delta S_t}{\delta \phi} - c_{n,l} \lambda \, \frac {\delta S_t}{\delta U_{2n+l-1} } +  \frac{ \delta{S_t}}{ \delta Z_l} \simeq 0 \, ,
  \ee
  neglecting contributions of ${\rm O}(U,\, Z, \, E)$. This is assumed to extend to the finite 1PI generating functional $\Gamma[\vphi,\lambda , U,Z,E]$, which is
  linear in $(U,\, Z,\, E)$, so that
  \be
   \frac{\delta\hskip 0.5pt\Gamma}{\delta U_l}\, \frac {\delta \hskip 0.5pt \Gamma}{\delta \vphi} - 
   c_{n,l} \lambda \, \frac {\delta \hskip 0.5pt \Gamma}{\delta U_{2n+l-1} } +  \frac{ \delta \hskip 0.5pt \Gamma}{ \delta Z_l} \simeq 0 \, .
   \label{Gamid}
   \ee
This replaces the linear equation \eqref{LSG}.
  
Since
\begin{align}
\Big [ \D_\vphi , \, \frac{\delta}{\delta \vphi} \Big ] = {}& {\hat \gamma}_\vphi \, \frac{\delta} {\delta \vphi} \, , \qquad \Big [ \D_\beta , \, \frac{\delta}{\delta U_l} \Big ] = 
- {\hat \gamma}_l \, \frac{\delta} {\delta U_l}  \, , \ \ l<2n-1 \, , \nn \\
\Big [ \D_\beta , \, \frac{\delta}{\delta U_{2n+l-1} } \Big ] = {}&
- {\hat \gamma}_{2n+l-1}  \, \frac{\delta} {\delta U_{2n+l-1} }  - {\gamma}_{ZU,2n+l-1}  \, \frac{\delta} {\delta Z_l }  - {\gamma}_{EU,2n+l-1}  \, \frac{\delta} {\delta E_{l+1}} \, , \nn \\
\Big [ \D_\beta , \, \frac{\delta}{\delta {Z_l} } \Big ] = {}&
- {\gamma}_{UZ,l}  \, \frac{\delta} {\delta U_{2n+l-1} }  - {\hat \gamma}_{Z,l}  \, \frac{\delta} {\delta Z_l }  - {\gamma}_{EZ,l}  \, \frac{\delta} {\delta E_{l+1}} \, ,
\end{align}
we may obtain consistency conditions
\begin{align} 
{\gamma}_{UZ,l} = {}&  - c_{n,l} \, {\hat  \beta}_\lambda + c_{n,l} \, \lambda\big  ( {\hat \gamma}_{2n+l-1} - {\hat \gamma_l} + {\hat \gamma}_\phi  +\vep\big  )\, , \nn \\
{\hat \gamma}_{Z,l}={}& c_{n,l}\, \lambda \, \gamma_{ZU,2n+l-1} + {\hat \gamma}_l - {\hat \gamma}_\phi -\vep  \, , \qquad
{\gamma}_{EZ,l}  =  c_{n,l}\, \lambda \, \gamma_{EU,2n+l-1} \, .
\end{align} 
 The anomalous dimension matrix at the fixed point \eqref{fixp} then has the form
\be
\begin{pmatrix}  {\hat \gamma}_{2n+l-1} & c_{n,l} \lambda_* \big ( {\hat \gamma}_{2n+l-1} - {\hat \gamma}_l +  {\hat \gamma}_\phi +\vep \big )   & 0 \\
\noalign{\vskip 2pt}
 \gamma_{ZU,2n+l-1} & c_{n,l} \lambda_* {\gamma}_{ZU,2n+l-1}  +  {\hat \gamma}_l  -  {\hat \gamma}_\phi -\vep  & 0 \\
 \noalign{\vskip 2pt}
 {\gamma}_{EU,2n+l-1 } &   c_{n,l} \lambda_* {\gamma}_{EU,2n+l-1} \, & {\hat \gamma}_{l+1}  \end{pmatrix} \, , \quad l= 1, \dots, n-1 \, .
\ee
This has an eigenvector and eigenvalue
\be
u_{l,2}= \begin{pmatrix} c_{n,l}\lambda_* \\ -1 \\ 0 \end{pmatrix} \, , \qquad \omega_{l,2} = {\hat \gamma}_l - {\hat \gamma}_\phi  -\vep \ \ \Rightarrow \ \
\Delta_{l,2} = d + \Delta_l - \Delta_\phi   \, , 
\label{Drel}
\ee
corresponding to the equation of motion scaling operator 
\be 
E_{\phi,l} = \frac{1}{l!} \phi^l \, E_\phi \, ,
\ee
 with $E_\phi$ as in \eqref{Ephi}. For $l=1$,  $ \omega_{1,2} =0$, $\Delta_{1,2}=d$.  In terms of the usual generating functional 
 for connected correlation functions,
 $W[J] = \int J \vphi - \Gamma[\vphi] , \, \frac{\delta}{\delta \vphi} \Gamma[\vphi] = J$,
 \eqref{Gamid} translates, since $\frac {\delta}{\delta X} W \big |_J =  - \frac {\delta}{\delta X} \Gamma \big |_\vphi$, into
 \be
 \big \langle  \tfrac{1}{l!}\phi^l(x) \big \rangle_J \,  J(x) = \big \langle  E_{\phi,l} (x) \big \rangle_J \, ,
 \ee  
 expressing this in terms of composite operators
 and then setting $U,\, Z,\, E$  to zero. By differentiating with respect to $J$
 \be
\big  \langle E_{\phi,l} (x) \, \phi(x_1 ) \dots \phi(x_r) \dots  \big \rangle = {\ts \sum_r }\,  \delta^d(x-x_r) 
\langle \dots \phi(x_{r-1}) \, \tfrac{1}{l!}\phi^l(x_r) \,  \phi(x _{r+1} ) \dots  \big \rangle \, ,
\label{Ephi1}
\ee
which reduces to \eqref{Ephi0} for $l=0$.
The relations for the scaling dimensions in \eqref{Drel} are essential for consistency.

For $l=1$ we may also consider constant rescaling of  the coupling $\lambda$ generated by
\be
 \L_{\lambda,\alpha} = - \alpha \, \lambda \frac{\pr}{\pr \lambda} + \alpha  \int \rmd^dx\; \frac{\delta}{\delta U_{2n}} \, .
\ee
In this case
\begin{align}
 \big  [ \L_{\lambda,\alpha}, \, \D_\phi + \D_\beta & \big  ] =   \L_{\lambda,\alpha \F_\lambda} - \L_{\alpha \, \gamma_{ZU,2n},1} \nn \\
& {} +
 \int \rmd^dx\; \alpha \bigg (\lambda \big (  {\hat \gamma}_\phi{\!}'  - \gamma_{ZU,2n} \big ) \,  \phi \frac{\delta}{\delta \phi} 
 + \big ( \lambda \, {\hat \gamma}_{2n}  - {\hat \beta}_\lambda +  2n\lambda^2\,  \gamma_{ZU,2n }  + \F_{\lambda} \big ) 
 \frac{\delta}{\delta U_{2n}} \bigg )\, , \nn \\
&\quad  \F_{\lambda} =  - \lambda \, {\hat \beta}_\lambda {\!}' + {\hat \beta}_\lambda \, .
\end{align}
Here we have discarded an apparent contribution involving $\gamma_{EU,2n} \frac{\delta}{\delta U_{2n}}$  since
this depends on there being non constant modes in $U_{2n}$ and these are absent for a  constant rescaling in \eqref{Llambda}.
The  consistency conditions determine ${\hat \gamma}_{2n}$ and $\gamma_{ZU,2n }$
\begin{align}
{\hat \gamma}_{2n}   ={}&   {\hat \beta}_\lambda {\!}' - 2n\lambda\,
 {\hat \gamma}_\phi{\!}'  \, ,  \qquad 
   \gamma_{ZU,2n } =   {\hat \gamma}_\phi{\!}'  \, .
\end{align}
At a fixed point the anomalous dimension matrix becomes
\be
\begin{pmatrix} {\hat \gamma}_{2n,*} & 2n \lambda_*\,    {\hat \gamma}_{2n,*} & 0 \\
{\hat \gamma}_\phi{\!}' (\lambda_*) & 2n\lambda_* \, {\hat \gamma}_\phi{\!}' (\lambda_*) & 0 \\
  \gamma_{EU,2n} (\lambda_*) & 2n\lambda_* \,  \gamma_{EU,2n} (\lambda_*) & {\hat \gamma}_2(\lambda_*) \end{pmatrix} \, , \quad
   {\hat \gamma}_{2n,*} =  {\hat \beta}_\lambda {\!}' (\lambda_* )- 2n\lambda_*\, {\hat \gamma}_\phi{\!}' (\lambda_*)  \, .
  \ee
  The scaling operators and associated scaling dimensions are then
  \begin{align}
  & \tfrac12 \, \pr^2 \phi^2 \, ,&& \Delta_{\pr^2 \phi^2} = d +  \gamma_{2,*}  = 2 +\Delta_{\phi^2} \, , \nn \\
  &  E_{\phi,1} \, , && \Delta_{E_{\phi,1}} = d \, , \nn \\
  & \tfrac{1}{(2n)!} \phi^{2n} - {\hat \gamma}_\phi{\!}' E_{\phi,1}
  + \gamma_{EU,2n} (\lambda_*)  \tfrac{ {\hat \beta}_\lambda {\!}' (\lambda_* )}
  { {\hat \beta}_\lambda {\!}' (\lambda_* ) - \gamma_{2,*}}\, \tfrac12\, \pr^2 \phi^2\, ,
 &&   \Delta_{\phi^{2n}}= d+{\hat \beta}_\lambda {\!}' (\lambda_* ) \, .
  \end{align}

\end{appendices}

\bibliographystyle{utphys}
\bibliography{3D.bib}

@book{KleinertB,
    author = "Kleinert, H. and Schulte-Frohlinde, V.",
    title = "{Critical properties of $\phi^4$-theories}",
    publisher={World Scientific, Singapore},
    year = "2001"
}

@article{Kotikov1,
      author         = "Kotikov, A.V.",
      title          = "{The Gegenbauer Polynomial Technique: the evaluation of a
                        class of Feynman diagrams}",
      journal        = "Phys. Lett.",
      volume         = "B375",
      year           = "1996",
      pages          = "240-248",
      doi            = "10.1016/0370-2693(96)00226-2",
      eprint         = "hep-ph/9512270",
      archivePrefix  = "arXiv",
      primaryClass   = "hep-ph",
      reportNumber   = "ENSLAPP-A-568-95",
      SLACcitation   = "%%CITATION = HEP-PH/9512270;%%"   
      }

@article{Broadhurst,
    author = "Broadhurst, David J. and Gracey, J. A. and Kreimer, D.",
    title = "{Beyond the triangle and uniqueness relations: Nonzeta counterterms at large N from positive knots}",
    eprint = "hep-th/9607174",
    archivePrefix = "arXiv",
    reportNumber = "OUT-4102-46, LTH-360, MZ-TH-95-28",
    doi = "10.1007/s002880050500",
    journal = "Z. Phys. C",
    volume = "75",
    pages = "559--574",
    year = "1997"
}

@inproceedings{Kotikov2,
    author = "Kotikov, A. V.",
    title = "{The Gegenbauer polynomial technique: The Evaluation of complicated Feynman integrals}",
    booktitle = "{15th International Workshop on High-Energy Physics and Quantum Field Theory (QFTHEP 2000)}",
    eprint = "hep-ph/0102177",
    archivePrefix = "arXiv",
    pages = "211--217",
    month = "7",
    year = "2000"
}

@article{Kotikov3,
    author = "Kotikov, A. V. and Teber, S.",
    title = "{New Results for a Two-Loop Massless Propagator-Type Feynman Diagram}",
    doi = "10.1134/S0040577918020083",
    journal = "Theor. Math. Phys.",
    volume = "194",
    number = "2",
    pages = "284--294",
    year = "2018"
}

@article{Kleinert,
    author = "Kleinert, Hagen and Schulte-Frohlinde, Verena",
    title = "{Critical exponents from five-loop strong coupling $\phi^4$ theory in 4 - epsilon dimensions}",
    eprint = "cond-mat/9907214",
    archivePrefix = "arXiv",
    doi = "10.1088/0305-4470/34/5/308",
    journal = "J. Phys. A",
    volume = "34",
    pages = "1037--1050",
    year = "2001"
}

@article{GrozinH,
    author = "Grozin, A. G.",
    title = "{Massless two-loop self-energy diagram: Historical review}",
    eprint = "1206.2572",
    archivePrefix = "arXiv",
    primaryClass = "hep-ph",
    reportNumber = "TTP12-019",
    doi = "10.1142/S0217751X12300189",
    journal = "Int. J. Mod. Phys. A",
    volume = "27",
    pages = "1230018",
    year = "2012"
}

@article{Panzer,
    author = "Kompaniets, Mikhail V. and Panzer, Erik",
    title = "{Minimally subtracted six loop renormalization of $O(n)$-symmetric $\phi^4$ theory and critical exponents}",
    eprint = "1705.06483",
    archivePrefix = "arXiv",
    primaryClass = "hep-th",
    doi = "10.1103/PhysRevD.96.036016",
    journal = "Phys. Rev. D",
    volume = "96",
    number = "3",
    pages = "036016",
    year = "2017"
}

@article{Bednyakov,
    author = "Bednyakov, A. and Pikelner, A.",
    title = "{Six-loop beta functions in general scalar theory}",
    eprint = "2102.12832",
    archivePrefix = "arXiv",
    primaryClass = "hep-ph",
    doi = "10.1007/JHEP04(2021)233",
    journal = "JHEP",
    volume = "04",
    pages = "233",
    year = "2021"
}

@article{Schnetz7,
    author = "Schnetz, Oliver",
     title = "{$\phi^4$ theory at seven loops}",
    eprint = "2212.03663",
    archivePrefix = "arXiv",
    primaryClass = "hep-th",
    doi = "10.1103/PhysRevD.107.036002",
    journal = "Phys. Rev. D",
    volume = "107",
    number = "3",
    pages = "036002",
    year = "2023"
}

@article{Schnetz6,
    author = "Schnetz, Oliver",
    title = "{$\phi^3$ theory at six loops}",
    eprint = "2505.15485",
    archivePrefix = "arXiv",
    primaryClass = "hep-th",
    doi = "10.1103/sp5q-km2c",
    journal = "Phys. Rev. D",
    volume = "112",
    number = "1",
    pages = "016028",
    year = "2025"
}

@article{SchnetzPeriod,
    author = "Schnetz, Oliver",
    title = "{Quantum periods: A Census of $\phi^4$-transcendentals}",
    eprint = "0801.2856",
    archivePrefix = "arXiv",
    primaryClass = "hep-th",
    reportNumber = "FAU-TP3-07-9",
    doi = "10.4310/CNTP.2010.v4.n1.a1",
    journal = "Commun. Num. Theor. Phys.",
    volume = "4",
    pages = "1--48",
    year = "2010"
}

@article{Gudmundsdottir1A,
    author = "Gudmundsdottir, R. and Rydnell, G. and Salomonson, P.",
    title = "{More on $O(N)$ symmetric \ensuremath{\phi^6} in three-dimensions Theory}",
    doi = "10.1103/PhysRevLett.53.2529",
    journal = "Phys. Rev. Lett.",
    volume = "53",
    pages = "2529--2531",
    year = "1984"
}

@article{Gudmundsdottir2,
    author = "Gudmundsdottir, Ragnheidur and Rydnell, Gunnar and Salomonson, Per",
    title = "{On $1/N$ Expansion in  $ (\phi^2)^3$ in Three-dimensions Field Theory}",
    reportNumber = "GOTEBORG-84-25",
    doi = "10.1016/0003-4916(85)90228-3",
    journal = "Annals Phys.",
    volume = "162",
    pages = "72--84",
    year = "1985"
}

@article{Townsend0,
    author = "Townsend, P. K.",
    title = "{Spontaneous Symmetry Breaking in $O(n)$  Symmetric $\phi^6$ Theory in the $1/n$ Expansion}",
    reportNumber = "Print-75-0617 (BRANDEIS)",
    doi = "10.1103/PhysRevD.12.2269",
    journal = "Phys. Rev. D",
    volume = "12",
    pages = "2269",
    year = "1975",
    note = "[Erratum: Phys.Rev.D 16, 533 (1977)]"
}

@article{Townsend1,
    author = "Townsend, Paul K.",
    title = "{The Global Ground State of $\phi^6$ Theory in Three-Dimensions}",
    reportNumber = "Print-76-0245 (BRANDEIS)",
    doi = "10.1103/PhysRevD.14.1715",
    journal = "Phys. Rev. D",
    volume = "14",
    pages = "1715",
    year = "1976"
}

@article{Townsend2,
    author = "Townsend, P. K.",
    title = "{Consistency of the $1/n$ Expansion for Three-Dimensional $\phi^6$ Theory}",
    reportNumber = "Print-76-0244 (BRANDEIS)",
    doi = "10.1016/0550-3213(77)90306-6",
    journal = "Nucl. Phys. B",
    volume = "118",
    pages = "199--217",
    year = "1977"
}

@article{Giombi2,
    author = "Giombi, Simone and Klebanov, Igor R. and Popov, Fedor and Prakash, Shiroman and Tarnopolsky, Grigory",
    title = "{Prismatic Large $N$ Models for Bosonic Tensors}",
    eprint = "1808.04344",
    archivePrefix = "arXiv",
    primaryClass = "hep-th",
    reportNumber = "PUPT-2568",
    doi = "10.1103/PhysRevD.98.105005",
    journal = "Phys. Rev. D",
    volume = "98",
    number = "10",
    pages = "105005",
    year = "2018"
}

@article{Badel,
    author = "Badel, Gil and Cuomo, Gabriel and Monin, Alexander and Rattazzi, Riccardo",
    title = "{Feynman diagrams and the large charge expansion in $3-\varepsilon$ dimensions}",
    eprint = "1911.08505",
    archivePrefix = "arXiv",
    primaryClass = "hep-th",
    doi = "10.1016/j.physletb.2020.135202",
    journal = "Phys. Lett. B",
    volume = "802",
    pages = "135202",
    year = "2020"
}

@article{Sakhi,
    author = "Sakhi, S.",
    title = "{Renormalization functions of the tricritical $O(N)$-symmetric $\Phi^6$ model 
    beyond the next-to-leading order in $1/N$}",
    doi = "10.1088/2399-6528/abfe4b",
    journal = "J. Phys. Comm.",
    volume = "5",
    number = "5",
    pages = "055011",
    year = "2021"
}

@article{Kharuk,
    author = "Kharuk, N. V.",
    title = "{Four-loop renormalization with a cutoff in a sextic model}",
    eprint = "2504.07688",
    archivePrefix = "arXiv",
    primaryClass = "hep-th",
    doi = "10.1088/1751-8121/ae0798",
    journal = "J. Phys. A",
    volume = "58",
    number = "39",
    pages = "395401",
    year = "2025"
}

@article{Shalaby,
    author = "Shalaby, Abouzeid M.",
    title = "{Accurate critical exponents from the optimal truncation of the $\varepsilon $-expansion within the O(N)-symmetric field theory for large N}",
    eprint = "2409.00271",
    archivePrefix = "arXiv",
    primaryClass = "hep-th",
    doi = "10.1140/epjc/s10052-025-14473-7",
    journal = "Eur. Phys. J. C",
    volume = "85",
    number = "7",
    pages = "751",
    year = "2025"
}

@article{Fei3loop,
	Archiveprefix = {arXiv},
	Author = {Fei, Lin and Giombi, Simone and Klebanov, Igor R. and Tarnopolsky, Grigory},
	Doi = {10.1103/PhysRevD.91.045011},
	Eprint = {1411.1099},
	Journal = {Phys. Rev.},
	Pages = {045011},
	Primaryclass = {hep-th},
	Reportnumber = {PUPT-2474},
	Slaccitation = {%%CITATION = ARXIV:1411.1099;%%},
	Title = {{Three loop analysis of the critical $O(N)$ models in $6-\epsilon$ dimensions}},
	Volume = {D91},
	Year = {2015}}

@article{Gracey4loop,
    author = "Gracey, J. A.",
    title = "{Four loop renormalization in six dimensions using forcer}",
    eprint = "2405.00413",
    archivePrefix = "arXiv",
    primaryClass = "hep-th",
    reportNumber = "LTH 1369",
    doi = "10.1103/PhysRevD.110.045015",
    journal = "Phys. Rev. D",
    volume = "110",
    number = "4",
    pages = "045015",
    year = "2024"
}

@article{Kompaniets,
    author = "Kompaniets, Mikhail and Pikelner, Andrey",
    title = "{Critical exponents from five-loop scalar theory renormalization near six-dimensions}",
    eprint = "2101.10018",
    archivePrefix = "arXiv",
    primaryClass = "hep-th",
    doi = "10.1016/j.physletb.2021.136331",
    journal = "Phys. Lett. B",
    volume = "817",
    pages = "136331",
    year = "2021"
}

@article{Borinsky,
    author = "Borinsky, M. and Gracey, J. A. and Kompaniets, M. V. and Schnetz, O.",
    title = "{Five-loop renormalization of \ensuremath{\phi^3} theory with applications to the Lee-Yang edge singularity and percolation theory}",
    eprint = "2103.16224",
    archivePrefix = "arXiv",
    primaryClass = "hep-th",
    reportNumber = "Nikhef-2021-006, LTH-1254",
    doi = "10.1103/PhysRevD.103.116024",
    journal = "Phys. Rev. D",
    volume = "103",
    number = "11",
    pages = "116024",
    year = "2021"
}

@article{Bednyakov2,
    author = "Bednyakov, A. V. and Kompaniets, M. V. and Trenogin, A. V.",
    title = "{On the six-loop scaling dimensions of the $(\phi^2)^n$ operators in $d=3$}",
    eprint = "2512.05059",
    archivePrefix = "arXiv",
    primaryClass = "hep-th",
    doi = "10.1016/j.nuclphysb.2026.117331",
    journal = "Nucl. Phys. B",
    volume = "1024",
    pages = "117331",
    year = "2026"
}

@article{Kompaniets2,
    author = "Adzhemyan, L. Ts. and Kompaniets, M. V. and Trenogin, A. V.",
    title = "{Six-loop renormalization group analysis of the $\varphi^4 + \varphi^6$ model}",
    eprint = "2601.21515",
    archivePrefix = "arXiv",
    primaryClass = "cond-mat.stat-mech",
    doi = "10.1134/S0040577926060012",
    journal = "Theor. Math. Phys.",
    volume = "227",
    number = "3",
    pages = "925--936",
    year = "2026"
}

@article{WallaceG,
	Author = {Wallace, D. J. and Zia, R. K. P.},
	Doi = {10.1016/0003-4916(75)90267-5},
	Journal = {Annals Phys.},
	Pages = {142},
	Reportnumber = {THEP 73-4/5},
	Slaccitation = {%%CITATION = APNYA,92,142;%%},
	Title = "{Gradient Properties of the Renormalization Group Equations in Multicomponent Systems}",
	Volume = {92},
	Year = {1975}}

@article{Analogs,
	Author = {Jack, I. and Osborn, H.},
	Doi = {10.1016/0550-3213(90)90584-Z},
	Journal = {Nucl. Phys.},
	Pages = {647-688},
	Reportnumber = {DAMTP-90-02},
	Slaccitation = {%%CITATION = NUPHA,B343,647;%%},
	Title = {{Analogs for the $c$ Theorem for Four-dimensional Renormalizable Field Theories}},
	Volume = {B343},
	Year = {1990}}

@article{Weyl,
      author         = "Osborn, H.",
      title          = "{Weyl consistency conditions and a local renormalization
                        group equation for general renormalizable field theories}",
      journal        = "Nucl. Phys.",
      volume         = "B363",
      year           = "1991",
      pages          = "486-526",
      doi            = "10.1016/0550-3213(91)80030-P",
      reportNumber   = "DAMTP-91-1",
      SLACcitation   = "%%CITATION = NUPHA,B363,486;%%"
}

@article{Kapoor,
    author = "Kapoor, Samarth and Prakash, Shiroman",
    title = "{Bifundamental multiscalar fixed points in \ensuremath{d=3-\varepsilon}}",
    eprint = "2112.01055",
    archivePrefix = "arXiv",
    primaryClass = "hep-th",
    doi = "10.1103/PhysRevD.108.026002",
    journal = "Phys. Rev. D",
    volume = "108",
    number = "2",
    pages = "026002",
    year = "2023"
}

@article{JackJones,
    author = "Jack, I. and Jones, D. R. T.",
    title = "{Anomalous dimensions for $\phi^n$ in scale invariant $d=3$ theory}",
    eprint = "2007.07190",
    archivePrefix = "arXiv",
    primaryClass = "hep-th",
    reportNumber = "LTH1239",
    doi = "10.1103/PhysRevD.102.085012",
    journal = "Phys. Rev. D",
    volume = "102",
    number = "8",
    pages = "085012",
    year = "2020"
}

@article{Seeking,
      author         = "Osborn, Hugh and Stergiou, Andreas",
      title          = "{Seeking fixed points in multiple coupling scalar
                        theories in the $\epsilon$ expansion}",
      journal        = "JHEP",
      volume         = "05",
      year           = "2018",
      pages          = "051",
      doi            = "10.1007/JHEP05(2018)051",
      eprint         = "1707.06165",
      archivePrefix  = "arXiv",
      primaryClass   = "hep-th",
      reportNumber   = "DAMTP-2017-30, CERN-TH-2017-149, DAMTP-2017-30,
                        CERN-TH-2017-149",
      SLACcitation   = "%%CITATION = ARXIV:1707.06165;%%"
}

@article{Sixloop,
    author = "Adzhemyan, L. Ts. and Ivanova, E. V. and Kompaniets, M. V. and Kudlis, A. and Sokolov, A. I.",
    title = "{Six-loop $\varepsilon$ expansion study of three-dimensional $n$-vector model with cubic anisotropy}",
    eprint = "1901.02754",
    archivePrefix = "arXiv",
    primaryClass = "cond-mat.stat-mech",
    doi = "10.1016/j.nuclphysb.2019.02.001",
    journal = "Nucl. Phys. B",
    volume = "940",
    pages = "332--350",
    year = "2019"
}

@article{Appelquist2,
      author         = "Appelquist, Thomas and Heinz, Ulrich W.",
      title          = "{Vacuum Stability in Three-dimensional $O(N)$ Theories}",
      journal        = "Phys. Rev.",
      volume         = "D25",
      year           = "1982",
      pages          = "2620-2633",
      doi            = "10.1103/PhysRevD.25.2620",
      reportNumber   = "YTP-82-01",
      SLACcitation   = "%%CITATION = PHRVA,D25,2620;%%"
}

@article{Appelquist,
      author         = "Appelquist, Thomas and Heinz, Ulrich W.",
      title          = "{Three-Dimensional $O(N)$ Theories at Large Distances}",
      journal        = "Phys. Rev.",
      volume         = "D24",
      year           = "1981",
      pages          = "2169-2181",
      doi            = "10.1103/PhysRevD.24.2169",
      reportNumber   = "YTP-81-16",
      SLACcitation   = "%%CITATION = PHRVA,D24,2169;%%"
}

@article{Giombi,
    author = "Giombi, Simone and Klebanov, Igor R. and Tarnopolsky, Grigory",
    title = "{Bosonic tensor models at large $N$ and small $\epsilon$}",
    eprint = "1707.03866",
    archivePrefix = "arXiv",
    primaryClass = "hep-th",
    reportNumber = "PUPT-2528",
    doi = "10.1103/PhysRevD.96.106014",
    journal = "Phys. Rev. D",
    volume = "96",
    number = "10",
    pages = "106014",
    year = "2017"
}

@article{Bardeen,
  title = {Spontaneous Breaking of Scale Invariance and the Ultraviolet Fixed Point in $\mathrm{O}(N)$-Symmetric (${\ensuremath{\varphi}}_{3}^{6}$) Theory},
  author = "Bardeen, William A. and {Moshe}, Moshe and {Bander}, Myron",
  journal = {Phys. Rev. Lett.},
  volume = {52},
  pages = {1188--1191},
  numpages = {0},
  year = {1984},
  publisher = {American Physical Society},
  doi = {10.1103/PhysRevLett.52.1188},
}

@article{David,
  title = {Bardeen-{Moshe}-{Bander} Fixed Point and the Ultraviolet Triviality of $ {({\vec \Phi}}{}^{2})_{3}{\!}^{3}$},
  author = {David, Francois and Kessler, David A. and Neuberger, Herbert},
  journal = {Phys. Rev. Lett.},
  volume = {53},
  pages = {2071--2074},
  numpages = {0},
  year = {1984},
  publisher = {American Physical Society},
  doi = {10.1103/PhysRevLett.53.2071},
}

@article{Semenoff,
      author         = "Omid, Hamid and Semenoff, Gordon W. and Wijewardhana, L. C. R.",
      title          = "{Light dilaton in the large $N$ tricritical $O(N)$ model}",
      journal        = "Phys. Rev.",
      volume         = "D94",
      year           = "2016",
      pages          = "125017",
      doi            = "10.1103/PhysRevD.94.125017",
      eprint         = "1605.00750",
      archivePrefix  = "arXiv",
      primaryClass   = "hep-th",
      SLACcitation   = "%%CITATION = ARXIV:1605.00750;%%"
}

@article{Tricrit,
	Author = {Stephen, M.J. and McCauley Jr., J.L.},
	Doi = {10.1016/0375-9601(73)90799-8},
	Journal = {Phys. Letters},
	Pages = {89-90},
	Title = {{Feynman Graph Expansion for Tricritical Exponents}},
	Volume = {44A},
	Year = {1973}}

@article{Lewis,
    author = "Lewis, A. L. and Adams, F. W.",
    title = "{Tricritical behavior in two dimensions. 2. Universal quantities from the epsilon expansion}",
    doi = "10.1103/PhysRevB.18.5099",
    journal = "Phys. Rev. B",
    volume = "18",
    pages = "5099--5111",
    year = "1978"
}

@article{Dwyer,
	Archiveprefix = {arXiv},
	Author = {O'Dwyer, J. and Osborn, H.},
	Doi = {10.1016/j.aop.2007.10.005},
	Eprint = {0708.2697},
	Journal = {Annals Phys.},
	Pages = {1859-1898},
	Primaryclass = {hep-th},
	Reportnumber = {DAMTP-07-77},
	Slaccitation = {%%CITATION = ARXIV:0708.2697;%%},
	Title = {{Epsilon Expansion for Multicritical Fixed Points and Exact Renormalisation Group Equations}},
	Volume = {323},
	Year = {2008}}

@article{Basu,
    author = "Basu, Pallab and Krishnan, Chethan",
    title = "{$\epsilon$-expansions near three dimensions from conformal field theory}",
    eprint = "1506.06616",
    archivePrefix = "arXiv",
    primaryClass = "hep-th",
    doi = "10.1007/JHEP11(2015)040",
    journal = "JHEP",
    volume = "11",
    pages = "040",
    year = "2015"
}

@article{Hager0,
  title = {\ensuremath{\Theta}-point behavior of diluted polymer solutions: Can one observe the universal logarithmic corrections predicted by field theory?},
  author = {Hager, Johannes and Sch\"afer, Lothar},
  journal = {Phys. Rev. E},
  volume = {60},
  pages = {2071--2085},
  numpages = {0},
  year = {1999},
  publisher = {American Physical Society},
  doi = {10.1103/PhysRevE.60.2071}
}

@article{Hager,
	Author = {Hager, J. S.},
	Doi = {10.1088/0305-4470/35/12/301},
	Journal = {J. Phys.},
	Pages = {2703-2711},
	Slaccitation = {%%CITATION = JPAGA,A35,2703;%%},
	Title = {{Six-loop renormalization group functions of $O(n)$-symmetric $\phi^6$-theory and $\epsilon$-expansions of tricritical exponents up to $\epsilon^3$}},
	Volume = {A35},
	Year = {2002}}

@article{Pisarski,
	Author = {Pisarski, R. D.},
	Doi = {10.1103/PhysRevLett.48.574},
	Journal = {Phys. Rev. Lett.},
	Pages = {574-576},
	Slaccitation = {%%CITATION = PRLTA,48,574;%%},
	Title = {{Fixed Point Structure of $\phi^6$ in three-dimensions at large $N$}},
	Volume = {48},
	Year = {1982}}

@article{Pisarski2,
      author         = "Pisarski, Robert D.",
      title          = "{On the fixed points  of  $\phi^6$ in three-dimensions and
                        $\phi^4$ in four-dimensions}",
      journal        = "Phys. Rev.",
      volume         = "D28",
      year           = "1983",
      pages          = "1554-1556",
      doi            = "10.1103/PhysRevD.28.1554",
      reportNumber   = "NSF-ITP-83-44",
      SLACcitation   = "%%CITATION = PHRVA,D28,1554;%%"
}

@article{Pisarski3A,
    author = "Pisarski, Robert D.",
    title = "{Asymptotically Free Fluids}",  
    reportNumber = "NSF-ITP-82-45",
    doi = "10.1103/PhysRevD.26.3543",
    journal = "Phys. Rev. D",
    volume = "26",
    pages = "3543",
    year = "1982"
}

@article{Jack6d,
	Archiveprefix = {arXiv},
	Author = {Gracey, J. A. and Jack, I. and Poole, C.},
	Doi = {10.1007/JHEP01(2016)174},
	Eprint = {1507.02174},
	Journal = {JHEP},
	Pages = {174},
	Primaryclass = {hep-th},
	Reportnumber = {LTH1045},
	Slaccitation = {%%CITATION = ARXIV:1507.02174;%%},
	Title = "{The $a$-function in six dimensions}",
	Volume = {01},
	Year = {2016}}

@article{Jack3d,
      author         = "Jack, I. and Jones, D. R. T. and Poole, C.",
      title          = "{Gradient flows in three dimensions}",
      journal        = "JHEP",
      volume         = "09",
      year           = "2015",
      pages          = "061",
      doi            = "10.1007/JHEP09(2015)061",
      eprint         = "1505.05400",
      archivePrefix  = "arXiv",
      primaryClass   = "hep-th",
      reportNumber   = "LTH1043",
      SLACcitation   = "%%CITATION = ARXIV:1505.05400;%%"
}

@article{Jack3d2,
      author         = "Jack, I. and Poole, C.",
      title          = "{$a$-function in three dimensions: Beyond the
                        leading order}",
      journal        = "Phys. Rev.",
      volume         = "D95",
      year           = "2017",
      pages          = "025010",
      doi            = "10.1103/PhysRevD.95.025010",
      eprint         = "1607.00236",
      archivePrefix  = "arXiv",
      primaryClass   = "hep-th",
      reportNumber   = "LTH-1082",
      SLACcitation   = "%%CITATION = ARXIV:1607.00236;%%"
}

@article{Jack3d3,
    author = {Gracey, J. A. and Jack, I. and Poole, C. and Schr\"oder, Y.},
    title = "{a-function for $N=$ 2 supersymmetric gauge theories in three dimensions}",
    eprint = "1609.06458",
    archivePrefix = "arXiv",
    primaryClass = "hep-th",
    reportNumber = "LTH1097",
    doi = "10.1103/PhysRevD.95.025005",
    journal = "Phys. Rev. D",
    volume = "95",
    number = "2",
    pages = "025005",
    year = "2017"
}

@article{Jack4d,
    author = "Jack, I. and Poole, C.",
    title = "{Scheme invariants in $\phi^4$ theory in four dimensions}",
    eprint = "1806.08598",
    archivePrefix = "arXiv",
    primaryClass = "hep-th",
    reportNumber = "LTH1168; CP\textasciicircum{}3-Origins-2018-24 DNRF90, LTH1168, CP3-ORIGINS-2018-24",
    doi = "10.1103/PhysRevD.98.065011",
    journal = "Phys. Rev. D",
    volume = "98",
    number = "6",
    pages = "065011",
    year = "2018"
}

@article{Jepsen,
    author = "Jepsen, Christian and Oz, Yaron",
    title = "{RG flows and fixed points of~$O(N)^r$ models}",
    eprint = "2311.09039",
    archivePrefix = "arXiv",
    primaryClass = "hep-th",
    doi = "10.1007/JHEP02(2024)035",
    journal = "JHEP",
    volume = "02",
    pages = "035",
    year = "2024"
}

@article{Pannell1,
    author = "Pannell, William H. and Stergiou, Andreas",
    title = "{Gradient properties of perturbative multiscalar RG flows to six loops}",
    eprint = "2402.17817",
    archivePrefix = "arXiv",
    primaryClass = "hep-th",
    doi = "10.1016/j.physletb.2024.138701",
    journal = "Phys. Lett. B",
    volume = "853",
    pages = "138701",
    year = "2024"
}

@article{Pannell2,
    author = "Pannell, William H. and Stergiou, Andreas",
    title = "{Gradient flows and the curvature of theory space}",
    eprint = "2502.06940",
    archivePrefix = "arXiv",
    primaryClass = "hep-th",
    doi = "10.1007/JHEP09(2025)117",
    journal = "JHEP",
    volume = "09",
    pages = "117",
    year = "2025"
}

@article{Delamotte,
    author = "Yabunaka, Shunsuke and Delamotte, Bertrand",
    title = "{Surprises in $O(N)$ Models: Nonperturbative Fixed Points, Large $N$ Limits, and Multicriticality}",
    eprint = "1707.04383",
    archivePrefix = "arXiv",
    primaryClass = "cond-mat.stat-mech",
    doi = "10.1103/PhysRevLett.119.191602",
    journal = "Phys. Rev. Lett.",
    volume = "119",
    number = "19",
    pages = "191602",
    year = "2017"
}

@article{HenrikssonRev,
    author = "Henriksson, Johan",
    title = "{The tricritical Ising CFT and conformal bootstrap}",
    eprint = "2501.18711",
    archivePrefix = "arXiv",
    primaryClass = "hep-th",
    reportNumber = "CERN-TH-2025-028",
    doi = "10.1007/JHEP08(2025)031",
    journal = "JHEP",
    volume = "08",
    pages = "031",
    year = "2025"
}

@article{TSteudtner,
    author = "Kvedarait{\.{e}}, Sandra and Steudtner, Tom and Uetrecht, Max",
    title = "{Revisiting the $\phi^6$ theory in three dimensions at large $N$}",
    eprint = "2502.07880",
    archivePrefix = "arXiv",
    primaryClass = "hep-th",
    doi = "10.1103/tnh4-7lnv",
    journal = "Phys. Rev. D",
    volume = "112",
    number = "5",
    pages = "056004",
    year = "2025"
}

@article{Brezin,
	doi = {10.1007/BF02819916},
	year = 1974,
        volume = {9},
	number = {12},
	pages = {483-486},
	author = {Brezin, E.  and De Dominicis, C. and Zinn-Justin, J.},
	title = {Anomalous Dimensions of Higher-Order operators in the $\varphi ^4 ${-Theory}},
	journal = {Lettere al Nuovo Cimento (1971-1985)}
}

@article{Zhang,
	doi = {10.1088/0305-4470/15/10/032},
	year = 1982,
	month = {oct},
	publisher = {{IOP} Publishing},
	volume = {15},
	number = {10},
	pages = {3303--3305},
	author = {F C Zhang and R K P Zia},
	title = {A correction-to-scaling critical exponent for fluids at order $\varepsilon^3$},
	journal = {Journal of Physics A: Mathematical and General}
}

@article{Nicoll2,
  title = "{Critical phenomena of fluids: Asymmetric Landau-Ginzburg-Wilson mode}l",
  author = {Nicoll, J. F.},
  journal = {Phys. Rev. A},
  volume = {24},
  issue = {4},
  pages = {2203--2220},
  numpages = {0},
  year = {1981},
  month = {Oct},
  publisher = {American Physical Society},
  doi = {10.1103/PhysRevA.24.2203},
}

@article{Nicoll,
  title = {Fluid-magnet universality: Renormalization-group analysis of ${\ensuremath{\varphi}}^{5}$ operators},
  author = {Nicoll, J. F. and Zia, R. K. P.},
  journal = {Phys. Rev. B},
  volume = {23},
  issue = {11},
  pages = {6157--6163},
  numpages = {0},
  year = {1981},
  month = {Jun},
  publisher = {American Physical Society},
  doi = {10.1103/PhysRevB.23.6157},
}

@article{Behan,
    author = "Behan, Connor and Rastelli, Leonardo and Rychkov, Slava and Zan, Bernardo",
    title = "{A scaling theory for the long-range to short-range crossover and an infrared duality}",
    eprint = "1703.05325",
    archivePrefix = "arXiv",
    primaryClass = "hep-th",
    reportNumber = "CERN-TH-2017-052, YITP-SB-17-13, CERN-PH-TH-2017-052",
    doi = "10.1088/1751-8121/aa8099",
    journal = "J. Phys. A",
    volume = "50",
    number = "35",
    pages = "354002",
    year = "2017"
}

@book{Andrews,
  title={Special Functions},
  author={Andrews, G.E. and Askey, R. and Roy, R.},
  year={1999},
  publisher={Cambridge University Press, Cambridge}
}

\end{document}